\documentclass[a4paper,12pt]{scrartcl}
\DeclareOldFontCommand{\rm}{\normalfont\rmfamily}{\mathrm}
\usepackage[utf8]{inputenc}				
\usepackage{graphicx}						
\usepackage{fancyhdr}						
\usepackage{units}							
\usepackage{cite}								
\usepackage{hyperref}						
\usepackage{psfrag}
\usepackage{amsmath}
\usepackage{latexsym} 
\usepackage{psfrag}
\usepackage{color}
\usepackage{longtable}
\usepackage{chngcntr}
\numberwithin{equation}{section}
\numberwithin{table}{section}

\definecolor{lblue}{RGB}{225,232,242}

\usepackage{amsmath}
\usepackage{tikz}
\usepackage{subcaption}

\usepackage{mathtools}
\DeclarePairedDelimiter\abs{\lvert}{\rvert}

\allowdisplaybreaks

\begin{document}
\parindent 0pt
\parskip 1.1ex

\newcommand{\X}{\overrightarrow{X}}
\newcommand{\Y}{\overrightarrow{Y}}
\newcommand{\nv}{\overrightarrow{n}}
\newcommand{\Phic}{\underline{p}}
\newcommand{\vc}{\underline{v}}
\newcommand{\kc}{\underline{k}}
\newcommand{\Fc}{\underline{F}}
\newcommand{\rc}{\underline{r}}
\newcommand{\Green}{\underline{G}}
\newcommand{\parta}[1]{\frac{\partial}{\partial #1}}

\begin{titlepage}

\begin{figure}[t]
	\centering
		\includegraphics[width=0.25\textwidth]{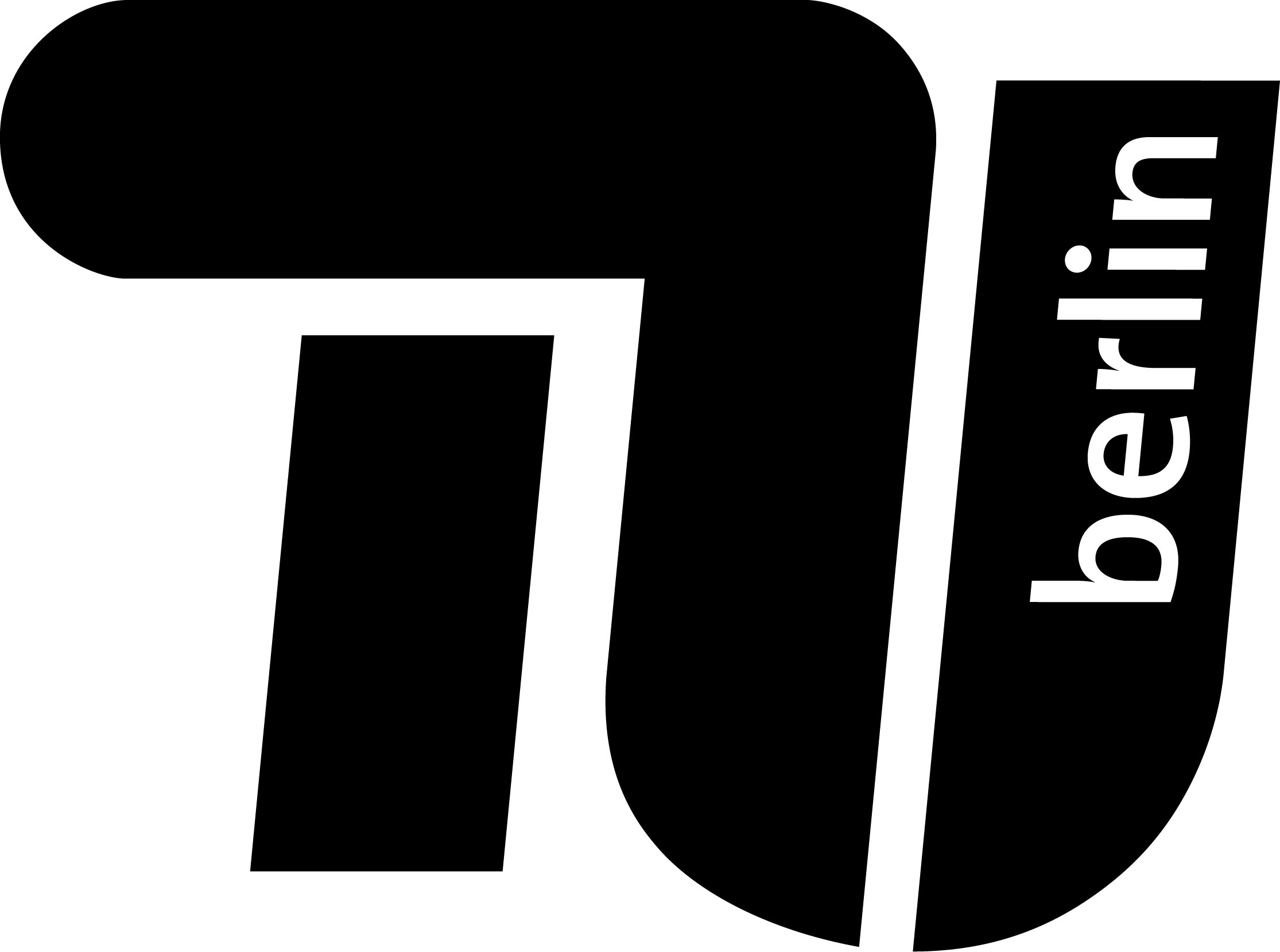}
\end{figure}

\vspace*{0,5cm}

\begin{center}
	\large{\textbf{Technische Universität Berlin}\\
	Fakultät V - Verkehrs- und Maschinensysteme\\
	Institut für Strömungsmechanik und Technische Akustik\\
	Fachgebiet Technische Akustik\\
	Sekr. TA 7  -  Einsteinufer 25  -  10587 Berlin}\\
\end{center}

\vspace*{1cm}

\begin{center}
	\Large{\textbf{Room Acoustics (lecture notes)\\
	}}
	\bigskip
	\bigskip
	\bigskip
	\bigskip
	\bigskip
	\bigskip
	\bigskip
	\bigskip
	\bigskip
	\bigskip
	\bigskip
	\bigskip
	\bigskip
	\bigskip
	\bigskip
	\bigskip
	\bigskip
	\bigskip
	\bigskip
	\bigskip
	\bigskip
	\bigskip
	\bigskip
	\bigskip
	\bigskip
	\bigskip
	\bigskip
	\bigskip
	\bigskip
	\bigskip
	\bigskip
	\bigskip
	\bigskip
	\bigskip
	\bigskip
	\bigskip
	\bigskip
	\bigskip
\end{center}
\begin{flushleft}
	\large Drasko Masovic, TU Berlin, 2018 \quad\quad (last update: \today)
\end{flushleft}
\vfill

\end{titlepage}

\pagestyle{fancy}
\addtolength{\headheight}{20pt}
\lhead{TU Berlin}
\chead{\scriptsize{Institut für Strömungsmechanik und Technische Akustik \\
				Fachgebiet Technische Akustik}}
\rhead{}

\tableofcontents
\section*{List of frequently used symbols}

\addcontentsline{toc}{section}{List of symbols}
	
\begin{longtable}{ l l }
	$A_s$ & equivalent absorption area \\
	$AL_{cons}$ & articulation loss of consonants \\
	$a$ & radius \\
	$BQI$ & binaural quality index \\
	$BR$ & bass ratio \\
	$C$ & clarity \\
	$C_p$ & specific heat capacity at constant pressure \\
	$c$ & speed of sound \\
	$D$ & damping, definition \\
	$D_i$ & directivity \\
	$d$ & distance, thickness, depth \\
	$E$ & acoustic energy (density) \\
	$EDT$ & early decay time \\
	$\boldsymbol e$ & unit vector \\
	$F$ & force (one-dimensional) \\
	$f$ & frequency, generic function \\
	$f_{Schroed}$ & Schroeder frequency \\
	$G$ & Green's function, strength factor \\
	$g$ & impulse response \\
	$H$ & height \\
	$H_e$ & Helmholtz number \\
	$\boldsymbol I$ & intensity vector \\
	$IACC$ & interaural cross correlation coefficient \\
	$IACF$ & interaural cross correlation function \\
	$I_s$ & irradiation strength \\
	$K$ & heat conductivity \\
	$k$ & wave number \\
	$\boldsymbol k$ & wave vector \\
	$L$ & length, sound pressure level \\
	$L_W$ & sound power level \\
	$LEF$ & lateral energy fraction \\
	$l$ & length, mean free path length \\
	$M$ & mass \\
	$m$ & attenuation constant, modulation transfer function \\
	$\boldsymbol n$ & unit vector normal to a surface \\
	$P$ & sound power \\
	$p$ & pressure, received signal \\
	$Q$ & source function, Q-factor \\
	$q$ & source function (per unit volume) \\
	$\boldsymbol q$ & heat flux \\
	$R_s$ & reflection coefficient \\
	$RASTI$ & rapid speech transmission index \\
	$RP$ & radiation pattern \\
	$\mathcal{R}$ & specific gas constant \\
	$r$ & distance, magnitude of radius vector \\
	$r_c$ & critical distance \\
	$S$ & stiffness, surface area \\
	$SC$ & Schroeder curve \\
	$SNR$ & signal-to-noise ratio \\
	$ST$ & stage support factor \\
	$STI$ & speech transmission index \\
	$s$ & signal, sequence \\
	$s_s$ & scattering coefficient \\
	$T$ & temperature, period, duration, reverberation time \\
	$T_s$ & transmission coefficient \\
	$t$ & time \\
	$t_s$ & centre time \\
	$V$ & volume \\
	${\boldsymbol v}$ & velocity \\
	${\boldsymbol x}$ & position vector \\
	${\boldsymbol y}$ & position vector (for a source or surface) \\
	$Z$ & impedance \\
	$\mathcal{Z}$ & specific impedance \\
	$z$ & random number between 0 and 1 \\
	$\alpha_s$ & absorption coefficient \\
	$\beta$ & volume fraction \\
	$\gamma$ & heat capacity ratio, directivity factor (gain) \\
	$\Delta t_{init}$ & initial time delay gap \\
	$\delta$ & Dirac delta function, end correction, boundary layer thickness \\
	$\epsilon$ & infinitesimal \\
	$\zeta$ & damping constant \\
	$\theta$ & polar angle \\
	$\iota$ & polar angle (for a source or surface) \\
	$\lambda$ & wavelength \\
	$\mu$ & dynamic viscosity \\
	$\nu$ & kinematic viscosity \\
	$\Xi$ & flow resistivity \\
	$\xi$ & displacement (one-dimensional) \\
	$\rho$ & density \\
	$\sigma$ & porosity, perforation ratio \\
	$\tau$ & viscous stress tensor, time (for a source or surface) \\
	$\Phi$ & phase, phase shift \\
	$\phi$ & azimuthal angle \\
	$\chi$ & azimuthal angle (for a source or surface) \\
	$\psi$ & mode (eigenfunction) \\
	$\Omega$ & solid angle, modulation frequency \\
	$\omega$ & angular frequency
\end{longtable}
\addtocounter{table}{-1}

\quad Note: Bold symbols denote vectors. Subscript $_0$ is used for steady background values and $\hat{}$ above a symbol indicates a complex amplitude. The functions $\mathcal{R}_e(\text{ })$ and $\mathcal{I}_m(\text{ })$ give real and imaginary part of a complex number, respectively, $^*$ denotes a complex conjugate value, and $j$ is used as the imaginary unit. Symbol $\partial$ is used for partial derivatives, $d$ for total derivatives, and $\nabla$ is the nabla operator ($\nabla$ is typically gradient of a scalar and $\nabla \cdot$ divergence of a vector). Symbol $\propto$ stands for ``proportional to'', $\sim$ for ``scales as'', and $\sim \mathcal{O}(\text{ })$ for order of magnitude. Angle brackets $\langle \text{ } \rangle$ are used for averaging (for example, over time: $\langle \text{ } \rangle_t$). $| \text{ } |$ gives magnitude of a vector or an absolute value (modulus). If a vector, say $\boldsymbol v$, has complex components, the expression $|\boldsymbol v|$ gives its magnitude $|\boldsymbol v| = \sqrt{\boldsymbol v ^* \cdot \boldsymbol v}$, which is a scalar (not a vector of the moduli of its components).

\section{Introduction}\label{ch:introduction}
\setcounter{footnote}{0}

\begin{figure}[h]
	\centering
	\begin{subfigure}{.5\textwidth}
		\centering
		\includegraphics[width=.95\linewidth]{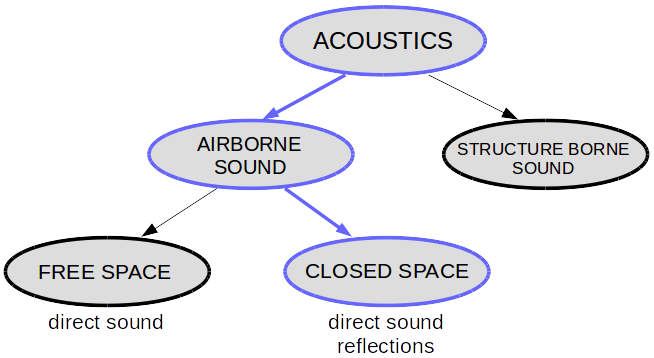}
		\label{fig:room_acoustics_object}
	\end{subfigure}%
	\begin{subfigure}{.5\textwidth}
		\centering
		\includegraphics[width=.95\linewidth]{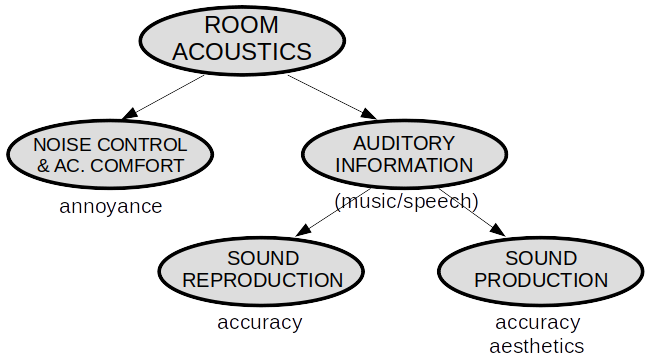}
		\label{fig:room_acoustics_applications}
	\end{subfigure}
	\caption{Room acoustics: (left) the object of study indicated with blue colour, (right) general types of applications.}
	\label{fig:room_acoustics_object_applications}
\end{figure}

Room acoustics is a sub-discipline of classical acoustics in fluids dealing with sound fields in closed (or partly open) spaces (see Fig. \ref{fig:room_acoustics_object_applications}, left). Although most commonly associated with relatively large rooms filled with air, it can also include small spaces (cavities) and other fluids in which sound propagates. Boundedness of the observed space implies that besides the direct sound, which in the absence of obstacles reaches a receiver directly from the source, a series of reflections usually takes place at the boundary surfaces of the room, which substantially contribute to the sound field in it.

\begin{figure}[h]
	\centering
	\includegraphics[width=0.4\textwidth]{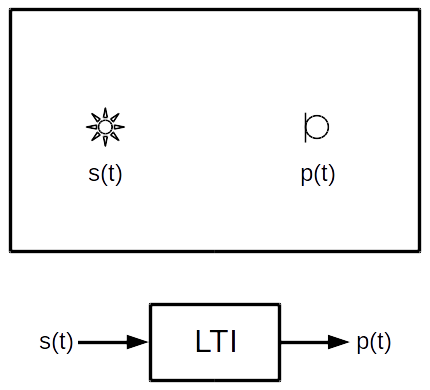}
	\caption{Room as a linear time-invariant (LTI) system.}
	\label{fig:room_LTI_system}
\end{figure}

Under normal circumstances, a room can be treated as a physical linear time-invariant (LTI) transmission system\footnote{As we will see in section~\ref{LTI_systems_and_impulse_response}, such systems are entirely characterized with their impulse responses. The conditions for linearity will be discussed shortly, while the time invariance holds whenever the acoustically relevant conditions in the room do not change significantly over time, for example, there are no moving objects in the room and the thermodynamic properties of the medium are essentially stable.} between a source emitting a certain signal $s(t)$ and a receiver which receives its possibly modified version $p(t)$. This is illustrated in Fig. \ref{fig:room_LTI_system}. Room acoustics provides means for assessment, estimation, measurement, and optimization of the system -- the direct sound and reflections -- such that the resulting sound field at the receiver location satisfies certain objective or subjective criteria, which depend on the type of the room.

As shown in Fig. \ref{fig:room_acoustics_object_applications} (right), the applications of room acoustics spread in two main directions. The first group of applications involves primarily \textbf{noise control} of various sources of unwanted sound which may be present inside or, less commonly, outside the room. The acoustics of such rooms should ensure a sufficient acoustic comfort to people in them or protection of sensitive equipment against high noise levels. Moreover, even if a potentially disturbed listener or sensitive equipment are in a separate room, knowledge of the sound field in the source room can be used to efficiently suppress noise closer to the source.

For this type of applications, sound field is usually well described with basic acoustic quantities, such as sound pressure level (SPL) or acoustic energy (often supplemented with reverberation time), averaged over the entire room volume, without a detailed analysis of the field. Furthermore, such applications are most common in engineering practice, frequently in combination with sound insulation and building acoustics (which will not be covered here\footnote{Room acoustics and building acoustics together are commonly referred to as architectural acoustics.}). The reason for this are various regulations concerned with environmental noise and acoustic comfort which exist in many countries and which set the upper limits for noise levels in different spaces. Very wide range of rooms in which people live or work can belong to this group of applications. Some examples are factories, machine rooms and workshops, open-plan offices, railway stations, airports, and other public places.

Quite different is the second group of applications, in which sound fields in rooms carry a specific auditory information, typically music or speech, which is of interest to its receivers. The listeners are often located in the same room, although the sound can also be first recorded with a microphone placed in the room. According to the type of the information, rooms can be simply but crudely divided into rooms for music (concert halls, opera houses, music studios, etc.), rooms for speech (lecture halls, theatres, conference rooms, etc.), and rooms with general purposes (including both speech and music). However, such a division becomes impractical inasmuch as speech and music appear together, often with similar significance (for example, in an opera house). A more appropriate division can be made according to the origin of sound emitted by the source in the room, which leads to two sub-cases.

In the first case, the main goal is accurate \textbf{reproduction} of already complete and given sound information, which has been previously stored on a medium. The reproduction is achieved with the aid of an audio system and the loudspeakers as sources of sound in the rooms. Typical examples are control rooms in studios for music or speech as well as cinemas. Especially sensitive to the influence of room acoustics are stereo and multichannel recordings, which contain not only sound but also spatial information on the location of virtual sources. In general, the reproduced audio signal should be distorted as least as possible by the room as the transmission system.

In the second case, the auditory information is created (\textbf{produced}) directly in the room and the room is expected not only to transmit it accurately to the receiver, but also to enhance it in a desirable way. The usual aim is to support the source of sound by increasing sound energy at the receiver location (the room gain) and to meet more sophisticated subjective, even aesthetic criteria of the listeners. Such rooms ``actively'' contribute to the sound field and affect the perceived sound and, therefore, their acoustics becomes critical. The examples are concert halls and opera houses, as well as lecture halls and theatres. The acoustics of these rooms is commonly considered as part of the overall design of the room interior, thus interweaving with architectural aesthetic criteria. Architects and acoustic consultants are then expected to collaborate in order to achieve a successful design satisfying both visual and auditory criteria. The described general applications of room acoustics are briefly summarized in Table \ref{tab:room_acoustics_applicatons}.

\begin{table}[h]
	\caption{Applications of room acoustics in engineering practice.}
	\label{tab:room_acoustics_applicatons}
	\begin{tabular}{ | l | p{5.5cm} | p{5.3cm} | }
		\hline
		 & \multicolumn{2}{c|}{\textbf{room acoustics}} \\
		\hline
		 & \textbf{noise control, ac. comfort} & \textbf{auditory information} \\
		\hline
		main goal & noise suppression & accuracy/enhancement \\
		\hline
		criteria & objective & objective and subjective \\
		\hline
		target values & regulations & guidelines/suggested values \\
		\hline
		acoustic treatment & norms & best practice \\
		\hline
		related discipline & building acoustics & interior architecture \\
		\hline
		project collaborator & civil engineer & architect \\
		\hline
	\end{tabular}

\end{table}

\subsection{Types of rooms}

Several common types of rooms in which an auditory information is produced or reproduced are listed in Table \ref{tab:room_types}. Some of them, such as theatres, may involve both produced (by the performers on the stage) and reproduced (by the public address (PA) system) sound. The table indicates also the usual sources of sound for each type of the room, whether the receivers are mainly human listeners or microphones, as well as the expected spatial distribution of the sources and receivers. Note that a podium or stage can also be treated as a receiver location, since the speakers and performers are also listeners of the sound they produce and the room provides them with a necessary acoustic feedback. Type of the auditory information (music/speech/singing) is also relevant for the specific applications and given in the table.

\begin{table}[h]
	\caption{Types of rooms and common sources and receivers of sound in them. Abbreviations: room type -- (P) sound production, (R) sound reproduction; sound -- Mu music, Si singing, Sp speech; source -- MI musical instrument, Hu human, LS loudspeaker (sound reinforcement); receiver -- Hu human, Mi microphone.}
	\label{tab:room_types}
	\begin{tabular}{ | l | l | l | l | l | l |}
		\hline
		\textbf{room type} & \textbf{sound} & \textbf{source} & \textbf{source location} & \textbf{rec.} & \textbf{rec. location} \\
		\hline
		concert hall (P) & Mu, Si & MI, Hu & stage & Hu & auditorium \\
		\hline
		opera house (P) & Si, Mu, Sp & Hu, MI & stage, orch. pit & Hu & auditorium \\
		\hline
		theatre (P, R) & Sp, Mu & Hu, LS & stage & Hu & auditorium \\
		\hline
		rec. studio (P) & Mu, Si, Sp & MI, Hu & not fixed & Mi & not fixed \\
		\hline
		control room (R) & Mu, Si, Sp & LS & fixed & Hu & fixed, limited \\
		\hline
		cinema (R) & Sp, Mu & LS & fixed & Hu & auditorium \\
		\hline
		lecture hall (P) & Sp & Hu & podium & Hu & auditorium \\
		\hline
		classroom (P) & Sp & Hu & podium & Hu & auditorium \\
		\hline
		club (P, R) & Mu, Si & MI, LS & stage, fixed & Hu & not fixed \\
		\hline
	\end{tabular}
\end{table}

The table gives only a general overview of typical scenarios and various deviations from them are certainly possible, for example, use of PA systems in lecture halls or opera houses, a single room used as a control room and recording studio, and so on. Placement of sources and microphones in recording studios can also be controlled and predetermined, so that their preferred locations are more or less fixed. All these aspects should be carefully considered when analysing acoustics of a room.

Other types of spaces which are not included in Table \ref{tab:room_types} but can also be treated using the methods of room acoustics are churches, amphitheatres and stadiums (as partly open spaces), sport halls, offices (especially open-plan offices), various public spaces, small cabins (for example, in vehicles) and cavities, and many other. A special type of rooms which appears in practice quite frequently are multipurpose rooms. In general, multifunctionality of these rooms assumes different types of the auditory information (speech and music) or varying locations of sources and receivers in the room. Although certain flexibility can be achieved with changeable geometry of the room and adaptive acoustic elements, the outcome is always a compromise between the more general functionality of the room and reasonable but sub-optimal acoustic quality.

The ultimate receiver of the auditory information is usually human, whether directly, being present in the room, or indirectly, when listening the recordings made with the microphones located in the room. However, if the receiver in the room is a microphone, its directivity can be taken into consideration in order to relax the requirements for the room acoustics. For example, undesirable strong reflections in a recording studio can be largely avoided with a careful placement of directional microphones. Similar consideration holds for directional sources. For example, controllable directivity of line array loudspeaker systems is frequently used to circumvent acoustic deficiencies in large spaces (very large halls, stadiums, public spaces, etc.), whether they lack useful reflections which ''amplify`` the source without major distortion, or the undesired (late) reflections must be suppressed. In general, we can conclude that certain acoustic properties of sources and receivers (besides their location in the room) complement room acoustics and should be taken into consideration. In the rest of this section we study in more details the most important characteristics of sources which commonly appear in rooms, as well as subjective criteria of human listeners, which are all relevant for the appropriate room acoustics.

\subsection{Characteristics of sources}\label{ch:characteristics_of_sources}

Next we consider some of the basic acoustic properties of various sources of sound (music, singing, or speech) which affect the criteria for optimal room acoustics. These are dynamic range, frequency range, and directivity.

\subsubsection{Dynamics and ambient noise}\label{ch:sources_dynamics}

Dynamic range of a source of sound determines the maximum acceptable level of noise in a room, as well as the minimum gain of the room, which ensures sufficient sound energy at the receiver's location. Table~\ref{tab:common_SPLs} gives a very rough indication of relatively sustained maximum sound power levels of some common sources of sound. For easier comparison with everyday sounds, the values are also converted to sound pressure levels in free space, at the distance\footnote{According to eq. (\ref{eq:critical_distance_diffuse_T}), the distance of 1m matches quite closely the critical distance at middle frequencies in many common rooms. At larger distances from the source, sound pressure level in an ideally diffuse sound field remains approximately constant, so the given values can also be taken as very roughly estimated maximum values of sound pressure level in the room outside the zone of the direct sound dominance.} 1\,m, estimated using the equality\footnote{See eq. (\ref{eq:acoustic_power_level_complex_time_average_plane_wave_omni_source_pressure}). For simplicity, no directivity is taken into account and the relation for a small (point) source is used. Therefore, the estimated sound pressure levels are indeed only indicative, especially for large sources.} $L_{free,max,1m} = L_{W,max} - 11$\,dB. Time average values can be considerable lower, say, by 10\,dB or more. Including the attenuation due to sound propagation (6\,dB per doubling the distance from a point source in free space, as will be derived in eq. (\ref{eq:solution_tailored_Green_wave_eq_free_space_compact_source_emission_time})), it becomes evident that most sources of interest, such as human voice, cannot provide sufficient sound pressure levels for typical distances of the receivers without additional gain of the room due to the reflections. Especially critical are relatively high frequencies, at which the sources typically radiate less efficiently, but which are essential for intelligibility of speech and clarity of music.

\begin{table}[h]
	\caption{Maximum sound power levels ($L_{W,max}$) and approximate sound pressure levels at the distance 1m in free space ($L_{free,max,1m}$) of common sources of sound.}
	\label{tab:common_SPLs}
	\begin{tabular}{ | l | l | l |}
		\hline
		\textbf{source} & \textbf{$L_{W,max}$} & \textbf{$L_{free,max,1m}$} \\
		\hline
		normal voice & 85\,dB & 74\,dB \\
		\hline
		loud voice & 95\,dB & 84\,dB \\
		\hline
		loud singing & 100\,dB & 89\,dB \\
		\hline
		\hline
		plucked string instruments & 90\,dB & 79\,dB \\
		\hline
		bowed string instruments & 95\,dB & 84\,dB \\
		\hline
		woodwind instruments & 100\,dB & 89\,dB \\
		\hline
		piano & 110\,dB & 99\,dB \\
		\hline
		brass instruments & 115\,dB & 104\,dB \\
		\hline
		percussions (peaks) & 120\,dB & 109\,dB \\
		\hline
		organs & 130\,dB & 119\,dB \\
		\hline
		orchestra & 135\,dB & 124\,dB \\
		\hline
	\end{tabular}
\end{table}

With the exception of percussions, the values shown in Table \ref{tab:common_SPLs} correspond to relatively sustained achievable sound levels. Of course, sound pressure level is expected to vary in time. Extreme examples of time-dependence are short impacts of percussions. If the duration of such transient sounds is shorter than around 200\,ms, the natural inertia of human auditory system can lead to even lower perceived sound levels compared to the actual, objective ones. For instance, a transient sound with duration 10\,ms is perceived as around 10\,dB weaker than a sound with equal amplitude and duration 100\,ms. The perceived sound level of very short sounds increases nearly linearly per doubling the duration. However, slower variations of sound energy are also important for clear reception of sound information, especially speech (as will be discussed further in section~\ref{ch:descriptors_of_room_acoustics}). A good illustration of this are highly reverberant spaces, which tend to smooth out the time variations of sound level. Parts of the information which are contained in the dynamics of the signal, such as silent consonants or short quiet intervals between successive musical tones, are thus diminished. This results in poor speech intelligibility and inappropriate room response for most of the music styles (especially those which involve fast succession of tones). Consequently, not only sufficient sound pressure level must be ensured at the location of the receiver, but the dynamic range of the source should be preserved to a certain extent.

Dynamic range, which is the difference between maximum and minimum sound power level, varies between different sources. It is around 30\,dB for woodwind instruments and somewhat larger for brass instruments. Violins can achieve even larger dynamic range, up to around 50\,dB. Dynamics of human voice is around 20-30\,dB. These are also the approximate ranges of sound pressure levels which a room should provide at the location of the receiver.

\begin{figure}[h]
	\centering
	\includegraphics[width=0.55\textwidth]{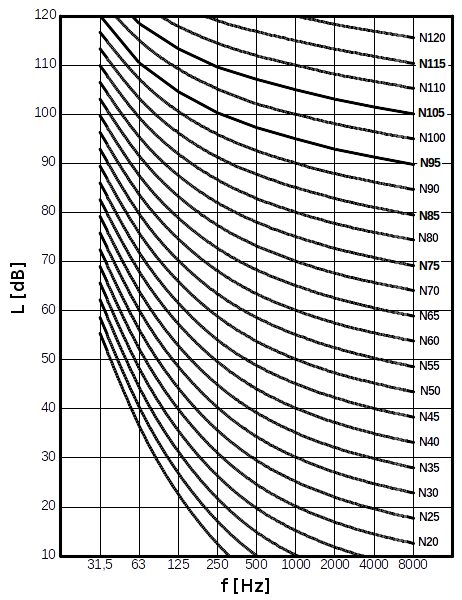}
	\caption{The NR-curves.}
	\label{fig:NR_curves}
\end{figure}

While the maximum sound pressure level depends on the sound power of the source and the room's gain, lower limit of the perceivable useful sound is most often determined by the \textbf{background noise} in the room. Its level sets a threshold for the quietest sounds which can be heard by a listener without being masked by the ambient noise. The sources of airborne and structure-borne noise can be located outside the room, such as traffic or many other sources of noise in adjacent rooms and corridors (when they are not efficiently suppressed by means of sound insulation), but also inside the room, such as ventilation systems, lighting, and other installations. The highest allowed levels of noise are often strictly defined in norms and regulations for different types of rooms. Since both the sensitivity of human hearing and ambient noise are frequency dependent, single-number values can be obtained after A-weighting and energy summation or by implementation of the noise rating curves. In the latter case, rating of broadband noise with a single number is done in the following way: octave noise spectrum is drawn on the top of the NR-curves shown in Fig.~\ref{fig:NR_curves}; it is then represented by the lowest NR-curve which has all its values (at all frequencies) larger than the values of the noise spectrum; number in the name of each NR-curve, which corresponds to its value at 1\,kHz, is the rated value in dB.

Suggested maximum values of the background noise level are given in Table~\ref{tab:NR_criteria} for the rooms from Table~\ref{tab:room_types} and several other types. These values vary somewhat between countries and their regulations. For easier comparison with Table~\ref{tab:common_SPLs} and assessment of the dynamic range, A-weighted equivalent sound pressure levels in the last column are estimated to be 10\,dB above the rated value of the NR-curve, which is a reasonable estimation for broadband sounds. Table~\ref{tab:NR_criteria} clearly shows that the requirements with regard to ambient noise are particularly stringent for the rooms for music listening or recording. This is primarily due to the large expected dynamics of music. They are followed by the rooms for speech. The highest levels of ambient noise are allowed when sound reinforcement is present.

\begin{table}[h]
	\caption{Suggested maximum ambient noise levels in rooms expressed in terms of NR-curves and approximate A-weighted sound pressure levels.}
	\label{tab:NR_criteria}
	\begin{tabular}{ | l | c | c | }
		\hline
		\textbf{room type} & \textbf{NR-curve} & \textbf{$L_A$} \\
		\hline
		\hline
		concert hall & 15 & 25\,dB \\
		\hline
		opera house & 15 & 25\,dB \\
		\hline
		theatre & 20 & 30\,dB \\
		\hline
		rec. studio & 15 & 25\,dB \\
		\hline
		control room & 15 & 25\,dB \\
		\hline
		cinema & 30 & 40\,dB \\
		\hline
		lecture hall & 25 & 35\,dB \\
		\hline
		classroom & 25 & 35\,dB \\
		\hline
		club (public spaces) & 35 & 45\,dB \\
		\hline
		\hline
		sleeping room & 25 & 35\,dB \\
		\hline
		living room & 30 & 40\,dB \\
		\hline
		church & 30 & 40\,dB \\
		\hline
		stadium & 45 & 55\,dB \\
		\hline
		library & 30 & 40\,dB \\
		\hline
		open-plan office & 35 & 45\,dB \\
		\hline
	\end{tabular}
\end{table}

\subsubsection{Frequency range}\label{ch:frequency_range}

Although the audible frequency range is approximately from 20\,Hz to 20\,kHz, the fact that not all its parts are equally relevant for particular applications of room acoustics can substantially simplify the treatment. At moderate sound pressure levels, around 80\,dB, human hearing is much less sensitive to the tones at low frequencies than at mid-to-high frequencies, 1-5kHz, where the sensitivity is highest (for example, around 20\,dB at 50\,Hz). On the other hand, useful auditory information is rarely contained in the sound at very high frequencies, above this range. Moreover, most sources of sound in rooms, such as musical instruments and human voice, radiate less sound energy at high frequencies above several kilohertz and the energy is strongly dissipated in air or absorbed at room surfaces. In practice, we are rarely interested in octaves above 4kHz.

Bulk of the energy of speech is contained in the octaves from 63\,Hz to 8\,kHz. The fundamental frequency of an adult male voice is between 80\,Hz and 180\,Hz and of a female voice around 150-250\,Hz. However, most of the spoken information is carried by the higher harmonics (formants). Sound energy of vowels is predominantly in the octaves 250-2000\,Hz. Consonants appear at even higher frequencies, in the octaves between 2\,kHz and 8\,kHz. While low and middle frequencies carry most of the energy of speech (the vowels), high-frequency contribution of consonants is critical for intelligibility, even though their energy can be 10-20\,dB weaker.

Frequency bands in which the fundamental frequencies of singing appear are 63-250\,Hz (bass), 125-500\,Hz (tenor), and 250-1000\,Hz (soprano). Another frequency region of special importance is around 3000\,Hz, which contains the so-called singer's formant. It allows trained singers to stand out from the orchestra, which is at lower frequencies often a challenging task. Table~\ref{tab:instruments_freq_range} summarizes the discussion above by listing the frequency ranges of fundamental frequencies of common musical instruments, or in which bulk of sound energy is contained, in the case of percussions and human voice. It should be mentioned that higher harmonics of musical tones as well as broadband, non-harmonic (stochastic) components can extend somewhat above these ranges.

\begin{table}[h]
	\caption{The most relevant frequency ranges (in octave bands) of musical instruments and human voice.}
	\label{tab:instruments_freq_range}
	\begin{tabular}{ | l | c | }
		\hline
		\textbf{source} & \textbf{freq. range} \\
		\hline
		\hline
		speech & 125\,Hz -- 4\,kHz \\
		\hline
		\hline
		bass & 63\,Hz -- 250\,Hz \\
		\hline
		tenor & 125\,Hz -- 500\,Hz \\
		\hline
		soprano & 250\,Hz -- 1\,kHz \\
		\hline
		\hline
		guitar & 63\,Hz -- 1\,kHz \\
		\hline
		harp & 31.5\,Hz -- 4\,kHz \\
		\hline
		double bass & 31.5\,Hz -- 250\,Hz \\
		\hline
		cello & 63\,Hz -- 1\,kHz \\
		\hline
		violin & 125\,Hz -- 4\,kHz \\
		\hline
		contrabassoon & 31.5\,Hz -- 250\,Hz \\
		\hline
		clarinet & 125\,Hz -- 2\,kHz \\
		\hline
		flute & 250\,Hz -- 4\,kHz \\
		\hline
		piano & 31.5\,Hz -- 4\,kHz \\
		\hline
		tuba & 31.5\,Hz -- 500\,Hz \\
		\hline
		trombone & 63\,Hz -- 500\,Hz \\
		\hline
		trumpet & 125\,Hz -- 1\,kHz \\
		\hline
		timpani & 63\,Hz -- 250\,Hz \\
		\hline
		drums & 125\,Hz -- 500\,Hz \\
		\hline
		xylophone & 250\,Hz -- 4\,kHz \\
		\hline
		triangle & 2\,kHz -- 8\,kHz \\
		\hline
		organs & 16\,Hz -- 8\,kHz \\
		\hline
	\end{tabular}
\end{table}

As a general outcome, in room acoustics we are typically interested in the frequency range of octave bands from \textbf{63\,Hz} to \textbf{4\,kHz}. This corresponds to the third-octave bands from 50\,Hz to 5\,kHz. If the room is intended only for speech, the lowest octave (63\,Hz) becomes less relevant, as far as useful sound information is concerned. In contrast to this, particularly significant are the octaves in the middle of the audible range, \textbf{500\,Hz}, \textbf{1\,kHz} and \textbf{2\,kHz}, in which the human auditory system is most sensitive and the information content is high.

\subsubsection{Directivity}

Practically all real sources of sound are directional, that is, they do not radiate sound equally in all directions, at least not at all frequencies. While pronounced directivity can cause too low sound levels in certain directions from the source, it can also help the acoustic optimization of rooms. For example, locations and orientations of musical instruments on the stage of a concert hall with regard to the auditorium are well defined by the tradition of music performance. Positions of the instruments and singers in a recording studio can be controlled. Location of a speaker on the podium of a lecture hall or a loudspeaker of a PA system is also fixed and known. Together with more or less known orientation and directivity of the sources, regions and directions of higher sound radiation can be identified and utilized for the optimization. This is, of course, much more difficult if the source is expected to move or change its orientation frequently, in which case its directivity becomes less relevant for the acoustic design.

Since it involves the dependence on both angle to the source and frequency, directivity is quite cumbersome to represent accurately graphically or numerically (see also the discussion in section~\ref{ch:directivity}). Table \ref{tab:instruments_coverage_angle} gives very approximate values of coverage angles (here angles of principal radiation within which the far-field sound pressure level varies less than $\pm 1.5$dB) of human voice and musical instruments. The values are given for the horizontal plane, which is, for example, plane of the bridge of string instruments or the bore of wind instruments. The directivity clearly increases with the size of the source. Almost all of the listed sources are omnidirectional only at frequencies below 250\,Hz, which means that they are directional in most of the frequency range of interest.

\begin{table}[h]
	\caption{Coverage angle (0 to -3\,dB) of musical instruments and human voice in the horizontal plane. Only one main radiation lobe is considered. Note: sound generation of flute in the octave band 125\,Hz is practically negligible (compare with Table \ref{tab:instruments_freq_range}).}
	\label{tab:instruments_coverage_angle}
	\begin{tabular}{ | l | l | l | l | l | l | l |}
		\hline
		\textbf{source} & \textbf{125\,Hz} & \textbf{250\,Hz} & \textbf{500\,Hz} & \textbf{1\,kHz} & \textbf{2\,kHz} & \textbf{4\,kHz} \\
		\hline
		\hline
		speech/voice & 300$^\circ$ & 280$^\circ$ & 160$^\circ$ & 160$^\circ$ & 120$^\circ$ & 90$^\circ$ \\
		\hline
		\hline
		harp & 360$^\circ$ & 360$^\circ$ & 180$^\circ$ & 80$^\circ$ & 60$^\circ$ & 30$^\circ$ \\
		\hline
		double bass & 320$^\circ$ & 270$^\circ$ & 220$^\circ$ & 100$^\circ$ & 90$^\circ$ & 80$^\circ$ \\
		\hline
		cello & 360$^\circ$ & 150$^\circ$ & 110$^\circ$ & 80$^\circ$ & 70$^\circ$ & 80$^\circ$ \\
		\hline
		violin & 360$^\circ$ & 360$^\circ$ & 220$^\circ$ & 180$^\circ$ & 140$^\circ$ & 90$^\circ$ \\
		\hline
		contrabassoon & 360$^\circ$ & 360$^\circ$ & 120$^\circ$ & 80$^\circ$ & 60$^\circ$ & 40$^\circ$ \\
		\hline
		clarinet & 360$^\circ$ & 360$^\circ$ & 360$^\circ$ & 220$^\circ$ & 100$^\circ$ & 40$^\circ$ \\
		\hline
		flute & / & 100$^\circ$ & 130$^\circ$ & 130$^\circ$ & 110$^\circ$ & 110$^\circ$ \\
		\hline
		tuba & 190$^\circ$ & 160$^\circ$ & 80$^\circ$ & 45$^\circ$ & 30$^\circ$ & 20$^\circ$ \\
		\hline
		trombone & 360$^\circ$ & 360$^\circ$ & 140$^\circ$ & 100$^\circ$ & 45$^\circ$ & 45$^\circ$ \\
		\hline
		trumpet & 360$^\circ$ & 360$^\circ$ & 360$^\circ$ & 90$^\circ$ & 80$^\circ$ & 40$^\circ$ \\
		\hline
	\end{tabular}
\end{table}

Coverage angle is a simple and convenient parameter when the source has a single direction of principal radiation. However, certain musical instruments exhibit more complicated directivity patterns. For example, a flute radiates sound both from the blow hole and the tone holes or the open end and, accordingly, its radiation pattern has maxima in these different directions, which makes the values in Table \ref{tab:instruments_coverage_angle} inaccurate. This is especially true at high frequencies, when complicated radiation patterns are expected in general.

Radiation patterns are also often asymmetric with respect to the direction of maximum radiation, which is not captured by the value of coverage angle. This is usually the case when the source itself is asymmetric and large, or the player of a musical instrument is located very close to the instrument thus affecting its radiation. For instance, head of a violinist presents an obstacle for sound radiation in a particular direction. Directivity patterns of piano and organs, which are not listed in Table~\ref{tab:instruments_coverage_angle}, are especially complicated due to their large sizes. Most of the sound energy of a grand piano is radiated from the soundboard upwards and downwards and certain fraction of the energy is reflected from the lid. Therefore, position of the lid can also affect the radiation pattern. Percussions are also omitted from the overview in the table above. In general, they radiate predominantly normal to the membrane or plate. Finally, directivity can vary even between the instruments of the same type, for instance, due to different physical properties of wooden boards of string instruments or their particular assembly.

A special type of the sources of sound are \textbf{loudspeakers}. Their directivity depends largely on their size and increases with frequency. Conveniently, the directivity patterns and other related parameters of loudspeakers are usually provided by the manufacturers and thus well known. Of special interest are line arrays of loudspeakers which consist of several loudspeakers clustered typically in the form of a straight or J-shaped array. Careful pre-processing of the input signals supplied to each of the loudspeakers allows a high flexibility and control over the directivity of the entire system. With larger number of loudspeakers in the array, higher directivity can be achieved and the direct sound energy can be distributed more uniformly over a large auditorium, compared to a single loudspeaker of equal sound power. In certain applications, this can be used to suppress the deficiencies of room acoustics, such as lack of useful reflected sound energy in (semi) open spaces, for example stadiums, or detrimental late reflections in highly reverberant rooms, sport halls, and churches. The pronounced and controllable directivity of line arrays is thus utilized both to avoid undesired reflections and minimize the losses of useful sound energy. Note that essentially the same applies to directional \textbf{microphones} and microphone arrays, when they are used as receivers.

\subsection{Subjective criteria}
\label{ch:subjective_criteria}

Since we mostly consider human listener as the receiver of sound information, sound field in a room is eventually assessed subjectively. With regard to that, Table \ref{tab:rooms_subjective_properties} lists the most relevant subjective criteria for the already discussed applications of room acoustics. It is important to note that these criteria must be satisfied only at those particular locations in rooms where the receivers are expected. This is especially critical when the locations of sources and listeners are known and fixed, for example at the stage or in the auditorium, or even spatially limited, such as in control rooms (locations of the loudspeakers and at the mixing desk). Acoustic design can then be focused on these key locations in the room, which makes the treatment somewhat easier and more efficient.

\begin{table}[h]
	\caption{The main subjective criteria for different applications of room acoustics.}
	\label{tab:rooms_subjective_properties}
	\begin{tabular}{ | l | p{10.3cm} |}
		\hline
		\textbf{application} & \textbf{subjective criteria} \\
		\hline
		noise control, ac. comfort & low loudness, appropriate reverberance \\
		\hline
		sound reproduction & suppressed effects of the room \\
		\hline
		sound production & appropriate loudness and dynamics, balance, reverberance, intelligibility (speech) or clarity (music), synchronization with the visual component, spaciousness, envelopment, apparent source width and localization, intimacy, absence of echo and coloration; \textit{on the stage}: ease of ensemble, support; \textit{*visual criterion} -- visibility of the source \\
		\hline
	\end{tabular}
\end{table}

Already at the first glance the table indicates that the subjective requirements are most complicated and diverse for rooms in which the sound information is created. In addition to the acoustic-related aspects which we consider next, in many cases, such as opera houses and theatres, visibility of the source (actor or performer) presents an important additional component. It is determined at the first place by the distance between a listener (viewer) and the source, as well as by the angle at which the source is observed. The question how the listed subjective criteria can be met in practice will be considered in section~\ref{ch:basic_strategies}. Here we will only give a few general remarks on each criterion and how they can be linked to certain objective physical properties of sound fields.

\textbf{Loudness} is closely related to the energy of sound and sound pressure level. Since human auditory system reacts to the excitation with certain inertia, it tends to ``integrate'' the sound energy over a certain finite time interval, which is of the order of 10\,ms. The total (rather than instantaneous) sound energy which is received in such short time windows determines the subjectively perceived loudness of sound events. Consequently, the perceived sound level should be well above the (supposedly steady) ambient noise level in the room (see Table \ref{tab:NR_criteria}). A room should also provide an appropriate \textbf{balance} between sound energy at different frequencies, especially rooms for music performances.

For good \textbf{intelligibility} of speech or \textbf{clarity} of music, sufficient loudness of the signal is not enough. Certain minimum dynamic range of the received sound has to be ensured, since much of the information in both speech and music is contained in the temporal fluctuations of sound energy between the spoken syllables or played musical tones as well as in the quieter parts. Characteristic time scale of such fluctuations has typically the order of 100\,ms or larger\footnote{This value can be directly compared with the modulation frequencies for speech which are given in section \ref{ch:descriptors_of_room_acoustics}.}. Accordingly, sound energy inside a room should decay fast enough during the quiet intervals not to mask the succeeding sounds\footnote{As a quick analysis, if we suppose that the sound energy should decay at least 10\,dB in 100\,ms, the reverberation time in the room (which will be defined later) below $T_{60} = 60\text{\,dB} \cdot 0.1\text{\,s} / 10\text{\,dB} = 0.6\text{\,s}$ is necessary for good speech intelligibility. For comparison, optimal reverberation time of lecture halls is typically around 1\,s and even below 0.5\,s in small classrooms.}. On the other hand, the early decaying energy which reaches the listener shortly after the direct sound, well within the first 100\,ms, does not compromise clarity but even contributes positively by increasing the loudness. 

\textbf{Synchronization} between the perceived auditory and visual components depends mainly on the distance between the source and receiver. Since the visual information is received practically instantaneously, time delay with which the direct sound reaches the listener (and the attenuation during propagation) should not be too large. Maximum distance of around 30\,m is recommendable for concert halls, when the visual component is of secondary importance. When good \textbf{visibility} of the source is more or less equally important as the auditive component, as in theatres and opera houses, the maximum distance should be smaller. For a proper distinction of facial expressions of actors on the stage, the viewer should not be placed more than around 20\,m from the stage.

The propagation time of the direct sound affects the synchronization. On the other hand, the delay (with respect to the direct sound) and energy of the earliest reflections which reach the listener determine the perceived distance from the source. If the earliest reflections arrive relatively long after the direct sound and with low energy, the source appears to be closer (both source and receiver appear to be far from the reflecting surfaces and close to each other). This quality of the perceived sound is often called \textbf{intimacy}. It decreases when strong reflections immediately follow the direct sound.

\textbf{Reverberance} makes the sound more natural and subjectively more appealing than a pure ``dry'' sound. However, optimal reverberance (and the associated objective physical phenomenon -- reverberation) for music performance depends on the music style and has partly historical roots. For example, Baroque music is usually associated with longer reverberation times than Romantic music. Based on this, the halls for music performances (the first two rows of Table \ref{tab:room_types}) can be further subdivided according to the type of the music which is produced in them. A compromise is often inevitable, which should make the room's reverberance satisfactory for most of the music performed in it. While reverberation time of 2 seconds or longer is appropriate for church music, shorter reverberation times are more suitable for contemporary music genres, and even shorter (around and below 1\,s) for speech.

\textbf{Spaciousness} is often stated to consist of two components: \textbf{listener envelopment} (impression of being immersed in the sound field, or being surrounded by it) and \textbf{apparent source width}. It depends mostly on the amount of sound energy which reaches the listener laterally. Hence, appropriate room acoustics, especially for music performances, should provide not only sufficient level of sound and clarity (which can in principle be achieved with localized sound reinforcement even in an anechoic environment), but also appropriate angular distribution of sound energy with respect to the listener. Uniform angular distribution of the incoming sound energy approaches the idealization of a diffuse field (see section \ref{ch:statistical_theory}), in which many reflections with equal energy reach the listener from all directions.

During the performance, the players and singers have to be able to hear well the sound of their own instrument or voice, as well as the other performers. This introduces additional requirements for room acoustics. In fact, in the absence of other listeners (for example, in recording studios), these requirements and associated subjective components become essential. They are often divided into ease of ensemble and support. \textbf{Ease of ensemble} rates the ability of the performers to hear well each other, which allows easier common performance. When it is low, the result can be poor synchronization and balance (relative strength) of the sound produced by different performers. In contrast to this, \textbf{support} relates to the feedback which the performers obtain from the room. Lack of support means that the performers might not hear their instruments or voices loud enough. As a consequence, they will probably tend to play or sing louder than actually necessary. Both ease of ensemble and support can be low in overdamped rooms (such as in poor recording studios) or at stages lacking local early reflections (typically from the closest rear wall or ceiling).

Presence of strong (not more than around 20\,dB weaker than the direct sound\footnote{In certain cases, reflected sound can even be several decibels stronger than the direct sound, for example, when a concave surface focuses the reflections to a receiver or when the direct sound is significantly attenuated due to the grazing propagation over an absorbing surface, such as auditorium.}) and distinct reflections or resonances of the room can cause undesired \textbf{echo} or \textbf{coloration} (change of the spectral content of the sound). Both of these phenomena are quite easily perceived and therefore the room acoustic design should prevent or at least minimize them. The major difference between echo and coloration is in the length of the delay of the reflections with respect to the direct sound. A rough border between the two is the delay time of 100\,ms. The reflections which reach the receiver with shorter delays are difficult to be perceived as an echo, a replica of the direct sound, but rather as a change of timbre\footnote{The delay of 100\,ms corresponds to the difference in sound path lengths of around 34\,m. This is one of the reasons why loudspeakers of public address systems at large events are normally placed at shorter distances from each other.}.

If the reflections immediately follow the direct sound, with the delays even less than around 5\,ms, they may cause an apparent \textbf{shift} of the location of the source or change of its \textbf{width}. While this may be less problematic or even desirable in rooms for music production, since it contributes to the spaciousness (see above), it can be extremely detrimental for a stereo or multichannel sound image. Especially interfering are reflections which reach a listener at relatively low angles, around 30-50$^\circ$, in the horizontal plane, where spatial resolution of the human auditory system is particularly high.

It should be bear in mind that although the subjective components of a sound field can be associated with certain physical parameters, these relations are by no means direct and simple. For example, minimum energy of the reflected sound which is required for the perception of echo, coloration, or apparent shift of the source depends strongly on the character of the sound (tonal sound, random noise, transient, etc.) and its angle of incidence. Correlation between objective and subjective parameters is the object of many studies, based on which widely used numerical descriptors of room acoustics are derived, which will be presented in section \ref{ch:descriptors_of_room_acoustics}.

\section{Basic equations}\label{ch:basic_equations}

As already stated, room acoustics is a sub-discipline of acoustics in fluids, in particular air. Accordingly, in this section we repeat some of the most important equations of classical acoustics and derive them from the governing equations of fluid dynamics.

\subsection{Conservation laws}\label{conservation_laws}

Fluid dynamics is governed at the first place by the \textbf{conservation laws}. Sound waves in fluids involve oscillatory motion of the fluid particles and therefore obey the same equations. These are:
\begin{itemize}
	\item conservation of mass (continuity equation)
	\begin{equation}\label{eq:Navier_Stokes_mass}
	\frac{\partial \rho_f}{\partial t} + \nabla \cdot (\rho_f \boldsymbol v_f) = 0,
	\end{equation}
	\item conservation of momentum (equation of motion)
	\begin{equation}\label{eq:Navier_Stokes_momentum}
	\frac{\partial}{\partial t}(\rho_f \boldsymbol v_f) + \nabla \cdot (\rho_f \boldsymbol v_f \boldsymbol v_f) + \nabla p_f - \nabla \cdot \tau_f = 0.
	\end{equation}
\end{itemize}
The third common governing equation, conservation of energy, is rarely used in the context of room acoustics, since heat exchange, change of entropy, and similar thermodynamic phenomena rarely have a significant role\footnote{These phenomena will be only briefly treated in relation to porous absorbers and thermal boundary layer in section~\ref{viscous_and_thermal_losses}.}. The physical quantities which depend on time ($t$) and location (vector $\boldsymbol x$) are: $\rho$ density, $\boldsymbol{v}$ velocity, $p$ pressure, and $\tau$ viscous stress tensor (which will be mostly neglected later). The subscript $_f$ stands for fluid, because the quantities are associated not only with sound waves, but with the fluid motion in general. Notice that both $\tau_f$ and $\boldsymbol v_f \boldsymbol v_f$ in the equation of motion are second-order tensors and that in their most general form, the conservation equations are non-linear and without a closed-form analytical solution.

The calculations are greatly simplified when sound waves are the \textbf{only perturbation} of an otherwise uniform (constant in space) and steady (constant in time) fluid, as is normally the case with air in rooms. We can then write $\rho_f = \rho_0 + \rho(\boldsymbol x, t)$, $p_f = p_0 + p(\boldsymbol x, t)$, and $\boldsymbol v_f = \boldsymbol v_0 + \boldsymbol v(\boldsymbol x, t)$, where the subscript $_0$ is used for the constant values of the background fluid as the medium in which sound waves propagate and $\rho$, $p$, and $\boldsymbol v$ are purely acoustic quantities (acoustic pressure, density, and particle velocity). In the following, we will consider mostly still air at the room temperature ($T_0 = 293$\,K), for which $\boldsymbol v_0 = 0$ and $\rho_0 = 1.2 \text{\,kg/m\textsuperscript{3}}$. Neglecting viscosity, the two conservation equations become, respectively,
\begin{equation}\label{eq:Navier_Stokes_mass_perturbations}
	\frac{\partial \rho}{\partial t} + \nabla \cdot (\rho_0 \boldsymbol v + \rho \boldsymbol v) = 0 \text{ and}
\end{equation}
\begin{equation}\label{eq:Navier_Stokes_momentum_perturbations}
	\frac{\partial}{\partial t}(\rho_0 \boldsymbol v + \rho \boldsymbol v) + \nabla \cdot (\rho_0 \boldsymbol v \boldsymbol v + \rho \boldsymbol v \boldsymbol v) + \nabla p = 0.
\end{equation}

Another reasonable (and crucial) assumption is that the sound waves present only a \textbf{small perturbation} of the background fluid. By that, we mean that $p \ll p_0$. The static pressure is given by the equation of state of an ideal gas,
\begin{equation}\label{eq:static_pressure}
p_0 = \rho_0 \mathcal{R} T_0,
\end{equation}
where $\mathcal{R}$ is specific gas constant ($\mathcal{R} = 287 \text{\,J/kgK}$ for air) and $T_0$ is temperature in kelvins. This gives approximately $p_0 = 101\,\text{kPa}$ for air at room temperature. If we take roughly one percent of this value, 1010\,Pa, to be the upper limit for the amplitude of weak acoustic pressure fluctuations, the inequality $p \ll p_0$ holds whenever the sound pressure level\footnote{Sound pressure level will be defined in eq.~(\ref{eq:SPL}). The factor of $\sqrt{2}$ in the equation represents the ratio of amplitude and root mean square value, which is in the definition of sound pressure level. The factor applies for simple oscillations, as will be shown later.} is below
\begin{align*}
20\log_{10} \left(\dfrac{1010\text{\,Pa}/\sqrt{2}}{2 \cdot 10^{-5}\text{\,Pa}}\right) \approx 151 \text{\,dB}.
\end{align*}
This is a very high value which is normally surpassed only in very close proximity of the loudest sources of sound, so $p \ll p_0$ will be indeed satisfied in practically all applications of room acoustics. In addition to this, as equations (\ref{eq:p_rho}) and (\ref{eq:rho0_c0_p0}) will show for a homentropic (inviscid and adiabatic) flow of a perfect gas, $\rho/\rho_0 \sim \mathcal{O}( p/p_0 )$, so $\rho \ll \rho_0$ also holds. The conservation laws can thus be written approximately, without major loss of accuracy as
\begin{equation}\label{eq:Navier_Stokes_mass_small_perturbations}
\frac{\partial \rho}{\partial t} + \rho_0 \nabla \cdot \boldsymbol v = 0 \text{ and}
\end{equation}
\begin{equation}\label{eq:Navier_Stokes_momentum_small_perturbations}
\rho_0 \frac{\partial \boldsymbol v}{\partial t} + \rho_0 \nabla \cdot (\boldsymbol v \boldsymbol v) + \nabla p = 0.
\end{equation}

As we will see in section \ref{ch:spherical_and_plane_waves}, at the points which are not very close to the source of sound $|\boldsymbol v| \sim \mathcal{O}( p/(\rho_0 c_0) ) \sim \mathcal{O}( \rho c_0/\rho_0 ) \ll c_0$, where $c_0$ is the speed of sound in the fluid. Since sound waves are the only perturbation, we can also scale $\partial /\partial t \sim \omega$ and $\nabla \sim k = \omega / c_0$, where $\omega$ is angular frequency and $k$ is wave number of the waves. Therefore, the first two terms in eq.~(\ref{eq:Navier_Stokes_momentum_small_perturbations}) scale as $\rho_0 \omega |\boldsymbol v|$ and $\rho_0 \omega |\boldsymbol v|^2/c_0$, respectively, and we can neglect the second term in favour of the first one. This finally gives the two conservation laws of classical acoustics:
\begin{itemize}
	\item conservation of mass
	\begin{equation}\label{eq:Euler_mass}
	\boxed{ \frac{\partial \rho}{\partial t} + \rho_0 \nabla \cdot {\boldsymbol v} = 0 }
	\end{equation}
	\item conservation of momentum
	\begin{equation}\label{eq:Euler_momentum}
	\boxed{ \rho_0 \frac{\partial {\boldsymbol v}}{\partial t} + \nabla p = 0 }.
	\end{equation}
\end{itemize}
These equations are \textbf{linear} and the linearization is achieved based on the assumption of a weak acoustic perturbation (for sound pressure levels in air below around 150\,dB). The background fluid is otherwise uniform, motionless, and homentropic perfect gas. All acoustic quantities, $p$, $\rho$, and $\boldsymbol v$, depend on time $t$ and location $\boldsymbol x$, while the background (static) quantities are constants.

\subsection{Viscous and thermal losses}\label{viscous_and_thermal_losses}

In the derivation above we neglected the viscosity (the fluid was homentropic, that is, adiabatic and inviscid), which is usually acceptable for bulk of the volume of rooms or cavities of usual sizes. However, viscous effects can dominate inside thin boundary layers at solid surfaces. An important example when such effects can become significant are porous materials, which are commonly used as sound absorbers and which, indeed, rely on viscous (and thermal) dissipation of sound energy in their fine inner structure. These absorbers will be considered in more details in section~\ref{ch:basic_elements}. Here we will only give a rough estimation of thickness of the boundary layers in which dissipation takes place.

First we consider a \textbf{viscous boundary layer} (also called Stokes layer) with thickness $\delta_\tau$ located at a motionless rigid surface. The rate of increase of the momentum from eq.~(\ref{eq:Navier_Stokes_momentum}) scales as
\begin{equation}\label{eq:rate_of_increase_momentum}
\frac{\partial}{\partial t}(\rho_f \boldsymbol v_f) \sim \rho_0 \omega |\boldsymbol v|.
\end{equation}
In the viscous boundary layer the momentum is balanced by the viscous term which follows from Stokes' hypothesis for Newtonian fluids\footnote{It states: $\tau_f = \mu \nabla \boldsymbol v + \mu (\nabla \boldsymbol v)^T - \frac{2}{3} \mu (\nabla \cdot \boldsymbol v) I$, where $I$ is the unit tensor and $^T$ denotes transposition.},
\begin{equation}\label{eq:viscous_stress_tensor_Stokes}
\nabla \cdot \tau_f \sim \frac{1}{\delta_\tau} \mu \frac{1}{\delta_\tau} |\boldsymbol v| = \rho_0 \nu \frac{1}{\delta_\tau^2} |\boldsymbol v|,
\end{equation}
where $\mu = \rho_0 \nu$ is dynamic viscosity and $\nu$ is kinematic viscosity of the fluid. The spatial derivatives scale as $1/\delta_\tau$. Equations (\ref{eq:rate_of_increase_momentum}) and (\ref{eq:viscous_stress_tensor_Stokes}) then give
\begin{equation}\label{eq:viscous_thermal_boundary_layer_thickness}
\delta_\tau(\omega) \sim \mathcal{O} \left( \sqrt{\frac{\nu}{\omega}} \right).
\end{equation}
The viscous boundary layer thickness is thus inversely proportional to the square root of frequency. For air at room temperature and atmospheric pressure, $\mu \approx 1.8 \cdot 10^{-5}$\,kg/(ms) and $\nu \approx 1.5 \cdot 10^{-5}$\,m$^2/$s. As an example, for frequencies around 1\,kHz, the boundary layer thickness is of the order of 0.1\,mm. 

In order to estimate the thickness of \textbf{thermal boundary layer}, $\delta_T$, we will use Fourier's law,
\begin{equation}\label{eq:Fourier_law}
\boldsymbol q = -K \nabla T,
\end{equation}
where $\boldsymbol q$ denotes heat flux vector (with the unit W/m$^2$) describing heat conduction, $K$ is thermal conductivity ($K \approx 0.025$\,W/(Km) in air at the room temperature and normal atmospheric pressure), and $T$ is (unsteady) temperature in kelvins. The rate of increase of energy scales as
\begin{equation}\label{eq:internal_energy_rate}
\rho_0 C_p \frac{\partial T}{\partial t} \sim \rho_0 C_p \omega T,
\end{equation}
where $C_p$ is the specific heat capacity at constant pressure, which is for air $C_p \approx 1005$\,J/(kgK), and the time scale of fluctuations is again dictated by the acoustic perturbation. This term is balanced by the rate of energy due to the heat conduction,
\begin{equation}\label{eq:energy_heat_conduction}
- \nabla \cdot \boldsymbol q = K \nabla \cdot \nabla T \sim \frac{K T}{\delta_T^2},
\end{equation}
where the spatial derivatives now scale as $1/\delta_T$. From the last two equations it follows that the order of magnitude of the thermal boundary layer thickness is
\begin{equation}\label{eq:thermal_boundary_layer_thickness}
\delta_T(\omega) \sim \mathcal{O} \left( \sqrt{\frac{K}{\rho_0 C_p \omega}} \right),
\end{equation}
which is again inversely proportional to the square root of frequency. Moreover, thickness of both boundary layers is of the same order of magnitude. The dissipation effects at such small length scales are irrelevant for sound propagation in rooms with many orders of magnitude larger dimensions. However, thickness of the boundary layers matches well the characteristic length scales of the inner structures of porous materials. As a consequence, a significant sound absorption in such materials takes place in the thermal and viscous boundary layers at the solid skeleton, which build most of the fluid phase of a porous absorber.

\subsection{Wave equation}

Since all static values are assumed to be known, the two conservation equations (scalar eq.~(\ref{eq:Euler_mass}) and vector eq.~(\ref{eq:Euler_momentum})) contain three unknowns -- two scalars (acoustic pressure and density) and the particle velocity vector. An additional equation is necessary in order to close the system of equations and make it solvable. This is the thermodynamic \textbf{equation of state}, which in its linearized form for homentropic flows (with constant entropy, that is, inviscid and adiabatic flows) and with ${\boldsymbol v_0} = 0$ reads
\begin{equation}\label{eq:p_rho}
p = c_0^2 \rho.
\end{equation}
In a perfect gas,
\begin{equation}\label{eq:rho0_c0_p0}
\rho_0 c_0^2 = \gamma p_0,
\end{equation}
where $\gamma$ is the heat capacity ratio, which can be taken as constant ($\gamma = 1.4$ for a diatomic gas like air). This gives the speed of sound in air at the room temperature around $c_0 = 343 \text{\,m/s}$.

The equations (\ref{eq:Euler_mass}), (\ref{eq:Euler_momentum}), and (\ref{eq:p_rho}) now form a closed system of three equations which can be solved in general. For example, we can combine the equations (\ref{eq:Euler_mass}) and (\ref{eq:p_rho}) to remove the density\footnote{Alternatively, we can remove the pressure by inserting eq.~(\ref{eq:p_rho}) into eq.~(\ref{eq:Euler_momentum}). This would lead to the wave equation with density, rather than pressure, as the acoustic variable. In general, compressible (longitudinal) waves, such as sound waves in fluids, are completely described with a single scalar, such as pressure, density, or scalar potential, and they all satisfy the wave equation.}. The two remaining equations are
\begin{equation}\label{eq:Euler_mass_pressure}
	\frac{1}{c_0^2} \frac{\partial p}{\partial t} + \rho_0 \nabla \cdot {\boldsymbol v} = 0
\end{equation}
and
\begin{equation}\label{eq:Euler_momentum1}
	\rho_0 \frac{\partial {\boldsymbol v}}{\partial t} + \nabla p = 0.
\end{equation}
These can be brought to a single scalar equation with acoustic pressure as the only dependent variable. This is achieved by differentiating eq.~(\ref{eq:Euler_mass_pressure}) with respect to time and subtracting the divergence of eq.~(\ref{eq:Euler_momentum1}) (so its vector contracts to a scalar) from it. The result is
\begin{align*}
\begin{split}
\frac{1}{c_0^2} \frac{\partial^2 p}{\partial t^2} + \rho_0 \frac{\partial}{\partial t}(\nabla \cdot {\boldsymbol v}) - \rho_0 \nabla \cdot \left( \frac{\partial {\boldsymbol v}}{\partial t}\right) - \nabla^2 p = 0.
\end{split}
\end{align*}
Since the time derivative and divergence can switch the order, the second and third terms cancel, leaving \textbf{the wave equation} for sound pressure,
\begin{equation}\label{eq:wave_eq}
\boxed{ \frac{1}{c_0^2} \frac{\partial^2 p(\boldsymbol{x}, t)}{\partial t^2} - \nabla^2 p(\boldsymbol{x}, t) = 0 }.
\end{equation}

The wave equation is the central equation of classical acoustics in fluids. It holds under the same set of assumptions as in section~\ref{conservation_laws} and all sound fields under those conditions have to satisfy it. Importance of the wave operator in physics of various fields is so large that it is often abbreviated as
\begin{equation}
\frac{1}{c_0^2} \frac{\partial^2}{\partial t^2} - \nabla^2 = \Box, 
\end{equation}
so the wave equation above reads simply
\begin{equation}
\Box p = 0.
\end{equation}
However, the wave equation admits many different particular solutions and hence different values of $p(\boldsymbol x, t)$ in rooms. Each particular solution depends on the source of sound and boundary conditions, which we consider next.

\subsection{Sources and boundary conditions}

The sound field in practically entire volume of a room satisfies the wave equation. However, a complete solution requires additional information on the sources of the sound field, as well as the initial\footnote{The initial conditions can be thought of as a temporal analogue of the boundary conditions, where the ``boundary'' is some initial time $t_0$. However, it is usually assumed that there is no sound field before some considered source in the room is switched on at $t = t_0$, so we will neglect the trivial initial condition $p(t<t_0) = 0$ most of the time.} and boundary conditions. A source of sound is most easily included in eq.~(\ref{eq:wave_eq}) by introducing a scalar function $q(\boldsymbol y,\tau)$ at its right-hand side, which is non-zero only at the location of the source $\boldsymbol y$ (thus, $\boldsymbol x = \boldsymbol y$ in the source region, which can also be spatially distributed) and only at time $\tau$ when the source is active. Alternatively, sources can also be introduced as unsteady boundary conditions, that is, active surfaces which excite the medium in the room.

Wave equation with the \textbf{source} term reads
\begin{equation}\label{eq:wave_eq_source}
\boxed{ \frac{1}{c_0^2} \frac{\partial^2 p(\boldsymbol{x}, t)}{\partial t^2} - \nabla_x^2 p(\boldsymbol{x}, t) = q(\boldsymbol{y}, \tau) }.
\end{equation}
Note that we use the subscript $_x$ in the nabla operator (Laplacian) to emphasize that the differentiation is performed with respect to the receiver's location $\boldsymbol x$ (for example, $\nabla_x = (\partial / \partial x_1, \partial / \partial x_2, \partial / \partial x_3)$ in three-dimensional Cartesian coordinates), not $\boldsymbol y$. The explicit distinction between $\boldsymbol{x}$ and $\boldsymbol{y}$ (although they belong to the same physical space inside the room), as well as between $t$ and $\tau$, is made for mathematical clarity, which will become especially useful later\footnote{See for example the derivation of eq. (\ref{eq:solution_Green_Helmholtz}).}. In the source region, however,
\begin{equation}\label{eq:wave_eq_source_y}
\frac{1}{c_0^2} \frac{\partial^2 p(\boldsymbol{y}, \tau)}{\partial \tau^2} - \nabla_y^2 p(\boldsymbol{y}, \tau) = q(\boldsymbol{y}, \tau).
\end{equation}
The source term has the unit kg/(m$^3$s$^2$) and can physically represent the rate of volume injection as the mechanism of sound generation,
\begin{equation}\label{eq:source_mass_injection}
q(\boldsymbol y,\tau) = \rho_0 \frac{\partial^2 \beta(\boldsymbol y,\tau)}{\partial \tau^2},
\end{equation}
where $\beta$ is the injected volume fraction (a dimensionless quantity). This is particularly appropriate for an essentially incompressible ($\rho_f = \rho_0$) fluid around a compact monopole source. Similarly, divergence of a vector or double divergence of a second-order tensor are suitable for a physical description of dipole and quadrupole sources, respectively.

Appropriate \textbf{boundary conditions} have to be specified for all physical surfaces of the room (walls, ceiling, and floor) or solid bodies in it, or (less commonly) imaginary surfaces which bound the interior volume of the room in which the field is assessed. Various geometric and acoustic properties of surfaces introduce potentially a great complexity and one of the major obstacles for high accuracy of both analytical and numerical calculations in room acoustics. In general, a boundary condition for sound pressure as the dependent acoustic variable can be formulated with the expression\footnote{Here we do not consider moving objects in the room. However, the surfaces are allowed to oscillate around some fixed location, even without being excited by the acoustic field, which is captured by the ``source'' function $c(\boldsymbol y, \tau)$ on the right-hand side.}
\begin{equation}\label{eq:wave_eq_p_BC}
\boxed{ a p(\boldsymbol y, \tau) + b \nabla_y p(\boldsymbol y, \tau) \cdot \boldsymbol n(\boldsymbol y) = c(\boldsymbol y, \tau) },
\end{equation}
for any $\boldsymbol y$ at the surface\footnote{We use the same symbol $\boldsymbol y$ for locations inside the source region and at the boundaries. This will not cause a confusion, because we are primarily interested in the sound field in the rest of the room, where $\boldsymbol x \neq \boldsymbol y$, and the sources and boundaries will be treated quite similarly (as, for example, in eq.~(\ref{eq:solution_Green_Helmholtz})). The same holds for time $\tau$.}. Vector $\boldsymbol n(\boldsymbol y)$ is a unit vector which is normal to the boundary at $\boldsymbol y$ and points into the room, $a$ is a dimensionless constant, $b$ is a constant in meters, and $c$ is a time-dependent function (in Pa), although depending on the function on the right-hand side, other units can be used, as well. The function is non-zero if the surface is active (in which case it behaves like a source). For simplicity, we will practically always assume that the constants $a$ and $b$ and the function $c$ do not vary over a single surface and that the surface is locally reacting -- its motion at any point $\boldsymbol y$ depends on the field only at that point (for further details, see the definition of impedance in eq.~(\ref{eq:impedance})).

Particularly common are acoustically hard (rigid and motionless) surfaces. At every location at such surfaces normal component of the velocity vector is zero,
\begin{equation}\label{eq:rigid_wall_BC_v}
\boldsymbol v(\boldsymbol y, \tau) \cdot \boldsymbol n(\boldsymbol y) = 0.
\end{equation}
From the scalar product of eq.~(\ref{eq:Euler_momentum}) with the time-independent $\boldsymbol n(\boldsymbol y)$, it follows (after replacing $\boldsymbol x$ with $\boldsymbol y$ and $t$ with $\tau$):
\begin{equation}\label{eq:rigid_wall_BC_p}
\rho_0 \frac{\partial }{\partial t}(\boldsymbol v(\boldsymbol y, \tau) \cdot \boldsymbol n(\boldsymbol y)) + \nabla_y p(\boldsymbol y, \tau) \cdot \boldsymbol n(\boldsymbol y) = \nabla_y p(\boldsymbol y, \tau) \cdot \boldsymbol n(\boldsymbol y) = 0.
\end{equation}
Therefore, the coefficient $a$ in eq. (\ref{eq:wave_eq_p_BC}) is zero and $c(\boldsymbol y, \tau) = 0$ for any $b \neq 0$ (for example, $b = 1$\,m) in this particular case. Several other types of surfaces will be considered later. However an appropriate (physically justified) yet feasible mathematical formulation of the boundary conditions of real surfaces is a highly non-trivial task. On the other hand, inappropriate boundary conditions and models of actual sources of sound can lead to large calculation errors. A certain compromise between the accuracy and computational efforts is inevitable in practice. In addition to this, an irregular geometry can substantially increase the complexity or even prohibit analytical calculations of the sound field, even with simple boundary conditions.

\subsection{Sound energy and sound pressure level}

As already mentioned, we usually consider human listener as the receiver in a room. Roughly speaking, human auditory system perceives the sound energy averaged over certain finite time interval, rather than the instantaneous values of sound pressure or its amplitude. For this reason, the oscillating sound pressure of a sustained pure tone is not perceived as time-varying, but as a continuous sound with constant loudness. This also shows that it is not the sound pressure which is averaged over time (which, as any other first-order acoustic quantity, would always yield zero for simple oscillations), but its squared value, a second-order quantity closely related to sound energy (as will be shown shortly).

Such averaging can be mathematically expressed as
\begin{equation}\label{eq:RMS}
\sqrt{\frac{1}{T_{avg}} \int_{-T_{avg}/2}^{T_{avg}/2} p^2 dt} = \sqrt{\langle p^2 \rangle_{T_{avg}}},
\end{equation}
where $T_{avg}$ is the time interval in which averaging is performed. The right-hand side of the equation merely introduces a more compact notation. This is the definition of a \textbf{root mean square} (RMS) value. The square root in it recovers pascal as the unit of RMS, but its value is strictly positive and does not correspond to the instantaneous sound pressure or its absolute value (amplitude). For simple oscillations, the averaging interval can be conveniently set to one period of the oscillations in analytical considerations, or, what is more common in practice, many periods, so that the obtained RMS value does not depend significantly on the interval. For short transient (impulse) sounds, it is difficult to define a single universally applicable averaging interval. However, it is usually of the order of $10$\,ms or longer (compare with the discussion on loudness in section~\ref{ch:subjective_criteria}), for example, S(low) 1\,s, F(ast) 125\,ms, or I(mpulse) 35\,ms.

\textbf{Sound pressure level} (SPL) is defined essentially as the logarithm with base 10 of the RMS value:
\begin{equation}\label{eq:SPL}
\begin{aligned}
\boxed{ L = 20 \log_{10} \left( \frac{\sqrt{\langle p^2 \rangle_{T_{avg}}}}{2 \cdot 10^{-5} \text{\,Pa}} \right) = 10 \log_{10}\left( \frac{1}{T_{avg}} \int_{-T_{avg}/2}^{T_{avg}/2} \frac{p^2}{4 \cdot 10^{-10} \text{\,Pa}^2} dt \right) }.
\end{aligned}
\end{equation}
The reference value $2 \cdot 10^{-5}$\,Pa (which should not be confused with the static pressure $p_0$) is conventionally chosen, since it matches closely the threshold of human hearing at middle frequencies. The unit of sound pressure level is decibel\footnote{Some authors add the unit explicitly in the definition of sound pressure level. We will leave it implicit that $20\log_{10}(\text{ })$ of a first-order quantity and $10\log_{10}(\text{ })$ of a second-order quantity give quantities with the unit dB. Notice, however, that the expression in the logarithm is always dimensionless.} (dB). It should, therefore, be noted that it is actually not the sound pressure, but its non-negative squared value, which determines the sound pressure level. By definition, SPL also involves time-averaging, which is not the case with sound pressure.

Squared values of the basic (first-order) acoustic quantities ($p$, $\rho$, $\boldsymbol v$) are tightly related with sound energy and power. In order to show this, we can multiply eq.~(\ref{eq:Euler_mass_pressure}) with $p/\rho_0$ and eq.~(\ref{eq:Euler_momentum1}) with $\boldsymbol v$ (the scalar product). Sum of the two equations gives then the conservation law for the acoustic energy:
\begin{equation}\label{eq:enegy_governing_equations}
\begin{aligned}
\frac{p}{\rho_0 c_0^2} \frac{\partial p}{\partial t} &+ p \nabla \cdot {\boldsymbol v} + \boldsymbol v \cdot \rho_0 \frac{\partial {\boldsymbol v}}{\partial t} + \boldsymbol v \cdot \nabla p = \frac{1}{2 \rho_0 c_0^2} \frac{\partial p^2}{\partial t} + \frac{\rho_0}{2} \frac{\partial {(\boldsymbol v \cdot \boldsymbol v)}}{\partial t} + \nabla \cdot (p \boldsymbol v) & \\
& = \frac{\partial}{\partial t} \left[ \frac{p^2}{2 \rho_0 c_0^2} + \frac{\rho_0 |\boldsymbol v|^2}{2} \right] + \nabla \cdot (p \boldsymbol v) = \frac{\partial E}{\partial t} + \nabla \cdot \boldsymbol I = 0.
\end{aligned}
\end{equation}
Thereby, we defined \textbf{sound intensity} (rate of energy flux) as the vector
\begin{equation}\label{eq:intensity}
\boxed{ \boldsymbol I = p \boldsymbol v }
\end{equation}
with the unit W/m$^2$ and \textbf{sound energy} as
\begin{equation}\label{eq:energy}
\boxed{ E = E_{pot} + E_{kin} = \frac{p^2}{2 \rho_0 c_0^2} + \frac{\rho_0 |\boldsymbol v|^2}{2} },
\end{equation}
where the first term represents potential energy and the second term is kinetic energy. The unit of $E$ is J/m$^3$ and it is thus strictly speaking energy density, although it is commonly referred to simply as acoustic energy. Although both sound energy and intensity are obviously second-order acoustic quantities (the relation between acoustic pressure and velocity, eq.~(\ref{eq:Euler_momentum}), is linear), they do not involve any time averaging by the given definitions\footnote{However, many authors do include the time-averaging in the definitions of sound energy $E$ and intensity $\boldsymbol I$ and this is often implicit, so care is required when doing calculations. Moreover, some authors define sound intensity as the scalar quantity, $I = |\boldsymbol I|$. In order to avoid a confusion, we will always explicitly denote averaging with angle brackets and magnitudes of vectors with bars $|\text{ }|$.}. In order to relate them with sound pressure level or RMS value, averaging over time should be performed, which will be done in the next section.

Finally, we can also define \textbf{sound power} of a source. Instantaneous (in contrast to time-averaged or sustained) power output of a source is given by
\begin{equation}\label{eq:acoustic_power}
P_q(t) = \int_V \nabla \cdot \boldsymbol I (\boldsymbol x,t) d^3 \boldsymbol x = \oint_S \boldsymbol I(\boldsymbol x,t) \cdot \boldsymbol n(\boldsymbol x) d^2 \boldsymbol x = \oint_S p(\boldsymbol x, t) \boldsymbol v(\boldsymbol x, t) \cdot \boldsymbol n(\boldsymbol x) d^2 \boldsymbol x,
\end{equation}
where $S$ is a closed surface which encloses the volume $V$ containing the entire source placed in free space and $\boldsymbol n$ is unit vector normal to the surface pointing outwards. Here we used the divergence theorem to switch from the volume to the surface integral. Analogously to sound pressure level, sound power level is defined as
\begin{equation}\label{eq:sound_power_level}
\begin{aligned}
\boxed{ L_W = 10 \log_{10} \frac{\langle P_q \rangle_T}{10^{-12} \text{\,W}} }.
\end{aligned}
\end{equation}
The reference value $10^{-12}$\,W follows from the value $4 \cdot 10^{-10}$\,Pa$^2$ in eq.~(\ref{eq:SPL}) divided with $\rho_0 c_0 \approx 400$\,kg/(m$^2$s), which will be identified as the characteristic impedance of air in eq.~(\ref{eq:impedance_plane_wave}). Time-averaged sound power thus has a more practical relevance and will be calculated later. We should also emphasize that according to the given definition, sound power is an acoustic quantity related to a source (or any object radiating sound), not the sound field itself.

\section{Modal analysis}\label{ch:modal_analysis}

In the previous section we introduced the most important acoustic quantities, which are in general functions of time and location. We also derived the wave equation, the solution of which (for the given initial and boundary conditions, as well as sources of sound) represents the sound field. In many cases of interest, particularly when the sources generate a stationary sound and all other conditions are stable, propagation of sound waves in time is not relevant. Accordingly, we will first solve a time-independent version of the wave equation in this section. Further time-related phenomena will be treated in the next section.

Another reason for excluding time from the calculation is that all closed rooms or cavities\footnote{This holds also for partly open spaces, and, as will be discussed, in a more general sense even for open spaces. Adding to this, in room acoustics we do not explicitly treat interaction between the sound field inside the room and the exterior, which is the object of building acoustics. Any interface surface must be supplied with an appropriate boundary condition. Coupled rooms are in this sense observed as a single, partly divided space.} have more or less pronounced resonances at certain frequencies (the eigenfrequencies), which are of special concern, yet time-independent (depending only on the room geometry and other supposedly fixed properties, but not on the source or receiver). Moreover, linearity of the equations in question allows a very effective application of the Fourier analysis. The time-dependent fields can be decomposed into simple sine waves with different frequencies, allowing the time to be replaced by frequency as the independent variable. This is demonstrated next, which will serve as the basis for modal analysis.

\subsection{Complex sine waves}\label{ch:frequency_domain}

The goal is to solve the homogeneous (without the source term) wave equation~(\ref{eq:wave_eq}) (the time-independent effects of boundary conditions and sources will be added later). To this end we apply the usual method for solving partial differential equations -- the separation of variables. We suppose a solution in the form
\begin{equation}\label{wave_eq_solution_separation_of_variables}
	p(\boldsymbol x, t)=\hat{p}(\boldsymbol x) \widetilde{p}(t),
\end{equation}
where $\hat{p}(\boldsymbol x)$ depends only on location\footnote{More precisely, it is sufficient to assume that it changes over much larger time scale than $\widetilde{p}(t)$, so that the essential time dependence is inside the exponential term of $\widetilde{p}(t) \sim e^{j \omega t}$. See the comments after eq.~(\ref{eq:Fourier_time_derivative}) for more details. Note also that no complex quantities have been introduced yet and that the units of $\hat p$ and $\widetilde p$ are irrelevant as long as their product is in pascals.} and $\widetilde{p}(t)$ depends only on time. Then, if $p \neq 0$, the wave equation gives
\begin{equation}\label{eq:wave_eq_separated}
\frac{\hat{p}(\boldsymbol x)}{c_0^2} \frac{d^2 \widetilde{p}(t)}{d t^2} - \widetilde{p}(t) \nabla^2 \hat{p}(\boldsymbol x) = 0 \Rightarrow \frac{1}{c_0^2 \widetilde{p}(t)} \frac{d^2 \widetilde{p}(t)}{d t^2} = \frac{1}{\hat{p}(\boldsymbol x)} \nabla^2 \hat{p}(\boldsymbol x).
\end{equation}
The independent variables are separated on the two sides of the last equality -- time on the left-hand side and spatial coordinates on the right-hand side. The equality can thus be satisfied only if both sides are equal to some real constant $C$ (with the unit 1/m$^2$), since any temporal or spatial variation of one side cannot be followed by the opposite side of the equation.

The left-hand side gives
\begin{equation}\label{eq:wave_eq_time_separated}
\frac{d^2 \widetilde{p}(t)}{d t^2} = C c_0^2 \widetilde{p}(t),
\end{equation}
which has the solution
\begin{equation}\label{eq:wave_eq_solution_time}
\widetilde{p}(t) = A \cos(\omega t) + B \sin(\omega t),
\end{equation}
with $A$ and $B$ some real constants and, after replacing $\widetilde{p}(t)$ in eq.~(\ref{eq:wave_eq_time_separated}), $C = -(\omega/c_0)^2 = -k^2$. Here, $\omega$ is angular frequency, which is related to frequency $f$, period $T = 1/f$, and wavelength $\lambda = c_0 T$ by the equalities
\begin{equation}\label{eq:omega_f}
\omega = 2 \pi f = 2 \pi / T = 2 \pi c_0/ \lambda.
\end{equation}
We also introduced the \textbf{wave number},
\begin{equation}\label{eq:omega_k}
k = \omega/c_0 = 2 \pi f / c_0 = 2 \pi / (c_0 T) = 2 \pi / \lambda.
\end{equation}
All quantities which appear in eq.~(\ref{eq:omega_f}) and eq.~(\ref{eq:omega_k}) are strictly larger than zero and real. Inserting the value of $C$ in the location-dependent part of eq.~(\ref{eq:wave_eq_separated}) gives
\begin{equation}\label{eq:Helmholtz_eq_no_source}
	\nabla^2 \hat{p}(\boldsymbol x) + k^2 \hat{p}(\boldsymbol x) = 0.
\end{equation}
This has the same form as the homogeneous Helmholtz equation (which will be derived below in eq.~(\ref{eq:Helmholtz_source})), but for the real $\hat p(\boldsymbol x)$.

The two terms of the solution in eq. \ref{eq:wave_eq_solution_time} involve two real constants, $A$ and $B$, which in general depend on the source and initial conditions. They can be written more compactly if we let $p$ (and, consequently, all other first-order acoustic quantities\footnote{\label{ftn:complex_quantities}Since only the real part of the complex field is physical, the second-order quantities such as $E$, $\boldsymbol I$, and $P_q$ remain real-valued (see below), as well as all independent variables and constants: $t$, $\tau$, $\boldsymbol{x}$, $\boldsymbol{y}$, $\omega$, $f$, $T$, $\lambda$, $k$, $c_0$, $\rho_0$, $p_0$, etc., at least for now. The unit vector $\boldsymbol n$ and the constants $a$ and $b$ in the boundary condition in eq.~(\ref{eq:wave_eq_p_BC}) are also real, but the function $c$, like the source function $q$, can take complex values.} which characterize the sound field, sources, or boundary/initial conditions in the linearized theory) take complex values. First we can notice that the simple oscillatory functions $\sin(\omega t)$ and $\cos(\omega t)$ represent the same solution with the phase difference $\pi/2$ between them. Then we use Euler's formula,
\begin{equation}\label{eq:Euler}
e^{j \omega t} = \cos(\omega t) + j \sin(\omega t),
\end{equation}
to write a \textbf{complex sine wave}\footnote{\label{ftn:time_exponential_negative_frequencies} Some authors use $e^{-j \omega t} = \cos(\omega t) - j \sin(\omega t)$ for the time dependence, which also obeys eq.~(\ref{eq:wave_eq}), because the second-order time derivative cancels the factor -1 in the exponent. Such complex conjugate solutions could also be included with our notation, if we let $\omega$ take negative values (which will indeed reflect in the bounds of the Fourier integral in eq.~(\ref{eq:inv_Fourier_p})). However, this does not bring any new physical solutions, which are associated only with the real part of $p$ (as also discussed in footnote \footref{ftn:fft_negative_frequencies}) and $\mathcal{R}_e(e^{-j\omega t}) = \mathcal{R}_e(e^{j\omega t}) = \cos(\omega t)$, and we can still consider $\omega$ to be strictly positive. Still, the imaginary part is not equal and the adopted convention with regard to the complex exponent has to be clarified for accurate analytical treatment.}
\begin{equation}\label{eq:wave_eq_solution_time_complex}
\widetilde{p}(t) = D e^{j \omega t},
\end{equation}
where $D$ is now a complex number. Its real and imaginary part exactly capture the two degrees of freedom of the real constants $A$ and $B$ and eq.~(\ref{eq:wave_eq_solution_time_complex}) is merely a complex-valued reformulation of the solution in eq.~(\ref{eq:wave_eq_solution_time}). The imaginary part is added for mathematical efficiency, not the physics involved.

An important property of complex exponential functions is that they are \textbf{mutually orthogonal}\footnote{Since the exponential functions are complex, the integral contains product with a complex conjugate value in order to return a real value. It can be compared with the result of time averaging in eq.~(\ref{eq:intensity_complex_time_average}), the derivation of which also involves integration over time, or with eq.~(\ref{eq:modes_orthogonality}), which actually has the same meaning as eq.~(\ref{eq:complex_sine_wave_orthogonality}) only for room modes and thus involves integration over space.}:
\begin{equation}\label{eq:complex_sine_wave_orthogonality}
\int_{-\infty}^{\infty} e^{j \omega_1 t} (e^{j \omega_2 t})^* dt = \int_{-\infty}^{\infty} e^{j \omega_1 t} e^{-j \omega_2 t} dt = \int_{-\infty}^{\infty} e^{j (\omega_1-\omega_2) t} dt = 2 \pi \delta(\omega_1 - \omega_2),
\end{equation}
which is equal to zero whenever $\omega_1 \neq \omega_2$. Here we used the equality
\begin{equation}\label{eq:delta_function_integral_exponential}
\int_{-\infty}^{\infty}  e^{j \omega t} dt = 2 \pi \delta(\omega)
\end{equation}
with $\delta()$ denoting the Dirac delta function, which will be formally introduced in eq.~(\ref{eq:Dirac}) and its main properties will be discussed later, in the context of Green's functions. Here it is important to notice that the simple oscillatory functions with different frequencies are mutually orthogonal. This property will be shared by room modes. The functions $e^{j\omega t}$ can thus be seen as ``temporal'' modes with continuous frequency, since the time as considered here is (unlike space inside a room) unbounded.

Since the entire solution we are interested in is the product $\hat p(\boldsymbol x) \widetilde{p}(t)$ in eq.~(\ref{wave_eq_solution_separation_of_variables}), where $\hat p(\boldsymbol x)$ is now also complex in general, we can absorb the factor $D$ into the value of $\hat p(\boldsymbol x)$ and keep only the dimensionless exponential term as the time-dependent part of the solution. The solution of eq.~(\ref{eq:wave_eq}) is thus product of the time-independent (or very slowly varying) complex amplitude $\hat{p}(\boldsymbol x)$ and the factor $e^{j \omega t}$,
\begin{equation}\label{eq:complex_sine_wave}
\boxed{ p(\boldsymbol x, t) = \hat{p}(\boldsymbol x) e^{j \omega t} }.
\end{equation}
The other first-order quantities take the same form: $\boldsymbol{v}(\boldsymbol x, t) = \hat {\boldsymbol v}(\boldsymbol x) e^{j \omega t}$, $\rho(\boldsymbol x, t) = \hat \rho(\boldsymbol x) e^{j \omega t}$, and $q(\boldsymbol y, \tau) = \hat q(\boldsymbol y) e^{j \omega \tau}$, the last one being a complex source function. Magnitude of a complex vector is real, in this case $|\boldsymbol v| = \sqrt{\boldsymbol v^* \cdot \boldsymbol v}$, where $\boldsymbol v^* = \mathcal{R}_e(\boldsymbol v) - j \mathcal{I}_m(\boldsymbol v)$ is complex conjugate of $\boldsymbol v$. Note also that the complex amplitudes do carry an information on phase, so the listed acoustic quantities are not necessarily in phase.

The complex amplitude $\hat p(\boldsymbol x)$ (in pascals) will be calculated further below. In general, it depends on the sources and boundary conditions and by fixing the time dependence with a simple term $e^{j\omega t}$ we leave all the effects of boundary and initial conditions as well as the sources to be reflected in the complex amplitude. It should also be said that continuing the separation of spatial coordinates in eq.~(\ref{eq:Helmholtz_eq_no_source}) is feasible only when all location-dependent variables (including those in boundary conditions) can be decoupled in terms of the components of $\boldsymbol x$ analogously to $\hat p$ and $\tilde p$ in eq.~(\ref{eq:wave_eq_separated}). Unfortunately, in contrast to the unbounded time, this is not guaranteed for bounded spaces and it is usually possible only for simple geometries of rooms, such as a rectangular room, which will be treated in section~\ref{ch:rectangular_room_wave_theory}, and uniform boundaries.

We should also notice that starting with eq.~(\ref{eq:complex_sine_wave}) we use the same symbol $p$ for both complex sound pressure and actual physical sound pressure. This shall not cause a confusion, since it is understood that the latter can always be expressed as the real part of the former:
\begin{equation}\label{eq:real_complex_p}
\begin{aligned}
\mathcal{R}_e(p) &= \mathcal{R}_e(\hat{p}e^{j \omega t}) =  \mathcal{R}_e\left[ \left( \mathcal{R}_e(\hat{p}) + j \mathcal{I}_m(\hat{p}) \right) \left(\cos(\omega t) + j \sin(\omega t) \right) \right] &\\
&= \mathcal{R}_e\left[ \mathcal{R}_e(\hat{p}) \cos(\omega t) + j \mathcal{R}_e(\hat{p})\sin(\omega t) + j \mathcal{I}_m(\hat{p})\cos(\omega t) - \mathcal{I}_m(\hat{p})\sin(\omega t) \right] &\\
&= \mathcal{R}_e(\hat{p}) \cos(\omega t) - \mathcal{I}_m(\hat{p})\sin(\omega t).
\end{aligned}
\end{equation}
Hence, no particular notation is needed in order to distinguish between real and complex $p$, $\boldsymbol{v}$, $\rho$, or $q$. In fact, we (and many other authors) often refer to the complex sound pressure simply as sound pressure, complex amplitude as amplitude, complex sine wave as sine wave, and so on. Mathematical unambiguity is maintained as long as the equations are \textbf{linear} (which is the case with the equations with complex quantities which we have derived so far), since they never involve a product of two first-order quantities.

In contrast to this, the energy-related, \textbf{second-order} quantities require more care. For example, according to eq.~(\ref{eq:intensity}) the intensity vector equals
\begin{equation}\label{eq:intensity_complex}
\boldsymbol I = \mathcal{R}_e(p) \mathcal{R}_e(\boldsymbol v),
\end{equation}
which is in general not equal to the product $p \boldsymbol v$, if $p$ and $\boldsymbol v$ are complex. Fortunately, we are most often interested only in time-averaged second-order quantities (similarly as in the case of sound pressure level in eq.~(\ref{eq:SPL})), which simplifies the expressions. For example, sound intensity averaged over one period $T = 2 \pi/ \omega$ of a simple acoustic oscillation is\footnote{Assuming constant amplitudes during this time interval.}
\begin{align*}
\begin{aligned}
\langle\boldsymbol I\rangle_T &= \frac{1}{T} \int_{-T/2}^{T/2} \boldsymbol I dt = \frac{1}{T} \int_{-T/2}^{T/2} \mathcal{R}_e(p) \mathcal{R}_e(\boldsymbol v) dt &\\
&= \frac{1}{T} \int_{-T/2}^{T/2} \left[ \mathcal{R}_e(\hat{p}) \cos(\omega t) - \mathcal{I}_m(\hat{p})\sin(\omega t) \right]\left[ \mathcal{R}_e(\hat{\boldsymbol v}) \cos(\omega t) - \mathcal{I}_m(\hat{\boldsymbol v})\sin(\omega t) \right] dt &\\
&= \frac{1}{T} \int_{-T/2}^{T/2} \{ \mathcal{R}_e(\hat{p}) \mathcal{R}_e(\hat{\boldsymbol v}) \cos^2(\omega t) - \mathcal{I}_m(\hat{p}) \mathcal{R}_e(\hat{\boldsymbol v}) \sin(\omega t) \cos(\omega t) &\\
&-\mathcal{R}_e(\hat{p}) \mathcal{I}_m(\hat{\boldsymbol v}) \sin(\omega t) \cos(\omega t) + \mathcal{I}_m(\hat{p}) \mathcal{I}_m(\hat{\boldsymbol v})\sin^2(\omega t) \} dt &\\
&= \frac{1}{T} \mathcal{R}_e(\hat{p}) \mathcal{R}_e(\hat{\boldsymbol v}) \int_{-T/2}^{T/2} \cos^2(\omega t) dt + \frac{1}{T} \mathcal{I}_m(\hat{p}) \mathcal{I}_m(\hat{\boldsymbol v}) \int_{-T/2}^{T/2} \sin^2(\omega t) dt. &
\end{aligned}
\end{align*}
For the last equality we used the fact that $\sin(\omega t) \cos(\omega t)$ is an odd function of $t$, the integral of which over the symmetric interval $[-T/2,T/2]$ vanishes. Solving the integrals gives
\begin{align*}
\begin{aligned}
\langle\boldsymbol I\rangle_T &= \frac{1}{T} \mathcal{R}_e(\hat{p}) \mathcal{R}_e(\hat{\boldsymbol v}) \left( \frac{2 \omega T/2 + \sin(2 \omega T/2)}{4 \omega} - \frac{-2 \omega T/2 + \sin(-2 \omega T/2)}{4 \omega} \right) &\\
&+ \frac{1}{T} \mathcal{I}_m(\hat{p}) \mathcal{I}_m(\hat{\boldsymbol v}) \left( \frac{2 \omega T/2 - \sin(2 \omega T/2)}{4 \omega} - \frac{-2 \omega T/2 - \sin(-2 \omega T/2)}{4 \omega} \right) &\\
&= \frac{1}{T} \mathcal{R}_e(\hat{p}) \mathcal{R}_e(\hat{\boldsymbol v}) \frac{T}{2} + \frac{1}{T} \mathcal{I}_m(\hat{p}) \mathcal{I}_m(\hat{\boldsymbol v}) \frac{T}{2} = \frac{1}{2} \left[ \mathcal{R}_e(\hat{p}) \mathcal{R}_e(\hat{\boldsymbol v}) + \mathcal{I}_m(\hat{p}) \mathcal{I}_m(\hat{\boldsymbol v}) \right],&
\end{aligned}
\end{align*}
which can be written shorter as
\begin{equation}\label{eq:intensity_complex_time_average}
\boxed{ \langle\boldsymbol I\rangle_T = \frac{1}{2} \mathcal{R}_e(\hat{p}^* \hat{\boldsymbol v})= \frac{1}{4} (\hat{p}^* \hat{\boldsymbol v} + \hat{p} \hat{\boldsymbol v}^*) }.
\end{equation}

Sustained acoustic power of a source can be expressed from eq.~(\ref{eq:acoustic_power}):
\begin{equation}\label{eq:acoustic_power_complex_time_average}
\begin{aligned}
\langle P_q \rangle_T &= \frac{1}{T} \int_{-T/2}^{T/2} \oint_S \boldsymbol I(\boldsymbol x,t) \cdot \boldsymbol n(\boldsymbol x) d^2 \boldsymbol x dt = \oint_S \langle \boldsymbol I(\boldsymbol x,t) \rangle_T \cdot \boldsymbol n(\boldsymbol x) d^2 \boldsymbol x &\\
&= \frac{1}{2} \oint_S \mathcal{R}_e(\hat{p}^* \hat{\boldsymbol v}) \cdot \boldsymbol n(\boldsymbol x) d^2 \boldsymbol x = \frac{1}{4} \oint_S (\hat{p}^* \hat{\boldsymbol v} + \hat{p} \hat{\boldsymbol v}^*) \cdot \boldsymbol n(\boldsymbol x) d^2 \boldsymbol x,&
\end{aligned}
\end{equation}
for a fixed closed surface $S$. Sound energy follows from eq.~(\ref{eq:energy}):
\begin{equation}\label{eq:energy_complex}
E = \frac{\mathcal{R}_e(p)^2}{2 \rho_0 c_0^2} + \frac{\rho_0 |\mathcal{R}_e(\boldsymbol v)|^2}{2}
\end{equation}
and after averaging over one period:
\begin{align*}
\begin{aligned}
\langle E \rangle_T &= \frac{1}{2 \rho_0 c_0^2} \frac{1}{T} \int_{-T/2}^{T/2} \left[ \mathcal{R}_e(\hat{p}) \cos(\omega t) - \mathcal{I}_m(\hat{p}) \sin(\omega t) \right]^2 dt &\\
&+ \frac{\rho_0}{2} \frac{1}{T} \int_{-T/2}^{T/2} \left[ \mathcal{R}_e(\hat{\boldsymbol v}) \cos(\omega t) - \mathcal{I}_m(\hat{\boldsymbol v}) \sin(\omega t) \right]^2 dt &\\
&= \frac{\mathcal{R}_e(\hat{p})^2}{2 \rho_0 c_0^2 T} \int_{-T/2}^{T/2} \cos^2(\omega t) dt - \frac{2 \mathcal{R}_e(\hat{p}) \mathcal{I}_m(\hat{p})}{2 \rho_0 c_0^2 T} \int_{-T/2}^{T/2} \sin(\omega t) \cos(\omega t) dt &\\
&+ \frac{\mathcal{I}_m(\hat{p})^2}{2 \rho_0 c_0^2 T} \int_{-T/2}^{T/2} \sin^2(\omega t) dt + \frac{\rho_0 \mathcal{R}_e(\hat{\boldsymbol v})^2}{2 T} \int_{-T/2}^{T/2} \cos^2(\omega t) dt &\\
&- \frac{2 \rho_0  \mathcal{R}_e(\hat{\boldsymbol v}) \mathcal{I}_m(\hat{\boldsymbol v})}{2 T} \int_{-T/2}^{T/2} \sin(\omega t) \cos(\omega t) dt + \frac{\rho_0 \mathcal{I}_m(\hat{\boldsymbol v})^2}{2 T} \int_{-T/2}^{T/2} \sin^2(\omega t) dt &\\
&= \frac{\mathcal{R}_e(\hat{p})^2}{2 \rho_0 c_0^2 T} \frac{T}{2} + \frac{\mathcal{I}_m(\hat{p})^2}{2 \rho_0 c_0^2 T} \frac{T}{2} + \frac{\rho_0 \mathcal{R}_e(\hat{\boldsymbol v})^2}{2 T} \frac{T}{2} + \frac{\rho_0 \mathcal{I}_m(\hat{\boldsymbol v})^2}{2 T} \frac{T}{2} & \\
& = \frac{\mathcal{R}_e(\hat{p})^2 + \mathcal{I}_m(\hat{p})^2}{4 \rho_0 c_0^2} + \frac{\rho_0 \left( \mathcal{R}_e(\hat{\boldsymbol v})^2 + \mathcal{I}_m(\hat{\boldsymbol v})^2 \right)}{4}.&
\end{aligned}
\end{align*}
This can be written more compactly as
\begin{equation}\label{eq:energy_complex_time_average}
\boxed{ \langle E \rangle_T = \frac{|\hat{p}|^2}{4 \rho_0 c_0^2} + \frac{\rho_0 |\hat{\boldsymbol v}|^2}{4} = \frac{\hat{p} \hat{p}^*}{4 \rho_0 c_0^2} + \frac{\rho_0 \hat{\boldsymbol v} \cdot \hat{\boldsymbol v}^*}{4} }.
\end{equation}
Finally, sound pressure level from eq.~(\ref{eq:SPL}) becomes
\begin{equation}\label{eq:SPL_complex_real_pressure}
\begin{split}
L = 10 \log_{10}\left( \frac{1}{T} \int_{-T/2}^{T/2} \frac{\mathcal{R}_e(p)^2}{4 \cdot 10^{-10} \text{\,Pa}^2} dt \right),
\end{split}
\end{equation}
which gives
\begin{equation}\label{eq:SPL_complex}
\begin{split}
\boxed{ L = 10 \log_{10}\left( \frac{|\hat{p}|^2 / 2}{4 \cdot 10^{-10} \text{\,Pa}^2} \right) = 20 \log_{10}\left( \frac{|\hat{p}| / \sqrt{2}}{2 \cdot 10^{-5} \text{\,Pa}} \right) }.
\end{split}
\end{equation}
Indeed, the root mean square value from eq.~(\ref{eq:SPL}) equals $|\hat{p}|/\sqrt{2}$ for simple oscillations with amplitude $|\hat{p}|$.

The solution in eq.~(\ref{eq:complex_sine_wave}) is given for any fixed angular frequency $\omega$. More generally, any continuous complex function $p(\boldsymbol x, t)$ (or some other acoustic variable replacing $p$) can be written as a superposition of (in general infinite number of) complex sine functions. This holds because the functions $e^{j\omega t}$ with all possible values of $\omega$ form a \textbf{complete set} of mutually orthogonal (eigen)functions, the weighted sum of which can match any given function of time. Mathematically this follows from Fourier's analysis and it is expressed by the integral
\begin{equation}\label{eq:inv_Fourier_p}
p(\boldsymbol x, t) = \int_{-\infty}^{\infty} \hat{p}(\boldsymbol x, \omega) e^{j \omega t} d\omega = \mathcal{F}_\omega^{-1} \{ \hat{p} (\boldsymbol x, \omega) \},
\end{equation}
where $\hat{p}(\boldsymbol x, \omega) d\omega$ takes the role of $\hat{p}(\boldsymbol x)$ from eq.~(\ref{eq:complex_sine_wave})\footnote{Strictly speaking, the unit of $\hat{p}(\boldsymbol x, \omega)$ is Pa$\cdot$s and $\hat{p}(\boldsymbol x, \omega)$ is density of $\hat{p}(\boldsymbol x)$ over continuous $\omega$. However, this difference does not have far-reaching implications and the use of the same symbol ($\hat p$) is usual in literature. Furthermore, the integral becomes a simple sum of the sinusoidal components from eq.~(\ref{eq:complex_sine_wave}) when the frequency is discrete ($d\omega$ is finite).} and acts as the weighting factor. In other words, the acoustic quantity $p(\boldsymbol x,t)$ is entirely represented by its spectrum $\hat p(\boldsymbol x,\omega)$. The complex amplitude $\hat{p}(\boldsymbol x, \omega)$ can thus vary with frequency but does not depend on time. It can also incorporate an arbitrary phase $\Phi(\omega)$: $p = \hat{p} e^{j (\omega t + \Phi(\omega))} = (\hat{p} e^{j \Phi(\omega)}) e^{j \omega t}$, where $\Phi(\omega)$ can be specified, for example, by the initial conditions or source function.

The integral in eq.~(\ref{eq:inv_Fourier_p}) is known as Fourier integral. In fact, eq.~(\ref{eq:inv_Fourier_p}) represents the inverse Fourier transform of $\hat{p}$ with $\omega$ as the parameter (hence the notation $\mathcal{F}_\omega^{-1} \{ \hat{p} \}$). The associated \textbf{Fourier transform} of $p$ is
\begin{equation}\label{eq:Fourier_p}
\hat{p}(\boldsymbol x, \omega) = \mathcal{F}_t \{ p (\boldsymbol x, t) \} = \frac{1}{2 \pi} \int_{-\infty}^{\infty} p(\boldsymbol x, t) e^{-j \omega t} dt,
\end{equation}
with time $t$ as the parameter of the transform\footnote{\label{ftn:fft_negative_frequencies}The Fourier transform introduces negative angular frequencies to $\hat p(\boldsymbol x, \omega)$. However, since we are interested in real $p(\boldsymbol x, t)$, then $\hat{p}(\boldsymbol x, \omega) = \mathcal{F}_t \{ p (\boldsymbol x, t) \} = \hat{p}^*(\boldsymbol x, -\omega)$ and no new information on the sound field is brought by these frequencies.}. Notice the appearance of $e^{-j\omega t}$, the complex conjugate of $e^{j\omega t}$.
Additional factor $1/(2 \pi)$ is necessary in order to satisfy the equality
\begin{align*}
	\begin{aligned}
	p(\boldsymbol x, t) &= \int_{-\infty}^{\infty} \hat{p}(\boldsymbol x, \omega) e^{j \omega t} d\omega = 
	\int_{-\infty}^{\infty} \frac{1}{2 \pi} \left( \int_{-\infty}^{\infty} p(\boldsymbol x, t') e^{-j \omega t'} dt' \right) e^{j \omega t} d\omega &\\ &= \int_{-\infty}^{\infty} \frac{1}{2 \pi} p(\boldsymbol x, t') \left( \int_{-\infty}^{\infty}  e^{j \omega (t-t')} d\omega \right) dt' = 
	\int_{-\infty}^{\infty} \frac{1}{2 \pi} p(\boldsymbol x, t') 2 \pi \delta(t-t') dt' &\\
    &= \int_{-\infty}^{\infty} p(\boldsymbol x, t') \delta(t'-t) dt' = p(\boldsymbol x, t).&
	\end{aligned}
\end{align*}
The independent time variable $t'$ is introduced here only to distinguish it from $t$. We also used eq.~(\ref{eq:delta_function_integral_exponential}) with $t$ and $\omega$ switched,
\begin{equation}\label{eq:delta_function_integral_exponential_Fourier_transform}
\int_{-\infty}^{\infty}  e^{j \omega t} d\omega = 2 \pi \delta(t),
\end{equation}
and the sampling property of the delta function (see eq.~(\ref{eq:Dirac_sampling})) for the last equality.

\subsection{Acoustics in frequency domain}

Since all equations involving first-order quantities are linear  (Fourier transform is also a linear operation), we can, without any loss of generality, observe a single complex sine wave from eq.~(\ref{eq:complex_sine_wave}), with a generic angular frequency $\omega$, as a solution of the wave equation. This is possible because any given angular frequency $\omega$ does not change in linear systems:
\begin{equation}\label{eq:linearity}
A\hat{p}_1(\boldsymbol x) e^{j \omega t} + B\hat{p}_2(\boldsymbol x) e^{j \omega t} = (A\hat{p}_1(\boldsymbol x)+B\hat{p}_2(\boldsymbol x)) e^{j \omega t},
\end{equation}
where $A$ and $B$ are some constant complex factors. Spectral components at different frequencies do not affect each other or any other component and do not generate new spectral components. Dependence of the complex amplitude and other quantities on frequency and, when necessary, integration over all frequencies of interest according to eq.~(\ref{eq:inv_Fourier_p}) are understood and left implicit.

Conveniently, $p(\boldsymbol x, t)$ and $\hat{p}(\boldsymbol x)$ have the same physical unit -- pascal. The symbol $\hat{}$ thus only implies that the variable is observed in frequency (rather than time) domain. For the specific frequency it depends only on the location $\boldsymbol x$. Equation~(\ref{eq:complex_sine_wave}) clearly separates time and location dependence and shows how $p$ (the real part of which is physical sound pressure) is straightforwardly obtained after multiplying $\hat p$ with $e^{j\omega t}$. If the amplitudes are allowed to (slowly) vary in time, it should be clearly stated.

\textbf{Time-independent} equations are obtained simply by dividing the corresponding time-dependant equations with the term $e^{j\omega t}$, which is common to all unsteady quantities and not affected by linear operations. For example, boundary condition for sound pressure amplitude $\hat{p}$ follows from eq.~(\ref{eq:wave_eq_p_BC}) (with complex $p$ and $c$) divided with $e^{j \omega \tau}$:
\begin{equation}\label{eq:Helmholtz_p_BC}
\boxed{ a \hat{p}(\boldsymbol y) + b \nabla_y \hat{p}(\boldsymbol y) \cdot \boldsymbol n(\boldsymbol y) = \hat{c}(\boldsymbol y) },
\end{equation}
where $\boldsymbol y$ is location on the surface and $\boldsymbol n(\boldsymbol y)$ is real, as before. Obviously, $a$, $b$, and $c$, as well as $\hat{p}$ can depend on frequency.

An important property of complex sine functions such as $p = \hat p e^{j\omega t}$, which further simplifies the mathematical treatment, is that their time derivative is very simple:
\begin{equation}\label{eq:Fourier_time_derivative}
\frac{\partial p}{\partial t} = \left( \frac{\partial \hat{p}}{\partial t} \right) e^{j \omega t} + \hat{p} \left( \frac{\partial}{\partial t}e^{j \omega t} \right) = \hat{p} \left( \frac{\partial}{\partial t}e^{j \omega t} \right) = j \omega \hat{p} e^{j \omega t} = j \omega p.
\end{equation}
Replacing time derivative with a product with $j\omega$ is one of the main reasons for introducing complex quantities at the first place. However, we have to suppose that $|e^{j \omega t} \partial \hat{p} / \partial t| \ll |\hat{p} \partial e^{j \omega t} / \partial t| = |j \omega \hat{p} e^{j \omega t}|$ for the second equality to hold. This is justified when any characteristic time scale $|2 \pi \hat{p} / ( \partial \hat{p} / \partial t)|$ over which $\hat{p}$ might vary appreciably (if we indeed allow $\hat p$ from eq.~(\ref{wave_eq_solution_separation_of_variables}) to vary with time) is much larger than the period of oscillations $T$, which is the time scale of $e^{j\omega t}$ ($|2 \pi e^{j\omega t} / ( \partial e^{j\omega t} / \partial t)| = |2\pi/(j\omega)| = 1/f = T$).

We can use the derivative from eq. (\ref{eq:Fourier_time_derivative}) to completely remove the time dependence from the wave equation. The second-order time derivative equals
\begin{equation}\label{eq:Fourier_double_time_derivative}
\frac{\partial^2 p}{\partial t^2} = j \omega \frac{\partial p}{\partial t} = (j \omega)^2 p = - \omega^2 p.
\end{equation}
Inserting this in the wave equation (\ref{eq:wave_eq_source}) gives
\begin{equation}\label{eq:Helmholtz_withExp_source}
k^2 p(\boldsymbol x, t) + \nabla_x^2 p(\boldsymbol x, t) = -q(\boldsymbol y, \tau),
\end{equation}
where $k$ is defined in eq.~(\ref{eq:omega_k}). Writing the source term as $q(\boldsymbol y, \tau)=\hat{q}(\boldsymbol y) e^{j \omega \tau}$ and dividing both sides with the factor $e^{j \omega \tau}$, which is not a function of $\boldsymbol x$, we obtain \textbf{the Helmholtz equation} with the source term,
\begin{equation}\label{eq:Helmholtz_source}
\boxed{ k^2 \hat{p}(\boldsymbol x) + \nabla_x^2 \hat{p}(\boldsymbol x) = -\hat{q}(\boldsymbol y) }.
\end{equation}
Notice that the complex amplitude $\hat p(\boldsymbol x)$ absorbed also the phase shift due to the wave propagation from the source to the receiver, $e^{j \omega (t - \tau)}$, which does not vary with time because we assume that the source, receiver, and medium are motionless. According to eq.~(\ref{eq:source_mass_injection}), the complex source function of a compact monopole can be modelled as
\begin{equation}\label{eq:source_mass_injection_Helmholtz}
\hat{q}(\boldsymbol y) = -\rho_0 \omega^2 \hat{\beta}(\boldsymbol y).
\end{equation}
Similarly as with equations~(\ref{eq:wave_eq_source}) and (\ref{eq:wave_eq_source_y}), eq.~(\ref{eq:Helmholtz_source}) can be formally written in the source region:
\begin{equation}\label{eq:Helmholtz_source_y}
k^2 \hat{p}(\boldsymbol y) + \nabla_y^2 \hat{p}(\boldsymbol y) = -\hat{q}(\boldsymbol y).
\end{equation}

The Helmholtz equation is thus time-independent wave equation. Together with initial conditions and boundary conditions (eq.~(\ref{eq:Helmholtz_p_BC})), which all contribute to the value of the complex amplitude, it defines the acoustic problem in frequency domain. For completeness, we can also write the conservation of mass and momentum in frequency domain, following directly from equations~(\ref{eq:Euler_mass}) and (\ref{eq:Euler_momentum}):
\begin{equation}\label{eq:Euler_mass_sine_wave}
\frac{\partial \rho}{\partial t} + \rho_0 \nabla_x \cdot {\boldsymbol v} = j \omega \rho + \rho_0 \nabla_x \cdot {\boldsymbol v} = 0
\end{equation}
and
\begin{equation}\label{eq:Euler_momentum_sine_wave}
\rho_0 \frac{\partial {\boldsymbol v}}{\partial t} + \nabla_x p = j \omega \rho_0 \boldsymbol v + \nabla_x p = 0.
\end{equation}

\subsection{Damping and reverberation time}\label{ch:damping_and_reverberation_time}

Before continuing with the solution of the Helmholtz equation it is worthwhile to give a few general remarks on damping and very closely related reverberation time. In order to obtain the simple expression for time derivative in eq.~(\ref{eq:Fourier_time_derivative}), we had to suppose that the sound amplitude varies slowly over time (if at all). If we neglect usual decay of the direct sound with increasing distance from the source, as well as sources of non-stationary sound and similar active elements, amplitude of a sound wave in a room decreases over time due to damping. The acoustic energy losses can be due to absorption at surfaces, transmission of sound outside the room, or dissipation in the medium (air).

If we suppose that $\hat{p}$ decays \textbf{exponentially} in time\footnote{A reasonable assumption in many cases of practical importance, as it will turn out later (see, for example, section~\ref{statistical_theory_energy_decay}). Nevertheless, in this way we obviously model different physical mechanisms of sound attenuation, because we do not derive eq.~(\ref{eq:p_amplitude_exp_decay}) from any governing equation.} with the value $\hat{p}_{t=0}$ at $t=0$, that is
\begin{equation}\label{eq:p_amplitude_exp_decay}
\hat{p}(\boldsymbol x) = \hat{p}_{t=0}(\boldsymbol x) e^{-\zeta t},
\end{equation}
with $\zeta$ real and non-negative \textbf{damping constant}\footnote{Which can, of course, like $\hat p(\boldsymbol x)$ vary with frequency.} with the unit 1/s, then the condition from above becomes $|e^{j \omega t} \partial \hat{p} / \partial t| = |e^{j \omega t} (- \zeta) \hat{p}_{max}e^{-\zeta t}| \ll |j \omega \hat{p}_{max}e^{-\zeta t} e^{j \omega t}|$. In other words,
\begin{equation}\label{eq:damping_omega}
\zeta \ll \omega.
\end{equation}
We call this the condition for \textbf{weak damping} and a room satisfying it a weakly damped room. It says that the sound amplitude at each frequency of interest does not decay dramatically due to damping during one period (or along one wavelength) of the sine wave (because the attenuation in dB for one period is $20 \log_{10} e^{\zeta T} \approx 54.6 \zeta/\omega \text{\,dB} \ll 54.6$\,dB). The condition is practically always satisfied for dissipation in air and low to moderate damping in absorbing materials, as will also be shown shortly. Importantly, this justifies ignoring time-dependence of the complex amplitude in eq.~(\ref{eq:complex_sine_wave}).

Largely for historical reasons which date back to the works of W. C. Sabine (the founder of room acoustics), the quantity which is most frequently used for expressing damping in rooms is not the damping constant $\zeta$, but the \textbf{reverberation time}\footnote{The number 60 in the index indicates the sound pressure level drop of 60\,dB in the definition of reverberation time. It is usually omitted and only the symbol $T$ is used. We will keep it, though, to distinguish between the value of $T_{60}$ ``per definition'' and the usually measured $T_{30}$, $T_{20}$, etc., which will be introduced in section~\ref{ch:descriptors_of_room_acoustics}.} $T_{60}$. It is defined as the time interval during which sound pressure level in a room drops 60\,dB after its source of stationary sound is switched off. As such, reverberation time can be easily related to the damping constant. If $\hat{p}$ is given in eq.~(\ref{eq:p_amplitude_exp_decay}), then at any $\boldsymbol x$
\begin{align*}
\begin{aligned}
10 \log_{10} \left(\frac{|\hat p(t)|^2}{|\hat p(t + T_{60})|^2} \right) = 20 \log_{10} \left(\frac{|\hat p(t)|}{|\hat p(t) e^{-\zeta T_{60}}|} \right) = 20 \log_{10} \left(\frac{1}{e^{-\zeta T_{60}}} \right) = 60\text{\,dB}.
\end{aligned}
\end{align*}
The (frequency-dependent) reverberation time is inversely proportional to the damping constant,
\begin{equation}\label{eq:T60_damping}
\boxed{ T_{60} = \frac{1}{\zeta} \ln \left( 10^{60/20} \right) \approx \frac{6.91}{\zeta} }.
\end{equation}

Reverberation time in ordinary rooms is between around 0.3\,s and several seconds. Hence, $\zeta < 23$\,s$^{-1}$ for most of the rooms. Since we are interested in audible frequencies, $\omega > 2 \pi \cdot 20\text{\,Hz} \approx 125.7$\,rad/s and the assumption of low damping ($\zeta \ll \omega$) is justified practically for the entire audible frequency range. Moreover, the shortest reverberation times, around 0.3\,s and below, occur usually only at relatively high frequencies, well above 20\,Hz.

Here we introduced the reverberation time as a function of continuous frequency. In practice, its values averaged over some finite frequency ranges (bands) are usually considered (estimated and measured), because they rarely vary substantially between close frequencies. A possible exception worth mentioning are coupled rooms (rooms connected by relatively small apertures between them, which thus act as interfaces). If only weakly damped, the eigenmodes\footnote{The room eigenmodes will be discussed in section~\ref{ch:room_eigenmodes}.} of the coupled rooms with similar frequencies can cause larger narrow-band fluctuations of energy decay, which are known as beats. Such occurrences can make the estimation of reverberation time difficult and the frequency-averaged values less informative.

For future analytical treatment, it is of interest to note that according to equations~(\ref{eq:complex_sine_wave}) and (\ref{eq:p_amplitude_exp_decay})
\begin{equation}\label{eq:p_exp_decay}
p(\boldsymbol x, t)=\hat{p}(\boldsymbol x) e^{j \omega t}=\hat{p}_{t=0}(\boldsymbol x) e^{-\zeta t} e^{j \omega t} = \hat{p}_{t=0}(\boldsymbol x) e^{j (\omega + j \zeta) t}.
\end{equation}
Hence, the imaginary part of phase, $\Phi = j \zeta t$, accounts for the exponential decay of the sound amplitude in time. This is very often utilized for modelling damping by promoting $\omega$ to a \textbf{complex angular frequency},
\begin{equation}\label{eq:complex_omega}
\omega \rightarrow \omega + j \zeta,
\end{equation}
while keeping the wave number $k$ and other associated quantities real (also the time and other independent variables, the imaginary parts of which would also interfere with the imaginary part of angular frequency; recall the footnote~\footref{ftn:complex_quantities}). Alternatively, the same can be achieved for a forward propagating wave with a complex wave number
\begin{equation}\label{eq:complex_wave_number}
k \rightarrow k - j \zeta / c_0,
\end{equation}
and real angular frequency. The sign of the imaginary part is changed because the phase of a forward propagating wave (say, in the direction of $x$-axis) has the form\footnote{See for example the sound field from a point source in eq.~(\ref{eq:solution_tailored_Green_wave_eq_free_space_compact_source_emission_time})} $\omega t - kx = (\omega/c_0 - k) x$. Accordingly, $k \rightarrow k + j\zeta/c_0$ would be valid for a wave travelling in the opposite direction, which makes this approach less suitable when both propagation directions are possible, for example, due to reflected waves.

The two ways of introducing damping with imaginary parts of angular frequency or wave number may seem to have different physical interpretations -- in the first case the amplitude decay occurs in time and in the second case in space. However, this is only a formal difference, because the time and distance are coupled via the constant $c_0$ in the phase $(\omega/c_0 \pm k) x$ of the wave during its propagation through a medium or absorbing material. On the other hand, even energy losses at absorbing (or transmitting) surfaces can be modelled this way, if we are interested only in the cumulative effect of damping over a longer time interval. The actual abrupt drop of amplitude after a reflection takes place at a surface appears then to be distributed in space and time, but the overall decay of sound energy in the room is the same (as long as the decay is exponential). For this reason, complex angular frequency will be used in the modal analysis below for expressing damping of modes, even though it usually occurs mainly at surfaces. We will be interested there in the time-independent solution and overall damping\footnote{Alternatively, damping at surfaces can be introduced by allowing the constants $a$ and $b$ in eq.~(\ref{eq:Helmholtz_p_BC}) to take complex values, but, when appropriate, the approach with complex angular frequency is much easier.}.

Sound attenuation due to \textbf{dissipation} in air is approximately exponential and it is commonly expressed in terms of the \textbf{attenuation constant} $m_{air}$. It is the factor by which the sound energy decreases per unit length and therefore has the unit 1/m. From eq.~(\ref{eq:p_amplitude_exp_decay}) it follows:
\begin{equation}\label{eq:p_energy_exp_decay}
\hat{p}^2 \propto e^{-2 \zeta t} = e^{-2 \zeta x/c_0}
\end{equation}
and therefore
\begin{equation}\label{eq:attenuation_constant}
m_{air} = \frac{2\zeta_{air}}{c_0}.
\end{equation}
Attenuation in dB/m is then $10 \log_{10}e^{m_{air}} = m_{air} \cdot 10 \log_{10}e \approx 4.34 m_{air}$. As always, exponential decays of amplitude and energy imply linear decay of sound pressure level. Table~\ref{tab:air_atttenuation_constant} shows values of the attenuation constant in air in 1/m and the attenuation in dB/m for the room temperature, atmospheric pressure, and relative humidity 50\%. We can see that the attenuation is appreciable only for frequencies well above 1\,kHz and long propagation paths, around 100\,m or longer. In room acoustics, this can be relevant only in very large and weakly damped spaces, such as churches or large halls. In most of the rooms, especially at relatively low frequencies or in the presence of other causes of damping (absorbing materials, transmitting walls or openings), the contribution of dissipation in air is negligible.

\begin{table}[h]
	\caption{Attenuation constant in air, $m_{air}$, and attenuation in dB/m at the room temperature, normal atmospheric pressure, and relative humidity 50\%.}
	\label{tab:air_atttenuation_constant}
	\begin{tabular}{ | p{2.8cm} | p{2cm} | p{2cm}| p{2cm}| p{2cm}| p{2cm}|}
		\hline
		& \textbf{500\,Hz} & \textbf{1000\,Hz} & \textbf{2000\,Hz} & \textbf{4000\,Hz} & \textbf{8000\,Hz}\\
		\hline
		$m_{air}$\,[1/m] & 0.6 $\cdot 10^{-3}$ & 1.1 $\cdot 10^{-3}$ & 2.3 $\cdot 10^{-3}$ & 6.8 $\cdot 10^{-3}$ & 24.3 $\cdot 10^{-3}$ \\
		\hline
		$4.34 m_{air}$\,[dB/m] & 0.003 & 0.005 & 0.010 & 0.030 & 0.105 \\
		\hline
	\end{tabular}
\end{table}

\subsection{Green's function in frequency domain}\label{ch:Greens's_function}

In the previous sections we reduced the acoustic problem to solving the time-independent Helmholtz equation~(\ref{eq:Helmholtz_source}) supplied with boundary conditions\footnote{We neglect the initial conditions and allow the complex sine waves to have an infinite duration. Their phase is determined by the location and phase of the source function. On the other hand, as we are about to see, the boundary conditions affect critically the solution of the equation -- either in a tailored Green's function in eq.~(\ref{eq:solution_tailored_Green_Helmholtz}) or as the surface integral in eq.~(\ref{eq:solution_Green_Helmholtz}).}. This can be done by introducing a \textbf{Green's function}, $\hat{G}(\boldsymbol{x}|\boldsymbol{y})$, which satisfies an equation with the same operator on the left-hand side,
\begin{equation}\label{eq:Helmholtz_Green}
k^2 \hat{G}(\boldsymbol{x}|\boldsymbol{y}) + \nabla_x^2 \hat{G}(\boldsymbol{x}|\boldsymbol{y}) = -\hat{\delta}(\boldsymbol x - \boldsymbol y).
\end{equation}
We will see shortly that the Green's function captures entirely sound propagation from $\boldsymbol y$ to $\boldsymbol x$, which is indicated by its argument $\boldsymbol x|\boldsymbol y$. The source term is replaced by the delta function\footnote{Physical description of room acoustics alone should not depend on a particular source of sound in the room. Therefore, replacement of the source function with delta function, as an ideal point source, makes sense. The hat symbol above the delta function is actually unnecessary. It only indicates that we are working in frequency domain and time is not its argument.}, a \textbf{generalized function} which satisfies (we remove the hat symbol, since we are listing properties for some generic argument $x$):
\begin{equation}\label{eq:Dirac}
\int_{-\infty}^{\infty} \delta(x) dx = \int_{-\epsilon}^{\epsilon} \delta(x) dx = 1,
\end{equation}
where $\epsilon \rightarrow 0^+$ is a positive infinitesimal, while its value is zero for every $x \neq 0$. Therefore, delta function is properly defined with finite values only in the integral sense\footnote{It is often loosely defined as a function which equals zero for every $x \neq 0$ and infinity for $x=0$ (which is necessary in order to satisfy eq.~(\ref{eq:Dirac})). However, it becomes physically meaningful only after integration over its argument, which is the reason why it is sometimes referred to as a distribution rather than a function. In fact, a pointwise definition of such generalized functions is not necessary.}. Since the integral gives a dimensionless value, the product $\delta(x)dx$ is also dimensionless and the unit of $\hat{\delta}(\boldsymbol x - \boldsymbol y) = \hat{\delta}(x_1-y_1) \hat{\delta}(x_2-y_2) \hat{\delta}(x_3-y_3)$ is 1/m$^3$ in a three-dimensional space. According to eq.~(\ref{eq:Helmholtz_Green}), the unit of $\hat{G}(\boldsymbol{x}|\boldsymbol{y})$ is equal to the unit of $\hat{\delta}(\boldsymbol x - \boldsymbol y)$ multiplied with m$^2$, which is 1/m.

The most important properties of delta function are \textbf{symmetry},
\begin{equation}\label{eq:Dirac_symmetry}
\delta(x-y) = \delta(y-x),
\end{equation}
and \textbf{sampling} (selectivity),
\begin{equation}\label{eq:Dirac_sampling}
\int_{-\infty}^{\infty} f(x) \delta(x-y) dx = f(y).
\end{equation}
Thus, the integral with the delta function over all values of $x$ extracts the value (sample) of a continuous function $f(x)$ only at $x$ for which the argument of $\delta$ is zero, that is, at $x=y$. An important property of Green's function, which we will not prove, is \textbf{reciprocity}\footnote{Hence, the order of $\boldsymbol x$ and $\boldsymbol y$ in the argument of $\hat G$ is important if the Green's function is complex. Nevertheless, similarly as with complex amplitude $\hat p(\boldsymbol x)$, we often omit the arguments altogether for brevity, when it does not lead to a confusion.}, $\hat{G}(\boldsymbol{x}|\boldsymbol{y}) = \hat{G}^*(\boldsymbol{y}|\boldsymbol{x})$. Since delta function is a generalized function, so is the Green's function.

Due to the symmetry of delta function and the reciprocity of Green's function, we can switch $\boldsymbol x$ and $\boldsymbol y$ in eq.~(\ref{eq:Helmholtz_Green}) and obtain from its complex conjugate
\begin{equation}\label{eq:Helmholtz_Green_y}
k^2 \hat{G} + \nabla_y^2 \hat{G} = -\hat{\delta}(\boldsymbol x - \boldsymbol y).
\end{equation}
The only change is in the variable of the Laplacian. If we now multiply eq.~(\ref{eq:Helmholtz_source_y}) with $\hat{G}(\boldsymbol{x}|\boldsymbol{y})$ and subtract eq.~(\ref{eq:Helmholtz_Green_y}) multiplied with $\hat{p}(\boldsymbol y)$ from it, we obtain
\begin{align*}
\begin{aligned}
\hat{G} k^2 \hat{p}(\boldsymbol y) &+ \hat{G} \nabla_y^2 \hat{p}(\boldsymbol y) + \hat{G} \hat{q}(\boldsymbol y) -\hat{p}(\boldsymbol y) k^2 \hat{G} -\hat{p}(\boldsymbol y) \nabla_y^2 \hat{G} -\hat{p}(\boldsymbol y) \hat{\delta}(\boldsymbol x - \boldsymbol y) &\\
&= \hat{G} \nabla_y^2 \hat{p}(\boldsymbol y) + \hat{G} \hat{q}(\boldsymbol y)
-\hat{p}(\boldsymbol y) \nabla_y^2 \hat{G} -\hat{p}(\boldsymbol y) \hat{\delta}(\boldsymbol x - \boldsymbol y) = 0.&
\end{aligned}
\end{align*}
Integrating the result with respect to the source location $\boldsymbol y$ over some volume V gives
\begin{align*}
\begin{aligned}
\hat{p} (\boldsymbol x) &= \int_V \hat{p}(\boldsymbol y) \hat{\delta}(\boldsymbol x - \boldsymbol y) d^3 \boldsymbol y &\\
&= \int_V  \hat{G} \hat{q}(\boldsymbol y) d^3 \boldsymbol y + \int_V \left[ \hat{G} \nabla_y^2 \hat{p}(\boldsymbol y) - \hat{p}(\boldsymbol y) \nabla_y^2 \hat{G} \right] d^3 \boldsymbol y &\\
&= \int_V  \hat{G} \hat{q}(\boldsymbol y) d^3 \boldsymbol y + \int_V \{ \nabla_y \cdot ( \hat{G} \nabla_y \hat{p}(\boldsymbol y) ) - (\nabla_y \hat{G} ) \cdot ( \nabla_y \hat{p}(\boldsymbol y) ) &\\
&- \nabla_y \cdot ( \hat{p}(\boldsymbol{y}) \nabla_y \hat{G} ) + (\nabla_y \hat{p}(\boldsymbol{y}) ) \cdot ( \nabla_y \hat{G} ) \}  d^3 \boldsymbol y &\\
&= \int_V  \hat{G} \hat{q}(\boldsymbol y) d^3 \boldsymbol y + \int_V \nabla_y \cdot \{ \hat{G} \nabla_y \hat{p}(\boldsymbol y) - \hat{p}(\boldsymbol{y}) \nabla_y \hat{G} \}  d^3 \boldsymbol y,&
\end{aligned}
\end{align*}
where for the first equality we used the sampling property of delta function. Finally, we can apply the divergence theorem on the last integral:
\begin{equation}\label{eq:solution_Green_Helmholtz}
\boxed{ \hat{p} (\boldsymbol x) = \int_V  \hat{q}(\boldsymbol y) \hat{G} d^3 \boldsymbol y + \oint_S \{ \hat{G} \nabla_y \hat{p}(\boldsymbol y) - \hat{p}(\boldsymbol{y}) \nabla_y \hat{G} \} \cdot \boldsymbol n d^2 \boldsymbol y },
\end{equation}
where $S$ is the surface which encloses the volume $V$ and $\boldsymbol n$ is the unit vector normal to $S$ and pointing outwards. If the volume $V$ does not contain any sources of sound, $\hat q(\boldsymbol y) = 0$ in $V$ and the first integral vanishes. However, the volume is usually chosen such that it matches the interior of the room and thus contains all sources in the room. The surface $S$ coincides then with the boundary surfaces of the room and the locations $\boldsymbol y$ refer both to the source region and boundaries. This justifies using the same variable $\boldsymbol y$ on the right-hand side of the wave equation~(\ref{eq:wave_eq_source}) and in the boundary condition, eq.~(\ref{eq:wave_eq_p_BC}). Indeed, the Green's function in eq.~(\ref{eq:solution_Green_Helmholtz}) describes the propagation of sound to the receiver at $\boldsymbol x$ either from the source or surface, which can thus be seen as a virtual or secondary source, even when not radiating sound on its own.

Equation~(\ref{eq:solution_Green_Helmholtz}) is called \textbf{Kirchhoff's (integral) equation} (sometimes also called Kirchhoff-Helmholtz or Rayleigh equation). It shows that for known sources, the Green's function indeed provides the solution of the entire acoustic problem. Unfortunately, it is only by means of the integral equation with the acoustic variable $\hat p$ appearing on its both sides. Calculation of the integral on the right-hand side requires knowledge of the field on the surface $S$, which is part of the solution we are calculating. Still, eq.~(\ref{eq:solution_Green_Helmholtz}) has important implications.

For a source located at $\boldsymbol y$ emitting sound waves with angular frequency $\omega = k c_0$ into \textbf{free space} (without any obstacles or boundaries), it can be shown that the Green's function equals
\begin{equation}\label{eq:Green_Helmholtz_free_space}
\boxed{ \hat{G}_{free}(\boldsymbol{x}|\boldsymbol{y}) = \frac{e^{-j k |\boldsymbol x-\boldsymbol y|}}{4 \pi |\boldsymbol x-\boldsymbol y|} = \frac{e^{-j k r}}{4 \pi r} },
\end{equation}
where $r = |\boldsymbol x-\boldsymbol y|$ is the distance between the source and receiver points. It indeed describes the propagation from $\boldsymbol y$ to $\boldsymbol x$ in free space -- decay of sound with the increasing distance from the source (even without dissipation and other damping mechanisms, because the sound energy is distributed over the expanding wavefront) and the associated change of phase in the numerator. Moreover, it is spherically symmetric and, as expected, satisfies the reciprocity condition, $\hat{G}(\boldsymbol{x}|\boldsymbol{y}) = \hat{G}^*(\boldsymbol{y}|\boldsymbol{x})$, since it depends only on the radial coordinate of the spherical coordinate system, not on the polar or azimuthal angle, and when $\boldsymbol x$ and $\boldsymbol y$ are switched, the outgoing spherical wave becomes incoming, so $-r \rightarrow r$ in the exponent (similarly as $e^{-jkx}$ and $e^{jkx}$ represent two plane waves propagating in opposite directions parallel to the $x$-axis).

In this form $\hat G_{free}$ describes sound propagation from a point monopole in free space. However, its value is not defined at the source location ($r = 0$) (recall that it is a generalized function, which is physically meaningful only after integration). This fact allows an efficient analytical procedure for including angularly dependent radiation (directivity) even for an infinitesimally small (point) source. The so-called  multipole expansion is achieved by expanding $r=|\boldsymbol x-\boldsymbol y|$ into a Taylor series around $\boldsymbol y = 0$. The additional terms of the expansion capture sound propagation from a dipole, quadrupole, etc. Rather than applying this procedure, we will just bear in mind that any solution which is obtained with the spherically symmetric form in eq.~(\ref{eq:Green_Helmholtz_free_space}) has to separately account for the directivity of the source. This is reasonable since room acoustics is most neutrally assessed for an omnidirectional source and, when required, given directivity of the source can be included explicitly (as, for example, in section~\ref{ch:directivity} in eq.~(\ref{eq:solution_tailored_Green_wave_eq_free_space_compact_directed_source_emission_time})). Still, it should be noted that a Green's function can ``add'' the directivity to the otherwise non-directional scalar source function. Alternatively, a directional source can be represented with a vector (dipole) or higher-order tensor (quadrupole etc.) source function, or with two (dipole) or more scalar functions.

In certain cases (for example, if we calculate only the reflected sound $\hat p(\boldsymbol x)$ from an isolated body in free space for the given incident sound wave and therefore known $\hat p(\boldsymbol y)$, as we do in section~\ref{ch:rectangular_surface}), the simple free space Green's function can be used in eq.~(\ref{eq:solution_Green_Helmholtz}). Another important case, especially for closed spaces, is when the Green's function $\hat{G}(\boldsymbol{x}|\boldsymbol{y})$ satisfies both eq.~(\ref{eq:Helmholtz_Green}) and the same boundary conditions as the initial variable $\hat{p}$ at all boundaries. It is then called \textbf{tailored Green's function} (adjusted to the boundary conditions). A general boundary condition is given in eq.~(\ref{eq:Helmholtz_p_BC}) and a tailored Green's function satisfies accordingly
\begin{equation}\label{eq:tailored_Green_Helmholtz_BC}
a \hat{G}_{tail} + b \nabla_y \hat{G}_{tail} \cdot \boldsymbol n(\boldsymbol y) = \hat{c}(\boldsymbol y)
\end{equation}
at every location $\boldsymbol y$ at each boundary surface and with the same values of $a$, $b$, and function $\hat c$ (with the adapted unit, for example, multiplied with 1/(Pa$\cdot$m)). In this case, the two terms inside the curly brackets in eq.~(\ref{eq:solution_Green_Helmholtz}) cancel at $S$ and the equation simplifies to
\begin{equation}\label{eq:solution_tailored_Green_Helmholtz}
\boxed{ \hat{p} (\boldsymbol x) = \int_V \hat{q}(\boldsymbol y) \hat{G}_{tail} d^3 \boldsymbol y }.
\end{equation}

For a known $\hat G_{tail}$, this equation is evidently much easier to solve than eq.~(\ref{eq:solution_Green_Helmholtz}), since it does not include the pressure amplitude $\hat p$ on the right-hand side nor the integral and geometry of the surfaces. The tailored Green's function completely describes acoustic behaviour of the room (without the source function), that is sound propagation between the source and receiver, including the effects of the room geometry and boundary conditions. Unfortunately, it is very difficult to obtain its exact form for arbitrary geometries and boundary conditions. We will be able to calculate it analytically in section~\ref{ch:rectangular_room_wave_theory} for a simple rectangular room with hard walls and, apart from a few similarly idealized scenarios, it has to be estimated numerically (for example, using the finite element method). In the trivial case when the boundaries are absent, $\hat{G}_{tail} = \hat{G}_{free}$.

If the source is an ideal \textbf{point source} located at $\boldsymbol y$, we can write its function as the distribution $\hat{q}(\boldsymbol y') = \hat{Q}(\boldsymbol y') \delta(\boldsymbol y' - \boldsymbol y)$, where $\hat{Q}$ has the unit kg/s$^2$ in a three-dimensional space. It can correspond to the unsteady volume injection from eq.~(\ref{eq:source_mass_injection_Helmholtz}). Notice, however, that the function $\hat q$ becomes a generalized function, which does not pose a problem because the pressure amplitude is expressed as an integral in eq.~(\ref{eq:solution_tailored_Green_Helmholtz}). Indeed, integral over a volume containing the source is finite
\begin{equation}\label{eq:tailored_Green_Helmholtz_compact_source}
\int_V \hat{q}(\boldsymbol y') d^3 \boldsymbol y' = \int_V \hat{Q}(\boldsymbol y') \delta(\boldsymbol y' - \boldsymbol y) d^3 \boldsymbol y' = \hat{Q}(\boldsymbol y),
\end{equation}
following from the sampling property of delta function. Similarly,
\begin{equation}\label{eq:solution_tailored_Green_Helmholtz_point_source}
\begin{aligned}
\hat{p} (\boldsymbol x) &= \int_V \hat{q}(\boldsymbol y') \hat{G}_{tail}(\boldsymbol x|\boldsymbol y') d^3 \boldsymbol y' = \int_V \hat{Q}(\boldsymbol y') \delta(\boldsymbol y' - \boldsymbol y) \hat{G}_{tail}(\boldsymbol x|\boldsymbol y') d^3 \boldsymbol y' &\\
&= \hat{Q}(\boldsymbol y) \hat{G}_{tail}(\boldsymbol x|\boldsymbol y),&
\end{aligned}
\end{equation}
which gives a very simple relation between the pressure amplitude, point source, and tailored Green's function. In fact, tailored Green's function represents \textbf{frequency response} of a room defined as\footnote{Alternatively, frequency response can be defined in terms of the velocity component normal to the surface of the source as $\hat p/(\hat{\boldsymbol v} \cdot \boldsymbol n) \sim \rho_0 S \omega \hat p / \hat Q$, where $S$ is surface area of the source, which involves additional multiplication with frequency. While this definition is more convenient for sources with frequency-independent velocity, the definition above is more appropriate for frequency-independent acceleration.} $\hat p/\hat Q$, for the given $\boldsymbol x$ and $\boldsymbol y$. Still, the last equality makes physical sense only when $\hat{G}_{tail}$ is an ordinary function at $\boldsymbol x$ outside the source region (because it leaves the integral). For example, for a point \textbf{monopole} in free space
\begin{equation}\label{eq:solution_tailored_Green_Helmholtz_point_source_free_space}
\begin{aligned}
\hat{p}_{free} (\boldsymbol x) = \hat{Q}(\boldsymbol y) \hat{G}_{free}(\boldsymbol x|\boldsymbol y) = \hat{Q}(\boldsymbol y) \frac{e^{-j k |\boldsymbol x-\boldsymbol y|}}{4 \pi |\boldsymbol x-\boldsymbol y|}.
\end{aligned}
\end{equation}
The pressure amplitude at $\boldsymbol x \neq \boldsymbol y$ is merely a scaled version of $\hat Q$ with a phase shift. The complete solution is obtained after integration over all wave numbers $k$, analogously to eq.~(\ref{eq:inv_Fourier_p}) ($e^{-jk|\boldsymbol x - \boldsymbol y|}/(4\pi |\boldsymbol x - \boldsymbol y|)$ thus has practically the same role as $e^{j\omega t}$ for spherical outgoing waves in frequency domain).

\subsection{Room eigenmodes}\label{ch:room_eigenmodes}

Next we treat a room completely generally, as a closed cavity with an arbitrary shape and volume $V$ in which an omnidirectional point source generates a sound field (the Green's function does not account for the directivity). Tailored Green's function of closed spaces can always be written as a weighted sum of an infinite number of distinct \textbf{eigenmodes} (complex eigenfunctions, or simply modes):
\begin{equation}\label{eq:tailored_Green_Helmholtz_modes}
\hat{G}_{tail}(\boldsymbol x | \boldsymbol y) = \sum_{n} A_n(\boldsymbol y) \psi_n (\boldsymbol x),
\end{equation}
where $\psi_n$ denotes mode $n$ ($n = 1,2,3...$) and $A_n$ is the complex weighting factor, which does not depend on $\boldsymbol x$. Since $\hat{G}_{tail} (\boldsymbol x | \boldsymbol y)$ depends on both $\boldsymbol x$ and $\boldsymbol y$, we expect $A_n$ to depend on $\boldsymbol y$, location of the source (and, of course, frequency). Thus, we separate the two variables representing the source and receiver locations.

This is completely analogous to the expansion in section~\ref{ch:frequency_domain}, which separated time and location and lead to eq.~(\ref{eq:inv_Fourier_p}). The Fourier integral is replaced by the series due to the spatial boundedness (there was no ``boundary'' in time). While $\omega$ (or $k$, see the free-space Green's function in eq.~(\ref{eq:Green_Helmholtz_free_space})) can take any (real and positive) value in an unbounded time (or space), we will see shortly that each $\psi_n(\boldsymbol x)$ is associated with a value $k_n$ from a discrete set of wave numbers, so the integral over $k$ is replaced by the sum over $n$. Eigenfrequencies of the modes depend entirely on the geometry and boundary conditions. Modes in unbounded time and space also have well-defined forms: $e^{j \omega t}$ and $e^{-jkx}$ for forward propagating plane waves or $e^{-j k r}/(4\pi r)$ for spherical waves radiated from the point $r=0$ (eq.~(\ref{eq:Green_Helmholtz_free_space})), etc. A more general form of eigenmodes, $\psi_n(\boldsymbol x)$, is necessary because of the variety of possible boundary conditions in bounded spaces.

Just like $e^{j\omega t}$ in eq.~(\ref{eq:complex_sine_wave}) or $e^{-jkr}$ in eq.~(\ref{eq:Green_Helmholtz_free_space}), the modes provide a complete set of mutually orthogonal functions. The orthogonality is expressed with
\begin{equation}\label{eq:modes_orthogonality}
\int_V \psi_m^* (\boldsymbol x) \psi_n (\boldsymbol x) d^3 \boldsymbol x = \begin{cases} K_n &\text{for $m = n$}\\
0 &\text{otherwise}.
\end{cases}
\end{equation}
The integral over the entire room volume is zero for any two non-equal modes. For $m=n$, it gives a real (due to the complex conjugate value, as in eq.~(\ref{eq:complex_sine_wave_orthogonality})) number $K_n$. Finite volume $V$ results in a finite value of $K_n$, in contrast to the delta function in eq.~(\ref{eq:complex_sine_wave_orthogonality}) for the unbounded time. Since a mode can be multiplied with an arbitrary constant without changing its eigenvalue (that is, frequency), the value of $K_n$ is not uniquely determined. However, we will see that this constant vanishes in the expressions for measurable physical quantities.

Outside the source region, Green's function has to satisfy eq.~(\ref{eq:Helmholtz_Green}) with the right-hand side equal to zero. Integrating the equation over $k$ and using the expansion from eq.~(\ref{eq:tailored_Green_Helmholtz_modes}) which turns the integral into a sum, we obtain
\begin{equation}\label{eq:G_tail_Helmholtz}
\begin{aligned}
k^2 \hat{G}_{tail} + \nabla_x^2 \hat{G}_{tail} &= \sum_{n} k_n^2 A_n(\boldsymbol y) \psi_n (\boldsymbol x) + \nabla_x^2 \sum_{n} A_n(\boldsymbol y) \psi_n (\boldsymbol x) &\\
&= \sum_{n} A_n(\boldsymbol y) \left[ k_n^2 \psi_n (\boldsymbol x) + \nabla_x^2 \psi_n (\boldsymbol x) \right]= 0.&
\end{aligned}
\end{equation}
Therefore, for every non-trivial $A_n \neq 0$,
\begin{equation}\label{eq:Helmholtz_Green_no_source}
k_n^2 \psi_n (\boldsymbol x) + \nabla_x^2 \psi_n (\boldsymbol x) = 0.
\end{equation}
Every mode satisfies the homogeneous Helmholtz equation (supplied with boundary conditions) with the corresponding \textbf{wave number}.

In order to evaluate the coefficients $A_n(\boldsymbol y)$, we use the property of orthogonality of the eigenmodes. We again insert the expansion from eq.~(\ref{eq:tailored_Green_Helmholtz_modes}) into eq.~(\ref{eq:Helmholtz_Green}). This time, however, we seek for the coefficients $A_n(\boldsymbol y)$ which do not \textit{a priori} depend on the modes and their frequencies. Therefore, we have to allow a general continuous $k$. Furthermore, the delta function on the right-hand side of eq.~(\ref{eq:Helmholtz_Green}) brings the dependence on $\boldsymbol y$. This gives
\begin{align*}
\begin{aligned}
k^2 \hat{G}_{tail} + \nabla_x^2 \hat{G}_{tail} &= k^2 \sum_{n} A_n(\boldsymbol y) \psi_n (\boldsymbol x) + \nabla_x^2 \sum_{n} A_n(\boldsymbol y) \psi_n (\boldsymbol x) &\\
&= \sum_{n} A_n(\boldsymbol y) \left[ k^2 \psi_n (\boldsymbol x) + \nabla_x^2 \psi_n (\boldsymbol x) \right] = 
\sum_{n} A_n(\boldsymbol y) \left[ k^2 \psi_n (\boldsymbol x) -k_n^2 \psi_n (\boldsymbol x) \right] &\\
&= \sum_{n} A_n(\boldsymbol y) (k^2-k_n^2) \psi_n (\boldsymbol x) =
-\hat{\delta}(\boldsymbol x - \boldsymbol y),&
\end{aligned}
\end{align*}
where we also used eq.~(\ref{eq:Helmholtz_Green_no_source}.

Next we multiply the last equality with the complex conjugate of an arbitrary eigenfunction $\psi_m (\boldsymbol x)$ and integrate over the entire room volume:
\begin{align*}
\begin{aligned}
\int_V \sum_{n} A_n(\boldsymbol y) &(k^2-k_n^2) \psi_n (\boldsymbol x) \psi_m^* (\boldsymbol x) d^3 \boldsymbol x = \sum_{n} A_n(\boldsymbol y) (k^2-k_n^2) \int_V \psi_n (\boldsymbol x) \psi_m^* (\boldsymbol x) d^3 \boldsymbol x &\\
&= \left( A_n(\boldsymbol y) (k^2-k_n^2) K_n \right)_{n=m} = A_m(\boldsymbol y) (k^2-k_m^2) K_m &\\
&=-\int_V \psi_m^* (\boldsymbol x) \hat{\delta}(\boldsymbol x - \boldsymbol y) d^3 \boldsymbol x = -\psi_m^* (\boldsymbol y).&
\end{aligned}
\end{align*}
Orthogonality of the modes from eq.~(\ref{eq:modes_orthogonality}) left only one member of the sum, for $n = m$, and the sampling property of delta function replaced $\boldsymbol x$ with $\boldsymbol y$ as the argument of the eigenfunction. Since this is valid for any mode $m$, it also holds for any mode $n$ and we can simply change the index:
\begin{equation}\label{eq:Helmholtz_Green_An}
A_n(\boldsymbol y) = \frac{\psi_n^* (\boldsymbol y)}{ K_n(k_n^2-k^2)}.
\end{equation}
Therefore, the tailored Green's function equals
\begin{equation}\label{eq:tailored_Green_Helmholtz_modes_An}
\boxed{ \hat{G}_{tail}(\boldsymbol x | \boldsymbol y) = \sum_{n} A_n(\boldsymbol y) \psi_n (\boldsymbol x) = \sum_{n} \frac{\psi_n (\boldsymbol x) \psi_n^* (\boldsymbol y)}{ K_n(k_n^2-k^2)}
	= c_0^2\sum_{n} \frac{\psi_n (\boldsymbol x) \psi_n^* (\boldsymbol y)}{ K_n(\omega_n^2-\omega^2)} }.
\end{equation}
It satisfies the reciprocity: $\hat{G}_{tail}(\boldsymbol x | \boldsymbol y) = \hat{G}_{tail}^*(\boldsymbol y | \boldsymbol x)$. Like eigenmodes, it is also a regular function, well defined for $\omega \neq \omega_n$, so eq.~(\ref{eq:solution_tailored_Green_Helmholtz_point_source}) holds. We can also see that multiplying the mode $\psi_n$ with any constant would not affect the value of $\hat G_{tail}$, because $K_n$ is then multiplied with its squared value, according to eq.~(\ref{eq:modes_orthogonality}). Even though the values of eigenfunctions are not uniquely defined, the frequency response in eq.~(\ref{eq:solution_tailored_Green_Helmholtz_point_source}) is.

Hence, if we are able to determine the eigenmodes $\psi_n$ (up to the arbitrary constants) and their wave numbers $k_n$, which satisfy both eq.~(\ref{eq:Helmholtz_Green_no_source}) and the boundary conditions (since we started with the tailored Green's function in eq.~(\ref{eq:tailored_Green_Helmholtz_modes})), we can also determine the constants $K_n$ from eq.~(\ref{eq:modes_orthogonality}) followed by the coefficients $A_n$ from eq.~(\ref{eq:Helmholtz_Green_An}). These coefficients together with the eigenmodes determine the tailored Green's function in eq.~(\ref{eq:tailored_Green_Helmholtz_modes}), as in eq.~(\ref{eq:tailored_Green_Helmholtz_modes_An}). The tailored Green's function can then be used in eq.~(\ref{eq:solution_tailored_Green_Helmholtz}) together with a known source function to obtain the complete solution in terms of the frequency-dependent complex amplitude of sound pressure. The time dependence is then simply attached by multiplication with $e^{j\omega t}$, according to eq.~(\ref{eq:complex_sine_wave}), and the physical sound pressure is the real part of the result.

However, we are still far from a closed-form solution, since the boundary conditions at all surfaces of the room determine the eigenmodes and their associated wave numbers, making their calculation a challenging task. Since tailored Green's function is a weighted sum of eigenmodes (eq.~(\ref{eq:tailored_Green_Helmholtz_modes})), the eigenmodes have to satisfy the same boundary conditions as the tailored Green's function, which are, as already discussed, also satisfied by the original variable $\hat{p}$. From equations~(\ref{eq:tailored_Green_Helmholtz_BC}) and (\ref{eq:tailored_Green_Helmholtz_modes}) it follows:
\begin{align*}
a \hat{G}_{tail} + b \nabla_y \hat{G}_{tail} \cdot \boldsymbol n(\boldsymbol y) &= \sum_{n} a_n A_n \psi_n (\boldsymbol y) + \sum_{n} b_n [ \nabla_y (A_n \psi_n (\boldsymbol y)) \cdot \boldsymbol n(\boldsymbol y) ] &\\
&= \sum_{n} A_n \left[ a_n \psi_n (\boldsymbol y) + b_n \nabla_y \psi_n (\boldsymbol y) \cdot \boldsymbol n(\boldsymbol y) \right] = \hat c(\boldsymbol y),&
\end{align*}
where we left only $\boldsymbol y$ at the boundary as the argument, in order to avoid confusion ($A_n$ is not a function of $\boldsymbol y$, but on the source location). Consequently,
\begin{equation}\label{eq:modes_Helmholtz_BC}
a_n \psi_n(\boldsymbol y) + b_n \nabla_y \psi_n(\boldsymbol y) \cdot \boldsymbol n(\boldsymbol y) = \frac{\hat c_n(\boldsymbol y)}{A_n},
\end{equation}
with $\sum_{n} \hat{c}_n(\boldsymbol y) = \hat c(\boldsymbol y)$.

Still, even without knowing the exact forms of eigenmodes, certain important conclusions can be made based on the result in eq.~(\ref{eq:tailored_Green_Helmholtz_modes_An}). For more generality, we can also include (in real rooms always present) \textbf{damping} as in equations~(\ref{eq:p_exp_decay}) and (\ref{eq:complex_omega}), assuming thereby exponential decay of the amplitude, by adding the imaginary part $j \zeta_n$ to the real angular frequency $\omega_n$ of mode $n$. The tailored Green's function becomes
\begin{equation}\label{eq:tailored_Green_Helmholtz_modes_An_damping_full}
\begin{split}
\hat{G}_{tail}(\boldsymbol x | \boldsymbol y) = c_0^2 \sum_{n} \frac{\psi_n (\boldsymbol x) \psi_n^* (\boldsymbol y)}{ K_n(\omega_n^2 + 2j \zeta_n \omega_n - \zeta_n^2-\omega^2)},
\end{split}
\end{equation}
which is for $\zeta_n \neq 0$ well defined even at $\omega = \omega_n$. From eq.~(\ref{eq:damping_omega}), $\zeta_n \ll \omega_n$ in weakly damped rooms, so we can approximate
\begin{equation}\label{eq:tailored_Green_Helmholtz_modes_An_damping}
\hat{G}_{tail}(\boldsymbol x | \boldsymbol y) = c_0^2 \sum_{n} \frac{\psi_n (\boldsymbol x) \psi_n^* (\boldsymbol y)}{ K_n(\omega_n^2 + 2j \zeta_n \omega_n -\omega^2)}.
\end{equation}
Inserting this into eq.~(\ref{eq:solution_tailored_Green_Helmholtz}), the formal solution of eq.~(\ref{eq:Helmholtz_source}) is
\begin{equation}\label{eq:solution_tailored_Green_Helmholtz_damped}
\hat{p} (\boldsymbol x) = c_0^2 \int_V \hat{q}(\boldsymbol y) \sum_{n} \frac{\psi_n (\boldsymbol x) \psi_n^* (\boldsymbol y)}{ K_n(\omega_n^2+2j \zeta_n \omega_n -\omega^2)} d^3 \boldsymbol y.
\end{equation}
The solution for a point monopole source is from eq.~(\ref{eq:solution_tailored_Green_Helmholtz_point_source})
\begin{equation}\label{eq:solution_tailored_Green_Helmholtz_modes}
\boxed{ \hat{p} (\boldsymbol x) = \hat Q(\boldsymbol y) \hat G_{tail}(\boldsymbol x | \boldsymbol y) = \hat{Q}(\boldsymbol y) c_0^2 \sum_{n} \frac{\psi_n (\boldsymbol x) \psi_n^* (\boldsymbol y)}{ K_n(\omega_n^2+2j \zeta_n \omega_n -\omega^2)} }.
\end{equation}
It is important to notice the difference between the measurable $\hat p(\boldsymbol x)$ and unmeasurable (and not uniquely defined) $\psi (\boldsymbol x)$. The latter does affect the former, but only in conjunction with the source and its location. Sound field of the point monopole at the location $\boldsymbol y$ in the room emitting stationary sound is thus a superposition of all modes which are excited ($\psi_n(\boldsymbol y) \neq 0$) by the source, normalized with the constant $K_n$. 

\begin{figure}[h]
	\centering
	\begin{subfigure}{.48\textwidth}
		\centering
		\includegraphics[width=.93\linewidth]{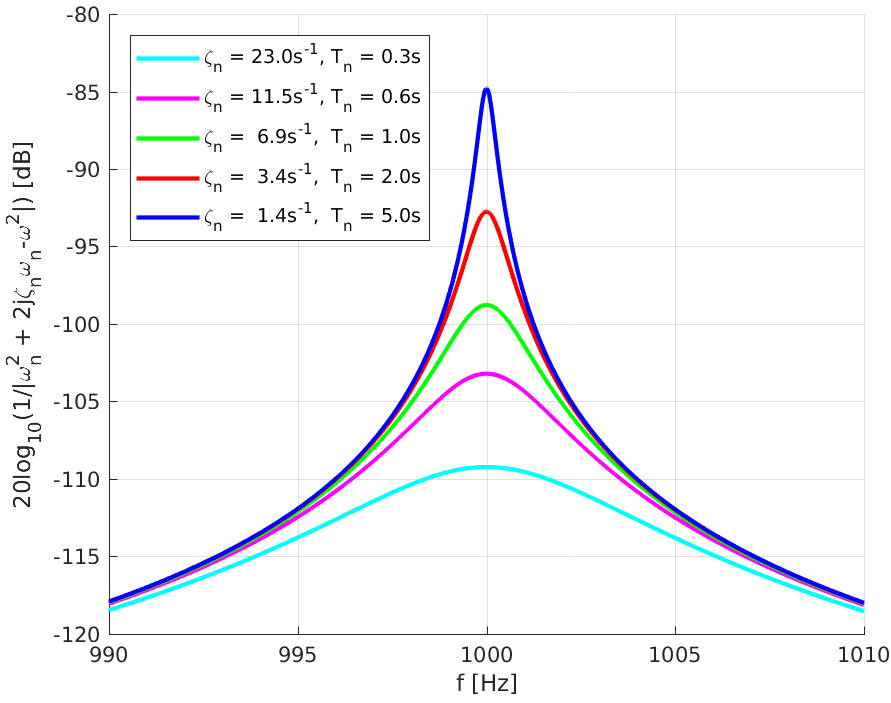}
		\label{fig:resonance_1000Hz}
	\end{subfigure}
	\begin{subfigure}{.50\textwidth}
		\centering
		\includegraphics[width=.98\linewidth]{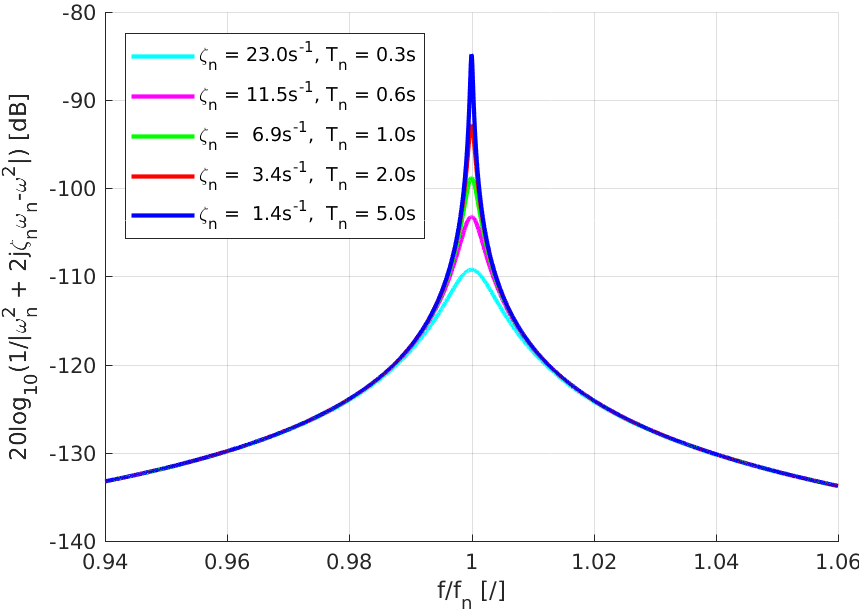}
		\label{fig:resonance_norm_freq}
	\end{subfigure}%
	\caption{Amplitude of a damped eigenmode with the eigenfrequency 1000\,Hz as a function of (left) frequency and (right) normalized eigenfrequency.}
	\label{fig:resonance}
\end{figure}

Each member of the sum corresponds to one mode of the room and its contribution to the complex sound pressure amplitude is a continuous function of frequency (not only $\omega_n$)
\begin{equation}\label{eq:solution_tailored_Green_Helmholtz_mode}
\hat{p}_n (\boldsymbol x) = \hat{Q}(\boldsymbol y) c_0^2 \frac{\psi_n (\boldsymbol x) \psi_n^* (\boldsymbol y)}{ K_n(\omega_n^2+2j \zeta_n \omega_n -\omega^2)}.
\end{equation}
Different complex terms in this expression can be interpreted as follows:
\begin{itemize}
	\item $\hat{Q}(\boldsymbol y)$ depends only on the (point) source -- its strength and initial phase
	\item$\psi_n (\boldsymbol x) \psi_n^* (\boldsymbol y) / K_n$ indicates how strongly the mode $\psi_n$ is excited by the source. It depends on the geometry of the problem -- shape and size of the room, boundary conditions, as well as the locations of both the source and receiver. (For example, if $\psi_n(\boldsymbol y)=0$ at the source location, the theoretical value of sound pressure due to the mode $n$ will be zero regardless of the mode shape $\psi_n(\boldsymbol x)$ and power of the source.)
	\item $1/ (\omega_n^2+2j \zeta_n \omega_n -\omega^2)$ is frequency dependence of the contribution of mode $n$ to the total sound pressure. It determines how prominent the mode can be in the frequency response of the room, whether it adds constructively (with equal sign) or destructively (with opposite sign) to the other modes. More damped rooms exhibit less pronounced resonances. Logarithm of the term's modulus, $20 \log_{10} (1/ |\omega_n^2+2j \zeta_n \omega_n -\omega^2|)$, is shown in Fig.~\ref{fig:resonance} for several realistic values of the damping constant $\zeta_n$ and the eigenfrequency of the mode $f_n = 1$\,kHz.
\end{itemize}
Since $\zeta_n \ll \omega_n$ and thus $2 \zeta_n \omega_n \ll \omega_n^2$, it follows that the damping term in the frequency dependence is important only when $\omega_n^2$ and $\omega^2$ largely cancel, that is, when $\omega \approx \omega_n$. Therefore, the damping is critical close to the eigenfrequency of the mode and the entire frequency dependence can also be approximated as $1/ (\omega_n^2+2j \zeta_n \omega -\omega^2)$.

The curves in Fig.~\ref{fig:resonance} depict typical behaviour of a \textbf{damped linear oscillator}, such as a mass-spring system. This follows from the analogy between eq.~(\ref{eq:Helmholtz_source}) and the equation of an undamped\footnote{We introduced damping first in eq.~(\ref{eq:tailored_Green_Helmholtz_modes_An_damping_full}).} oscillator, with the only essential difference being in the independent variable (location instead of time). The ordinary (time is the only independent variable, so we can write $d/dt$ instead of $\partial / \partial t$) differential equation of a simple weakly damped mass-spring system exhibiting forced oscillatory motion reads
\begin{equation}\label{eq:oscillator_time}
M \frac{d^2 \xi(t)}{dt^2} + D \frac{d \xi(t)}{dt} + S \xi(t) = F(\tau),
\end{equation}
where $\xi$ is one-dimensional displacement (m), $M$ is mass (kg), $D$ is damping (kg/s), $S$ is stiffness (kg/s$^2$), and $F$ is one-dimensional force (N) acting on the mass $M$. By comparing this equation with the Helmholtz equation~(\ref{eq:Helmholtz_source}), we see that the second-order spatial derivative (in a three-dimensional space) is replaced with the second-order time derivative and the following substitutions apply: $\hat{p}(\boldsymbol x) \rightarrow \xi(t)$, $k^2 \rightarrow S/M$, and $-\hat{q}(\boldsymbol y) \rightarrow F(\tau)/M$. The remaining term $(D/M)(d \xi(t) / dt)$ is due to the damping, which is not present in eq.~(\ref{eq:Helmholtz_source}), but modelled in eq.~(\ref{eq:tailored_Green_Helmholtz_modes_An_damping_full}).

If we write the displacement and force as $\xi(t) = \hat{\xi} e^{j \omega t}$ and $F(\tau) = \hat{F} e^{j \omega \tau}$, respectively, and divide eq.~(\ref{eq:oscillator_time}) with $e^{j \omega t}$ (since the force acts directly on the moving lumped mass, the excitation is instantaneous and $e^{j\omega t} = e^{j\omega \tau}$), we obtain a simple algebraic equation
\begin{equation}\label{eq:oscillator_freq}
-M \omega ^2 \hat{\xi} + j \omega D \hat{\xi} + S \hat{\xi} = \hat{F}.
\end{equation}
The solution is
\begin{equation}\label{eq:oscillator_freq_solution}
\hat{\xi} = \frac{\hat{F}/M}{S/M + j \omega D/M -\omega^2}.
\end{equation}
For low damping, resonance of the mass-spring system occurs at the angular frequency $\omega_0 = \sqrt{S/M}$. Damping affects the Q-factor of the resonance, which is by definition $Q = \omega_0 M/D = \sqrt{S M}/D$. The Q-factor is tightly related with the half-width of the resonance curve, $\Delta \omega$, which is angular frequency range centred at $\omega_0$ within which amplitude of the oscillations drops by the factor not larger than $\sqrt{2}$ and thus energy by the factor not larger than $2$, compared to their peak values. The relation is $Q = \omega_0 /\Delta \omega$, or $\Delta \omega = \omega_0/Q = D/M$. \textbf{Mechanical impedance} of the system is defined as
\begin{equation}\label{eq:oscillator_impedance}
Z = \frac{F}{v} = \frac{F}{d \xi/dt} = \frac{\hat F}{j \omega \hat \xi} = j\omega M + D + \frac{S}{j\omega}.
\end{equation}

After comparing the denominator in eq.~(\ref{eq:oscillator_freq_solution}) with $S/M = \omega_0^2$ and the term $\omega_n^2+2j \zeta_n \omega -\omega^2$ from above, we can conclude that every room (so far we have only assumed that the damping is low and that the room is excited by a point omnidirectional source, with no other restrictions on the geometry or boundary conditions) behaves as an infinite series of simple harmonic oscillators with the resonances at frequencies $f_n = \omega_n/(2\pi)$, half-widths $\Delta \omega_n = 2 \zeta_n$, and Q-factors $Q_n = \omega_n/(2 \zeta_n)$. The damping constants $\zeta_n = \Delta \omega_n/2$ correspond to $D/(2M)$. The excitation ``force'' of the oscillators is, of course, introduced by the source of sound $\hat{Q}$. However, unlike the lumped mass-spring system, the level of excitation of the room resonances (and, for instance, the actual peaks of SPL like those in Fig.~\ref{fig:resonance}) is also spatially dependent, due to the remaining term $\psi_n (\boldsymbol x) \psi_n^* (\boldsymbol y) / K_n$ in eq.~(\ref{eq:solution_tailored_Green_Helmholtz_mode}). As discussed there, it depends on the source and receiver locations, as well as the shapes of the modes dictated by the geometry of the room and boundary conditions. Moreover, the total sound field in eq.~(\ref{eq:solution_tailored_Green_Helmholtz_modes}) is a sum of the contributions of many modes with different (or even equal) frequencies. Depending on the signs of their terms $\psi_n(\boldsymbol x) \psi_n^*(\boldsymbol y)$, the modes can add constructively or (partly) cancel, increasing further the peaks from Fig.~\ref{fig:resonance} or creating dips in the frequency response.

Summarizing the methodology we have used so far for solving the room acoustic problems, Table~\ref{tab:acoustic_quantities_freq_domain} lists the main quantities which appeared, the key equations (and associated boundary conditions) which they satisfy, and the quantities on which they depend. It should clearly demonstrate the gradual decomposition of the initial problem (defined by the governing equations and the wave equation), which was achieved by separating various variables and decoupling their contributions. In order to obtain a more specific solution, we need to calculate eigenmodes $\psi_n$ of a room and their wave numbers $k_n$, which satisfy eq.~(\ref{eq:Helmholtz_Green_no_source}) and boundary conditions, eq.~(\ref{eq:modes_Helmholtz_BC}). Since we removed the time dependence and work in the frequency domain, all boundary conditions have to be satisfied simultaneously by every mode. This presents a very difficult calculation task and the main challenge in modal analyses, which can be handled analytically only in very few idealized cases. These are the rooms with very simple geometries and boundary conditions, for which it is possible to further separate the spatial coordinates in a suitable coordinate system. For a rectangular room with uniform boundary conditions, this is most easily done with Cartesian coordinates\footnote{Similarly, cylindrical coordinates would be more convenient for cylindrical cavities and spherical coordinates for spherical cavities. Such geometries have much less practical relevance in room acoustics, so we will cover only rectangular rooms in more detail in the following.}. In any case, we should recall that the full solution of a sound field in terms of sound pressure values depends also on the source function, which thus has to be given. Alternatively, frequency response of a room (the tailored Green's function) requires only the source location.

\begin{table}[h]
	\caption{Main quantities and their dependencies. Notes: BC -- boundary conditions; initial condition in time domain is $p = 0$ before the source $q$ is switched on; in frequency domain the duration of sine waves is assumed to be infinite; frequency dependence is always implied.}
	\label{tab:acoustic_quantities_freq_domain}
	\begin{tabular}{ | p{2cm} | c | p{5.5cm} |}
		\hline
		\textbf{quantity} & \textbf{equation} & \textbf{depends on} \\
		\hline
		$p$ & eq.~(\ref{eq:wave_eq_source}) & $\boldsymbol x$, $t$, $q(\boldsymbol y, \tau)$, BC in eq.~(\ref{eq:wave_eq_p_BC}) \\
		\hline
		$\hat p$ & eq.~(\ref{eq:Helmholtz_source}) & $\boldsymbol x$, $\hat q(\boldsymbol y)$, BC in eq.~(\ref{eq:Helmholtz_p_BC}) \\
		\hline
		$\hat G_{tail}$ & eq.~(\ref{eq:Helmholtz_Green}) & $\boldsymbol x$, $\boldsymbol y$, BC in eq.~(\ref{eq:tailored_Green_Helmholtz_BC}) \\
		\hline
		$\psi_n$ & eq.~(\ref{eq:Helmholtz_Green_no_source}) & $\boldsymbol x$, BC in eq.~(\ref{eq:modes_Helmholtz_BC}) \\
		\hline
		$A_n$ & eq.~(\ref{eq:Helmholtz_Green_An}) & $\boldsymbol y$, $\psi_n$ \\
		\hline
	\end{tabular}
\end{table}

\subsection{Rectangular room with hard walls} \label{ch:rectangular_room_wave_theory}

Introducing axes of a Cartesian coordinate system parallel to the edges of a \textbf{rectangular room}, we can separate the three spatial coordinates by supposing the following form of the solution of eq.~(\ref{eq:Helmholtz_Green_no_source}):
\begin{equation}\label{eq:modes_rect_room_product}
\psi_n (\boldsymbol x) = \psi_n(x_1) \psi_n(x_2) \psi_n(x_3) = \psi_{n1} \psi_{n2} \psi_{n3}.
\end{equation}
In the last equality we placed the components of $\boldsymbol x$ in the subscripts of $\psi_n$ for easier notation. (The same will be done with the components of $\boldsymbol{y}$ below.) The modal orders are non-negative integers $n_i = 0,1,2...$ for $i = 1,2,3$ and each combination $(n_1,n_2,n_3)$ corresponds to one mode $n$, where $n=1,2,3...$. The mode $\psi_{n =0}$ for $(n_1,n_2,n_3) = (0,0,0)$ is not acoustic and therefore omitted.

Continuing with the separation of variables, eq.~(\ref{eq:Helmholtz_Green_no_source}) becomes
\begin{equation}\label{eq:Helmholtz_Green_no_source_separation}
\begin{aligned}
\nabla_x^2 (\psi_{n1} \psi_{n2} \psi_{n3}) &+ k_n^2 \psi_{n1} \psi_{n2} \psi_{n3} = \psi_{n2} \psi_{n3} \frac{d^2 \psi_{n1}}{dx_1^2} + \psi_{n1} \psi_{n3} \frac{d^2 \psi_{n2}}{dx_2^2} &\\
&+ \psi_{n1} \psi_{n2} \frac{d^2 \psi_{n3}}{dx_3^2}+ k_n^2 \psi_{n1} \psi_{n2} \psi_{n3} = 0&
\end{aligned}
\end{equation}
and after dividing with $\psi_{n1} \psi_{n2} \psi_{n3}$ for any $\psi_n \neq 0$
\begin{equation}\label{eq:Helmholtz_Green_no_source_separation2}
\begin{aligned}
\frac{1}{\psi_{n1}} \frac{d^2 \psi_{n1}}{dx_1^2} + \frac{1}{\psi_{n2}} \frac{d^2 \psi_{n2}}{dx_2^2} + \frac{1}{\psi_{n3}} \frac{d^2 \psi_{n3}}{dx_3^2}+ k_n^2 = 0.
\end{aligned}
\end{equation}
Since $\psi_{n1}$, $\psi_{n2}$, and $\psi_{n3}$ are functions of different components of $\boldsymbol x$, it is convenient to treat $k_n$ as the magnitude of a \textbf{wave vector} $\boldsymbol k_n$ with components $k_{n1}$, $k_{n2}$, and $k_{n3}$ in the three-dimensional $k$-space, that is
\begin{equation}\label{eq:modes_rect_room_kn}
k_n^2 = |\boldsymbol k_n|^2 = k_{n1}^2 + k_{n2}^2 + k_{n3}^2.
\end{equation}
In this way eq.~(\ref{eq:Helmholtz_Green_no_source_separation2}) gives three decoupled ordinary differential equations analogue to eq.~(\ref{eq:wave_eq_time_separated}):
\begin{equation}\label{eq:Helmholtz_Green_no_source_separated_system}
\begin{aligned}
\frac{d^2 \psi_{n1}}{dx_1^2} + k_{n1}^2 \psi_{n1} &= 0, \\
\frac{d^2 \psi_{n2}}{dx_2^2} + k_{n2}^2 \psi_{n2} &= 0, \\
\frac{d^2 \psi_{n3}}{dx_3^2} + k_{n3}^2 \psi_{n3} &= 0.
\end{aligned}
\end{equation}
General solutions of equations~(\ref{eq:Helmholtz_Green_no_source_separated_system}) are, respectively,
\begin{equation}\label{eq:modes_rect_room}
\begin{aligned}
\psi_{n1} &= C_{11}e^{-j k_{n1} x_1} + C_{21}e^{j k_{n1} x_1}
&\\&= (C_{11} + C_{21}) \cos(k_{n1} x_1) - j (C_{11}-C_{21}) \sin(k_{n1} x_1), &\\
\psi_{n2} &= C_{12}e^{-j k_{n2} x_2} + C_{22}e^{j k_{n2} x_2} 
&\\&= (C_{12} + C_{22}) \cos(k_{n2} x_2) - j (C_{12}-C_{22}) \sin(k_{n2} x_2), &\\
\psi_{n3} &= C_{13}e^{-j k_{n3} x_3} + C_{23}e^{j k_{n3} x_3}
&\\&= (C_{13} + C_{23}) \cos(k_{n3} x_3) - j (C_{13}-C_{23}) \sin(k_{n3} x_3),&
\end{aligned}
\end{equation}
where $C_{ji}$ ($j = 1,2$) are complex constants which depend on the boundary conditions in eq.~(\ref{eq:modes_Helmholtz_BC}) and we also used Euler's formula. Notice that we have to consider both complex conjugate solutions, $e^{jk_{ni}x_i}$ and $e^{-jk_{ni}x_i}$, since $\psi_{ni}$ are complex, in contrast to eq.~(\ref{eq:wave_eq_solution_time_complex}), where $e^{j\omega t}$ was sufficient to cover all physically relevant cases for the real sound field (see also footnote \footref{ftn:time_exponential_negative_frequencies}). More specifically, for our convention with $e^{j\omega t}$ as the time dependence, the components with $e^{-jk_{ni}x_i}$ propagate forwards (in the direction of the $x_i$-axis) and the components with $e^{jk_{ni}x_i}$ propagate backwards (in the opposite direction). Indeed, the solution in eq.~(\ref{eq:modes_rect_room}) suffices as the ansatz for plane waves in free-space when no other spatial constraints, such as boundaries, are given.

The last equations have 9 unknowns -- 6 complex constants $C_{ji}$ and three components $k_{ni}$ (as already mentioned, complex $a_n$ and $b_n$ in the boundary conditions can introduce damping and accordingly complex values of the wave number). They depend on the particular boundary conditions at the six plane surfaces of the rectangular room, which are in general given by eq.~(\ref{eq:modes_Helmholtz_BC}). If we assume that all surfaces are passive, with $\hat c_n(\boldsymbol y) = 0$, then
\begin{equation}\label{eq:modes_Helmholtz_BC_rect_room}
\begin{aligned}
a_n \psi_n(\boldsymbol y) + b_n \nabla_y \psi_n(\boldsymbol y) \cdot \boldsymbol n(\boldsymbol y) = 0
\end{aligned}
\end{equation}
for locations $\boldsymbol y$ at the surfaces and $\boldsymbol n$ pointing into the room. Notice that even though we can separate the variables (and then the equations) $y_1$, $y_2$, and $y_3$ in the eigenmodes like $x_1$, $x_2$, and $x_3$ above, the same might not hold for the vector $\boldsymbol n$, which is also a function of $\boldsymbol y$. The scalar product with $\boldsymbol n$ may still couple the variables.

Conveniently, we set the coordinate axes such that the room extends from $0$ to $L_1$, $0$ to $L_2$, and $0$ to $L_3$ along the axes $x_1$, $x_2$, and $x_3$, respectively. The component $n_1$ of the vector $\boldsymbol n$ is then non-zero only if the two surfaces which are normal to the $x_1$-axis, that is only for $y_1 = 0$ or $L_1$. Similarly, $n_2 \neq 0$ for $y_2 = 0$ or $L_2$ and $n_3 \neq 0$ for $y_3 = 0$ or $L_3$. This fact will allow us to separate the variables also in the boundary conditions and it is directly due to the rectangular shape of the room and the choice of Cartesian coordinates. More precisely:
\begin{equation}
\begin{aligned}
\boldsymbol n(y_1=0) &= (1,0,0)^T,\\
\boldsymbol n(y_1=L_1) &= (-1,0,0)^T,\\
\boldsymbol n(y_2=0) &= (0,1,0)^T,\\
\boldsymbol n(y_2=L_2) &= (0,-1,0)^T,\\
\boldsymbol n(y_3=0) &= (0,0,1)^T,\\
\boldsymbol n(y_3=L_3) &= (0,0,-1)^T,
\end{aligned}
\end{equation}
where the argument $y_1=0$ denotes simply the boundary which is normal to the $x_1$-axis and shares the point $x_1 = 0$ with it, and similarly for the other boundaries. In this way, a uniform boundary condition ($a_n$ and $b_n$ are assumed to be constant over each surface) can be defined for each of the six surfaces with separate variables. For more complicated (and realistic) geometries or non-uniform boundary conditions, the separation can be much more difficult. 

Next we insert the values of the unit vectors in eq.~(\ref{eq:modes_Helmholtz_BC_rect_room}), substitute $\psi_n = \psi_{n1}\psi_{n2}\psi_{n3}$ and neglect the trivial cases when $\psi_{ni} = 0$ for any $i=1,2,3$, so the six boundary conditions together with eq.~(\ref{eq:modes_rect_room}) give
\begin{align*}\label{eq:modes_Helmholtz_BC_rect_room_surfaces}
\begin{aligned}
a_{n10} \psi_{n1} + b_{n10} \frac{d \psi_{n1}}{dy_1} &= a_{n10} C_{11}e^{-j k_{n1} y_1} + a_{n10} C_{21}e^{j k_{n1} y_1} &\\ & + b_{n10} C_{11}(-j k_{n1})e^{-j k_{n1} y_1} + b_{n10} C_{21}(j k_{n1})e^{j k_{n1} y_1} = 0 \text{ for } y_1 = 0, &\\
a_{n1L} \psi_{n1} - b_{n1L} \frac{d \psi_{n1}}{dy_1} &= a_{n1L} C_{11}e^{-j k_{n1} y_1} + a_{n1L} C_{21}e^{j k_{n1} y_1} &\\ & - b_{n1L} C_{11}(-j k_{n1})e^{-j k_{n1} y_1} - b_{n1L} C_{21}(j k_{n1})e^{j k_{n1} y_1} = 0 \text{ for } y_1 = L_1, &\\
a_{n20} \psi_{n2} + b_{n20} \frac{d \psi_{n2}}{dy_2} &= a_{n20} C_{12}e^{-j k_{n2} y_2} + a_{n20} C_{22}e^{j k_{n2} y_2} &\\ & + b_{n20} C_{12}(-j k_{n2})e^{-j k_{n2} y_2} + b_{n20} C_{22}(j k_{n2})e^{j k_{n2} y_2} = 0 \text{ for } y_2 = 0, &\\
a_{n2L} \psi_{n2} - b_{n2L} \frac{d \psi_{n2}}{dy_2} &= a_{n2L} C_{12}e^{-j k_{n2} y_2} + a_{n2L} C_{22}e^{j k_{n2} y_2} &\\ & - b_{n2L} C_{12}(-j k_{n2})e^{-j k_{n2} y_2} - b_{n2L} C_{22}(j k_{n2})e^{j k_{n2} y_2} = 0 \text{ for } y_2 = L_2, &\\
a_{n30} \psi_{n3} + b_{n30} \frac{d \psi_{n3}}{dy_3} &= a_{n30} C_{13}e^{-j k_{n3} y_3} + a_{n30} C_{23}e^{j k_{n3} y_3} &\\ & + b_{n30} C_{13}(-j k_{n3})e^{-j k_{n3} y_3} + b_{n30} C_{23}(j k_{n3})e^{j k_{n3} y_3} = 0 \text{ for } y_3 = 0, &\\
a_{n3L} \psi_{n3} - b_{n3L} \frac{d \psi_{n3}}{dy_3} &= a_{n3L} C_{13}e^{-j k_{n3} y_3} + a_{n3L} C_{23}e^{j k_{n3} y_3} &\\ & - b_{n3L} C_{13}(-j k_{n3})e^{-j k_{n3} y_3} - b_{n3L} C_{23}(j k_{n3})e^{j k_{n3} y_3} = 0 \text{ for } y_3 = L_3. &\\
\end{aligned}
\end{align*}
The coefficients $a_{n10}$ and $b_{n10}$ are defined for the surface $y_1 = 0$, $a_{n1L}$ and $b_{n1L}$ for the surface $y_1 = L_1$, $a_{n20}$ and $b_{n20}$ for the surface $y_2 = 0$, and so on. In general, the coefficients $a_{n}$ (as well as $b_{n}$) are not all equal. After inserting the values of $y_i$,
\begin{equation}\label{eq:modes_Helmholtz_BC_rect_room_surfaces_normal}
\begin{aligned}
C_{11}(a_{n10}-j b_{n10} k_{n1}) + C_{21}(a_{n10}+j b_{n10} k_{n1}) &= 0,\\
C_{11}(a_{n1L}+j b_{n1L} k_{n1})e^{-j k_{n1} L_1} + C_{21}(a_{n1L} - j b_{n1L} k_{n1})e^{j k_{n1} L_1} &= 0, \\
C_{12}(a_{n20}-j b_{n20} k_{n2}) + C_{22}(a_{n20}+j b_{n20} k_{n2}) &= 0,\\
C_{12}(a_{n2L}+j b_{n2L} k_{n2})e^{-j k_{n2} L_2} + C_{22}(a_{n2L} - j b_{n2L} k_{n2})e^{j k_{n2} L_2} &= 0, \\
C_{13}(a_{n30}-j b_{n30} k_{n3}) + C_{23}(a_{n30}+j b_{n30} k_{n3}) &= 0,\\
C_{13}(a_{n3L}+j b_{n3L} k_{n3})e^{-j k_{n3} L_3} + C_{23}(a_{n3L}-j b_{n3L} k_{n3})e^{j k_{n3} L_3} &= 0.
\end{aligned}
\end{equation}
This gives 6 equations for the 9 unknowns $C_{ji}$ and $k_{ni}$. In fact, these are 3 systems of two equations. In order to have non-trivial solutions for the coefficients $C_{ji}$, their determinants have to equal zero, that is,
\begin{align*}
\begin{aligned}
(a_{n10}-j b_{n10} k_{n1})(a_{n1L} - j b_{n1L} k_{n1})e^{j k_{n1} L_1} &= (a_{n10}+j b_{n10} k_{n1})(a_{n1L}+j b_{n1L} k_{n1})e^{-j k_{n1} L_1}, \\
(a_{n20}-j b_{n20} k_{n2})(a_{n2L} - j b_{n2L} k_{n2})e^{j k_{n2} L_2} &= (a_{n20}+j b_{n20} k_{n2})(a_{n2L}+j b_{n2L} k_{n2})e^{-j k_{n2} L_2}, \\
(a_{n30}-j b_{n30} k_{n3})(a_{n3L} - j b_{n3L} k_{n3})e^{j k_{n3} L_3} &= (a_{n30}+j b_{n30} k_{n3})(a_{n3L}+j b_{n3L} k_{n3})e^{-j k_{n3} L_3},
\end{aligned}
\end{align*}
or after multiplication with $e^{j k_{ni} L_i}$,
\begin{align*}
\begin{aligned}
e^{2j k_{n1} L_1} = \frac{(a_{n10}+j b_{n10} k_{n1})(a_{n1L}+j b_{n1L} k_{n1})}{(a_{n10}-j b_{n10} k_{n1})(a_{n1L} - j b_{n1L} k_{n1})}, & \\
e^{2j k_{n2} L_2} = \frac{(a_{n20}+j b_{n20} k_{n2})(a_{n2L}+j b_{n2L} k_{n2})}{(a_{n20}-j b_{n20} k_{n2})(a_{n2L} - j b_{n2L} k_{n2})}, & \\
e^{2j k_{n3} L_3} = \frac{(a_{n30}+j b_{n30} k_{n3})(a_{n3L}+j b_{n3L} k_{n3})}{(a_{n30}-j b_{n30} k_{n3})(a_{n3L} - j b_{n3L} k_{n3})}. &
\end{aligned}
\end{align*}

Hence, we have derived three equations for the three components of the wave vector, $k_{ni}$, which can be solved for any given values of the coefficients $a_n$ and $b_n$ (although the equations are transcendental, which prohibits a closed-form analytical solution in general). The obtained values of $k_n$ can even be complex, if damping is introduced by the boundary conditions (complex $a_n$ and $b_n$), with the same physical interpretation as in eq.~(\ref{eq:complex_wave_number}). In any case, the values can be used back in eq.~(\ref{eq:modes_Helmholtz_BC_rect_room_surfaces_normal}) to find the values of the coefficients $C_{ji}$, which also determine the eigenmodes, according to eq.~(\ref{eq:modes_rect_room}).

For simplicity, we will assume that all surfaces are equal, that is, $a_{n10} = a_{n1L} = a_{n20} = a_{n2L} = a_{n30} = a_{n3L} = a_{n}$ and similarly for $b_{n}$. The last equations reduce to
\begin{align*}
\begin{aligned}
e^{2j k_{n1} L_1} = \frac{(a_{n}+j b_{n} k_{n1})^2}{(a_{n}-j b_{n} k_{n1})^2} \Rightarrow e^{j k_{n1} L_1} = \pm \frac{a_{n}+j b_{n} k_{n1}}{a_{n}-j b_{n} k_{n1}}, & \\
e^{2j k_{n2} L_2} = \frac{(a_{n}+j b_{n} k_{n2})^2}{(a_{n}-j b_{n} k_{n2})^2} \Rightarrow e^{j k_{n2} L_2} = \pm \frac{a_{n}+j b_{n} k_{n2}}{a_{n}-j b_{n} k_{n2}}, & \\
e^{2j k_{n3} L_3} = \frac{(a_{n}+j b_{n} k_{n3})^2}{(a_{n}-j b_{n} k_{n3})^2} \Rightarrow e^{j k_{n3} L_3} = \pm \frac{a_{n}+j b_{n} k_{n3}}{a_{n}-j b_{n} k_{n3}}, &
\end{aligned}
\end{align*}
with arbitrary signs $\pm$. If in addition to this all surfaces are rigid and motionless (acoustically \textbf{hard} and passive), normal component of the pressure gradient at them is zero (eq.~(\ref{eq:rigid_wall_BC_p})) and the boundary condition in eq.~(\ref{eq:modes_Helmholtz_BC_rect_room}) reads (with $a_n = 0$ and $b_n = 1$\,m)
\begin{equation}\label{eq:modes_Helmholtz_BC_hard_wall}
\nabla_y \psi_n(\boldsymbol y) \cdot \boldsymbol n(\boldsymbol y) = 0.
\end{equation}
Hence,
\begin{align*}
\begin{aligned}
e^{j k_{n1} L_1} = \pm 1, & \\
e^{j k_{n2} L_2} = \pm 1, & \\
e^{j k_{n3} L_3} = \pm 1, &
\end{aligned}
\end{align*}
which is satisfied for
\begin{equation}\label{eq:modes_rect_room_hard_wall_kn_values}
\begin{aligned}
k_{n1} &= n_{L_1} \pi/L_1, \\
k_{n2} &= n_{L_2} \pi/L_2, \\
k_{n3} &= n_{L_3} \pi/L_3,
\end{aligned}
\end{equation}
where\footnote{Negative values $n_{L_i}=-1, -2...$ with $k_{ni} < 0$ would also satisfy the equalities. However, these solutions are also included by eq.~(\ref{eq:modes_rect_room}), since eq.~(\ref{eq:modes_Helmholtz_BC_rect_room_surfaces_normal_hard_walls}) will show that $C_{11} = C_{21}$, $C_{12} = C_{22}$, and $C_{13} = C_{23}$. Therefore, we can consider only non-negative $k_{n1}$, $k_{n2}$, and $k_{n3}$ and still obtain all modes.} $n_{L_i} = n_i = 0, 1, 2...$ for $i = 1,2,3$ match the modal orders. Components of the wave vector are thus real (the surfaces are not absorbing).

The dimensionless quantity $kL$, where $L$ is some characteristic length scale of the acoustic problem (here room dimension) and $k$ is wave number (here component of a wave vector), is called \textbf{Helmholtz number}, $H_e$. According to eq.~(\ref{eq:omega_k}), it equals
\begin{equation}\label{eq:Helmholtz_number}
\boxed{ H_e = k L = \frac{\omega L}{c_0} = \frac{2\pi f L}{c_0} = \frac{2 \pi L}{\lambda} }.
\end{equation}
Hence, it expresses ratio of the characteristic length scale of the geometry of the problem ($L$) and the sound waves ($1/k = \lambda/(2\pi)$). It allows us do describe something as acoustically very small (compact), when $H_e \ll 1$, or very large when $H_e \gg 1$ (or neither, if $H_e \sim \mathcal{O}(1)$). This has such strong practical implications, that whenever we say that something is small or large (a room, an object, a distance, etc.) in the context of classical acoustics, we do not mean it in terms of absolute lengths in meters, but in terms of the the value of the Helmholtz number (with an appropriate length scale and for the given frequency). For example, eq.~(\ref{eq:modes_rect_room_hard_wall_kn_values}) suggests here (recalling that not all $n_{L_i}$ can be zero) that room modes cannot occur for Helmholtz number values below around one, with the rectangular room dimensions as the characteristic length scales. The same applies for rooms of any shapes, as long as some characteristic room dimension can be determined. Much below this value, the low-frequency sound waves cannot propagate inside the room's volume and their amplitudes and phases are not location-dependent. Sound pressure is uniform in the room (but does vary in time). The non-acoustic mode $(0,0,0)$ has the same behaviour. 

Each mode $n$ is associated with one particular combination of $n_{L_1}$, $n_{L_2}$, and $n_{L_3}$. Its wave number $k_n$ can be calculated from equations~(\ref{eq:modes_rect_room_kn}) and (\ref{eq:modes_rect_room_hard_wall_kn_values}). In order to determine shapes of the eigenmodes for the rectangular room with hard walls, we notice that eq.~(\ref{eq:modes_Helmholtz_BC_rect_room_surfaces_normal}) gives for the zero coefficients $a_n$
\begin{equation}\label{eq:modes_Helmholtz_BC_rect_room_surfaces_normal_hard_walls}
\begin{aligned}
-C_{11} + C_{21} &= 0,\\
C_{11}e^{-j k_{n1} L_1} - C_{21}e^{j k_{n1} L_1} &= 0, \\
-C_{12} + C_{22} &= 0,\\
C_{12}e^{-j k_{n2} L_2} - C_{22}e^{j k_{n2} L_2} &= 0, \\
-C_{13}+ C_{23} &= 0,\\
C_{13}e^{-j k_{n3} L_3} - C_{23}e^{j k_{n3} L_3} &= 0.
\end{aligned}
\end{equation}
For $k_{ni}$ from eq.~(\ref{eq:modes_rect_room_hard_wall_kn_values}), these are only 3 linearly independent equations,
\begin{equation}\label{eq:modes_Helmholtz_BC_rect_room_surfaces_normal_hard_walls_k_n}
\begin{aligned}
C_{11} - C_{21} &= 0, \\
C_{12} - C_{22} &= 0, \\
C_{13} - C_{23} &= 0.
\end{aligned}
\end{equation}
We can use them in eq.~(\ref{eq:modes_rect_room}) to obtain
\begin{align*}
\begin{aligned}
\psi_{n1} &= 2 C_{11} \cos(k_{n1} x_1), \\
\psi_{n2} &= 2 C_{12} \cos(k_{n2} x_2), \\
\psi_{n3} &= 2 C_{13} \cos(k_{n3} x_3).
\end{aligned}
\end{align*}
Simple (and real) cosine functions are thus characteristic for hard surfaces. Equation~(\ref{eq:modes_rect_room_product}) becomes
\begin{equation}\label{eq:modes_rect_room_hard_wall_shape}
\begin{aligned}
\psi_n (\boldsymbol x) = \psi_{n1} \psi_{n2} \psi_{n3} &= 8 C_{11} C_{12} C_{13} \cos(k_{n1} x_1) \cos(k_{n2} x_2) \cos(k_{n3} x_3) \\ &= 8 C_{11} C_{12} C_{13} \prod_{i=1}^{3} \cos(k_{ni} x_i).
\end{aligned}
\end{equation}
As expected, three coefficients, $C_{11}$, $C_{12}$, and $C_{13}$, remain unknown. Their product represents an arbitrary constant which can multiply an eigenfunction without affecting its eigenfrequency.

We can now write the tailored Green's function with damping, given in eq.~(\ref{eq:tailored_Green_Helmholtz_modes_An_damping}):
\begin{align*}
\begin{aligned}
\hat{G}_{tail}(\boldsymbol x | \boldsymbol y) = 64 c_0^2 C_{11}^2 C_{12}^2 C_{13}^2 &\sum_{n} \frac{\prod_{i=1}^{3} \cos(k_{ni} x_i) \prod_{i=1}^{3} \cos(k_{ni} y_i)}{ K_n(\omega_n^2 + 2j \zeta_n \omega_n -\omega^2)}.&
\end{aligned}
\end{align*}
In doing so, as already discussed, we actually model the weak damping, either at the surfaces or in the medium inside the room, while keeping the boundary conditions and eigenfunctions simple. The constant $K_n$ can also be determined from eq.~(\ref{eq:modes_orthogonality}):
\begin{align*}
\begin{aligned}
K_n &= \int_V \psi_n^* (\boldsymbol x) \psi_n (\boldsymbol x) d^3 \boldsymbol x &\\
&= 64 C_{11}^2 C_{12}^2 C_{13}^2 \int_{0}^{L1} \int_{0}^{L2} \int_{0}^{L3} \cos^2(k_{n1} x_1) \cos^2(k_{n2} x_2) \cos^2(k_{n3} x_3) dx_1 dx_2 dx_3 &\\
&= 64 C_{11}^2 C_{12}^2 C_{13}^2 \int_{0}^{L1} \cos^2(k_{n1} x_1) dx_1 \int_{0}^{L2} \cos^2(k_{n2} x_2) dx_2  \int_{0}^{L3} \cos^2(k_{n3} x_3) dx_3.&
\end{aligned}
\end{align*}
The first integral equals
\begin{equation}\label{spatial_integal_of_squared_cos}
\begin{aligned}
\int_{0}^{L1} \cos^2(k_{n1} x_1) dx_1 &= \frac{1}{k_{n1}}\int_{0}^{k_{n1} L1} \cos^2(k_{n1} x_1) d(k_{n1}x_1) &\\
&= \frac{1}{2 k_{n1}} [ (k_{n1} L_1) + \sin(k_{n1} L_1) \cos(k_{n1} L_1)] = \frac{L_1}{2}&
\end{aligned}
\end{equation}
for $k_{n1} \neq 0$ or $L_1$ for $k_{n1} = 0$. For the last equality in eq.~(\ref{spatial_integal_of_squared_cos}), we used $\sin(k_{n1} L_1) = 0$, which follows from eq.~(\ref{eq:modes_rect_room_hard_wall_kn_values}). We can express the remaining two integrals similarly, which gives
\begin{align*}
K_n = 8 C_n C_{11}^2 C_{12}^2 C_{13}^2 L_1 L_2 L_3 = 8 C_n C_{11}^2 C_{12}^2 C_{13}^2 V,
\end{align*}
where $V = L_1 L_2 L_3$ is volume of the room and
\begin{equation}
C_n =
\begin{cases}
4 &\text{for single $k_{ni} \neq 0$ (axial modes)}\\
2 &\text{for two $k_{ni} \neq 0$ (tangential modes)}\\
1 &\text{for all three $k_{ni} \neq 0$ (oblique modes)}.
\end{cases} 
\end{equation}
Hence,
\begin{equation}\label{eq:tailored_Green_Helmholtz_modes_damping_rect_room_hard_wall}
\begin{aligned}
\hat{G}_{tail}(\boldsymbol x | \boldsymbol y) =  \frac{8 c_0^2}{C_n L_1 L_2 L_3} \sum_{n} \frac{\prod_{i=1}^{3} \cos(k_{ni} x_i) \prod_{i=1}^{3} \cos(k_{ni} y_i)}{ \omega_n^2 + 2j \zeta_n \omega_n -\omega^2}.
\end{aligned}
\end{equation}
Importantly, the three unknown coefficients cancel out after the normalization with $K_n$, which makes them irrelevant for the tailored Green's function. Non-uniqueness of the modes is thus not reflected in the Green's function.

Finally, complex sound pressure amplitude can be calculated from eq.~(\ref{eq:solution_tailored_Green_Helmholtz}) or, if the sound source is compact and omnidirectional, from eq.~(\ref{eq:solution_tailored_Green_Helmholtz_modes}):
\begin{equation}\label{eq:solution_tailored_Green_Helmholtz_modes_rect_room_hard_wall}
\begin{aligned}
\boxed{ \hat{p} (\boldsymbol x) = \frac{8c_0^2 \hat{Q}(\boldsymbol y)}{C_n L_1 L_2 L_3} \sum_{n} \frac{\prod_{i=1}^{3} \cos(k_{ni} x_i) \prod_{i=1}^{3} \cos(k_{ni} y_i)}{ \omega_n^2+2j \zeta_n \omega_n -\omega^2} }.
\end{aligned}
\end{equation}

\begin{figure}[h]
	\centering
	\includegraphics[width=0.685\textwidth]{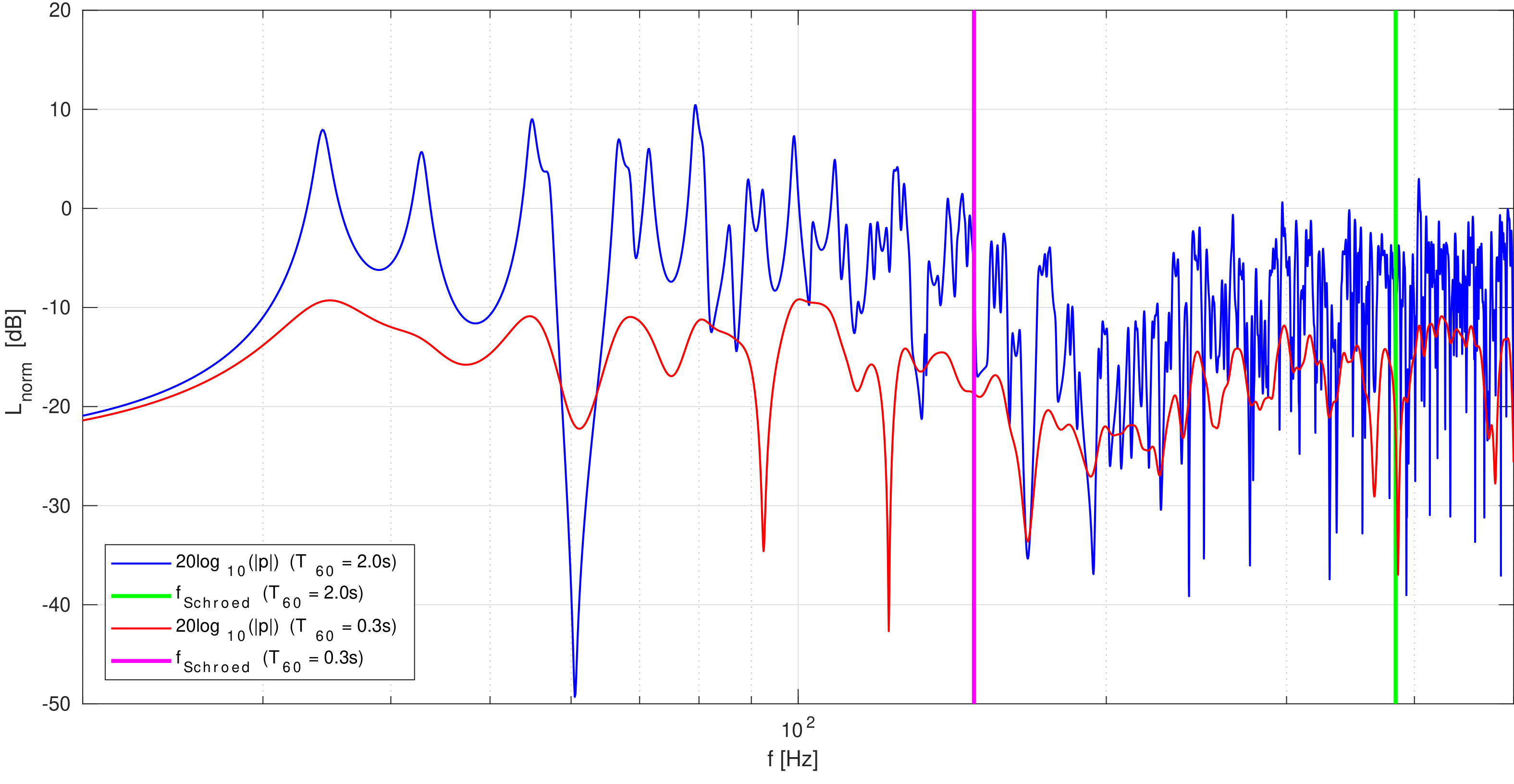}
	\includegraphics[width=0.69\textwidth]{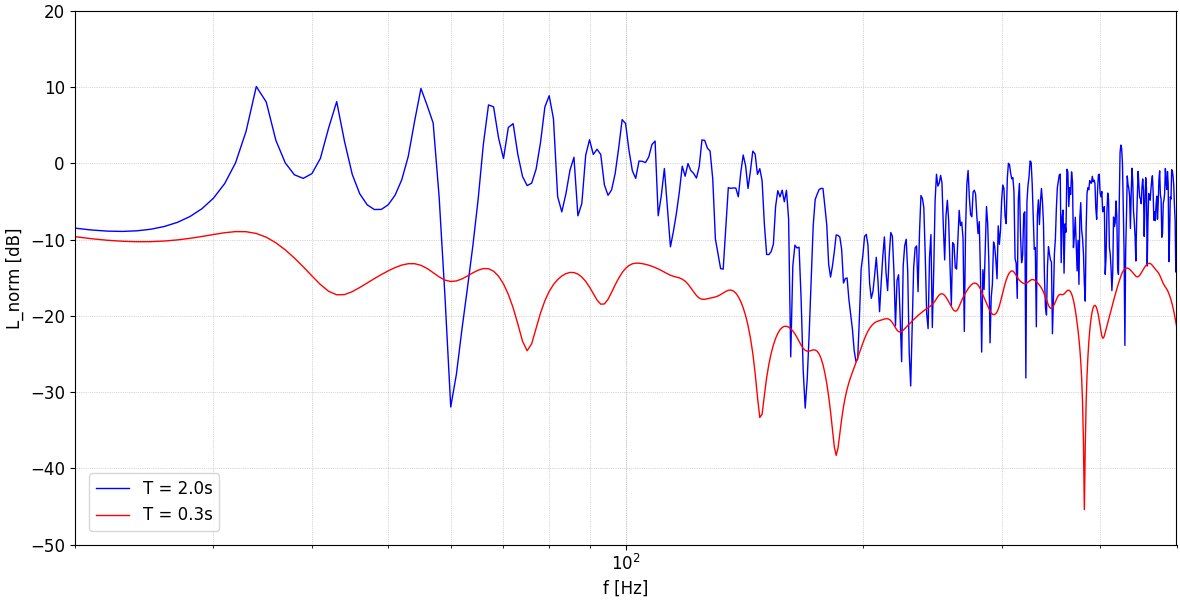}
	\caption{Frequency response of a rectangular room with hard walls for a point monopole source with constant $\hat Q = 1$\,kg/s$^2$; the room dimensions are $(L_1, L_2, L_3) = (5\text{\,m}, 4\text{\,m}, 3\text{\,m})$, receiver location is $(x_1, x_2, x_3)^T = (4.5\text{\,m}, 0.5\text{\,m}, 0.5\text{\,m})^T$, and source location is $(y_1, y_2, y_3)^T = (0.1\text{\,m}, 0.1\text{\,m}, 0.1\text{\,m})^T$; speed of sound is $c_0 = 343$\,m/s and reverberation time ($T_{60} = 2$\,s or $T_{60} = 0.3$\,s) is frequency-independent; the frequency response is calculated (above) analytically from the lowest 1000 modes (with eigenfrequencies up to 512\,Hz), with the frequency resolution 0.1\,Hz and (below) numerically, with the frequency resolution 1\,Hz.}
	\label{fig:rect_room_freq_response}
\end{figure}

The upper part of Fig.~\ref{fig:rect_room_freq_response} shows amplitude of the frequency response (in dB) calculated using eq.~(\ref{eq:tailored_Green_Helmholtz_modes_damping_rect_room_hard_wall}) (which is equivalent to eq.~(\ref{eq:solution_tailored_Green_Helmholtz_modes_rect_room_hard_wall}) with $\hat Q = 1$\,kg/s$^2$) for two values of frequency-independent reverberation time, 0.3\,s and 2\,s. The damping constant is calculated with eq.~(\ref{eq:T60_damping}) and the frequency resolution is 0.1\,Hz. Figure~\ref{fig:rect_room_freq_response} (below) shows the same frequency responses calculated numerically, with lower frequency resolution of 1\,Hz (for shorter computation time).

Notice that the overlapping modes can add constructively or (partly) cancel each other, depending on the signs of the cosines in eq.~(\ref{eq:tailored_Green_Helmholtz_modes_damping_rect_room_hard_wall}), which represent the modes $\psi_n$. This causes higher peaks or deeper drops of the amplitude of the response or sound pressure, $|p| = |\hat{p}|$. The figure also demonstrates strong fluctuations of the response at all frequencies. These are most often undesired, particularly at low frequencies, when the eigenfrequencies are not close enough to be indistinguishable by human auditory system. The fluctuations can be smoothed out by increasing damping of the modes, that is, decreasing their Q-factor (compare with Fig.~\ref{fig:resonance}). Additional consequence of higher damping is drop of the total energy of the sound field in the room, which can be desirable for noise control, but also undesirable, if the sound carries a useful information and the source is weak. Notice also that the overall increase at low frequencies largely vanishes, if we define the frequency response as $\hat p/(\hat{\boldsymbol v} \cdot \boldsymbol n) \propto \omega \hat p/\hat Q$.

\subsection{Modal density}\label{ch:density_of_modes}

It is clear from Fig.~\ref{fig:rect_room_freq_response} that the density of modes on the frequency axis and their overlapping increase with frequency. This inevitably leads to higher computational burden of both analytical and numerical calculations of sound fields. On the other hand, human auditory system has a limited frequency resolution which decreases with frequency, which naturally averages out fluctuations in the spectra of the received sounds. It averages sound energy within certain finite frequency bands in a similar manner as it performs averaging over time. This renders detailed calculations of each mode at high frequencies unnecessary for most of the applications. Therefore, it is of interest to inspect how the modal density and their overlapping depend on frequency. Further statistical treatment of sound fields in frequency domain (in particular the result in eq.~(\ref{eq:solution_tailored_Green_Helmholtz_modes_rect_room_hard_wall})) will be considered in section~\ref{diffuse_field_modal_analysis}. 

According to eq.~(\ref{eq:modes_rect_room_kn}), eigenfrequencies correspond to the points in $k$-space scaled with the factor $f/k = c_0/(2\pi)$. For the rectangular room, using eq.~(\ref{eq:modes_rect_room_hard_wall_kn_values}), 
\begin{equation}\label{eq:modes_rect_room_fn}
\boxed{ f_n^2 = \left( \frac{c_0}{2 \pi} \right)^2 k_n^2 = \frac{c_0^2}{4 \pi^2} \sum_{i=1}^3 k_{ni}^2 = \left(\frac{c_0 n_{L_1}}{2L_1}\right)^2 + \left(\frac{c_0 n_{L_2}}{2L_2}\right)^2 + \left(\frac{c_0 n_{L_3}}{2L_3}\right)^2 }.
\end{equation}
The number of frequencies $f_n$ below some frequency $f$ is approximately equal to the ratio of volume $(4 f^3 \pi/3)/8$ of the spherical sector in the first octant of the scaled $k$-space and its smallest volume element $(c_0/2L_1)(c_0/2L_2)(c_0/2L_3) = c_0^3/(8V)$, which follows from eq.~(\ref{eq:modes_rect_room_fn}) with $n_{L_1}=n_{L_2}=n_{L_3}=0$. The ratio equals
\begin{equation}\label{eq:modes_rect_room_Nfn}
N_{f_n<f} \approx \frac{(4 f^3 \pi/3)/8}{c_0^3/(8V)} = \frac{4 \pi V f^3}{3c_0^3}.
\end{equation}
This is only approximate for two reasons. First, the $k$-space is discretized with cuboids, which cannot exactly fit the curvature of one eight of a sphere. The inaccuracy is especially large for small spherical sectors, with volumes not much larger than the volume element, in which cases $N_{f_n<f}$ cannot even be approximated with any continuous function of frequency. Second, points on the flat boundaries and edges of the octant are not counted equally as the inner points. In particular, only half of the modes with exactly one of the values $n_{L_i}$ equal to zero (the surface modes) are included (the other half left to the adjacent octants) and only one quarter of the modes with two out of three $n_{L_i}$ equal to zero (the edge modes). Consequently, the approximation is satisfactory only when $k L \gg 1$, where $L$ is the smallest characteristic length scale of the room, so the number of the omitted modes on the boundaries and edges as well as on the curved surface of the spherical sector is small compared to the total number of points.

\begin{figure}[h]
	\centering
	\begin{subfigure}{.5\textwidth}
		\centering
		\includegraphics[width=1.05\linewidth]{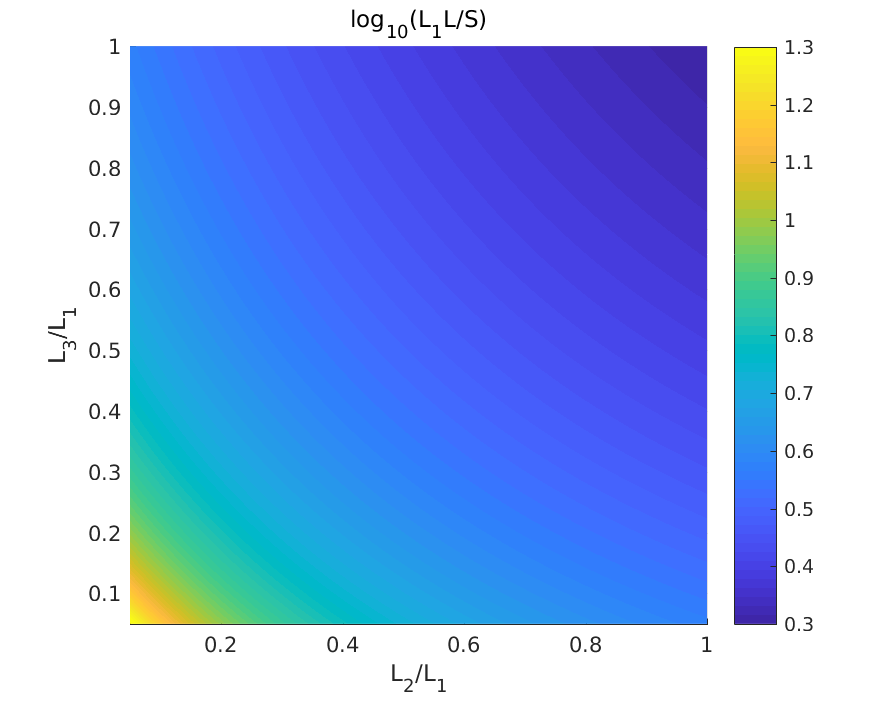}
		\label{fig:L_S_ratio_rect_room}
	\end{subfigure}%
	\begin{subfigure}{.5\textwidth}
		\centering
		\includegraphics[width=1.05\linewidth]{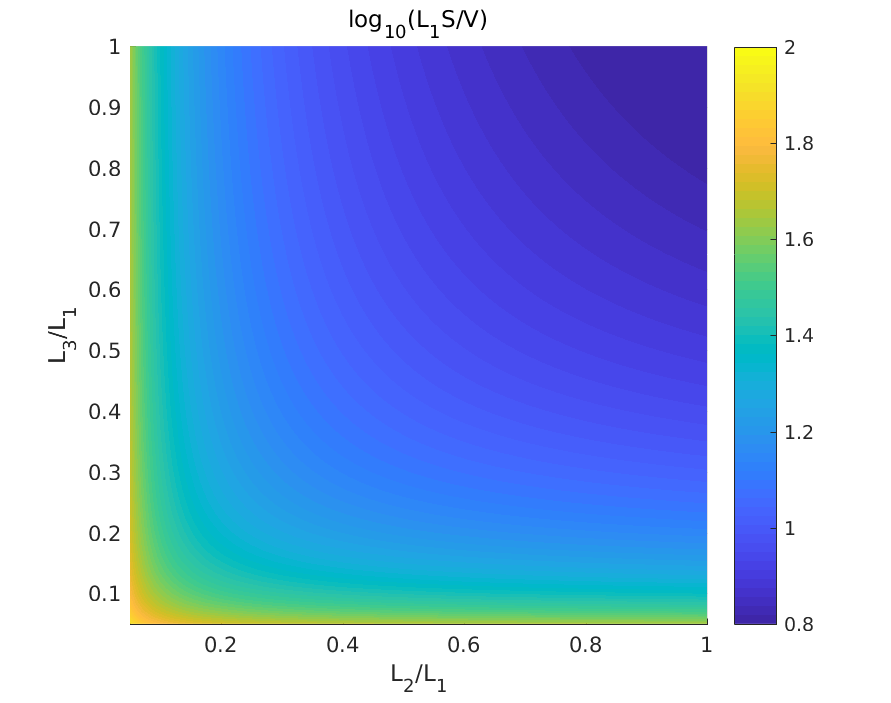}
		\label{fig:S_V_ratio_rect_room}
	\end{subfigure}
	\caption{Logarithm of the ratio (left) $L/S$ and (right) $S/V$ in a rectangular room, scaled with the longest dimension of the room, $L_1$.}
	\label{fig:L_S_V_ratios_rect_room}
\end{figure}

The expression which includes the missing half of the surface modes and three quarters of the edge modes can also be derived and it introduces two additional terms:
\begin{equation}\label{eq:modes_rect_room_Nfn_corrected}
N_{f_n<f} = \frac{4 \pi V f^3}{3c_0^3} + \frac{\pi S f^2}{4c_0^2} + \frac{L f}{8c_0},
\end{equation}
where $S = 2(L_1 L_2 + L_1 L_3 + L_2 L_3)$ is total area of the boundary surfaces of the room and $L = 4 (L_1 + L_2 + L_3)$ is total length of the edges. For later reference, Fig.~\ref{fig:L_S_V_ratios_rect_room} shows logarithms of the ratios $L/S$ and $S/V$ for a rectangular room scaled with the longest dimension of the room, $L_1$. As we can see, the contribution of the last two terms in eq.~(\ref{eq:modes_rect_room_Nfn_corrected}) becomes small in comparison to the first term for high $kL \sim k S^{1/2} \sim k V^{1/3} \gg 1$ (assuming all room dimensions have lengths of the same order). It is also interesting to note that the last two and the following equations hold even for arbitrary room shapes, when $kL \gg 1$ (or, as often simply written, but more restrictive, $f \rightarrow \infty$). 

The number of modes increases rapidly with frequency, $N_{f_n<f} \propto f^3$. At high enough values of $kL$, when $N_{f_n<f}$ is approximately continuous function of frequency, \textbf{modal density} (the number of eigenfrequencies per hertz) equals the derivative
\begin{equation}\label{eq:modes_rect_room_fn_density}
\frac{dN_{f_n<f}}{df} = \frac{4 \pi V f^2}{c_0^3} + \frac{\pi S f}{2 c_0^2} + \frac{L}{8c_0} \approx \frac{4 \pi V f^2}{c_0^3}.
\end{equation}
On the other hand, overlapping of modes in frequency domain depends also on their widths. We recall that the resonance curves, which follow from eq.~(\ref{eq:solution_tailored_Green_Helmholtz_mode}) for particular $\boldsymbol x$ and $\boldsymbol y$ and have the shape shown in Fig.~\ref{fig:resonance}, have the half-widths $\Delta f_n = \Delta \omega_n / (2\pi) = \zeta_n/\pi$. If the number of different modes within this small frequency range is relatively high, say larger than 3, the modes largely overlap and no single mode is expected to dominate in the frequency response of the room. This is the case when $dN_{f_n<f}/df > 3/\Delta f_n$, that is, at the frequencies
\begin{equation}
\begin{aligned}
f > f_{Schroed} &= \sqrt{\frac{3 c_0^3}{4 \pi V \Delta f_n}} = \sqrt{\frac{3 c_0^3}{4 V \zeta_n}} \approx 5500 \sqrt{\frac{1\text{m}^3/\text{s}^3}{V \zeta_n}},
\end{aligned}
\end{equation}
for $c_0 = 343$\,m/s, where $f_{Schroed}$ is \textbf{the Schroeder frequency}, named after M. Schroeder (one of the biggest names in room acoustics), who first proposed it as a measure of modal overlapping. Notice that the value of 3 is chosen quite arbitrarily.

Expressed in terms of reverberation time in eq.~(\ref{eq:T60_damping}), the Schroeder frequency equals
\begin{equation}\label{eq:Schroeder_freq}
\begin{aligned}
\boxed{ f_{Schroed} \approx 2100 \sqrt{\frac{T_{60,\Delta f_n} \cdot 1\text{m}^3/\text{s}^3}{V}} },
\end{aligned}
\end{equation}
where $T_{60,\Delta f_n}$ denotes the value of reverberation time in the small frequency range $\Delta f_n$. The overlapping modes are thus assumed to have similar values of the damping constant and reverberation time and, strictly speaking, $\Delta f_n$ should include $f_{Schroed}$, for example, by iteratively estimating $f_{Schroed}$ and observing the appropriate $\Delta f_n$. For simplicity, however, broadband reverberation time or its average value in the three middle octave bands (500\,Hz, 1\,kHz, and 2\,kHz) is most often used in practice for estimating the Schroeder frequency. The inaccuracy thereby is usually not critical, since the Schroeder frequency is only an indicator of high or low modal overlapping, not a clear limit between the two regimes.

Schroeder frequency indicates roughly the low-frequency range in which resonant behaviour of a room can be pronounced and perceivable. As an example, its value for the analysed rectangular room is shown in the upper part of Fig.~\ref{fig:rect_room_freq_response} for the two values of reverberation time. Above the Schroeder frequency the overlap of a large number of modes results in a much smoother frequency response of the room (assuming averaging in relatively narrow frequency bands, at least $\sim \Delta f_n$ or larger), which allows statistical treatment and the model of a diffuse field, as will be demonstrated in section~\ref{diffuse_field_modal_analysis}. Figure~\ref{fig:rect_room_freq_response} also shows that the Schroeder frequency is lower in more damped rooms, which agrees with the fact that their modes have larger half-widths and thus overlap more. Similarly, large rooms with lower eigenfrequencies also have lower Schroeder frequencies.

\begin{figure}[h]
	\centering
	\begin{subfigure}{.49\textwidth}
		\centering
		\includegraphics[width=1\linewidth]{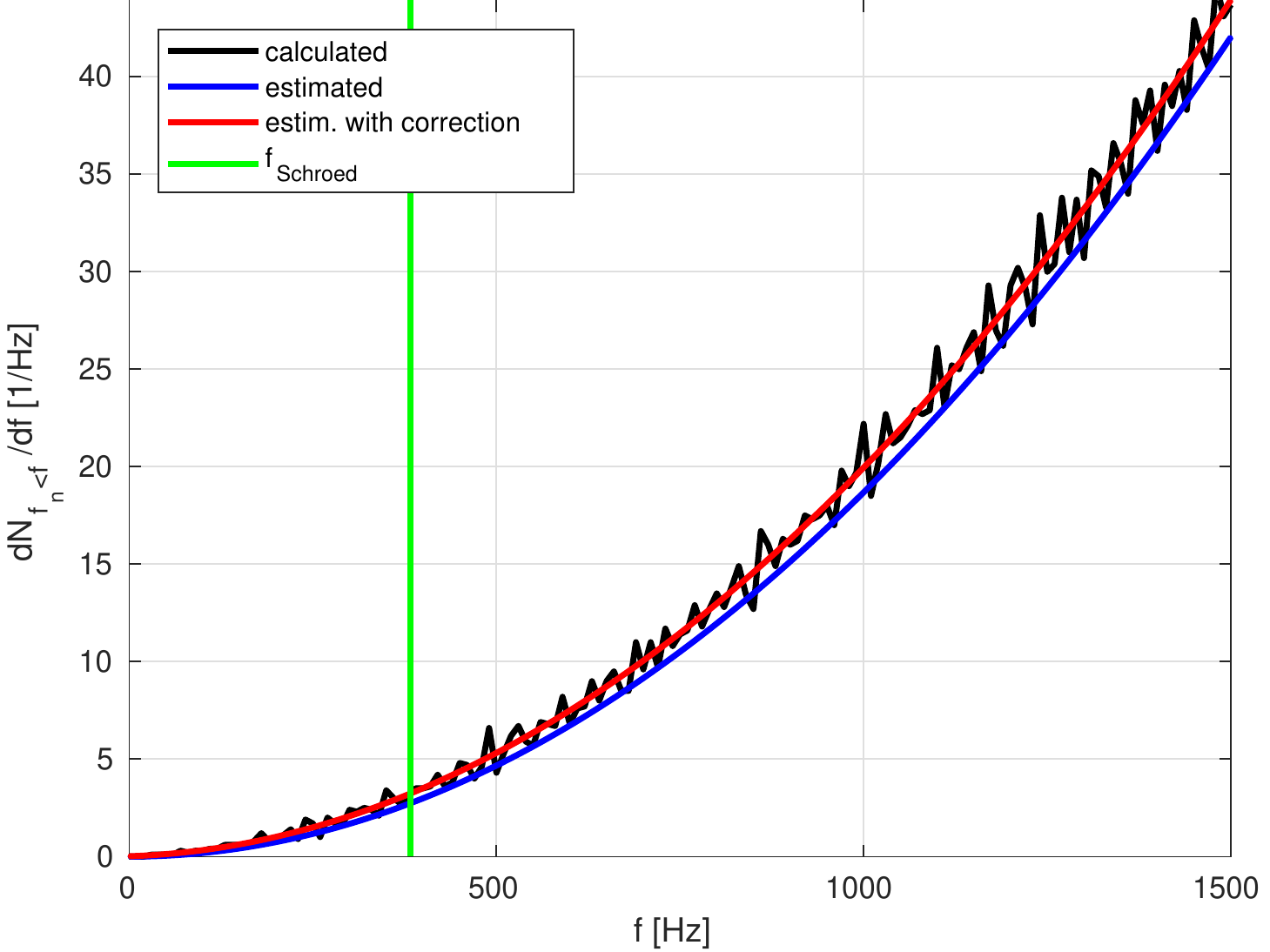}
		\label{fig:rect_room_modes_density}
	\end{subfigure}%
	\begin{subfigure}{.49\textwidth}
		\centering
		\includegraphics[width=1\linewidth]{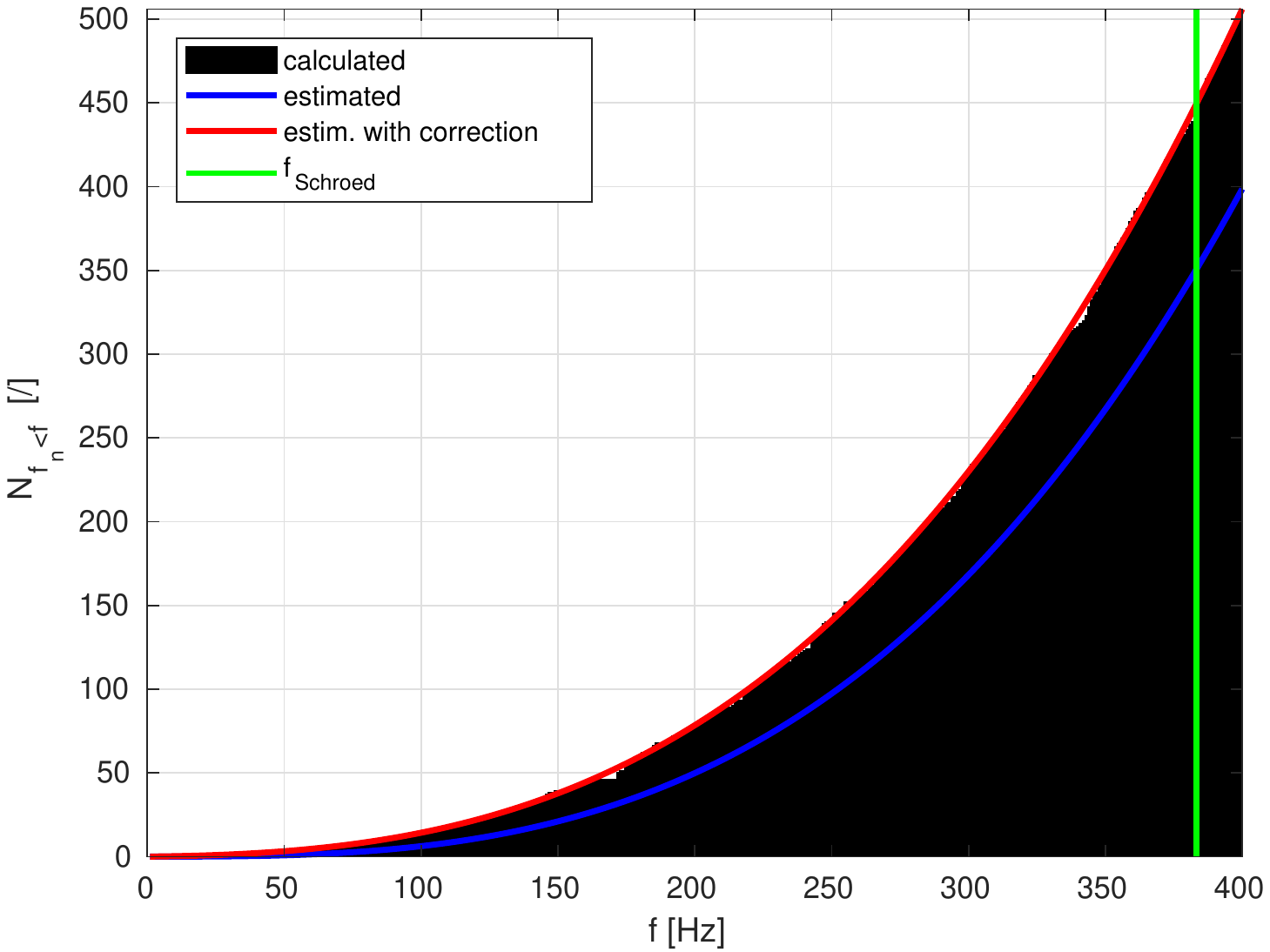}
		\label{fig:rect_room_modes_number}
	\end{subfigure}
	\caption{Modal density (left) and cumulative distribution (right) for a rectangular room with dimensions $(L_1, L_2, L_3) = (3\text{\,m}, 4\text{\,m}, 5\text{\,m})$. The exact values in black colour are obtained using eq.~(\ref{eq:modes_rect_room_fn}). For smoothness of the curve in the left diagram, the values are averaged in non-overlapping frequency bands of 10\,Hz. The estimated values (equations~(\ref{eq:modes_rect_room_Nfn})-(\ref{eq:modes_rect_room_fn_density})) are shown with and without the two last two terms in eq.~(\ref{eq:modes_rect_room_Nfn_corrected}). The Schroeder frequency is shown for the reverberation time 2\,s.}
	\label{fig:rect_room_modes}
\end{figure}

Figure~\ref{fig:rect_room_modes} shows density of the modes of the same room as in Fig.~\ref{fig:rect_room_freq_response} as well as their cumulative distribution, both exact (the eigenfrequencies are calculated using eq.~(\ref{eq:modes_rect_room_fn})) and estimated using equations~(\ref{eq:modes_rect_room_Nfn})-(\ref{eq:modes_rect_room_fn_density}) with and without the last two terms in eq.~(\ref{eq:modes_rect_room_Nfn_corrected}). The Schroeder frequency for $T_{60} = 2$s is also indicated, which is around 383\,Hz (this is more realistic value for the weakly damped room than for $T_{60} = 0.3$s, which is around 148\,Hz). The number of modes below Schroeder frequency can be quite large even for numerical calculations, especially in weakly damped rooms. The number of surface and edge modes which are included by the last two terms of eq.~(\ref{eq:modes_rect_room_Nfn_corrected}) increases with frequency and makes a relatively high fraction of the total number of modes even at the Schroeder frequency. However, the fraction drops with frequency as the number of volume modes increases more rapidly.

\bigskip

In this section we considered the modal theory of sound fields in rooms in frequency domain and obtained an analytical solution for a rectangular room with hard walls. Even such highly idealized case provided valuable information on the resonant behaviour of rooms and lead to some general conclusions. In practice, many rooms do actually have a rectangular shape or can be approximated so. The specific equations such as eq.~(\ref{eq:solution_tailored_Green_Helmholtz_modes_rect_room_hard_wall}) and eq.~(\ref{eq:modes_rect_room_fn}) can then be used for rough estimations of sound fields and resonance frequencies (sometimes even with surprising accuracy), especially when the deviations of the geometry from an ideal rectangular room are small compared to the considered wavelengths (at low frequencies, when the modal analysis is indeed relevant) and the rooms are weakly damped.

If complicated geometries and boundary conditions must be treated more accurately, solutions of the Helmholtz equation, tailored Green's functions, or room modes can be estimated \textbf{numerically}, typically using finite element method (FEM) or boundary element method (BEM). The former method relies on discretization of the (usually) three-dimensional computational domain, interior of a closed space. It can be used for solving the Helmholtz equation~(\ref{eq:Helmholtz_source}) directly, supplied with boundary conditions, such as the one in eq.~(\ref{eq:Helmholtz_p_BC}), as well as the source function, or, in a somewhat simpler case, for calculating the room modes for a given geometry and boundary conditions. The latter method is based on the discretization of surfaces. It can be used for solving the integral equation~(\ref{eq:solution_Green_Helmholtz}), especially in free space, with the known Green's function from eq.~(\ref{eq:Green_Helmholtz_free_space}). It is used less frequently for closed spaces.

The major challenge for obtaining accurate numerical (or analytical) solutions with modal analysis is including realistic boundary conditions. As already discussed, the modes have to satisfy the conditions at all boundaries simultaneously, since the time dependence is suppressed from the calculations. Moreover, complicated materials and micro geometries of surfaces cannot be easily expressed with equations such as eq.~(\ref{eq:Helmholtz_p_BC}). Apart from that, the applications of numerical solvers for room acoustics are still quite limited with reasonable computational costs. In practice, a relatively small number of modes can be calculated, which means that high accuracy can be achieved only at relatively low frequencies (that is, Helmholtz numbers, since the modal density depends also on the room dimensions), when the density of modes is low. Schroeder frequency can be seen as a very rough indicator of the highest frequencies which can be covered with such calculations. Calculations at higher frequencies are rarely feasible, especially if multiple scenarios (for example, when testing different geometries and boundary conditions) should be assessed in reasonable time. For these reasons, we shall reconsider acoustic analysis in time domain, in search of a more suitable methodology for high frequencies and large rooms. This will lead us to the two remaining general theories of room acoustics -- geometrical and statistical.

\section{Analysis in time domain}\label{ch:analysis_in_time_domain}

Modal analysis from the previous section point to the essential resonant behaviour of rooms and closed spaces in general, which is especially pronounced at relatively low Helmholtz numbers. However, calculations of fields or tailored Green's functions using modal analysis become rather impractical when many contributing modes have to be included. With the exception of acoustically small rooms or cavities, this is very often the case, if a substantial part of the audible frequency range is considered. In addition to this, all boundary conditions in the analysis had to be satisfied by the solutions at once, since time dependence of the acoustic quantities was suppressed. This makes the calculations computationally involved for all but the simplest room geometries and boundary conditions. With regard to this, we return in this section to the analysis in time domain, which, together with additional assumptions, will relax these requirements and lead to simpler solutions at high Helmholtz numbers.

\subsection{Green's function in time domain}\label{ch:Greens's_function_time}

Similarly as the Helmholtz equation~(\ref{eq:Helmholtz_source}), the wave equation~(\ref{eq:wave_eq_source}) can be formally solved with the aid of \textbf{Green's functions}. A Green's function $G(\boldsymbol{x}, t|\boldsymbol{y}, \tau)$ with the unit 1/(m$\cdot$s) has to satisfy the following equation (compare with eq.~(\ref{eq:Helmholtz_Green})):
\begin{equation}\label{eq:wave_eq_Green_x}
\frac{1}{c_0^2} \frac{\partial^2 G(\boldsymbol{x}, t|\boldsymbol{y}, \tau)}{\partial t^2} - \nabla_x^2 G(\boldsymbol{x}, t|\boldsymbol{y}, \tau) = \delta(\boldsymbol{x} - \boldsymbol{y}, t - \tau) = \delta(\boldsymbol{x} - \boldsymbol{y}) \delta(t - \tau).
\end{equation}
Due to the symmetry of delta function and the reciprocity of Green's function, $G(\boldsymbol{x}, t|\boldsymbol{y}, \tau) = G^*(\boldsymbol{y}, \tau |\boldsymbol{x}, t)$, we can switch $\boldsymbol x$ and $\boldsymbol y$ and $t$ and $\tau$ and obtain from the complex conjugate of eq.~(\ref{eq:wave_eq_Green_x})
\begin{equation}\label{eq:wave_eq_Green}
\frac{1}{c_0^2} \frac{\partial^2 G(\boldsymbol{x}, t|\boldsymbol{y}, \tau)}{\partial \tau^2} - \nabla_y^2 G(\boldsymbol{x}, t|\boldsymbol{y}, \tau) = \delta(\boldsymbol{x} - \boldsymbol{y}, t - \tau) = \delta(\boldsymbol{x} - \boldsymbol{y}) \delta(t - \tau).
\end{equation}

Mimicking the procedure in section~\ref{ch:Greens's_function}, we subtract the last equality multiplied with $p(\boldsymbol y, \tau)$ from eq.~(\ref{eq:wave_eq_source_y}) multiplied with $G(\boldsymbol{x}, t|\boldsymbol{y}, \tau)$ and obtain (leaving out the arguments of $G$ for brevity)
\begin{align*}
\begin{aligned}
G \frac{1}{c_0^2} \frac{\partial^2 p(\boldsymbol{y}, \tau)}{\partial \tau^2} &- G \nabla_y^2 p(\boldsymbol{y}, \tau) - p(\boldsymbol y, \tau) \frac{1}{c_0^2} \frac{\partial^2 G}{\partial \tau^2} + p(\boldsymbol y, \tau) \nabla_y^2 G - G q(\boldsymbol{y}, \tau) &\\
&= -p(\boldsymbol y, \tau) \delta(\boldsymbol{x} - \boldsymbol{y}) \delta(t - \tau).&
\end{aligned}
\end{align*}
We integrate the result with respect to $\boldsymbol y$ over some volume $V$ and with respect to $\tau$ over the time interval $(t_0^-,t)$, where $t_0^-<t_0$ is the time just before $t_0$, when the source $q$ is switched on. Due to causality in time domain, we also limit the integration up to the reception time $t>t_0$, since no event which happens after $t$ can influence the field at $t$. In a way, these are time constraints analogue to the boundedness of space. The integration gives
\begin{align*}
\begin{aligned}
\int_{t_0^-}^{t} \int_V G q(\boldsymbol{y}, \tau) d^3 \boldsymbol y d \tau &+ \frac{1}{c_0^2} \int_{t_0^-}^{t} \int_V \left( p(\boldsymbol y, \tau) \frac{\partial^2 G}{\partial \tau^2} - G \frac{\partial^2 p(\boldsymbol{y}, \tau)}{\partial \tau^2} \right) d^3 \boldsymbol y d \tau &\\ &+ \int_{t_0^-}^{t} \int_V \left( G \nabla_y^2 p(\boldsymbol{y}, \tau) - p(\boldsymbol y, \tau) \nabla_y^2 G \right) d^3 \boldsymbol y d \tau &\\ &= \int_{t_0^-}^{t} \int_V p(\boldsymbol y, \tau) \delta(\boldsymbol{x} - \boldsymbol{y}) \delta(t - \tau)d^3 \boldsymbol y d \tau &\\& = \int_{t_0^-}^{t} \left( \int_V p(\boldsymbol y, \tau) \delta(\boldsymbol{x} - \boldsymbol{y}) d^3 \boldsymbol y \right) \delta(t - \tau) d \tau &\\& = \int_{t_0^-}^{t} p(\boldsymbol x, \tau) \delta(t - \tau) d \tau = p(\boldsymbol x, t).&
\end{aligned}
\end{align*}

Next we notice that
\begin{align*}
p \frac{\partial^2 G}{\partial \tau^2} - G \frac{\partial^2 p}{\partial \tau ^2} = \frac{\partial}{\partial \tau} \left( p \frac{\partial G}{\partial \tau} \right) - \frac{\partial p}{\partial \tau} \frac{\partial G}{\partial \tau} - \frac{\partial}{\partial \tau} \left( G \frac{\partial p}{\partial \tau} \right) + \frac{\partial G}{\partial \tau} \frac{\partial p}{\partial \tau} = \frac{\partial}{\partial \tau} \left( p \frac{\partial G}{\partial \tau} - G \frac{\partial p}{\partial \tau} \right)
\end{align*}
and similarly
\begin{align*}
\begin{split}
G \nabla_y^2 p - p \nabla_y^2 G = \nabla_y \cdot (G \nabla_y p - p \nabla_y G).
\end{split}
\end{align*}
After also applying the divergence theorem, we obtain
\begin{align*}
\begin{aligned}
p(\boldsymbol x, t) &= \int_{t_0^-}^{t} \int_V G q(\boldsymbol{y}, \tau) d^3 \boldsymbol y d \tau + \int_{t_0^-}^{t} \oint_S \left( G \nabla_y p(\boldsymbol{y}, \tau) - p(\boldsymbol y, \tau) \nabla_y G \right) \cdot \boldsymbol n d^2 \boldsymbol y d \tau &\\&
+ \frac{1}{c_0^2} \int_{t_0^-}^{t} \int_V \frac{\partial}{\partial \tau} \left( p(\boldsymbol{y}, \tau) \frac{\partial G}{\partial \tau} - G \frac{\partial p(\boldsymbol{y}, \tau)}{\partial \tau} \right) d^3 \boldsymbol y d \tau,&
\end{aligned}
\end{align*}
where $\boldsymbol n(\boldsymbol y)$ is a unit vector normal to the surface $S$, which encloses the volume $V$, at $\boldsymbol y$ and pointing outwards. The last integral equals
\begin{align*}
\begin{aligned}
\int_{t_0^-}^{t} \int_V & \frac{\partial}{\partial \tau} \left( p(\boldsymbol{y}, \tau) \frac{\partial G}{\partial \tau} - G \frac{\partial p(\boldsymbol{y}, \tau)}{\partial \tau} \right) d^3 \boldsymbol y d \tau &\\
&= \int_V  \left[ \int_{t_0^-}^{t} \frac{\partial}{\partial \tau} \left( p(\boldsymbol{y}, \tau) \frac{\partial G}{\partial \tau} - G \frac{\partial p(\boldsymbol{y}, \tau)}{\partial \tau} \right) d \tau \right] d^3 \boldsymbol y &\\
&= \int_V \left[ \left( p(\boldsymbol{y}, \tau) \frac{\partial G}{\partial \tau} - G \frac{\partial p(\boldsymbol{y}, \tau)}{\partial \tau} \right)_{\tau = t} - \left( p(\boldsymbol{y}, \tau) \frac{\partial G}{\partial \tau} - G \frac{\partial p(\boldsymbol{y}, \tau)}{\partial \tau} \right)_{\tau = t_0^-} \right] d^3 \boldsymbol y &\\
&= \int_V \left( G \frac{\partial p(\boldsymbol{y}, \tau)}{\partial \tau} - p(\boldsymbol{y}, \tau) \frac{\partial G}{\partial \tau} \right)_{\tau = t_0^-} d^3 \boldsymbol y,&
\end{aligned}
\end{align*}
where $( \cdot )_{\tau = t}$ indicates that the expression in the parentheses is to be evaluated at time $t$ (that is, $( f(\tau) )_{\tau = t} = f(t)$) and $G(\boldsymbol{x}, t|\boldsymbol{y}, \tau) = 0$ for $\tau = t$ and any $\boldsymbol x \neq \boldsymbol y$ (we are concerned only with receivers outside the source region and no instantaneous propagation is allowed between $\boldsymbol y$ and $\boldsymbol x$). Therefore,
\begin{equation}\label{eq:solution_Green_wave_eq}
\begin{aligned}
p(\boldsymbol x, t) & = \int_{t_0^-}^{t^-} \int_V q(\boldsymbol{y}, \tau) G d^3 \boldsymbol y d \tau &\\& + \int_{t_0^-}^{t^-} \oint_S \left( G \nabla_y p(\boldsymbol{y}, \tau) - p(\boldsymbol y, \tau) \nabla_y G \right) \cdot \boldsymbol n d^2 \boldsymbol y d \tau &\\
&+ \frac{1}{c_0^2} \int_V \left( G \frac{\partial p(\boldsymbol{y}, \tau)}{\partial \tau} - p(\boldsymbol{y}, \tau) \frac{\partial G}{\partial \tau} \right)_{\tau = t_0^-} d^3 \boldsymbol y,&
\end{aligned}
\end{equation}
where $t^-$ denotes the time just before $t$.

We can compare this result with eq.~(\ref{eq:solution_Green_Helmholtz}). Besides the necessary integration over time because of the recovered time dependence, the additional term is due to the \textbf{initial conditions} at the time when the source is switched on. It vanishes if we assume that $p = 0$ for $t < t_0$ and we can replace $t_0^-$ with $t_0$ in the integration intervals. This leaves
\begin{equation}\label{eq:solution_Green_wave_eq_t0-infinity}
\boxed{ p(\boldsymbol x, t) = \int_{t_0}^{t^-} \int_V q(\boldsymbol{y}, \tau) G d^3 \boldsymbol y d \tau + \int_{t_0}^{t^-} \oint_S \left( G \nabla_y p(\boldsymbol{y}, \tau) - p(\boldsymbol y, \tau) \nabla_y G \right) \cdot \boldsymbol n d^2 \boldsymbol y d \tau }
\end{equation}
for any $\boldsymbol x \neq \boldsymbol y$. The first term on the right-hand side captures the direct sound of a source in $V$ and the second term captures effects of the boundaries, when $V$ matches the interior of the room. As before, the second term vanishes for a tailored Green's function satisfying the same boundary conditions as $p$,
\begin{equation}
a G_{tail} + b \nabla_y G_{tail} \cdot \boldsymbol n(\boldsymbol y) = c(\boldsymbol y, \tau)
\end{equation}
from eq.~(\ref{eq:wave_eq_p_BC}), and the equation simplifies further to
\begin{equation}\label{eq:solution_tailored_Green_wave_eq}
\boxed{ p(\boldsymbol x, t) = \int_{t_0}^{t^-} \int_V q(\boldsymbol{y}, \tau) G_{tail} d^3 \boldsymbol y d \tau = \int_{-\infty}^{t^-} \int_V q(\boldsymbol{y}, \tau) G_{tail} d^3 \boldsymbol y d \tau }.
\end{equation}
In the last equality we formally extended the integration interval to $-\infty$, which is allowed since $q = 0$ for $-\infty < t < t_0$. This equation is the time-domain counterpart of eq.~(\ref{eq:solution_tailored_Green_Helmholtz}).

\textbf{Free-space Green's function} of the wave equation~(\ref{eq:wave_eq_source}), which satisfies eq.~(\ref{eq:wave_eq_Green_x}) for an outgoing wave is
\begin{equation}\label{eq:Green_wave_eq_free_space}
\boxed{ G_{free}(\boldsymbol{x}, t|\boldsymbol{y}, \tau) = \frac{\delta(t-\tau-|\boldsymbol x-\boldsymbol y|/c_0)}{4 \pi |\boldsymbol x-\boldsymbol y|} = \frac{\delta(t-\tau-r/c_0)}{4 \pi r} }.
\end{equation}
It has the same denominator as $\hat G_{free}$ in eq.~(\ref{eq:Green_Helmholtz_free_space}), capturing the decay of sound with increasing distance from the source. It is also spherically symmetric (not containing the directivity of the source) and satisfies the reciprocity. However, its numerator is a delta function (for any $r$), a generalized function which is non-zero only when $\tau = t-r/c_0$, which is called emission (or retarded) time. In fact, the delta functions $\delta(t-\tau)$ for various $\tau$ are eigenfunctions in time domain (without further constraints on time) like the exponential functions $e^{j\omega t}$ or $e^{-jkr}$ in unbounded space in frequency domain. The functions $\delta(t-\tau)$ are mutually orthogonal and form a complete set for continuous $\tau$. Every function of time can be represented as an integral of weighted $\delta(t-\tau)$ with respect to $\tau$. This follows directly from the sampling property of the (symmetric) delta function from eq.~(\ref{eq:Dirac_sampling}). After replacing $x$ with $\tau$ and $y$ with $t$, it gives
\begin{equation}
p(\boldsymbol x, t) = \int_{-\infty}^{\infty} p(\boldsymbol x, \tau) \delta(t-\tau) d\tau.
\end{equation}
Evidently, $p(\boldsymbol x, \tau)$ play the same roles of weights as $\hat p(\boldsymbol x, \omega)$ in eq.~(\ref{eq:inv_Fourier_p}) and the sampling property in time domain corresponds to the Fourier integral in frequency domain. Complex sine waves are ``samples'' in the latter domain. Furthermore, since the delta function is real and symmetric, the inverse transform has the same form,
\begin{equation}
p(\boldsymbol x, \tau) = \int_{-\infty}^{\infty} p(\boldsymbol x, t) \delta(t-\tau) dt.
\end{equation}

In closed spaces, the tailored Green's function in time domain is a series of delta impulses from eq.~(\ref{eq:Green_wave_eq_free_space}) additionally scaled due to absorption and diffusion at the surfaces and with $r$ generalized to a total path length of (possibly reflected) sound. Analogously to such a superposition of weighted impulses with different delays (or time of arrival at the receiver location), $\hat{G}_{tail}$ in closed spaces in frequency domain is a weighted sum of eigenfunctions with different eigenfrequencies (eq.~(\ref{eq:tailored_Green_Helmholtz_modes})).

If a point source located at $\boldsymbol y$ in $V$ emits an ideal impulse at time $t_0$, we can write the source function as a generalized function $q(\boldsymbol y', \tau) = Q(\boldsymbol y',\tau) \delta(\boldsymbol y' - \boldsymbol y) \delta(\tau-t_0)$, with $Q$ in kg/s, so that
\begin{equation}\label{eq:tailored_Green_Helmholtz_compact_source_time_domain}
\int_{-\infty}^{\infty} \int_V q(\boldsymbol y',\tau) d^3 \boldsymbol y' d\tau = \int_{-\infty}^{\infty} \int_V Q(\boldsymbol y',\tau) \delta(\boldsymbol y' - \boldsymbol y) \delta(\tau-t_0) d^3 \boldsymbol y' d\tau = Q(\boldsymbol y,t_0).
\end{equation}
For a tailored Green's function, eq.~(\ref{eq:solution_tailored_Green_wave_eq}) gives
\begin{equation}\label{eq:solution_tailored_Green_wave_eq_impulse}
\begin{aligned}
p(\boldsymbol x, t) &= \int_{-\infty}^{t^-} \int_V q(\boldsymbol{y}', \tau) G_{tail}(\boldsymbol x, t| \boldsymbol y', \tau) d^3 \boldsymbol y' d \tau &\\
&= \int_{-\infty}^{t^-} \int_V Q(\boldsymbol y',\tau) \delta(\boldsymbol y' - \boldsymbol y) \delta(\tau-t_0) G_{tail}(\boldsymbol x, t| \boldsymbol y', \tau) d^3 \boldsymbol y' d \tau &\\
&= Q(\boldsymbol y, t_0) G_{tail}(\boldsymbol x, t| \boldsymbol y, t_0),&
\end{aligned}
\end{equation}
if $t_0 \leq t^-$ or zero otherwise. This is very similar to the result in eq.~(\ref{eq:solution_tailored_Green_Helmholtz_point_source}) with added time dependence. In this case, the tailored Green's function represents the \textbf{impulse response} (response to the impulse emitted by the point source $Q$) and contains all the information about sound propagation in the room between $\boldsymbol y$ and $\boldsymbol x$ in the time interval from $t_0$ until $t$. The tailored Green's function in frequency domain, which was considered in section~\ref{ch:modal_analysis}, represents the frequency response. In free space, $G_{tail} = G_{free}$ and for a point monopole source
\begin{equation}\label{eq:solution_tailored_Green_wave_eq_impulse_free_space}
\begin{aligned}
p_{free}(\boldsymbol x, t) = Q(\boldsymbol y, t_0) G_{free}(\boldsymbol x, t| \boldsymbol y, t_0) = Q(\boldsymbol y, t_0) \frac{\delta(t-t_0-|\boldsymbol x-\boldsymbol y|/c_0)}{4 \pi |\boldsymbol x-\boldsymbol y|}
\end{aligned}
\end{equation}
is merely a scaled and delayed (due to the propagation from $\boldsymbol y$ to $\boldsymbol x$ with the finite speed $c_0$) version of the radiated impulse\footnote{The point source function $q$ will be generalized to an arbitrary function of time $t$ in eq.~(\ref{eq:solution_tailored_Green_wave_eq_free_space_compact_source}).}. The delay $|\boldsymbol x-\boldsymbol y|/c_0$ corresponds to the phase shift $k|\boldsymbol x - \boldsymbol y|$ (divided with $\omega$) in eq.~(\ref{eq:solution_tailored_Green_Helmholtz_point_source_free_space}).

Although these interpretations are correct, additional care is necessary. The fact that $G_{tail}$ is a generalized function, a series of delta impulses, also outside the source region has another consequence. Sound pressure expressed by the last equality in eq.~(\ref{eq:solution_tailored_Green_wave_eq_impulse}) and in eq.~(\ref{eq:solution_tailored_Green_wave_eq_impulse_free_space}) is not physical. The reason is that the integrals in eq.~(\ref{eq:solution_tailored_Green_wave_eq_impulse}) did not act on the delta functions of $G_{tail}$. For example, a physically correct result in free space would be
\begin{equation}\label{eq:solution_tailored_Green_wave_eq_free_space}
\begin{aligned}
&p_{free}(\boldsymbol x, t) = \int_{-\infty}^{t^-} \int_V q(\boldsymbol{y}', \tau) \frac{\delta(t-\tau-r/c_0)}{4 \pi r} d^3 \boldsymbol y' d \tau &\\
&= \int_V \frac{1}{4 \pi r} \left( \int_{-\infty}^{t^-} q(\boldsymbol{y}', \tau) \delta(\tau-(t-r/c_0)) d \tau \right) d^3 \boldsymbol y' = \int_V \frac{q(\boldsymbol{y}', t-r/c_0)}{4 \pi r} d^3 \boldsymbol y',&
\end{aligned}
\end{equation}
since $t-r/c_0 < t$ and we used the symmetry of delta function and the fact that $r$ is not a function of time (the source and receiver are motionless). Hence, $p_{free}$ at time $t$ depends on the (arbitrary) source function evaluated at the emission time. If the omnidirectional source is compact, $r$ is essentially equal in the entire source region and we can write
\begin{equation}\label{eq:solution_tailored_Green_wave_eq_free_space_compact_source}
\begin{aligned}
p_{free}(\boldsymbol x, t) &= \frac{1}{4 \pi r} \int_V q(\boldsymbol{y}', t-r/c_0) d^3 \boldsymbol y' = \frac{1}{4 \pi r} \int_V Q(\boldsymbol y',t-r/c_0) \delta(\boldsymbol y' - \boldsymbol y) d^3 \boldsymbol y' \\
&= \frac{1}{4 \pi r} Q(\boldsymbol y, t-r/c_0),
\end{aligned}
\end{equation}
where now, similarly as in eq.~(\ref{eq:tailored_Green_Helmholtz_compact_source}), $q(\boldsymbol y', \tau) = Q(\boldsymbol y',\tau) \delta(\boldsymbol y' - \boldsymbol y)$, $Q$ has again the unit kg/s$^2$, and
\begin{equation}\label{eq:tailored_Green_wave_eq_compact_source}
Q(\boldsymbol y, \tau) = \int_V q(\boldsymbol y', \tau) d^3 \boldsymbol y' = \int_V Q(\boldsymbol y', \tau) \delta(\boldsymbol y' - \boldsymbol y) d^3 \boldsymbol y'.
\end{equation}
Sound pressure depends only on the source and the distance from it.

\subsection{LTI systems and impulse response}\label{LTI_systems_and_impulse_response}

A type of a system which is completely described with its impulse response is \textbf{linear time invariant (LTI) system}. Since we assume that all sources and receivers are motionless, all relevant conditions stable, and the theory is linear, a room can also be treated as an LTI system. The \textbf{impulse response} $g(t)$ is then the room's response at a fixed receiver location to the impulse emitted from a point monopole source at some other fixed location. The resulting signal $p$ at the receiver location is given by the \textbf{convolution} integral ($\delta$, $s$, $g$, and $p$ are here real functions of time only):
\begin{equation}\label{eq:convolution}
p(t+r/c_0) = \int_{-\infty}^{\infty} s(\tau) g(t-\tau) d\tau = \int_{-\infty}^{\infty} g(\tau) s(t-\tau) d\tau,
\end{equation}
where $s(t)$ is the input signal (emitted by the source), as in Fig.~\ref{fig:room_LTI_system}.

The first equality can be compared with the integral over time in eq.~(\ref{eq:solution_tailored_Green_wave_eq}), with $s$ acting as $q$ (that is, $Q$ for a point source $q(\boldsymbol y', \tau) = Q(\boldsymbol y',\tau) \delta(\boldsymbol y' - \boldsymbol y)$, after the remaining integration over $\boldsymbol y'$) and $g$ as $G_{tail}$. By inserting the free-space Green's function from eq.~(\ref{eq:Green_wave_eq_free_space}) in eq.~(\ref{eq:solution_tailored_Green_wave_eq}), we can see that the essential difference is that the emission time $t-r/c_0$ has been replaced by the reception time $t$ of the direct sound (the free-space component). This explains $t+r/c_0$ as the argument of the received sound pressure function on the left-hand side of eq.~(\ref{eq:convolution}). Moreover, the requirement of causality is met by the impulse responses starting at $t=0$, for which $g(t) = 0$ for $t<0$, so the time integration up to $\tau = t^-+r/c_0$ can be extended to $+\infty$, since for $\tau \geq t+r/c_0$, $g(t-\tau)=0$ for any $r>0$. Equation~(\ref{eq:convolution}) and thus the integral with respect to time in eq.~(\ref{eq:solution_tailored_Green_wave_eq}) are convolutions and the tailored Green's function is indeed an (ideal) impulse response of the room seen as an LTI system. However, it should be emphasized again that $g(t)$ does depend on the source and receiver locations ($\boldsymbol x$ and $\boldsymbol y$ in $G_{tail}$).

The second equality in eq.~(\ref{eq:convolution}) gives after replacing $t+r/c_0$ with $t$ (which does not affect the interval of integration)
\begin{equation}\label{eq:convolution_emission_time}
p(t) = \int_{-\infty}^{\infty} g(\tau) s(t-r/c_0-\tau) d\tau.
\end{equation}
For causal systems, the integral in eq.~(\ref{eq:convolution_emission_time}) becomes
\begin{equation}\label{eq:convolution_causal}
p(t) = \int_{0}^{\infty} g(\tau) s_e(t-\tau) d\tau,
\end{equation}
where $s_e(t) = s(t-r/c_0)$ is evaluated at the emission time. This form is often more convenient, since the emitted signal is usually also specified starting at $t=0$ (thus excluding the time delay $r/c_0$ from the analysis). More precisely, the input sequence $s_e$ is given as a function of the emission time, while the impulse response and sound pressure are expressed in the reception time. From the theory of LTI systems, we also know that in frequency domain convolution turns into product:
\begin{equation}\label{eq:convolution_freq}
\hat{p}(\omega) = \hat{s}_e(\omega) \hat{g}(\omega),
\end{equation}
with $\hat{g}(\omega)$ \textbf{frequency response} of a system. This has the same form as eq.~(\ref{eq:solution_tailored_Green_Helmholtz_point_source}) and, as already noted, $\hat G_{tail}$ represents frequency response of a room. However, in this form the information on the delay of the direct sound ($r/c_0$), which is contained in $\hat G_{tail}$, is not retained in $\hat g$. On the other hand, the delay is not essential (or can be easily estimated) in most of the applications and this is the form in which impulse responses of rooms are usually acquired and analysed (see section~\ref{ch:measurements_and_descriptors}).

\begin{figure}[h]
	\centering
	\begin{subfigure}{.5\textwidth}
		\centering
		\includegraphics[width=1\linewidth]{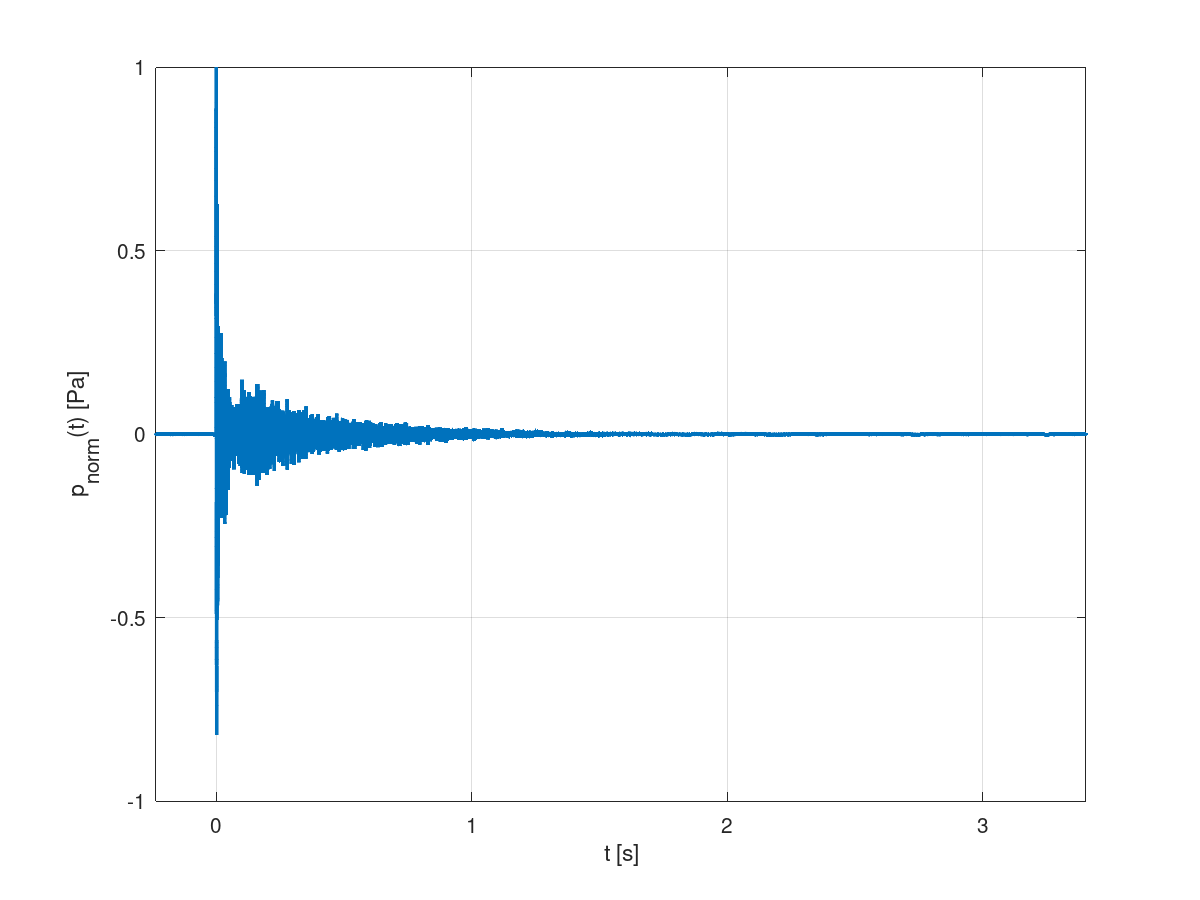}
		\label{fig:imp_response_norm}
	\end{subfigure}%
	\begin{subfigure}{.5\textwidth}
		\centering
		\includegraphics[width=1\linewidth]{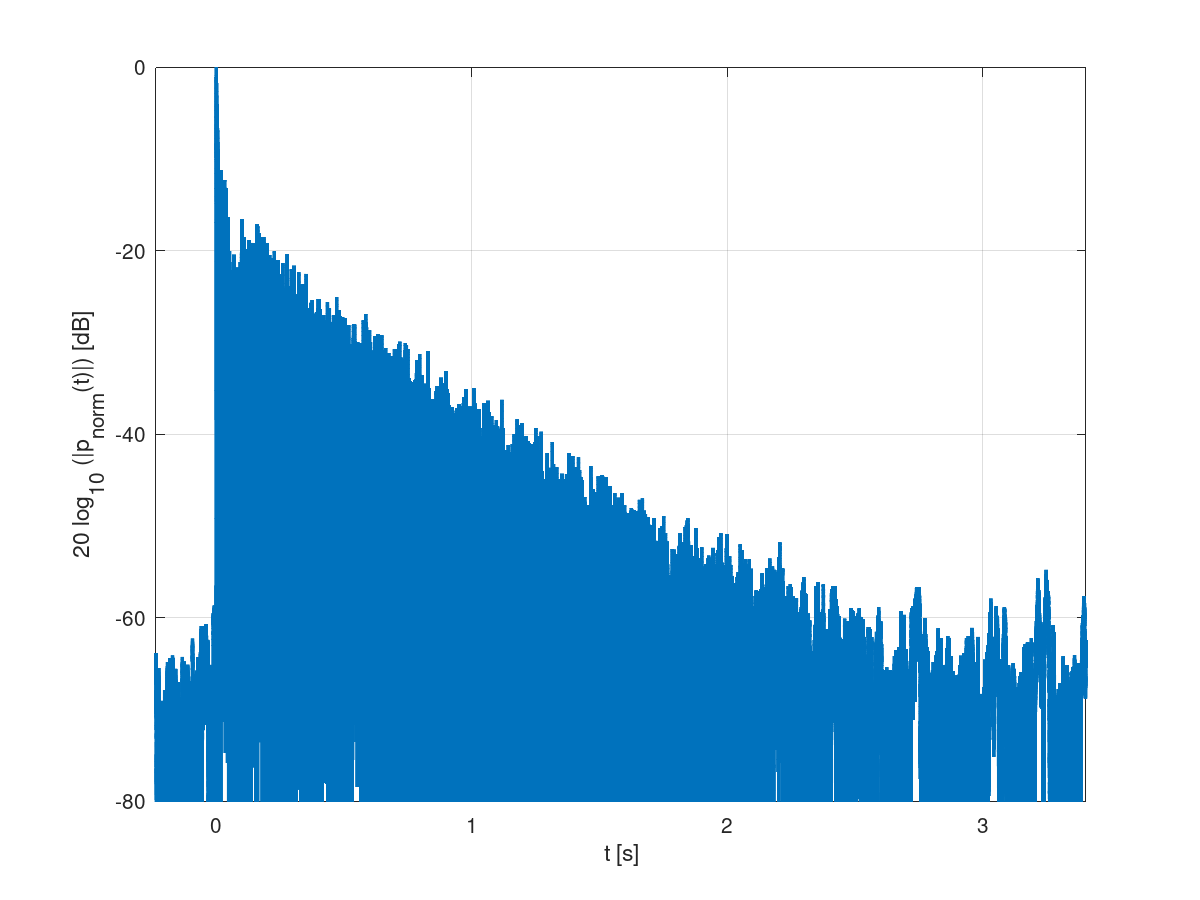}
		\label{fig:imp_response_norm_log}
	\end{subfigure}
	\caption{Example of a room impulse response: (left) normalized sound pressure $p_\text{norm}(t)$ and (right) $20 \log_{10} (|p_\text{norm}(t)|)$.}
	\label{fig:imp_response}
\end{figure}

\begin{figure}[h]
	\centering
	\includegraphics[width=0.7\linewidth]{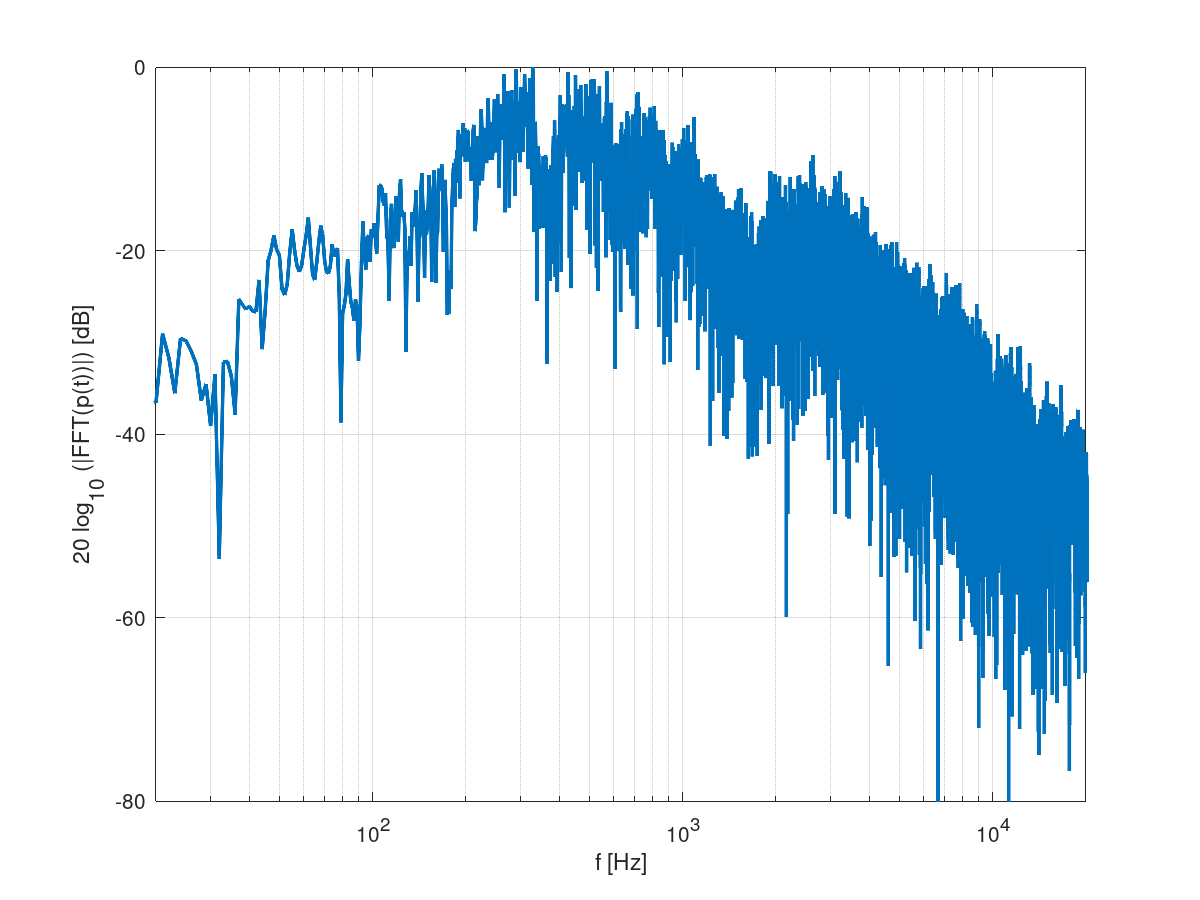}
	\caption{Normalized frequency response obtained from the impulse response in Fig.~\ref{fig:imp_response}.}
	\label{fig:freq_response}
\end{figure}

Figure~\ref{fig:imp_response} presents an example of a recorded broadband room impulse response. The room with the characteristic length scale $L \approx 20$\,m is excited with a sound impulse (by popping a balloon, see section~\ref{ch:acquisition_room_impulse_response}). Hence, $p(t) \approx g(t)$ from eq.~\ref{eq:convolution_causal} with $s_e(t) \approx \delta(t)$. The sound pressure is normalized so that the maximum of the direct sound at $t=0$ is 1\,Pa. Right-hand side of the figure shows logarithm of the absolute value of the signal. The direct sound is, as usually, followed by the strong early reflections, which precede a relatively smooth exponential decay of many late reflections. The noise level before and after the impulse response suggests that the broadband peak-to-noise ratio is around 65\,dB. Figure~\ref{fig:freq_response} shows normalized (to the maximum value 0\,dB) logarithm of the amplitude of the frequency response obtained from the same impulse response using the fast Fourier transform with resolution 1\,Hz. It has a form comparable to the exact frequency response in Fig.~\ref{fig:rect_room_freq_response}. However, the non-ideal sound impulse with the energy peak around 500\,Hz causes a gradual drop of the acquired frequency response at both low and high frequencies. The lowest modes of the large rooms lie at $f \sim \mathcal{O}(10\text{\,Hz})$ are also not captured well. In fact, most of the low-frequency energy of the response is likely due to the ambient noise, since the peak-to-noise ratio is much below the broadband value.

\subsection{Directivity}
\label{ch:directivity}

Working in time domain allows us to start the analysis of any field with an isolated source in free space, which is physically justified in a closed space before the emitted sound has reached an obstacle or boundary during its propagation. Therefore, in contrast to the modal analysis, we can consider the boundary conditions separately afterwards. Nevertheless, if we are interested in a stationary sound field, we can suppose for simplicity that the source emits a complex sine wave with generic angular frequency $\omega$:
\begin{equation}\label{eq:tailored_Green_wave_eq_compact_source_sine_wave}
Q(\boldsymbol y, \tau) = \hat{Q}(\boldsymbol y) e^{j \omega \tau}.
\end{equation}
Sound pressure from eq.~(\ref{eq:solution_tailored_Green_wave_eq_free_space_compact_source}) equals then
\begin{equation}\label{eq:solution_tailored_Green_wave_eq_free_space_compact_source_emission_time}
p(\boldsymbol x, t) = \frac{\hat{Q}(\boldsymbol y)}{4 \pi r} e^{j \omega (t-r/c_0)} = \frac{\hat{Q}(\boldsymbol y)}{4 \pi r} e^{j (\omega t - kr)}.
\end{equation}
We leave out the subscript $_{free}$ for brevity. The complex amplitude is thus
\begin{equation}\label{eq:solution_tailored_Green_wave_eq_free_space_compact_source_emission_time_amplitude}
\hat{p}(\boldsymbol x) = \frac{\hat{Q}(\boldsymbol y)}{4 \pi r} e^{- jkr},
\end{equation}
which is in full agreement with eq.~(\ref{eq:solution_tailored_Green_Helmholtz_point_source_free_space}).

Like $\hat{G}_{free}$ in eq.~(\ref{eq:Green_Helmholtz_free_space}), the free-space Green's function in eq.~(\ref{eq:Green_wave_eq_free_space}) is spherically symmetric, which leaves sound pressure amplitude dependent only on the source and distance $r$ from it. When necessary, directivity of the point source can be introduced explicitly, with a real and non-negative\footnote{The minus sign of a negative factor would be absorbed in the phase of $\hat{Q}(\boldsymbol y)$.} dimensionless factor $D_i$ (also called \textbf{directivity})\footnote{Directivity is by definition angular dependence of the acoustic far field (to be defined shortly) generated by the source. Therefore, the introduced directivity and all calculations which include it are valid only in the far field of the source, in which $kr \gg 1$.}:
\begin{equation}\label{eq:solution_tailored_Green_wave_eq_free_space_compact_directed_source_emission_time}
\boxed{ p(\boldsymbol x, t) = \frac{\hat{Q}(\boldsymbol y) D_i(\theta, \phi)}{4 \pi r} e^{j (\omega t - kr)} },
\end{equation}
for $\boldsymbol y = 0$ and $r = |\boldsymbol x|>0$. Here, $0 \leq \theta \leq \pi$ and $0 \leq \phi < 2 \pi$ are, respectively, polar and azimuthal angle of the receiver's location. The latter is given by $\boldsymbol x = (x_1, x_2, x_3)^T = (r \sin(\theta) \cos(\phi), r \sin(\theta) \sin(\phi), r \cos(\theta))^T$, since the source is located at the origin $\boldsymbol y = 0$ of the spherical coordinate system. The complex amplitude is accordingly
\begin{equation}\label{eq:solution_tailored_Green_wave_eq_free_space_compact_directed_source_emission_time_freq}
\hat{p}(\boldsymbol x) = \frac{\hat{Q}(\boldsymbol y) D_i(\theta, \phi)}{4 \pi r} e^{-jkr}.
\end{equation}
The field thus depends not only on the source and distance, but also on the directivity. Like $\hat{p}$ and $\hat{Q}$, $D_i$ is also a function of frequency in general, with the range of values over $\theta$ and $\phi$ which usually increases with frequency, resulting in a more pronounced directivity of the source. Consequently, the values of $D_i$ or any other exact descriptor of directivity should be given in the entire frequency range of interest, typically in octave bands.

\begin{figure}[h]
	\centering
	\begin{subfigure}{.3\textwidth}
		\centering
		\includegraphics[width=1\linewidth]{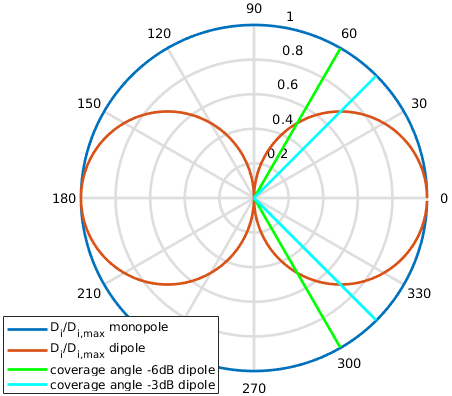}
		\label{fig:directivity_polar_monopole_dipole}
	\end{subfigure}%
	\begin{subfigure}{.3\textwidth}
		\centering
		\includegraphics[width=1\linewidth]{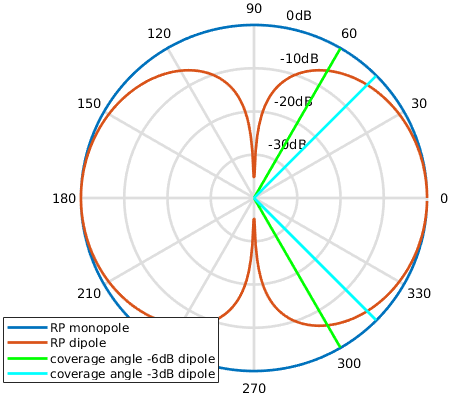}
		\label{fig:directivity_dB_polar_monopole_dipole}
	\end{subfigure}
	\begin{subfigure}{.35\textwidth}
		\centering
		\includegraphics[width=1\linewidth]{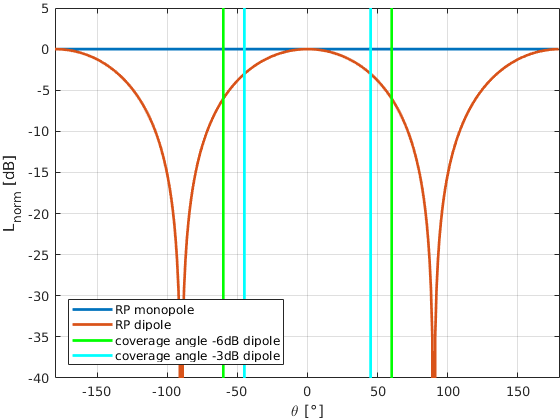}
		\label{fig:directivity_plot_monopole_dipole}
	\end{subfigure}
	\caption{Directivity of an ideal point monopole and dipole.}
	\label{fig:directivity_monopole_dipole}
\end{figure}

Directivity $D_i$ is most often \textbf{normalized} with its maximum value $D_{i,max}$ at the axis of principal radiation of the source (which typically matches the axis of symmetry of an axisymmetric source, for example a loudspeaker). In this way its values are limited to the range between (and including) 0 and 1. Logarithm of such an angularly dependent quantity is called the \textbf{radiation pattern}\footnote{Other normalizations in the definition of $RP$ are also possible, for example, with the sound power level of the source, such that $10 \log_{10} \left( \int_{0}^{2\pi} \int_{0}^{\pi} 10^{RP/10} \sin(\theta) d\theta d\phi \right) - L_W = 0$.}:
\begin{equation}\label{eq:radiation_pattern}
RP = 20 \log_{10} \left( \frac{D_i(\theta, \phi)}{D_{i,max}} \right) \leq 0 \text{\,dB}.
\end{equation}
As an example, Fig.~\ref{fig:directivity_monopole_dipole} shows different possible graphic representations of the directivity of an ideal monopole and dipole as a function of $\theta$ (the values do not depend on the azimuthal angle, so the curves in the first two polar plots can be rotated around the horizontal axis to obtain the three-dimensional radiation patterns). Notice how the eight-shaped curve of the dipole appears to be deformed in decibel units and the zero radiation is not entirely captured by the finite resolution of the graph.

For convenience, directivity of a source is often expressed as a (frequency-dependent) single number value, with inevitable loss of details of the radiation pattern. A simple but crude descriptor is \textbf{coverage angle}, which is the angular spread in a given plane around the direction of maximum radiation of a source, in which the radiation pattern remains above some prescribed value, typically -3\,dB, -6\,dB, or -10\,dB. It takes values between 0$^\circ$ and 360$^\circ$ (see, for example, Table~\ref{tab:instruments_coverage_angle}). For a point dipole it equals $90^\circ$ for -3\,dB and $120^\circ$ for -6\,dB, while a compact monopole has the full coverage angle, $360^\circ$. For more complex sources, the radiation patterns of which are not axisymmetric, the coverage angle is usually given in the horizontal and vertical plane.

Another approach for expressing directivity with a single number is using the \textbf{directivity factor (gain)} defined as
\begin{equation}\label{eq:directivity_factor}
\gamma = \frac{4 \pi}{\int_{0}^{4 \pi} \left( D_i(\theta, \phi)/D_{i,max} \right)^2 d\Omega} = \frac{4 \pi D_{i,max}^2}{\int_{0}^{4 \pi} D_i^2(\theta, \phi) d\Omega} \ge 1,
\end{equation}
where $d\Omega = \sin(\theta)d\theta d\phi$ is the solid angle differential. The full solid angle equals $\int d\Omega = \int_{0}^{2\pi} \int_{0}^{\pi} \sin(\theta)d\theta d\phi = 4\pi$, which is in the numerator in eq.~(\ref{eq:directivity_factor}). Hence, the directivity factor represents ratio of the maximum magnitude of time-averaged sound intensity (or maximum energy) produced by the source (in the direction of maximum radiation of the source) and its average value over all directions. If a source has sound power $P_q$, it radiates effectively with the power $\gamma P_q$ in the direction of maximum radiation. Information on the radiation in other directions is lost after the averaging. Directivity factor is larger than or equal to 1 and, of course, frequency dependent. It equals 1 for an ideal monopole (hence the name gain, $\gamma$ times more sound energy is radiated in the direction of maximum radiation than by a monopole with the same sound power) and for a dipole
\begin{equation}\label{eq:directivity_factor_dipole}
\begin{aligned}
\gamma &= \frac{4 \pi}{\int_{0}^{4 \pi} \cos^2(\theta) d\Omega} = \frac{4 \pi}{\int_{0}^{2 \pi} \int_{0}^{\pi} \cos^2(\theta) \sin(\theta) d\theta d\phi} &\\
&= \frac{2}{\int_{0}^{\pi} \cos^2(\theta) \sin(\theta) d\theta} = -\frac{6}{[\cos^3(\pi) - \cos^3(0)]} = 3.
\end{aligned}
\end{equation}
Associated logarithmic quantity is
\begin{equation}\label{eq:directivity_factor_log}
10 \log_{10} \gamma = 10 \log_{10} \left( \frac{4 \pi D_{i,max}^2}{\int_{0}^{4 \pi} D_i^2(\theta, \phi) d\Omega} \right) \ge 0\text{dB}.
\end{equation}
It equals 0\,dB for omnidirectional sources, around 4.8\,dB for a point dipole, and generally rarely exceeds 20\,dB. 

\subsection{Spherical and plane waves}
\label{ch:spherical_and_plane_waves}

Sound pressure of a point source in free space from eq.~(\ref{eq:solution_tailored_Green_wave_eq_free_space_compact_directed_source_emission_time}) decays proportionally to $1/r$, regardless of the directivity of the source. In rooms, however, additional change of amplitude of the propagating wave can occur when the wave hits a solid surface -- boundary surface of the room or any other solid object in it. As implied by eq.~(\ref{eq:wave_eq_p_BC}), boundary conditions are naturally expressed in terms of $p$ and/or its gradient. Moreover, the pressure gradient is also related to the acoustic particle velocity by the conservation of momentum, eq.~(\ref{eq:Euler_momentum}), which appears in the definitions of sound intensity and energy (equations~(\ref{eq:intensity}) and (\ref{eq:energy})) and therefore sound power. For these reasons, it is of interest to calculate the gradient of sound pressure obtained above.

If we suppose that the point source is omnidirectional\footnote{\label{ftn:directional_source_far_field}If the source is directional, the following analysis is still valid in the far field (with additional multiplication with the factor $D_i$). It will be shown shortly that sound waves act as plane waves in the acoustic far field and thus (locally) depend spatially essentially only on $r$, not on $\theta$ and $\phi$.} with the given source function $Q = \hat{Q} e^{j\omega \tau}$, the radiated waves are \textbf{spherical} and the pressure is spatially dependent only on $r$ (eq.~(\ref{eq:solution_tailored_Green_wave_eq_free_space_compact_source_emission_time})), distance from the source. We can replace $\nabla_x p$ with $(\partial p / \partial r)\boldsymbol e_r$ in spherical coordinates\footnote{Gradient of $p$ is in spherical coordinates $\nabla p = \frac{\partial p}{\partial r} \boldsymbol e_r + \frac{1}{r} \frac{\partial p}{\partial \theta} \boldsymbol e_\theta + \frac{1}{r \sin(\theta)} \frac{\partial p}{\partial \phi} \boldsymbol e_\phi$ and we neglect the last two terms in favour of the first one.}, where $\boldsymbol e_r$ is a unit vector pointing into the direction of sound propagation. Consequently,
\begin{equation}\label{eq:p_solution_tailored_Green_wave_eq_free_space_compact_source_emission_time}
\begin{aligned}
\nabla_x p(\boldsymbol x, t) &= \boldsymbol e_r \frac{\partial}{\partial r} \left( \frac{\hat{Q}(\boldsymbol y)}{4 \pi r} e^{j (\omega t - kr)} \right) &\\& = \frac{\hat{Q}(\boldsymbol y)}{4 \pi} \left[ \left( \frac{1}{r} \frac{\partial}{\partial r} e^{j (\omega t - kr)} \right) + \left( e^{j (\omega t - kr)} \frac{\partial}{\partial r} \frac{1}{r} \right) \right] \boldsymbol e_r&\\
&= \frac{\hat{Q}(\boldsymbol y)}{4 \pi} \left(-\frac{jk}{r} e^{j (\omega t - kr)} - \frac{1}{r^2} e^{j (\omega t - kr)} \right)  \boldsymbol e_r &\\& = -\frac{\hat{Q}(\boldsymbol y)}{4 \pi} \left(\frac{jk}{r} + \frac{1}{r^2} \right) e^{j (\omega t - kr)} \boldsymbol e_r&\\
&= -p(\boldsymbol x, t)\left( jk + \frac{1}{r} \right)  \boldsymbol e_r.&
\end{aligned}
\end{equation}
The gradient scales in the two terms with phase difference of $\pi/2$ as $\sim k$ (acoustic) and $\sim 1/r$ (non-acoustic). Particle velocity follows from the momentum equation~(\ref{eq:Euler_momentum_sine_wave}),
\begin{equation}\label{eq:v_solution_tailored_Green_wave_eq_free_space_compact_source_emission_time}
\begin{aligned}
\boldsymbol v(\boldsymbol x, t) &= -\frac{1}{j \omega \rho_0}\nabla_x p(\boldsymbol x, t) = \frac{1}{j \omega \rho_0} p(\boldsymbol x, t)\left( jk + \frac{1}{r} \right) \boldsymbol e_r &\\
&= \frac{1}{\rho_0 c_0} p(\boldsymbol x, t)\left( 1 - \frac{j}{kr} \right) \boldsymbol e_r = \frac{\hat{Q}(\boldsymbol y)}{4 \pi r \rho_0 c_0} \left( 1 - \frac{j}{kr} \right) e^{j (\omega t - kr)} \boldsymbol e_r.&
\end{aligned}
\end{equation}

Furthermore, from eq.~(\ref{eq:intensity_complex_time_average}), we can express time-averaged sound intensity of the spherical sound wave:
\begin{equation}\label{eq:intensity_complex_time_average_plane_wave}
\begin{aligned}
\langle\boldsymbol I\rangle_T &= \frac{1}{4} (\hat{p}^* \hat{\boldsymbol v} + \hat{p} \hat{\boldsymbol v}^*) &\\& = \frac{1}{4} \hat{p}^* \left[ \frac{1}{\rho_0 c_0} \hat{p}\left( 1 - \frac{j}{kr} \right) \boldsymbol e_r \right] + \frac{1}{4} \hat{p} \left[ \frac{1}{\rho_0 c_0} \hat{p}\left( 1 - \frac{j}{kr} \right) \boldsymbol e_r \right]^* &\\
&=  \frac{1}{4} \hat{p}\hat{p}^* \left[ \frac{1}{\rho_0 c_0} \left( 1 - \frac{j}{kr} \right) \boldsymbol e_r \right] + \frac{1}{4} \hat{p} \hat{p}^* \left[ \frac{1}{\rho_0 c_0} \left( 1 - \frac{j}{kr} \right) \boldsymbol e_r \right]^* &\\
&= \frac{1}{4} |\hat{p}|^2 \frac{1}{\rho_0 c_0} \left( 1 - \frac{j}{kr} + 1 + \frac{j}{kr} \right) \boldsymbol e_r = \frac{|\hat{p}|^2}{2 \rho_0 c_0} \boldsymbol e_r &\\
&= \frac{1}{2 \rho_0 c_0} \left| \frac{\hat{Q}(\boldsymbol y)}{4 \pi r} e^{- jkr} \right| ^2 \boldsymbol e_r = \frac{1}{2 \rho_0 c_0} \left| \frac{\hat{Q}(\boldsymbol y)}{4 \pi r} \right| ^2 |e^{- jkr}|^2 \boldsymbol e_r = \frac{|\hat{Q}(\boldsymbol y)|^2}{32 \rho_0 c_0 \pi^2 r^2} \boldsymbol e_r.
\end{aligned}
\end{equation}
Note that the two non-acoustic terms $\propto j/(kr)$ cancel out, leaving $\langle \boldsymbol I \rangle_T/|\hat p|^2$ independent of $kr$. This is not the case with time-averaged energy, which is according to eq.~(\ref{eq:energy_complex_time_average})
\begin{equation}\label{eq:energy_complex_time_average_spherical_wave}
\begin{aligned}
\langle E \rangle_T &= \frac{\hat{p} \hat{p}^*}{4 \rho_0 c_0^2} + \frac{\rho_0 \hat{\boldsymbol v} \cdot \hat{\boldsymbol v}^*}{4} = \frac{\hat{p} \hat{p}^*}{4 \rho_0 c_0^2} + \frac{\rho_0}{4} \frac{1}{\rho_0^2 c_0^2} \hat p \hat p^* \left( 1 + \frac{1}{k^2r^2} \right) \boldsymbol e_r \cdot \boldsymbol e_r &\\
&= \frac{|\hat{p}|^2}{2 \rho_0 c_0^2} \left( 1 + \frac{1}{2k^2r^2} \right) = \frac{1}{c_0} \langle\boldsymbol I\rangle_T \cdot \boldsymbol e_r \left( 1 + \frac{1}{2k^2r^2} \right) &\\
&= \frac{|\hat{Q}(\boldsymbol y)|^2}{32 \rho_0 c_0^2 \pi^2 r^2} \left( 1 + \frac{1}{2k^2r^2} \right).&
\end{aligned}
\end{equation}
On the other hand, sound pressure level is from eq.~(\ref{eq:SPL_complex})
\begin{equation}\label{eq:SPL_complex_spherical_wave}
\begin{aligned}
L &= 10 \log_{10}\left( \frac{|\hat{p}|^2 / 2}{4 \cdot 10^{-10} \text{\,Pa}^2} \right) = 10 \log_{10}\left( \frac{|\hat{Q}(\boldsymbol y)|^2/(32 \pi^2 r^2)}{4 \cdot 10^{-10} \text{\,Pa}^2} \right)&\\
&= 10 \log_{10}\left( \frac{\rho_0 c_0 \langle\boldsymbol I\rangle_T \cdot \boldsymbol e_r}{4 \cdot 10^{-10} \text{\,Pa}^2} \right) \approx 10 \log_{10}\left( \frac{\langle\boldsymbol I\rangle_T \cdot \boldsymbol e_r}{10^{-12} \text{\,W/m}^2} \right).&
\end{aligned}
\end{equation}
For the last approximation we used $\rho_0 c_0 \approx 400$\,kg/(m$^2$s), which gives the reference value for sound intensity, 10$^{-12}$\,W/m$^2$, in agreement with the reference value in eq.~(\ref{eq:sound_power_level}).

Sound power of an omnidirectional point source emitting the spherical wave is from eq.~(\ref{eq:acoustic_power_complex_time_average})
\begin{equation}\label{eq:acoustic_power_complex_time_average_spherical_wave}
\begin{aligned}
\langle P_q \rangle_T &= \oint_S \langle \boldsymbol I(\boldsymbol x,t) \rangle_T \cdot \boldsymbol n(\boldsymbol x) d^2 \boldsymbol x = 4 \pi r^2 \langle\boldsymbol I\rangle_T \cdot \boldsymbol e_r = \frac{2 \pi r^2}{\rho_0 c_0} |\hat{p}|^2 &\\
&= \frac{2 \pi r^2}{\rho_0 c_0} \frac{|\hat{Q}(\boldsymbol y)|^2}{16 \pi^2 r^2} = \frac{|\hat{Q}(\boldsymbol y)|^2}{8 \pi \rho_0 c_0},
\end{aligned}
\end{equation}
where we chose a spherical control surface $S$ over which $|\hat{p}|^2$ and $\boldsymbol I \cdot \boldsymbol e_r$ are constant, with the centre at the source location, radius $r$, and unit vector $\boldsymbol n(\boldsymbol x) = \boldsymbol e_r$ normal to it and pointing outwards. As expected, sound power depends on $\hat Q$ but not on $r$, since it is a property of the source, independent of any receiver location. Sound power level in eq.~(\ref{eq:sound_power_level}) equals then
\begin{equation}\label{eq:acoustic_power_level_complex_time_average_plane_wave_omni_source_pressure_derivation}
\begin{aligned}
L_W &= 10 \log_{10} \frac{\langle P_q \rangle_T}{10^{-12} \text{\,W}} = 10 \log_{10} \frac{|\hat{Q}(\boldsymbol y)|^2/(8 \pi \rho_0 c_0)}{10^{-12} \text{\,W}} = 10 \log_{10} \frac{4 \pi r^2 \langle\boldsymbol I\rangle_T \cdot \boldsymbol e_r}{10^{-12} \text{\,W}} &\\
&= 10 \log_{10} \frac{|\hat{p}|^2/2}{4 \cdot 10^{-10} \text{\,Pa}^2} + 10 \log_{10} \frac{4 \cdot 10^{-10}\text{\,Pa}^2 }{\rho_0 c_0 10^{-12} \text{\,W/m}^2} + 10 \log_{10} \left( \frac{4 \pi r^2}{1\text{\,m}^2} \right),&
\end{aligned}
\end{equation}
so with $\rho_0 c_0 \approx 400$\,kg/(m$^2$s) and the first equality in eq.~(\ref{eq:SPL_complex_spherical_wave}) sound pressure level can be approximated as
\begin{equation}\label{eq:acoustic_power_level_complex_time_average_plane_wave_omni_source_pressure}
\boxed{ L(r) = L_W - 10 \log_{10} \left( \frac{4 \pi r^2}{1\text{\,m}^2} \right) }.
\end{equation}
This expression is commonly used to relate directly sound power level of an omnidirectional point source and sound pressure level at certain distance from it. In all expressions above, $|\hat p|$ of the simple sine waves can also be replaced with the root mean square value multiplied with $\sqrt{2}$.

In the expressions for velocity, eq.~(\ref{eq:v_solution_tailored_Green_wave_eq_free_space_compact_source_emission_time}), and energy, eq.~(\ref{eq:energy_complex_time_average_spherical_wave}), we recognize again \textbf{the Helmholtz number} $kr$, this time with the distance from the source $r$ as the reference length scale. The associated terms followed from the gradient of sound pressure. If the Helmholtz number is much larger than 1, the second (non-acoustic) term in the brackets can be neglected compared to the first (acoustic) term, and we can write
\begin{equation}\label{eq:v_solution_tailored_Green_wave_eq_free_space_compact_source_emission_time_far_field}
\boxed{ \boldsymbol v(\boldsymbol x, t) = \frac{1}{\rho_0 c_0} p(\boldsymbol x, t) \boldsymbol e_r = \frac{\hat{Q}(\boldsymbol y)}{4 \pi r \rho_0 c_0} e^{j (\omega t - kr)} \boldsymbol e_r },
\end{equation}
or after dividing with $e^{j \omega t}$
\begin{equation}\label{eq:v_solution_tailored_Green_wave_eq_free_space_compact_source_emission_time_far_field_amlitude}
\hat{\boldsymbol v}(\boldsymbol x) = \frac{1}{\rho_0 c_0} \hat{p}(\boldsymbol x) \boldsymbol e_r = \frac{\hat{Q}(\boldsymbol y)}{4 \pi r \rho_0 c_0} e^{- j kr} \boldsymbol e_r.
\end{equation}
This is \textbf{acoustic far field}\footnote{We emphasize that it is acoustic far field to distinguish it from geometric far field, which uses size of a source as the reference length scale (see for example section \ref{ch:rectangular_surface}). However, due to its larger importance in practice, acoustic far field is often called simply far field.} at large enough (compared to the wavelength) distances from the source, more precisely for
\begin{equation}\label{eq:far_field_condition}
kr \gg 1 \Rightarrow r \gg \lambda/2\pi.
\end{equation}

In the far field, acoustic pressure and velocity are in phase as in a \textbf{plane wave} (which is the simplest solution of the wave equation). The sound intensity and energy from equations~(\ref{eq:intensity}) and (\ref{eq:energy}) are also simply related,
\begin{equation}\label{eq:energy_plane_wave}
E = \frac{p^2}{2 \rho_0 c_0^2} + \frac{|p \boldsymbol e_r|^2}{2 \rho_0 c_0^2} = \frac{p^2}{\rho_0 c_0^2} = \frac{\boldsymbol I \cdot \boldsymbol e_r}{c_0} = \frac{|\boldsymbol I|}{c_0}
\end{equation}
(potential and kinetic energies are equal), and the time-averaged energy from eq.~(\ref{eq:energy_complex_time_average_spherical_wave}) becomes
\begin{equation}\label{eq:intensity_energy_complex_time_average_plane_wave}
\begin{aligned}
\boxed{ \langle E \rangle_T = \frac{|\hat{p}|^2}{2 \rho_0 c_0^2} = \frac{\langle\boldsymbol I\rangle_T \cdot \boldsymbol e_r}{c_0} = \frac{|\langle\boldsymbol I\rangle_T|}{c_0} = \frac{|\hat{Q}(\boldsymbol y)|^2}{32 \rho_0 c_0^2 \pi^2 r^2} = \frac{\langle P_q \rangle_T}{4 \pi r^2 c_0} },
\end{aligned}
\end{equation}
where we used eq.~(\ref{eq:acoustic_power_complex_time_average_spherical_wave}) for the last equality. Sound pressure level from eq.~(\ref{eq:SPL_complex_spherical_wave}) equals
\begin{equation}\label{eq:SPL_complex_plane_wave}
\begin{aligned}
L = 10 \log_{10}\left( \frac{\rho_0 c_0 \langle\boldsymbol I\rangle_T \cdot \boldsymbol e_r}{4 \cdot 10^{-10} \text{\,Pa}^2} \right) = 10 \log_{10}\left( \frac{\rho_0 c_0^2 \langle E \rangle_T}{4 \cdot 10^{-10} \text{\,Pa}^2} \right)
\end{aligned}
\end{equation}
and if we again approximate $\rho_0 c_0 \approx 400$\,kg/(m$^2$s)
\begin{equation}\label{eq:SPL_complex_plane_wave_intensity}
\begin{aligned}
\boxed{ L = 10 \log_{10}\left( \frac{\langle\boldsymbol I\rangle_T \cdot \boldsymbol e_r}{10^{-12} \text{\,W/m}^2} \right) = 10 \log_{10}\left( \frac{c_0 \langle E \rangle_T}{10^{-12} \text{\,J}/(\text{m}^2\text{s})} \right) }.
\end{aligned}
\end{equation}
Still, although $p$ and $\boldsymbol v$ are in phase as in a plane wave, this does not change the fact that they both decay with the distance, $\propto 1/r$. This is the essential difference between a spherical and plane wave, as a result of different free-space Green's functions. The second-order quantities (energy and intensity) decay accordingly proportionally to $1/r^2$ in the far field, and the sound pressure level $20 \log_{10} (2) = 10 \log_{10} (4) \approx 6$\,dB per doubling the distance.

An important outcome is that in the far field time-averaged sound energy, rather than an RMS value of sound pressure, can be used for calculation of sound pressure level, even without a need to calculate particle velocity explicitly. As already discussed, sound pressure level is closely related to the human perception of sound (recall eq.~(\ref{eq:SPL}) and the discussion related to it). This will be used as the basis for both statistical theory and geometrical (ray) acoustics. In particular, the latter theory treats sound waves not as sound pressure fields, but as beams of energy radiated from a source in particular directions $\boldsymbol e_r$. Applied time averaging allows neglecting the phase differences between waves, which substantially simplifies the calculations. As will be shown, only absolute values of complex amplitudes remain relevant, as in eq.~(\ref{eq:intensity_energy_complex_time_average_plane_wave}).

In contrast to the far field, \textbf{acoustic (very) near field} (that is, part of the field in which the non-acoustic near field term dominates over the acoustic far field term) of a point source\footnote{If the source is not acoustically compact, the receiver can be in the far field with respect to some parts of the source region, even if it is close to the source, so that the field at its location is mixed, both near and far field. Some authors introduce the term geometric near field for such locations. In any case, the condition for acoustic far field, eq.~(\ref{eq:far_field_condition}), remains unchanged.} can be defined for distances from the source which satisfy $kr \ll 1$. Then, from eq.~(\ref{eq:v_solution_tailored_Green_wave_eq_free_space_compact_source_emission_time}),
\begin{equation}\label{velocity_pressure_near_field}
\boldsymbol v(\boldsymbol x, t) = \frac{-j}{\rho_0 c_0 kr} p(\boldsymbol x, t) \boldsymbol e_r.
\end{equation}
Pressure and velocity are out of phase by the value $\pi/2$ and the velocity magnitude is much larger (by the factor $1/(kr)$) than that of a plane wave, $|\hat{p}|/(\rho_0 c_0)$. In addition to this, from eq.~(\ref{eq:energy_complex_time_average_spherical_wave}), $\langle E \rangle_T = \langle\boldsymbol I\rangle_T \cdot \boldsymbol e_r / (2 c_0 k^2 r^2)$, which is by the factor $1/(2k^2 r^2)$ larger than the far-field value $\langle\boldsymbol I\rangle_T \cdot \boldsymbol e_r / c_0$. This shows that most of the kinetic energy in the acoustic near field does not propagate far from the source, into the far field, since the associated particle velocity magnitude decays $\propto 1/r^2$, while the sound pressure amplitude decays $\propto 1/r$. In fact, the acoustic near field is essentially incompressible\footnote{For $kr \ll 1$ the second term in the wave equation~(\ref{eq:wave_eq}) $\sim p/r^2$ dominates over the first term $\sim \omega^2 p/c_0^2 = k^2 p$, so the wave equation becomes Laplace's equation, $\nabla^2 p = 0$. The same holds for an incompressible medium, in which the excitation is transmitted instantaneously, with $c_0 \rightarrow \infty$. Acoustic compactness of the space prohibits propagation of waves, like in an acoustically compact cavity discussed in the context of eq.~(\ref{eq:Helmholtz_number}) or the small source region in eq.~(\ref{eq:source_mass_injection}), where $\rho_f = \rho_0$.} (hence, non-acoustic) and the medium (air) in it acts as an oscillating solid body, rather than a compressible fluid. Nevertheless, this energy does contribute to the value of sound level in eq.~(\ref{eq:SPL_complex_plane_wave_intensity}), which is why the expression holds only in the far field. A typical example of the (acoustically relevant) effect of such non-acoustic energy are in-ear headphones. The receiver (eardrum) is located in the very near field of the compact source of sound for practically entire audible frequency range. This allows efficient reception of low-frequency sound, which otherwise does not propagate far from the small source.

Acoustic near field has much less practical relevance in room acoustics than the far field, since inequality (\ref{eq:far_field_condition}) is usually satisfied, except very close to the sources (or other objects of interest, which might reflect or scatter the sound and thus behave as secondary sources) of relatively low-frequency sound. For example, a distance of $r = 5$\,m, satisfies the far-field condition for all frequencies above 100\,Hz, while for middle frequencies above 500\,Hz, the near field can be neglected already at 1\,m. Moreover, between $r \ll 1/k$ and $r \gg 1/k$ is a transition zone in which the near-field effects are still relatively small, so eq.~(\ref{eq:far_field_condition}) is often replaced in practice with less restrictive $r > \lambda$ or even $r > \lambda/2$.

In the far field of a compact directional source, magnitude $|\hat p(r)|$ in eq.~(\ref{eq:acoustic_power_level_complex_time_average_plane_wave_omni_source_pressure_derivation}) should be additionally multiplied with the angularly dependent factor $D_i(\theta, \phi)$ (see also footnote~\footref{ftn:directional_source_far_field}), which gives a generalization of eq.~(\ref{eq:acoustic_power_level_complex_time_average_plane_wave_omni_source_pressure}) for directional sources,
\begin{equation}\label{eq:acoustic_power_level_complex_time_average_plane_wave_directional_source_pressure}
\boxed{ L(r,\theta, \phi) = L_W - 10 \log_{10} \left( \frac{4 \pi r^2}{1\text{\,m}^2} \right) + 10 \log_{10} D_i^2(\theta, \phi) }.
\end{equation}
Similarly, power of a directional point source can be expressed with acoustic quantities in the far field by multiplying the angularly-independent intensity in the surface integral in eq.~(\ref{eq:acoustic_power_complex_time_average_spherical_wave}) with the second-order factor $D_i^2$. As a result,
\begin{equation}\label{eq:acoustic_power_complex_time_average_plane_wave}
\begin{aligned}
\langle P_q \rangle_T &= \oint_S D_i^2(\theta, \phi) \langle \boldsymbol I(r,t) \rangle_T \cdot \boldsymbol e_r(\boldsymbol x) d^2 \boldsymbol x
= |\langle \boldsymbol I(r, t) \rangle_T| \oint_S D_i^2(\theta, \phi) d^2 \boldsymbol x &\\
&= c_0 \langle E(r,t) \rangle_T \oint_S D_i^2(\theta, \phi) d^2 \boldsymbol x = \frac{|\hat{p}(r)|^2}{2 \rho_0 c_0} \oint_S D_i^2(\theta, \phi) d^2 \boldsymbol x &\\
&= \frac{|\hat{Q}(\boldsymbol y)|^2}{32 \rho_0 c_0 \pi^2 r^2} \oint_S D_i^2(\theta, \phi) d^2 \boldsymbol x,&
\end{aligned}
\end{equation}
where we also used eq.~(\ref{eq:intensity_energy_complex_time_average_plane_wave}). For an omnidirectional source with constant $D_i = 1$ (which does not affect sound power of the source) this reduces to eq.~(\ref{eq:acoustic_power_complex_time_average_spherical_wave}), since
\begin{equation}
\begin{aligned}
\oint_S 1 d^2 \boldsymbol x = 4 \pi r^2
\end{aligned}
\end{equation}
for the spherical control surface.

\subsection{Acoustic impedance}\label{ch:acoustic_impedance}

In the acoustic far field of a source of sound in free space, complex amplitude of sound pressure and particle velocity are simply related by means of eq.~(\ref{eq:v_solution_tailored_Green_wave_eq_free_space_compact_source_emission_time_far_field_amlitude}). Since pressure and velocity are in phase, the scaling factor is real. It defines the \textbf{characteristic impedance} of the fluid,
\begin{equation}\label{eq:impedance_plane_wave}
Z_0 = \frac{\hat p}{\hat {\boldsymbol v} \cdot \boldsymbol e_r} =  \rho_0 c_0.
\end{equation}
In air at room temperature $Z_0 \approx 412$\,kg/(m$^2$s), which is often conveniently approximated as 400\,kg/(m$^2$s). In general, ratio of $\hat p$ and $\hat {\boldsymbol v} \cdot \boldsymbol e_r$ can be complex even in a simple unbounded medium, as in the near field of a point source, where the waves are not plane (eq.~(\ref{velocity_pressure_near_field}) divided with $e^{j\omega t}$). However, characteristic impedance of a medium is given for plane waves and $\boldsymbol e_r$ matches the direction of wave propagation.

On an arbitrary real or imaginary surface, \textbf{impedance} at point $\boldsymbol y$ is defined similarly as
\begin{equation}\label{eq:impedance}
\boxed{ Z(\boldsymbol y) = \frac{\hat{p}(\boldsymbol y)}{\hat{\boldsymbol v}(\boldsymbol y) \cdot \boldsymbol n(\boldsymbol y)} },
\end{equation}
with the unit vector $\boldsymbol n(\boldsymbol y)$ normal to the surface at $\boldsymbol y$ and pointing into the surface\footnote{Notice that in the earlier formulations of boundary conditions we set $\boldsymbol n$ pointing into the room, which is exactly the opposite direction if $Z$ is defined for a boundary surface. Therefore some care is required (see, for example, the derivation of eq.~(\ref{eq:impedance2}) below). Here  we stick to the usual convention for impedance. For normal sound incidence $\boldsymbol n$ has the same direction as the incoming sound wave, $\boldsymbol n = \boldsymbol e_r$. The same direction was also chosen for closed control surfaces above, for example, in the definition of sound power, eq.~(\ref{eq:acoustic_power}).} (and not necessarily into the direction of wave propagation). It is infinite for a rigid motionless wall, since $\hat{\boldsymbol v} \cdot \boldsymbol n = 0$ for any finite $\hat{p}$ at its surface. Hence, $Z = \infty$ is an alternative form of the boundary condition for a rigid wall in eq.~(\ref{eq:Helmholtz_p_BC}). Indeed, impedance introduces another form for specification of boundary conditions. It is sometimes called a mixed boundary condition, because it involves two acoustic quantities, $\hat p$ and $\hat{\boldsymbol v}$. The scalar product with $\boldsymbol n$ turns the latter vector into scalar.

In the special case of normal incidence of an incoming plane wave, $\boldsymbol n = \boldsymbol e_r$ from above and the surface behaves effectively as a fluid with $Z_0 = Z$ in the half-space behind the surface, for the sound field in front of it. If this impedance also matches the impedance of the fluid in front of the surface, the surface is acoustically transparent, as if the fluid extended behind the surface to infinity. It does not affect the amplitude or phase of the sound wave, nor it produces a reflection. Its impedance is fully matched to the characteristic impedance of the fluid in front of it. Accordingly, surface impedance is often normalized with the value $Z_0$ of the fluid in front of it:
\begin{equation}\label{eq:specific_impedance}
\mathcal{Z}(\boldsymbol y) = \frac{Z(\boldsymbol y)}{Z_0} = \frac{Z(\boldsymbol y)}{\rho_0 c_0},
\end{equation}
which is a dimensionless quantity called \textbf{specific impedance}.

The impedance is defined pointwise and, as such, its use is meaningful only for \textbf{locally reacting surfaces}. The impedance of such surfaces at a certain point $\boldsymbol y$ does not depend on the sound field or associated motion of the surface at any other location and, within linear theory, this makes the impedance a property of the surface alone, regardless of the sound field. This does not hold in general. For example, excitation of solid plates or membranes at one point propagates through the structure and affects its motion at other locations. The ratio $\hat{p}/(\hat{\boldsymbol v} \cdot \boldsymbol n)$ can still be formally defined at any $\boldsymbol y$, but apart from the values of $\hat p$ and $\hat{\boldsymbol v}$ there, it also depends on their distributions over the surface. Hence, when we talk about impedance, we assume that the surface is essentially locally reacting. This is the case, for example, with porous materials as well as a rigid motionless wall.

As already mentioned, impedance presents a mixed type of boundary condition which can be related to the general boundary condition given in eq.~(\ref{eq:Helmholtz_p_BC}). To this end, we notice that the scalar product of the conservation of momentum in frequency domain, eq.~(\ref{eq:Euler_momentum_sine_wave}), with $\boldsymbol n(\boldsymbol y)$ divided with $e^{j\omega \tau}$ gives $j\omega \rho_0 (\hat{\boldsymbol v} \cdot \boldsymbol n) = - \nabla_y \hat p \cdot \boldsymbol n$ at the boundary. Then, from eq.~(\ref{eq:impedance}),
\begin{equation}\label{eq:impedance2}
Z(\boldsymbol y) \frac{\nabla_y \hat p(\boldsymbol y) \cdot \boldsymbol n(\boldsymbol y)}{j \omega \rho_0} - \hat{p}(\boldsymbol y) = 0.
\end{equation}
Since the unit vector $\boldsymbol n$ in eq.~(\ref{eq:Helmholtz_p_BC}) points in the direction opposite from the vector $\boldsymbol n$ in eq.~(\ref{eq:impedance}), we have replaced $\boldsymbol n$ with $-\boldsymbol n$. Multiplying both sides with $j\omega/c_0$ gives
\begin{equation}\label{eq:impedance3}
\frac{Z(\boldsymbol y)}{\rho_0 c_0} \nabla_y \hat p(\boldsymbol y) \cdot \boldsymbol n(\boldsymbol y) - \frac{j \omega}{c_0} \hat{p}(\boldsymbol y) = - jk \hat{p}(\boldsymbol y) + \mathcal{Z}(\boldsymbol y) \nabla_y \hat p(\boldsymbol y) \cdot \boldsymbol n(\boldsymbol y) = 0.
\end{equation}
Comparing this with eq.~(\ref{eq:Helmholtz_p_BC}) gives $a = -jk$ and $b = \mathcal{Z} = Z/Z_0$ and $c = 0$ for a passive locally reacting surface. Indeed, impedance is just an alternative form of the boundary condition for a locally reacting surface.

Impedance of a surface can also be related to the flow of acoustic energy through it. According to equations~(\ref{eq:intensity_complex_time_average}) and (\ref{eq:acoustic_power_complex_time_average}) (with $\boldsymbol n$ pointing outwards, in agreement with the definition of impedance), the associated power is expressed as the integral over the open surface:
\begin{equation}\label{eq:acoustic_power_complex_time_average1}
\begin{aligned}
\langle P \rangle_T &= \frac{1}{2} \int_S \mathcal{R}_e(\hat{p}^* \hat{\boldsymbol v} \cdot \boldsymbol n) d^2 \boldsymbol y&\\
&= \frac{1}{2} \int_S \mathcal{R}_e(Z^* (\hat{\boldsymbol v} \cdot \boldsymbol n)^* (\hat{\boldsymbol v} \cdot \boldsymbol n)) d^2 \boldsymbol y = \frac{1}{2} \int_S \mathcal{R}_e(Z) |\hat{\boldsymbol v} \cdot \boldsymbol n|^2 d^2 \boldsymbol y&\\
&= \frac{1}{2} \int_S \mathcal{R}_e \left(\hat p^* \frac{\hat p}{Z} \right) d^2 \boldsymbol y = \frac{1}{2} \int_S \frac{|\hat p|^2 }{\mathcal{R}_e(Z)} d^2 \boldsymbol y.&
\end{aligned}
\end{equation}
Hence, only real part of the impedance (the resistance) determines the flow of energy. When it is larger than zero, the surface is passive (the net flow of energy is into the surface), when it is negative, the surface is active (it radiates sound energy into the half-space in front of it). Notice that a passive surface may reflect ($\hat p$ is by definition total sound pressure amplitude, not only the incident sound component), but with the reflected energy not larger than the incoming energy.

Yet another option to define a boundary condition at a point $\boldsymbol y$ of the surface is with a relation between complex amplitudes of the incident sound pressure $\hat{p}_{inc}$ and reflected $\hat{p}_{refl}$ or total $\hat p = \hat{p}_{inc} + \hat{p}_{refl}$ sound pressure. If the incident wave reaches the surface from the direction specified by polar and azimuthal angle $(\iota_{inc},\chi_{inc})$ of the spherical coordinate system centred at $\boldsymbol y$, the relation can be written generally as
\begin{equation}\label{eq:surface_system}
\hat{p}_{refl}(\theta, \phi) = f(\hat{p}_{inc}(\iota_{inc},\chi_{inc})),
\end{equation}
where $0 \leq \iota_{inc}, \theta \leq \pi/2$, $0 \leq \chi_{inc}, \phi < 2 \pi$ in the half-space in front of the surface ($\iota_{inc} = 0$ for normal incidence) and $(\theta, \phi)$ specifies direction of the reflected wave. If the specular reflection ($\theta = \iota_{inc}$ and $\phi = \chi_{inc}+\pi$) dominates or direction of the reflected sound is irrelevant, the dependence on $\theta$ and $\phi$ can be omitted. Moreover, if a surface is locally reacting, the angle of incidence determined by $\iota_{inc}$ and $\chi_{inc}$ is also irrelevant for the field at a point $\boldsymbol y$. Indeed, boundary conditions are commonly expressed without angular dependence, like impedance above. Another example is \textbf{reflection coefficient} defined as $\hat{R}_s = \hat{p}_{refl}/\hat{p}_{inc}$, which will be used in section~\ref{ch:basic_elements}. This shows again diversity of representations of boundary conditions.

\subsection{Sound rays}\label{ch:sound_rays}

The concept of sound ray is essential for the geometrical theory of acoustics and we shall introduce it before discussing the theory in more details in the next section. \textbf{Sound ray} is an energy-based model of essentially plane sound wave in the far field of a (possibly directional) source. It models a very narrow section of the actual spherical wave (scaled by the directivity in the source) which is radiated within a small solid angle $d\Omega = \sin(\theta) d\theta d\phi$ of the spherical coordinate system $(r,\theta,\phi)$ centred at the location of the source. Surface element of a control sphere with the centre at the source location and radius $r \gg 1/k$ has area $dS = r^2d\Omega$, which for a fixed $d\Omega$ increases with $r$. Since $d\Omega$ is very small and $kr \gg 1$, the surface element is approximately flat and the sound field over it is nearly uniform (pressure gradient depends only on $r$), which justifies the plane-wave approximation.

Like energy of a plane wave, energy of a sound ray does not decay with $r$, because the solid angle $d\Omega$ is fixed, not $dS$. Integral of a generic function $f(\boldsymbol x)$ over the entire spherical control surface equals
\begin{equation}\label{eq:surface_integral}
\oint_S f(\boldsymbol x) d^2 \boldsymbol x = r^2 \int_{0}^{4\pi} f(r, \Omega) d\Omega = r^2 \int_{0}^{2\pi} \int_{0}^{\pi} f(r, \theta, \phi) \sin(\theta) d\theta d\phi.
\end{equation}
Using this in eq.~(\ref{eq:acoustic_power_complex_time_average_plane_wave}),
\begin{equation}\label{eq:acoustic_power_complex_time_average_plane_wave_differential}
\begin{aligned}
\langle P_q \rangle_T = \int d\langle P_q \rangle_T = \oint_S c_0 \langle E(r,t) \rangle_T D_i^2(\theta, \phi) d^2 \boldsymbol x,
\end{aligned}
\end{equation}
implies that the time-averaged sound energy which is radiated within the angle $d\Omega$ can be expressed in terms of sound power of the source as
\begin{align*}
d\langle P_q \rangle_T = c_0 D_i^2(\theta, \phi) \langle E (r,t) \rangle_T r^2 d\Omega.
\end{align*}
Therefore, the rate of energy (power) which is transferred by a sound ray with solid angle $d\Omega$ equals
\begin{equation}\label{eq:ray_power}
\langle P_{q,ray}(\theta, \phi, t) \rangle_T = \frac{d\langle P_q \rangle_T}{d\Omega} = c_0 D_i^2(\theta, \phi) \langle E (r,t) \rangle_T r^2 = D_i^2(\theta, \phi) \frac{\langle P_{q} \rangle_T}{4 \pi},
\end{equation}
which does not depend on $r$, since there is no energy decay due to the expansion of the wavefront ($d\Omega$ is fixed). For the last equality, we used eq.~(\ref{eq:intensity_energy_complex_time_average_plane_wave}), which also implies
\begin{equation}\label{eq:intensity_energy_complex_time_average_ray}
\begin{aligned}
\langle E_{ray}(\theta, \phi, t) \rangle_T &= \frac{|\langle\boldsymbol I_{ray}(\theta, \phi, t)\rangle_T|}{c_0} \\
&= \langle E (r,t) \rangle_T \frac{r^2}{1\text{\,m}^2} = D_i^2(\theta, \phi) \frac{\langle P_{q} \rangle_T/1\text{\,m}^2}{4 \pi c_0} = \frac{1}{c_0} \frac{\langle P_{q,ray}(\theta, \phi, t) \rangle_T}{1\text{\,m}^2}\\
&=\frac{|\hat p (r)|^2}{2 \rho_0 c_0^2} \frac{r^2}{1\text{\,m}^2}
\end{aligned}
\end{equation}
for the unit length of $r$. This is energy of a virtual \textbf{sound particle} propagating along the sound ray, which does not decay with $r$. It has a continuous angular distribution for the infinitesimal $d\Omega$. In numerical (ray tracing) simulations, the full solid angle is subdivided into finite number of sections, $N$, which then replaces $4\pi$ in the numerator in the last two equations.

The particles are thus carriers of sound energy, not sound pressure, and the energy does not decay with the distance. The ray model is valid only when the waves can be approximated as plane waves. Other energy losses, such as due to dissipation in air or at absorbing surfaces, can be included by additional multiplication of the second-order quantities with real positive factors taking values between 0 and 1. This is the case with absorption coefficient which will be introduced later, in eq.~(\ref{eq:conservation_of_energy}). After giving all the necessary preliminaries in this section we can proceed with the remaining two energy-based theories of sound fields in closed spaces.

\section{Geometrical and statistical theory}\label{ch:geometrical_statistical_theory}

After considering mostly free-space sound fields in time domain, we shall introduce statistical and geometrical theory of sound fields in closed spaces. Together with the modal analysis from section~\ref{ch:modal_analysis}, they are the major theories of room acoustics. Both of them are naturally expressed in terms of sound energy, rather than pressure, and in addition to this follow from the statistical treatment, which assumes averaging the energy over frequency, as we explain next.

\subsection{Summation of energy}\label{ch:energy_summation}

In the previous section we saw that sound energy suffices for a physical description of the acoustic far field, as long as we are interested in time-averaged values, such as sound pressure level in eq.~(\ref{eq:SPL_complex_plane_wave}), that is, when we can neglect phase of the waves. However, this was shown only for a single source in free space. If multiple sources or reflections in a bounded space are present, the field at a receiver location $\boldsymbol x$ is expected to be a \textbf{superposition} (the theory is linear) of two or more waves. In section~\ref{ch:modal_analysis}, the superposition of reflections in a room was treated in frequency domain, as a (weighted) sum of the room eigenmodes. In time domain, the sound field at $\boldsymbol x$ is a superposition of travelling waves, for example, direct sound from the source (the free-space component) and reflections from the boundary surfaces and other objects in the room. Before continuing building the theories based on sound energy as the main quantity, we have to make sure that superposition of waves in the far field can also be captured accurately enough.

For that purpose we can observe two simple real sine waves with frequency $f = \omega/(2 \pi)$. Two plane waves in the far field are most generally given with sound pressures $p_1 = A \cos(\omega t)$ and $p_2 = B \cos(\omega (t+\Delta t))$, where $A$ and $B$ are real amplitudes larger than one and $\omega \Delta t$ is an arbitrary phase shift between the two waves at the receiver location. Without a loss of generality, we can also let $t\geq 0$ and $\Delta t \geq 0$. Total energy of the superposition of these two waves is from eq.~(\ref{eq:energy_plane_wave})
\begin{align*}
\begin{aligned}
E_{sum} &= \frac{(p_1+p_2)^2}{\rho_0 c_0^2} = \frac{1}{\rho_0 c_0^2} \left[ A \cos(\omega t) + B \cos(\omega (t+\Delta t)) \right]^2 &\\
&= \frac{1}{\rho_0 c_0^2} \{ (A \cos(\omega t))^2 + [B \cos(\omega (t+\Delta t))]^2 + 2AB \cos(\omega t) \cos(\omega (t+\Delta t)) \} &\\
&= E_1 + E_2 + \frac{AB}{\rho_0 c_0^2} \{ \cos(-\omega \Delta t) + \cos(2 \omega t+ \omega \Delta t) \} &\\
&= E_1 + E_2 + \frac{AB}{\rho_0 c_0^2} \{ \cos(\omega \Delta t) + \cos(2 \omega (t + \Delta t)) \},&
\end{aligned}
\end{align*}
where $E_1 = p_1^2/(\rho_0 c_0^2)$ and $E_2 = p_2^2/(\rho_0 c_0^2)$ are energies of the two waves. In general, the energy $E_{sum}$ does not have to correspond to a single sound wave, if the two waves merely intersect at $\boldsymbol x$. It is simply the total (instantaneous) sound energy at the given receiver location.

\begin{figure}[h]
	\centering
	\begin{subfigure}{.47\textwidth}
		\centering
		\includegraphics[width=1\linewidth]{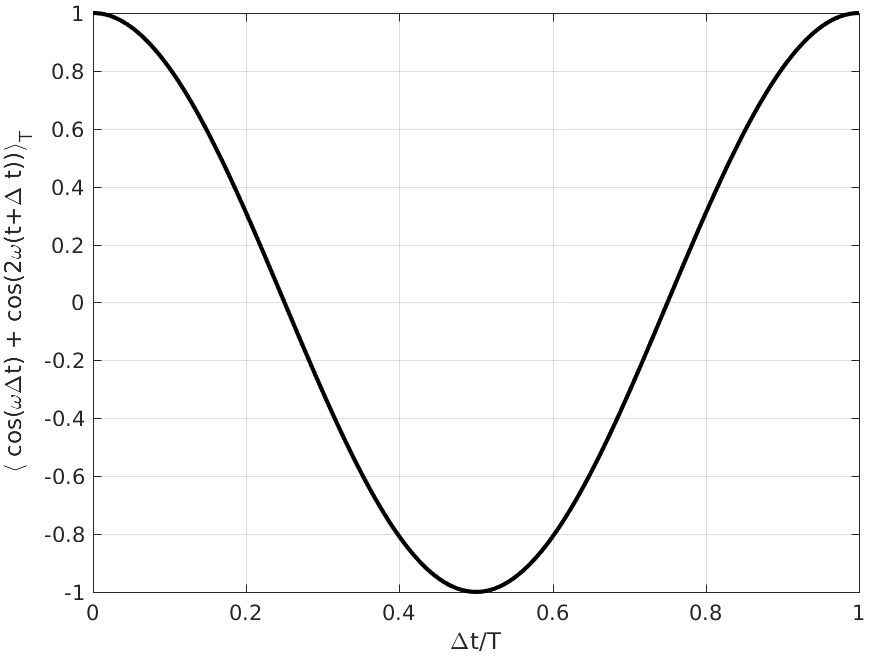}
		\label{fig:energy_sum_term}
	\end{subfigure}%
	\begin{subfigure}{.5\textwidth}
		\centering
		\includegraphics[width=1\linewidth]{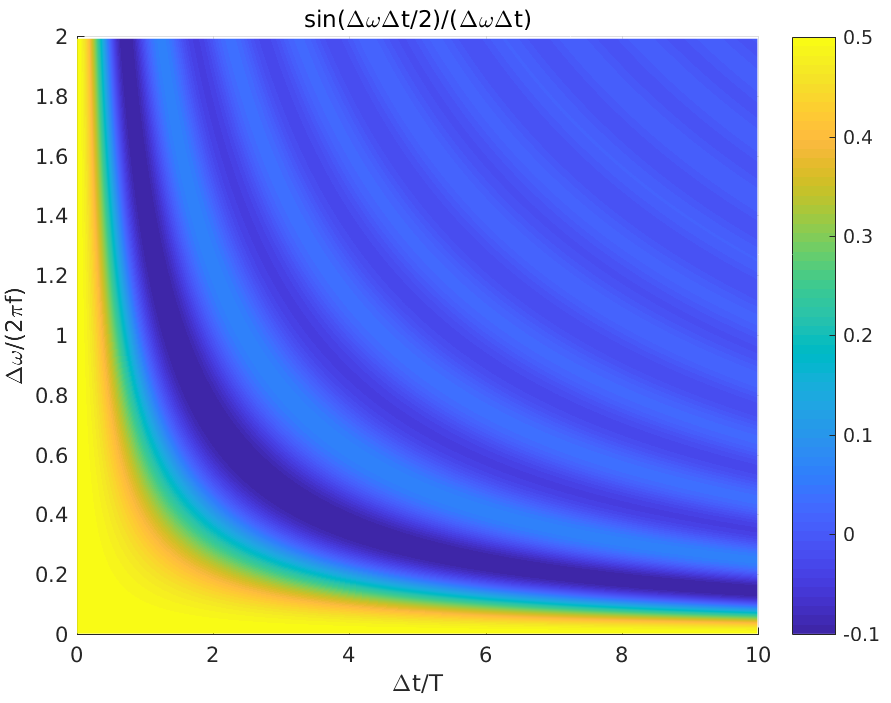}
		\label{fig:energy_sum_freq_average_term}
	\end{subfigure}
	\caption{Normalized correction of the sum of energy of two sine waves with frequency $f = \omega/(2\pi)$ averaged over (left) time and (right) frequency in the range $\Delta \omega/(2\pi)$ centred at $\omega$. Time difference $\Delta t$ of the waves is normalized with the period $T = 1/f$.}
	\label{fig:energy_sum_additional_terms}
\end{figure}

Time-averaged energy is accordingly
\begin{equation}\label{eq:energy_two_plane_waves_time_average}
\langle E_{sum} \rangle_T = \langle E_1 \rangle_T + \langle E_2 \rangle_T + \frac{AB}{\rho_0 c_0^2} \langle \cos(\omega \Delta t) + \cos(2 \omega (t + \Delta t)) \rangle_T
\end{equation}
and depends not only on the average energies of the two sound waves, but also on the time difference of their arrival at the receiver location, $\Delta t$. Depending on the phase shift between them, sine waves can add constructively or (partly) cancel. Moreover, any fixed value $\Delta t$ gives different phase shifts as a function of frequency, $\Delta \Phi = \omega \Delta t$. Value of the last term in eq.~(\ref{eq:energy_two_plane_waves_time_average}) without the amplitude-dependent factor $AB/(\rho_0 c_0^2)$ is shown in the left diagram in Fig.~\ref{fig:energy_sum_additional_terms}, where 1 corresponds to maximal constructive superposition and -1 to maximal cancellation (which does not necessarily leave zero total energy, because $A$ and $B$ do not have to be equal).

Evidently, simple sum of time-averaged energies does not quantify superposition of waves accurately enough. Fortunately, acoustic quantities are most commonly considered averaged in finite frequency bands, as well, typically octave or third-octave bands. As already discussed in the context of modal density in section~\ref{ch:density_of_modes}, human auditory system also averages sound energy over certain narrower, but finite frequency ranges. As we demonstrate next, it is exactly the \textbf{frequency averaging} which makes sound energy sufficient as a field quantity at relatively high frequencies. Still, the averaging is valid only if the observed frequency bands and the sound content in them are broad enough, meaning that the latter consists of many simple sinusoidal components which we analyse here. In general, we assume that the sound is at least as wide as the frequency band.

Frequency averaging of the total energy $E_{sum}$ from above in the range of angular frequencies $\Delta \omega$ centred at $\omega$ gives\footnote{Recall that the main reason for introducing complex sine functions was indeed to simplify the calculus.} (compare with time averaging in eq.~(\ref{eq:RMS})):
\begin{align*}
\begin{aligned}
\langle& E_{sum} \rangle_\omega = \langle E_1 \rangle_\omega + \langle E_2 \rangle_\omega + \frac{AB}{\rho_0 c_0^2} \langle \cos(\omega \Delta t) + \cos(2 \omega (t + \Delta t)) \rangle_\omega &\\
&= \langle E_1 \rangle_\omega + \langle E_2 \rangle_\omega + \frac{AB}{\rho_0 c_0^2} \frac{1}{\Delta \omega} \left[ \int_{\omega-\Delta \omega/2}^{\omega+\Delta \omega/2} \cos(\omega \Delta t) d\omega + \int_{\omega-\Delta \omega/2}^{\omega+\Delta \omega/2} \cos(2 \omega (t + \Delta t)) d\omega \right] &\\
&= \langle E_1 \rangle_\omega + \langle E_2 \rangle_\omega + \frac{AB}{\rho_0 c_0^2} \frac{1}{\Delta \omega} \frac{1}{\Delta t} \int_{(\omega-\Delta \omega/2) \Delta t}^{(\omega+\Delta \omega/2) \Delta t} \cos(\omega \Delta t) d(\omega \Delta t) &\\
&+ \frac{AB}{\rho_0 c_0^2} \frac{1}{\Delta \omega} \frac{1}{2(t+\Delta t)} \int_{2(\omega-\Delta \omega/2) (t+\Delta t)}^{2(\omega + \Delta \omega/2) (t+\Delta t)} \cos(2 \omega (t + \Delta t)) d(2\omega (t+\Delta t)) &\\
&= \langle E_1 \rangle_\omega + \langle E_2 \rangle_\omega + \frac{AB}{\rho_0 c_0^2} \frac{1}{\Delta \omega} \frac{1}{\Delta t} \left[ \sin((\omega+\Delta \omega/2)\Delta t) - \sin((\omega-\Delta \omega/2)\Delta t) \right] &\\
&+ \frac{AB}{\rho_0 c_0^2} \frac{1}{\Delta \omega} \frac{1}{2(t+\Delta t)} \left[ \sin(2(\omega + \Delta \omega/2) (t+\Delta t)) - \sin(2(\omega - \Delta \omega/2) (t+\Delta t)) \right] &\\
&= \langle E_1 \rangle_\omega + \langle E_2 \rangle_\omega + \frac{2AB}{\rho_0 c_0^2} \frac{1}{\Delta \omega} \frac{1}{\Delta t} \sin(\Delta \omega \Delta t/2) \cos(\omega \Delta t) &\\
&+ \frac{AB}{\rho_0 c_0^2} \frac{1}{\Delta \omega} \frac{1}{t+\Delta t} \sin(\Delta \omega (t+\Delta t)) \cos( 2\omega (t+\Delta t)).&
\end{aligned}
\end{align*}
The last two terms vanish in favour of the sum $\langle E_1 \rangle_\omega + \langle E_2 \rangle_\omega$ for any $\omega$ when $\Delta \omega \Delta t \gg 1$ and $\Delta \omega (t + \Delta t) \gg 1$, which is when $\Delta \omega \gg 1/ \Delta t$ (this automatically satisfies both conditions, since $t, \Delta t \geq 0$ and therefore $t+\Delta t \geq \Delta t$). This is also demonstrated in Fig.~\ref{fig:energy_sum_additional_terms} (right), which shows the value of $\sin(\Delta \omega \Delta t/2)/(\Delta \omega \Delta t)$ as a function of $\Delta \omega$ and $\Delta t$, normalized with the central angular frequency of the range, $\omega$, and the associated period $T = 2\pi/\omega$, respectively. As $\Delta \omega \Delta t \rightarrow \infty$, the value approaches zero. On the other hand, the factor $2AB/(\rho_0 c_0^2)$ is of the same order of magnitude as $\langle E_1 \rangle_\omega + \langle E_2 \rangle_\omega \sim (A^2 + B^2)/(\rho_0 c_0^2)$, if $A$ and $B$ are of the same order, or lower, if $A \gg B$ or $B \gg A$.

As a conclusion, sum of energies of sound waves (or sound rays as their representations, with which we will work later) matches accurately the energy of their superposition if the condition $\Delta \omega  \Delta t \gg 1$ is satisfied, even if the waves are mutually coherent\footnote{By coherent, we mean that spectra of the waves including the phases are very similar. Similarity of the spectra implies similarity of the waveforms. However, two waves with similar amplitude spectra can still be incoherent if their phase distributions are different.}, as the two sine waves here (but with some finite bandwidth, say, at least $\Delta \omega$). For example, if a delay between two reflections, replicas of the direct sound, is only 1\,ms, the condition becomes $\Delta f \gg 1/ (2 \pi \Delta t) \approx 160$\,Hz, so the sound and frequency range should be at least around 1\,kHz broad. However, if the delay is 10\,ms (which is roughly the initial time delay gap in large halls -- see section~\ref{ch:descriptors_of_room_acoustics}), the bandwidth should be $\Delta f \gg 16$\,Hz, which is satisfied by most of the octave bands.

Sound waves in rooms are rarely narrowband and coherent as the two sine waves above. If broadband sounds are incoherent (with different wave forms), their phase differences at different frequencies are random enough and summation of energies is justified regardless of $\Delta t$. In other words, energy summation is not applicable for narrowband (very small $\Delta \omega$) and coherent sounds (with equal freuquencies) or coherent broadband sounds with similar time of arrival (such that $\Delta \omega \Delta t \gg 1$ does not hold). In all other cases the approximation
\begin{equation}\label{eq:energy_two_plane_waves_time_average_incoherent}
\boxed{ \langle E_{sum} \rangle_{\omega} = \langle E_1 \rangle_{\omega} + \langle E_2 \rangle_{\omega} }
\end{equation}
is valid. In practice, short delays are quite common, so the condition is usually that the waves are \textbf{broadband} enough (not tonal) and \textbf{incoherent}. Fortunately, even the reflections originating from the same source are usually incoherent enough already after a few reflections from real surfaces, at least as long as the source does not emit pure tones.

Finally, we can allow again complex values of sound pressure. Additional averaging over time removes the phase dependence and leaves
\begin{equation}\label{eq:energy_two_plane_waves_time_freq_average_incoherent}
\begin{aligned}
\langle E_{sum} \rangle_{\omega, T} &= \langle E_1 \rangle_{\omega, T} + \langle E_2 \rangle_{\omega, T} = \frac{\langle |\hat{p}_1|^2 \rangle_{\omega} + \langle |\hat{p}_2|^2 \rangle_{\omega} }{2 \rho_0 c_0^2} &\\
&= \frac{\langle |\hat{Q}(\boldsymbol y)|^2 \rangle_{\omega}}{32 \rho_0 c_0^2 \pi^2} \left( \frac{1}{r_1^2} + \frac{1}{r_2^2} \right) = \frac{\langle P_q \rangle_{T,\omega}}{4 \pi c_0} \left( \frac{1}{r_1^2} + \frac{1}{r_2^2} \right).
\end{aligned}
\end{equation}
We used eq.~(\ref{eq:intensity_energy_complex_time_average_plane_wave}) and the last two equalities hold when both waves originate from the same point source (for example, two reflections with different paths to the receiver with total lengths $r_1$ and $r_2$). Comparing this with eq.~(\ref{eq:intensity_energy_complex_time_average_ray}) for unit $r$, we see that
\begin{equation}\label{rays_energy_summation}
\langle E_{ray, sum} \rangle_{\omega, T} = \langle E_{ray,1} \rangle_{\omega, T} + \langle E_{ray,2} \rangle_{\omega, T}.
\end{equation}
Two sound rays sum energetically under the same conditions. In the following we leave averaging over $\omega$ implicit for the rays.

As already indicated, energy summation may be inappropriate in certain situations involving narrowband or coherent sounds. For example, two simultaneous tones with slightly different frequencies may cause beating, a low-frequency time modulation of the sound. Oscillating energy of their sum cannot be captured by a simple sum of the two energies. Strong coherent early reflections (with short delays with respect to the direct sound) can cause perceivable coloration (see section \ref{ch:subjective_criteria}). Such effects are also ignored by the energy summation. Too long intervals of time averaging (longer than several tenths of milliseconds) may also neglect temporal variations of sound amplitude which is perceivable  by human listeners. Accordingly, energy summation is not advisable in frequency ranges smaller than, say, $\Delta f = 5/ (2 \pi \cdot 20\text{\,ms}) \approx 40$\,Hz. In general, frequency averaging does involve a loss of information, which might be relevant in certain situations, both objectively and subjectively. However, in spite of these deficiencies and mainly due to the substantial simplification of calculations which it allows, frequency averaging and energy summation are indispensable at high frequencies, when the modal analysis becomes tedious. They gives rise to the statistical theory, as well as ray tracing modelling, which, in contrast to the modal theory, consider only (relatively broadband) sound energy averaged in finite frequency bands.

\subsection{Statistical theory and diffuse field}\label{ch:statistical_theory}

\textbf{Statistical} treatment of sound fields in rooms is especially simple if the field are ideally \textbf{diffuse}, meaning that the sound waves reach every receiver location in the room uniformly (with equal energy) from all possible directions. In the language of geometrical (ray) acoustics\footnote{Although some (stationary) results of the statistical theory can be derived directly from the time-independent modal analysis (see section~\ref{diffuse_field_modal_analysis}), decaying fields in damped rooms are more easily observed in the geometrical theory in time domain. Therefore, we proceed with the statistical theory as a special case of geometrical acoustics in diffuse fields.}, every receiver point is hit by incoming sound rays from all directions and the received sound energy does not depend on the angle of incidence. Therefore, we can treat energy of the rays as independent of angle and according to eq.~(\ref{eq:intensity_energy_complex_time_average_ray}), it equals
\begin{equation}\label{eq:intensity_energy_complex_time_average_ray_diffuse}
\langle E_{ray,diff}(t) \rangle_T = \frac{1}{c_0} |\langle\boldsymbol I_{ray,diff}(t)\rangle_T|.
\end{equation}
Every ray in the room carries the same energy, regardless of its direction. 

This energy can be related to the total energy $\langle E \rangle_T$ at $\boldsymbol x$. Integration over all directions of the incoming rays and eq.~(\ref{eq:intensity_energy_complex_time_average_plane_wave}) give
\begin{equation}\label{eq:intensity_energy_complex_time_average_plane_wave_reciever}
\begin{aligned}
|\langle\boldsymbol I \rangle_T| &= c_0 \langle E \rangle_T = \int_{0}^{4\pi} \langle \boldsymbol I_{ray}(\Omega)\rangle_T \cdot \boldsymbol e_r d\Omega = \int_{0}^{4\pi} |\langle\boldsymbol I_{ray}(\Omega)\rangle_T| d\Omega,
\end{aligned}
\end{equation}
where $\Omega$ is, as before, solid angle and we left the time dependence inside $\langle \text{ } \rangle_T$ implicit. Notice that the integration implies simple summation of energies of the rays as in eq.~(\ref{rays_energy_summation}) and under the same assumptions -- broadband and incoherent sound waves and averaging over frequency. Like $E_{sum}$ in section~\ref{ch:energy_summation}, $\boldsymbol I$ and $E$ do not have to be associated with any resulting ray. They just quantify the superposition of many waves at the particular location in the room. For example, the resulting intensity vector, sum of the intensity vectors of all incoming rays, would be zero in a diffuse field, since all directions are covered with an equal incoming energy. The scalar quantity $|\boldsymbol I|$ on the left-hand side of the equation is simply magnitude of the (non-physical\footnote{Still, it is often found in literature, which is why we use it here, but it carries no more information than the scalar energy.}) intensity vector, which can be attributed to the total energy $E$ via eq.~(\ref{eq:intensity_energy_complex_time_average_plane_wave}), and eq.~(\ref{eq:intensity_energy_complex_time_average_plane_wave_reciever}) can be seen as its definition. In a diffuse field, the equation simplifies to
\begin{equation}\label{eq:intensity_energy_complex_time_average_plane_wave_reciever_diffuse}
\begin{aligned}
c_0 \langle E_{diff} \rangle_T = |\langle\boldsymbol I_{ray,diff}\rangle_T| \int_{0}^{4\pi} d\Omega = 4\pi |\langle\boldsymbol I_{ray,diff}\rangle_T|.
\end{aligned}
\end{equation}
and therefore
\begin{equation}\label{eq:intensity_energy_diff}
\boxed{ \langle E_{diff} \rangle_T = \frac{4 \pi |\langle\boldsymbol I_{ray,diff}\rangle_T|}{c_0} = 4 \pi \langle E_{ray,diff} \rangle_T },
\end{equation}
from eq.~(\ref{eq:intensity_energy_complex_time_average_ray_diffuse}). This is completely analogue to the uniform and continuous distribution of sound power of a source in eq.~(\ref{eq:ray_power}) (for $D_i=1$ of an omnidirectional source) over the full solid angle $4\pi$, merely at the receiver location.

If a field is diffuse, it can be shown that it is necessarily \textbf{uniform}, that is, $\langle E_{diff} \rangle_T$ is constant in space (does not depend on the location $\boldsymbol x$ in the room). Consequently, we can observe the field in a room as a whole and sound energy at any point is equal to its average over the room interior. Together with time and frequency averaging, which are also necessary for the geometrical theory, such spatial averaging is implied by the statistical theory. A simple intuitive proof of uniformity is that if energy in some part of the sound field were lower than in any other part, sound rays coming from the former would carry less energy than those from the latter and the field would not be diffuse (uniform over all angles of incidence) at the locations where the rays from the two regions intersect. Energy of rays $\langle E_{ray,diff} \rangle_T$ in eq.~(\ref{eq:intensity_energy_complex_time_average_ray_diffuse}) is constant in space. The opposite, however, does not hold in general. If a field is uniform, it is not necessarily diffuse. For example, time-averaged energy of simple plane waves is constant, but the field is obviously not diffuse -- the entire energy is transported in a single direction.

Even when it is uniformly distributed in the room, sound energy can vary over time. Neglecting dissipation in the medium, major losses of sound energy normally occur at surfaces, boundaries of the room. As before, in order to obtain a particular solution, general solution of the wave equation (which are sound rays as models of plane waves) and sources have to be complemented by boundary conditions. 

To this end, we first consider a small surface element $dS_s$ at $\boldsymbol x$ hit by a sound ray (accordingly we can use $\boldsymbol x$ for a location on the surface, as for a receiver, not $\boldsymbol y$) at the angle of incidence\footnote{Here and often in the following we use the term angle of incidence in a narrower sense, for the polar angle only. This is a very common usage, since the observed quantities rarely depend on the azimuthal angle (for example, when a surface is axisymmetric with respect to its normal).} $\theta$, which is, as before, angle to the normal of the surface. Relation between intensity of the ray, $\boldsymbol I_{ray}$, and the power received by the surface element, $dP_{s,ray}$, is
\begin{equation}\label{eq:received_power_surface_element_single_ray}
\begin{aligned}
d\langle P_{s,ray}(\Omega) \rangle_T &= \langle \boldsymbol I_{ray}(\Omega)\rangle_T \cdot (-\boldsymbol n(\boldsymbol x)) dS_s = |\langle\boldsymbol I_{ray}(\Omega)\rangle_T| \cos(\theta) dS_s.&
\end{aligned}
\end{equation}
The additional minus sign in front of $\boldsymbol n$ (the unit vector normal to the surface and pointing into the room) compared to the definition in eq.~(\ref{eq:acoustic_power}) is because we are expressing the received, not radiated power. The same could be achieved with $\boldsymbol n$ pointing in the opposite direction, into the surface. Total power received from all rays which hit the element $dS$ is an integral over all angles $\Omega$ in the half-space in front of the surface,
\begin{equation}\label{eq:received_power_surface_element_all_rays}
\begin{aligned}
d\langle P_s \rangle_T &= \int_{0}^{2\pi} d\langle P_{s,ray}(\Omega) \rangle_T d\Omega = dS_s \int_{0}^{2\pi} |\langle\boldsymbol I_{ray}(\Omega)\rangle_T| \cos(\theta) d\Omega.&
\end{aligned}
\end{equation}
The last integral is called \textbf{irradiation strength},
\begin{equation}\label{eq:irradiation_strength}
\langle I_s \rangle_T = \frac{d\langle P_s \rangle_T}{dS_s} = \int_{0}^{2\pi} |\langle\boldsymbol I_{ray}(\Omega)\rangle_T| \cos(\theta) d\Omega.
\end{equation}
It is a scalar quantity with the same physical unit as intensity, W/m$^2$. However, like $|\langle\boldsymbol I \rangle_T|$ in eq.~(\ref{eq:intensity_energy_complex_time_average_plane_wave_reciever}), it does not correspond to any wave nor it represents magnitude of some physical intensity vector. It simply quantifies total power incident to the surface per unit area.

In a diffuse field, the rays reaching the surface element from all angles have equal energy, so eq.~(\ref{eq:received_power_surface_element_all_rays}) simplifies to
\begin{equation}\label{eq:received_power_surface_element_all_rays_diffuse}
\begin{aligned}
d\langle P_{s,diff} \rangle_T &= |\langle\boldsymbol I_{ray,diff}\rangle_T| dS_s \int_{0}^{2\pi} \cos(\theta) d\Omega &\\
&= |\langle\boldsymbol I_{ray,diff}\rangle_T| dS_s \int_{0}^{\pi/2} \int_{0}^{2\pi} \cos(\theta) \sin(\theta) d\phi d\theta &\\
&= 2 \pi |\langle\boldsymbol I_{ray,diff}\rangle_T| dS_s \int_{0}^{\pi/2} \sin(\theta) \cos(\theta) d\theta &\\
&= 2 \pi |\langle\boldsymbol I_{ray,diff}\rangle_T| dS_s \left[ -\frac{1}{2} (\cos^2(\pi/2) - \cos^2(0)) \right] &\\& = \pi |\langle\boldsymbol I_{ray,diff}\rangle_T| dS_s&
\end{aligned}
\end{equation}
and the irradiation strength does not depend on the angle of incidence and equals
\begin{equation}\label{eq:irradiation_strength_diffuse}
\boxed{ \langle I_{s,diff} \rangle_T = \pi |\langle\boldsymbol I_{ray,diff}\rangle_T| = \frac{\langle E_{diff} \rangle_T c_0}{4} }.
\end{equation}
The last equality follows from eq.~(\ref{eq:intensity_energy_diff}). Like the energy $\langle E_{diff} \rangle_T$ in the bulk of the room, irradiation strength $\langle I_{s,diff} \rangle_T$ is also uniform over all surfaces and the relation between the two quantities is very simple.

In order to include energy losses at the surface, eq.~(\ref{eq:received_power_surface_element_single_ray}) can be multiplied with a real coefficient $\alpha_{s,\Omega}(\Omega)$ taking values between 0 (no energy losses, for fully reflecting surfaces, regardless of the direction of the reflected sound) and 1 (complete losses). In general, it depends on both angles of the spherical coordinate system with the origin at the location of the surface element, $\boldsymbol x$, in the half-space $0 \leq \theta \leq \pi/2$, $0\leq \phi < 2\pi$. The obtained quantity,
\begin{equation}\label{eq:energy_loss_ray}
d\langle P_{s,ray,loss}(\Omega) \rangle_T = \alpha_{s,\Omega}(\Omega) d\langle P_{s,ray}(\Omega) \rangle_T,
\end{equation}
represents fraction of the power delivered by the sound ray with the direction of arrival ($\theta$,$\phi$) which is not reflected back into the room (but either absorbed by the surface element or transmitted through it). The factor $\alpha_{s,\Omega}(\Omega)$ is (angularly-dependent) \textbf{absorption coefficient}. It is another form of boundary conditions suitable for energy-based acoustics. Notice that because it describes here boundary surfaces of rooms, it includes both true energy losses inside building elements (walls, porous absorbers, etc.) and apparent losses due to transmission outside the room. Total energy losses at the surface element from all incident rays are accordingly
\begin{equation}\label{eq:energy_loss_surface_element}
d\langle P_{s,loss} \rangle_T = \int_{0}^{2\pi} \alpha_{s,\Omega}(\Omega) d\langle P_{s,ray}(\Omega) \rangle_T d\Omega.
\end{equation}

In a diffuse field, eq.~(\ref{eq:received_power_surface_element_single_ray}) reads
\begin{equation}\label{eq:received_power_surface_element_single_ray_diff}
\begin{split}
d\langle P_{s,ray,diff}(\theta) \rangle_T = |\langle\boldsymbol I_{ray,diff}\rangle_T| \cos(\theta) dS_s
\end{split}
\end{equation}
The power $dP_{s,ray,diff}$ does not depend on $\phi$, but it does depend on $\theta$ due to the factor $\cos(\theta)$, even though the field is diffuse. After inserting this in eq.~(\ref{eq:energy_loss_surface_element}), we obtain
\begin{equation}\label{eq:energy_loss_surface_element_diffuse}
\begin{aligned}
d\langle P_{s,loss,diff} \rangle_T &=  |\langle\boldsymbol I_{ray,diff}\rangle_T| dS_s \int_{0}^{2\pi} \alpha_{s,\Omega}(\Omega) \cos(\theta) d\Omega &\\
&= \frac{d\langle P_{s,diff} \rangle_T}{\pi} \int_{0}^{2\pi} \alpha_{s,\Omega}(\Omega) \cos(\theta) d\Omega.&
\end{aligned}
\end{equation}
We used the result of eq.~(\ref{eq:received_power_surface_element_all_rays_diffuse}) for the last equality. The product $\alpha_{s,\Omega}(\Omega)dS_s$ is called \textbf{equivalent absorption area} of the surface element $dS_s$,
\begin{equation}\label{eq:equivalent_absoprtion_area_surface_element}
dA_s(\Omega) = \alpha_{s,\Omega}(\Omega)dS_s,
\end{equation}
and has the unit m$^2$. It is simply product of the absorption coefficient and surface area and matches (hence, ``equivalent'' in the name of the parameter) the area of a fully absorbing surface ($\alpha_{s,\Omega}=1$) with equal total absorption as the surface under consideration. As absorption coefficient, it can in general depend on the angle of incidence.

The last integral in eq.~(\ref{eq:energy_loss_surface_element_diffuse}) together with the factor $1/\pi$ represents absorption coefficient in a diffuse field,
\begin{equation}\label{eq:absorption_coeff_diffuse}
\alpha_{s,diff} = \frac{1}{\pi}\int_{0}^{2\pi} \alpha_{s,\Omega}(\Omega) \cos(\theta) d\Omega = 2 \int_{0}^{\pi /2} \alpha_{s,\Omega}(\theta) \sin(\theta) \cos(\theta) d\theta.
\end{equation}
The second equality holds in the most common case when $\alpha_{s,\Omega}$ does not depend on $\phi$ but only on the angle of incidence $\theta$. If $\alpha_{s,\Omega}(\Omega) = \alpha_s$ does not depend on $\theta$ either, its value is exactly equal to the absorption coefficient in a diffuse field:
\begin{equation}\label{eq:absorption_coeff_diffuse_const}
\alpha_{s,diff} = 2 \alpha_s \int_{0}^{\pi /2} \sin(\theta) \cos(\theta) d\theta = \alpha_s.
\end{equation}
The factor $1/\pi$ is necessary for this equality to hold. Note that the absorption coefficient $\alpha_{s,diff}$ of various absorbing materials is typically measured in reverberation chambers, for example, in octave bands, while $\alpha_{s,\Omega}(\theta = 0)$ for normal incidence is measured in a Kundt's tube. These two values do not necessarily match and neither one of them has to match the absorption coefficient value for some other particular angle of incidence\footnote{Although some simple relations can be derived under additional assumptions, eq.~(\ref{eq:absorption_coeff_diffuse}) suggests that in general the values of $\alpha_{s,\Omega}(\Omega)$ have to be known for all angles $\theta$ and $\phi$, in order to calculate $\alpha_{s,diff}$.}. It depends on a particular application, or, more specifically, type of the sound field at the surface, which absorption coefficient value is the most appropriate. For most of the applications in room acoustics, the field is approximately diffuse (at least the sound waves reach surfaces from many angles), so $\alpha_{s,diff}$ is commonly used.

Now that we expressed sound energy of a diffuse field in the bulk of the room, eq.~(\ref{eq:intensity_energy_diff}), as well as at the surfaces, eq.~(\ref{eq:irradiation_strength_diffuse}), and we know how to include energy losses at the surfaces, we can relate these by noticing that the total energy has to be conserved. Conservation of sound energy, eq.~(\ref{eq:enegy_governing_equations}), holds for any time scale. Therefore, it also holds for time-averaged energy, which, as discussed in section~\ref{ch:damping_and_reverberation_time}, can vary (decay) slowly over time in weakly damped rooms, such that its change over one period of the waves, $T$, is negligible\footnote{Otherwise, the time derivative in the conservation law would strongly couple with the integral over time of $\langle \text{ } \rangle_T$ from eq.~(\ref{eq:RMS}) and the two cannot be treated separately.}. Averaged over $T$ and integrated over the room volume $V$, with the losses at surfaces replacing the flux (sound intensity) term and introduced sources of sound with the total power $P_q$ on its right-hand side, the \textbf{conservation of energy} reads
\begin{equation}\label{eq:conservation_of_energy_integral}
\begin{aligned}
\int_{V} &\frac{\partial \langle E \rangle_T}{\partial t} d^3 \boldsymbol x + \int_{S} d\langle P_{s,loss} \rangle_T &\\
&= \int_{V} \frac{\partial \langle E \rangle_T}{\partial t} d^3 \boldsymbol x + \int_{S} \int_{0}^{2\pi} \alpha_{s,\Omega}(\Omega) d\langle P_{s,ray}(\Omega) \rangle_T d\Omega = \int_{V} d\langle P_q\rangle_T.&
\end{aligned}
\end{equation}
We used the divergence theorem to obtain the surface integral and inserted $d\langle P_{s,loss} \rangle_T$ from eq.~(\ref{eq:energy_loss_surface_element}). Energy rate in the first term is balanced by the losses at the boundaries with total surface area $S$ and total power of the sources in the room.

In a diffuse field, from equations (\ref{eq:energy_loss_surface_element_diffuse}), (\ref{eq:absorption_coeff_diffuse}), (\ref{eq:received_power_surface_element_all_rays_diffuse}), and (\ref{eq:irradiation_strength_diffuse}):
\begin{equation}\label{eq:conservation_of_energy_integral_diffuse}
\begin{aligned}
\int_{V} &\frac{\partial \langle E_{diff} \rangle_T}{\partial t} d^3 \boldsymbol x + \int_{S} d\langle P_{s,loss,diff} \rangle_T &\\
&= \int_{V} \frac{\partial \langle E_{diff} \rangle_T}{\partial t} d^3 \boldsymbol x + \int_{S} d\langle P_{s,diff} \rangle_T \alpha_{s,diff} &\\
&= \int_{V} \frac{\partial \langle E_{diff} \rangle_T}{\partial t} d^3 \boldsymbol x + \int_{S} |\langle\boldsymbol I_{ray,diff}\rangle_T| \pi \alpha_{s,diff} dS_s&\\
&= \int_{V} \frac{\partial \langle E_{diff} \rangle_T}{\partial t} d^3 \boldsymbol x + \int_{S} \alpha_{s,diff} \langle I_{s,diff} \rangle_T dS_s = \int_{V} d\langle P_q\rangle_T.&
\end{aligned}
\end{equation}
Since $\langle E_{diff} \rangle_T$ is uniform in $V$ and $\langle I_{s,diff} \rangle_T$ is uniform over $S$, they can leave the integrals and the equation simplifies to
\begin{equation}\label{eq:conservation_of_energy}
V \frac{d \langle E_{diff} \rangle_T}{d t} + \langle \alpha_{s,diff} \rangle_S S \langle I_{s,diff} \rangle_T = V \frac{d \langle E_{diff} \rangle_T}{d t} + \frac{\langle \alpha_{s,diff} \rangle_S S c_0}{4} \langle E_{diff} \rangle_T = \langle P_q\rangle_T,
\end{equation}
where we used eq.~(\ref{eq:irradiation_strength_diffuse}) and introduced average absorption coefficient of the surfaces,
\begin{equation}
\langle \alpha_{s,diff} \rangle_S = \frac{1}{S} \int_{S} \alpha_{s,diff} dS_s.
\end{equation}
Since the theory treats only spatially averaged and uniform energy, the exact distributions of absorption or sources are also irrelevant. The simple expression in eq.~(\ref{eq:conservation_of_energy}) is the central equation of the statistical theory. It describes two cases of great practical importance -- stationary state when the sources of stationary sound with the total time-averaged power $\langle P_q \rangle_T$ are on and the decay of sound energy after they have been switched off. These two regimes are considered next.

\subsubsection{Stationary state}\label{ch:stationary_state}

In a \textbf{stationary state}, after the sources of sustained sounds have been switched on and the field has reached a stationary regime, the energy is time-independent and eq.~(\ref{eq:conservation_of_energy}) becomes
\begin{align*}
\frac{\langle \alpha_{s,diff} \rangle_S S c_0}{4} \langle E_{diff} \rangle_T = \langle P_q\rangle_T,
\end{align*}
or
\begin{equation}\label{eq:energy_steady_diffuse}
\boxed{ \langle E_{diff} \rangle_T = \frac{4 \langle P_q\rangle_T}{\langle \alpha_{s,diff} \rangle_S S c_0} }.
\end{equation}
This important result relates uniform time and frequency averaged sound energy in the far diffuse field to acoustic power of the sources of stationary sound in the room, depending on the equivalent absorption area of the room. Sound pressure level can be expressed from equations~(\ref{eq:SPL_complex_plane_wave_intensity}) and (\ref{eq:energy_steady_diffuse}). It is, of course, constant in space and equals
\begin{equation}
\begin{aligned}
L = 10 \log_{10}\left( \frac{c_0 \langle E_{diff} \rangle_T}{10^{-12} \text{\,J}/(\text{m}^2\text{s})} \right) \approx 10 \log_{10}\left( \frac{\langle P_q\rangle_T /10^{-12}\text{\,W}}{\langle \alpha_{s,diff} \rangle_S S / 1 \text{\,m}^2} \right) + 6 \text{\,dB},
\end{aligned}
\end{equation}
that is,
\begin{equation}\label{eq:SPL_diffuse_field}
\boxed{ L \approx L_W - 10 \log_{10}\left( \frac{\langle \alpha_{s,diff} \rangle_S S}{1 \text{\,m}^2} \right) + 6 \text{\,dB} },
\end{equation}
where sound power level $L_W$ is defined in eq.~(\ref{eq:sound_power_level}). Under the assumption of a diffuse field and energy summation (recall the related discussion in section~\ref{ch:energy_summation}), one can easily estimate the sound pressure level in a room, based only on the sound power level of the sources and absorption coefficient values and areas of the surfaces.

However, close enough to a source, the field is neither diffuse nor uniform. One example is acoustic near field, but also acoustic far field is not diffuse, if the direct sound dominates over (approximately) diffuse reflected sound field. While energy of the diffuse field is constant in space, the direct sound energy decays with increasing distance from the source. This implies a certain distance $r_{c,diff}$ (called \textbf{critical distance} or reverberation distance), at which energies of the direct and reflected sound are equal. Equalizing eq.~(\ref{eq:energy_steady_diffuse}) with eq.~(\ref{eq:intensity_energy_complex_time_average_plane_wave}), which holds in the far-field of point monopole source with sound power $P_q$, we can express the critical distance from
\begin{equation}\label{eq:energy_complex_time_average_plane_wave_diffuse}
\langle E_{diff} \rangle_T = \frac{|\hat{p}_{diff}|^2}{2 \rho_0 c_0^2} = \frac{4 \langle P_q\rangle_T}{\langle \alpha_{s,diff} \rangle_S S c_0} = \frac{4}{\langle \alpha_{s,diff} \rangle_S S c_0} 2 \pi r_{c,diff}^2 \frac{|\hat{p}_{diff}|^2}{\rho_0 c_0}.
\end{equation}
It equals
\begin{equation}\label{eq:critical_distance_diffuse}
r_{c,diff} = \sqrt{\frac{\langle \alpha_{s,diff} \rangle_S S}{16\pi}}.
\end{equation}
Since it is derived for a far and diffuse field, the expression for critical distance is applicable only for frequencies which satisfy the condition $k r_{c,diff} \gg 1$ and if the reflected sound field is essentially diffuse.

\begin{figure}[h]
	\centering
	\begin{subfigure}{.46\textwidth}
		\centering
		\includegraphics[width=1\linewidth]{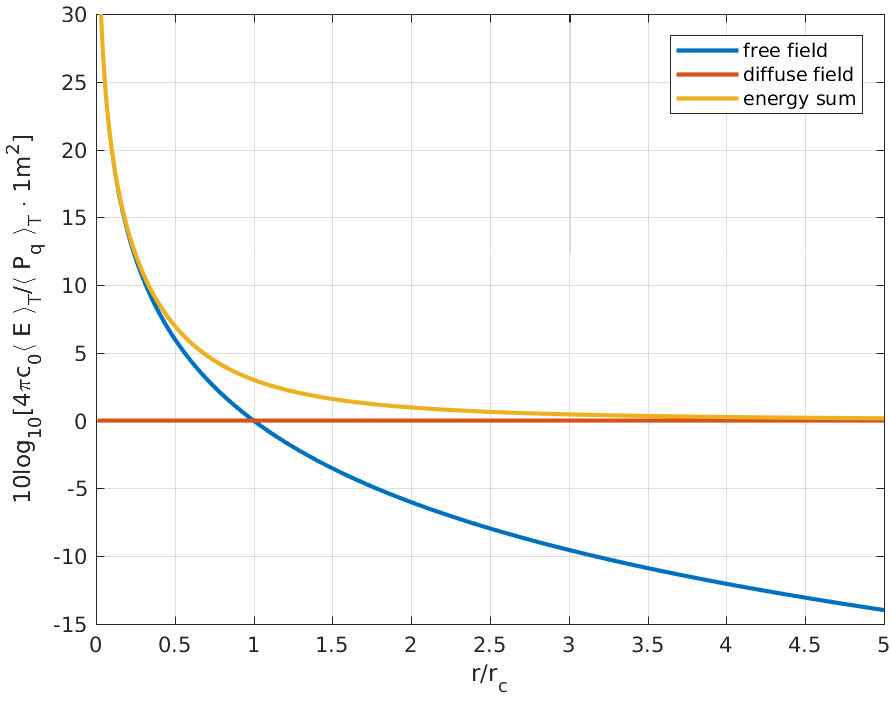}
		\label{fig:energy_rc}
	\end{subfigure}%
	\begin{subfigure}{.5\textwidth}
		\centering
		\includegraphics[width=1\linewidth]{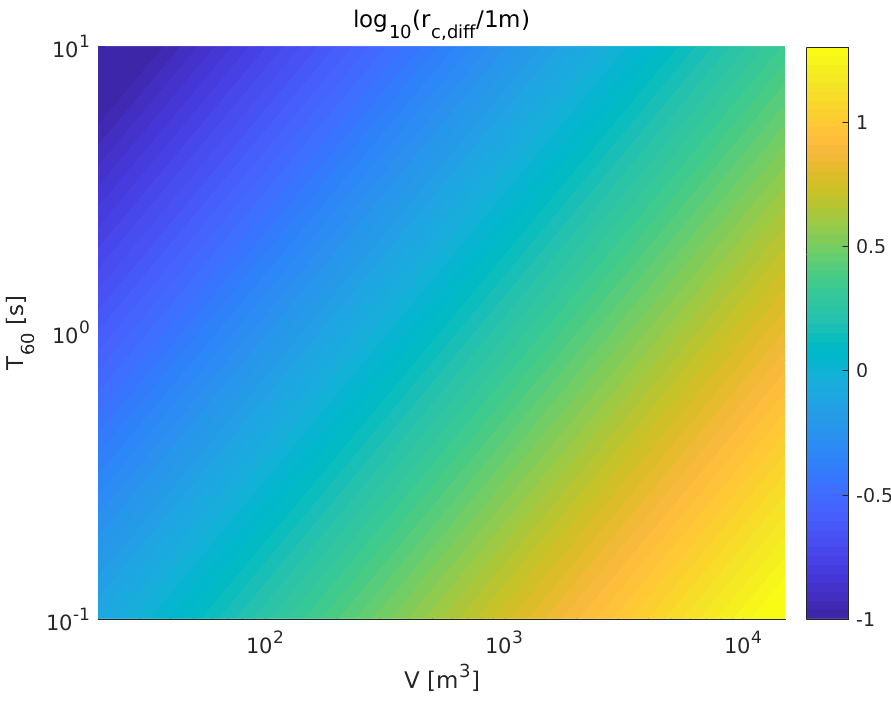}
		\label{fig:rc_V_T}
	\end{subfigure}
	\caption{Normalized sound pressure level in free and diffuse field as a function of distance from an omnidirectional point source normalized with the critical distance (left) and critical distance in a diffuse field from eq.~(\ref{eq:critical_distance_diffuse_T}) (right).}
	\label{fig:critical_distance}
\end{figure}

Only at distances from the source which are larger than the critical distance sound field can be diffuse (a necessary but not sufficient condition) and the statistical theory is valid. Conversely, direct sound can dominate over reflected sound only at distances smaller than the critical distance. Figure~\ref{fig:critical_distance} (left) shows the decay of sound energy of a point monopole in free space with distance from it, according to eq.~(\ref{eq:intensity_energy_complex_time_average_plane_wave}), and its constant value in a diffuse field generated by the same source, according to equations~(\ref{eq:energy_steady_diffuse}) and (\ref{eq:critical_distance_diffuse}). The blue and red lines thus represent the sound pressure levels from equations~(\ref{eq:acoustic_power_level_complex_time_average_plane_wave_omni_source_pressure}) and (\ref{eq:SPL_diffuse_field}) (with $\langle \alpha_{s,diff} \rangle_S S = 16\pi r_{c,diff}^2$ from eq.~(\ref{eq:critical_distance_diffuse})) and the yellow curve is their energy sum. The values in decibels are normalized such that sound pressure level in the diffuse field equals 0\,dB. For distances between roughly $r_{c,diff}/2$ and $2r_{c,diff}$, neither direct nor reflected sound dominates and for $r = r_{c,diff}$ the total energy is around 3\,dB higher than the energy of each one of the two components. It should also be mentioned that even when the direct sound energy can be neglected compared to the reflected energy for $r > r_{c,diff}$, this does not mean that it is of no importance, even in a diffuse field. For example, because it is the component which reaches a listener first, it is essential for source localization.

If the source is directional, critical distance in the direction of its maximum radiation can be estimated using the expression
\begin{equation}\label{eq:critical_distance_diffuse_directional_source}
r_{c,diff} = \sqrt{\gamma \frac{\langle \alpha_{s,diff} \rangle_S S}{16\pi}}.
\end{equation}
The gain $\gamma$ was defined in eq.~(\ref{eq:directivity_factor}) and its values larger than one can be utilized to extend the zone of the direct sound dominance around the axis of principal radiation of the source, when the effects of the room should be suppressed, as in control rooms and other rooms for sound reproduction. On the other hand, the field of a directional source is less likely to be diffuse, which makes the estimation of critical distance less accurate.

\subsubsection{Energy decay}\label{statistical_theory_energy_decay}

When the source is switched off (say, at time $t = 0$; more generally, but less commonly, several sources can be switched off at once), a transient regime begins, in which the sound \textbf{energy decays} until it vanishes. Following from eq.~(\ref{eq:conservation_of_energy}), this is expressed by the equality
\begin{equation}\label{eq:conservation_of_energy_decay_diffuse}
V \frac{d \langle E_{diff} \rangle_T}{d t} + \frac{\langle \alpha_{s,diff} \rangle_S S c_0}{4} \langle E_{diff} \rangle_T = 0.
\end{equation}
This ordinary differential equation has the solution\footnote{Compare the operators of equations~(\ref{eq:conservation_of_energy_decay_diffuse}) and (\ref{eq:wave_eq_time_separated}) and notice how the first-order time derivative gives real exponential function as the solution, while the second-order time derivative leads to oscillatory functions. Both effects are also captured by the left-hand side of eq.~(\ref{eq:oscillator_time}) of a damped oscillator.}
\begin{equation}\label{eq:conservation_of_energy_decay_diffuse_solution}
\langle E_{diff} \rangle_T = \langle E_{diff} \rangle_T^0 e^{-\frac{\langle \alpha_{s,diff} \rangle_S S c_0}{4 V}t}
\end{equation}
for any $t \geq 0$, where $\langle E_{diff} \rangle_T^0$ is sound energy in the room at the moment when the source is switched off, $\langle E_{diff} \rangle_T^0 = \langle E_{diff} \rangle_T$ at $t = 0$. We recognize exponential decay of energy considered in section~\ref{ch:damping_and_reverberation_time}. Hence, the statistical theory of decaying diffuse fields (with all its assumptions and averaging of quantities) justifies the exponential model of damping.

Comparing the last equation with the square of pressure amplitude in eq.~(\ref{eq:p_amplitude_exp_decay}), we find the damping constant
\begin{equation}\label{eq:damping_constant_diffuse}
\zeta = \frac{\langle \alpha_{s,diff} \rangle_S S c_0}{8 V},
\end{equation}
which is due to the absorbing surfaces of the room. The tightly related reverberation time from eq.~(\ref{eq:T60_damping}) equals
\begin{equation}\label{eq:T60_Sabine}
\boxed{ T_{60} \approx \frac{6.91}{\zeta} = 55.28 \frac{V}{\langle \alpha_{s,diff} \rangle_S S c_0} \approx 0.16 \frac{V \cdot 1\text{\,s/m} }{\langle \alpha_{s,diff} \rangle_S S} = 0.16\frac{V \cdot 1\text{\,s/m} }{A_{tot,diff}} },
\end{equation}
for $c_0 = 343$\,m/s and $A_{tot,diff} = \langle \alpha_{s,diff} \rangle_S S$ is total equivalent absorption area in the room, as introduced in eq.~(\ref{eq:equivalent_absoprtion_area_surface_element}). This is famous \textbf{Sabine's formula} which simply relates room volume, equivalent absorption area (product of average absorption coefficient and total surface area), and reverberation time. It is valid regardless of the room's geometry (the theory treats only spatially averaged values), as long as the far acoustic field is diffuse. In addition to this, the conditions for energy summation have to be fulfilled, which practically means that the superposed sound waves are broadband enough and mutually incoherent (see the comments with regard to eq.~(\ref{eq:energy_two_plane_waves_time_average_incoherent})) and time and frequency averaging are implied. It is interesting to note that the constant 0.16\,s/m (the unit is often omitted in literature) has been derived analytically, not empirically. It equals $8 \ln(10^{60/20})/c_0$. However, it can be rounded to 0.16\,s/m without a major loss of accuracy, since determining reverberation time down to a few percent of accuracy is futile. For further insight, the scaled ratio $S/V$ of a rectangular room is displayed in Fig.~\ref{fig:L_S_V_ratios_rect_room} (right). If necessary (see section~\ref{ch:damping_and_reverberation_time}), we can even include energy losses due to dissipation in air by adding the damping constant $\zeta_{air}$ from eq.~(\ref{eq:attenuation_constant}) to $\zeta$ from eq.~(\ref{eq:damping_constant_diffuse}). This gives a generalization of Sabine's formula
\begin{equation}\label{eq:T60_Sabine_air}
T_{60} \approx \frac{6.91}{\zeta + \zeta_{air}} = \frac{6.91}{\frac{\langle \alpha_{s,diff} \rangle_S S c_0}{8 V} + \frac{m_{air} c_0}{2}} \approx 0.16 \frac{V \cdot 1\text{\,s/m}}{\langle \alpha_{s,diff} \rangle_S S + 4 m_{air} V}.
\end{equation}
However, in most of the applications which do not involve very large reverberant spaces, $4 m_{air} V / (\langle \alpha_{s,diff} \rangle_S S) \sim m_{air} L^3/ (\langle \alpha_{s,diff} \rangle_S L^2) \sim m_{air} L/ \langle \alpha_{s,diff} \rangle_S \ll 1$, where $L$ is characteristic length scale of the room (see Table~\ref{tab:air_atttenuation_constant} for typical values of $m_{air}$).

Neglecting the dissipation in air, we can estimate the critical distance from eq.~(\ref{eq:critical_distance_diffuse}) in terms of reverberation time:
\begin{equation}\label{eq:critical_distance_diffuse_T}
\boxed{ r_{c,diff} = \sqrt{\frac{\langle \alpha_{s,diff} \rangle_S S}{16\pi}} = \sqrt{\frac{0.16 V \cdot 1\text{\,s/m}}{16\pi T_{60}}} \approx 0.056 \sqrt{\frac{V \cdot 1\text{\,s/m}}{T_{60}}} }.
\end{equation}
Logarithm of its value is shown in Fig.~\ref{fig:critical_distance} (right) for different values of $V$ and $T_{60}$. For $V \sim \mathcal{O}(100\text{\,m}^3)$ and $T_{60} \sim \mathcal{O}(1\text{\,s})$, critical distance is of the order of magnitude $r_{c,diff} \sim \mathcal{O}(1\text{\,m})$, increasing with the square root of volume and decreasing with the square root of reverberation time. This indicates that in most of the common rooms listeners are located in the zone where direct sound is much weaker than reflected sound and the room affects the field profoundly. Important exceptions (especially in large damped rooms) are microphones (as in recording studios or on stages) and mixing engineers in control rooms, which are located close to the sources, in order to receive strong direct sound of the source. Furthermore, as in eq.~(\ref{eq:critical_distance_diffuse_directional_source}), critical distance is larger around the axis of principal radiation of directional sources and equals
\begin{equation}\label{eq:critical_distance_diffuse_T_directional_source}
r_{c,diff} \approx 0.056 \sqrt{ \gamma \frac{V \cdot 1\text s/\text m}{T_{60}}}
\end{equation}
in the direction of maximum radiation. Analogue ``boost'' of the direct sound for the fixed distance from the source, due to the gain $\gamma$, can be achieved with directional receivers oriented towards the source.

The crucial assumption for obtaining the simple relations above is that the field is \textbf{diffuse} and therefore only spatially averaged values can be considered. With regard to that, it must be mentioned that truly diffuse field is very difficult to achieve in practice, even in the far field of omnidirectional sources. This can be due to specific geometry of the room (for example, the occurrence of acoustic shadows behind obstacles) or surfaces, which might reflect sound energy in different directions non-uniformly (for example, concave curved surfaces reflect bulk of energy in certain preferred directions), or due to unevenly distributed absorption inside the room, resulting in some surfaces reflecting significantly more energy than other. All this makes the incoming sound energy at the receiver location angularly dependent. Even in reverberation chambers with flat, non-parallel, and very low absorbing walls, additional diffusers are commonly used to make the sound field more diffuse. In spite of these complications, the statistical theory of diffuse fields remains robust enough and indispensable for quick and simple estimations of sound energy and reverberation time in rooms in relatively wide frequency bands, far enough from the source ($r > r_{c,diff}$ and $kr \gg 1$), using equations (\ref{eq:energy_steady_diffuse}) (or eq.~(\ref{eq:SPL_diffuse_field}) for sound pressure level) and (\ref{eq:T60_Sabine}). For more accurate (and computationally more involved) energy-based calculations of non-diffuse fields, geometrical acoustics and ray tracing simulations present a reasonable alternative.

\subsubsection{Diffuse field in modal analysis}\label{diffuse_field_modal_analysis}

Before we continue with the geometrical theory, it is very instructive to re-derive the expression for a stationary diffuse field in eq.~(\ref{eq:energy_steady_diffuse}) directly from the time-independent modal analysis of section~\ref{ch:modal_analysis}. This will show how the statistical model naturally emerges from the modal theory at high enough frequencies, above the Schroeder frequency, when modal analysis becomes unfeasible.

Together with equations~(\ref{eq:intensity_energy_complex_time_average_plane_wave}), (\ref{eq:T60_Sabine}), and (\ref{eq:acoustic_power_complex_time_average_spherical_wave}) and after recalling that $c_0/8 \approx 6.91/0.16\text{\,m}^{-1}\text{s}$, eq.~(\ref{eq:energy_steady_diffuse}) gives
\begin{equation}\label{mean_square_pressure_diffuse_field}
|\hat{p}_{diff}|^2 = 2 \rho_0 c_0^2 \langle E_{diff} \rangle_T = \frac{8 \rho_0 c_0 \langle P_q\rangle_T}{\langle \alpha_{s,diff} \rangle_S S} = \frac{8 \cdot 6.91 \rho_0 c_0 \langle P_q\rangle_T}{0.16\text{s/m} V \zeta} = \frac{|\hat{Q}|^2 c_0}{8\pi V \zeta},
\end{equation}
We shall derive the same result directly from the time-independent eq.~(\ref{eq:solution_tailored_Green_Helmholtz_modes_rect_room_hard_wall}), without reference to the ray acoustics in time domain. However, the statistical theory is energy-based, so we start with the square of eq.~(\ref{eq:solution_tailored_Green_Helmholtz_modes_rect_room_hard_wall}),
\begin{equation}\label{pressure_amplitude_modes_squared}
\begin{aligned}
|\hat{p}|^2 = \frac{64 c_0^4 |\hat{Q}|^2}{C_n^2 V^2} \left| \sum_{n} \frac{\prod_{i=1}^{3} \cos(k_{ni} x_i) \prod_{i=1}^{3} \cos(k_{ni} y_i)}{ \omega_n^2+2j \zeta_n \omega_n -\omega^2} \right|^2.
\end{aligned}
\end{equation}
As in section~\ref{ch:modal_analysis}, $L_1$, $L_2$, and $L_3$ are dimensions of the rectangular room with hard walls and $V = L_1 L_2 L_3$ its volume.

First we average eq.~(\ref{pressure_amplitude_modes_squared}) over all source and receiver locations in the volume $V$ (recall that the statistical theory implies that the energy distribution is uniform, giving its spatially averaged value for any distribution of sources):
\begin{equation}
\begin{aligned}
\frac{1}{V^2} \int_{V} &\int_{V} |\hat{p}|^2 d^3 \boldsymbol x d^3 \boldsymbol y \\
&= \frac{64 c_0^4 |\hat{Q}|^2}{C_n^2 V^4} \int_{V} \int_{V} \left| \sum_{n} \frac{\prod_{i=1}^{3} \cos(k_{ni} x_i) \prod_{i=1}^{3} \cos(k_{ni} y_i)}{ \omega_n^2 + 2j \zeta_n \omega_n -\omega^2} \right|^2 d^3 \boldsymbol x d^3 \boldsymbol y.
\end{aligned}
\end{equation}
At high enough frequencies, much above the Schroeder frequency, the modal density is high and many (mutually incoherent) modes overlap with different phases (section~\ref{ch:density_of_modes}). Accordingly, they sum energetically (as in section~\ref{ch:energy_summation}; averaging over finite frequency bands $\Delta \omega$ will be explicitly introduced shortly), so we can approximate
\begin{equation}
\begin{aligned}
\frac{1}{V^2} \int_{V} &\int_{V} |\hat{p}|^2 d^3 \boldsymbol x d^3 \boldsymbol y \\
&= \frac{64 c_0^4 |\hat{Q}|^2}{C_n^2 V^4} \int_{V} \int_{V} \sum_{n} \left| \frac{\prod_{i=1}^{3} \cos(k_{ni} x_i) \prod_{i=1}^{3} \cos(k_{ni} y_i)}{ \omega_n^2 + 2j \zeta_n \omega_n -\omega^2} \right|^2 d^3 \boldsymbol x d^3 \boldsymbol y \\
&= \frac{64 c_0^4 |\hat{Q}|^2}{C_n^2 V^4} \sum_{n} \frac{\int_{V} \int_{V} \prod_{i=1}^{3} \cos^2(k_{ni} x_i) \prod_{i=1}^{3} \cos^2(k_{ni} y_i) d^3 \boldsymbol x d^3 \boldsymbol y}{\left| \omega_n^2 + 2j \zeta_n \omega_n -\omega^2  \right|^2}.
\end{aligned}
\end{equation}
In the last equality we changed the order of the sum over modes with eigenfrequencies $\omega_n$ and the integrals over locations $\boldsymbol x$ and $\boldsymbol y$.

As in section~(\ref{ch:rectangular_room_wave_theory}), we can separate the three spatial variables in Cartesian coordinates and thereby split the two integrals into six,
\begin{equation}
\begin{aligned}
\int_{V} \int_{V} \prod_{i=1}^{3} &\cos^2(k_{ni} x_i) \prod_{i=1}^{3} \cos^2(k_{ni} y_i) d^3 \boldsymbol x d^3 \boldsymbol y \\
&= \prod_{i=1}^{3} \int_{0}^{L_i} \cos^2(k_{ni} x_i) dx_i \prod_{i=1}^{3} \int_{0}^{L_i} \cos^2(k_{ni} y_i) dy_i.
\end{aligned}
\end{equation}
At large Helmholtz numbers with respect to the room dimensions we can consider only the prevailing oblique modes, for which $k_{ni} \neq 0$ and $C_n = 1$. Equation~(\ref{spatial_integal_of_squared_cos}) gives the solution of the integrals,
\begin{equation}
\begin{aligned}
	\int_{0}^{L_i} \cos^2(k_{ni} x_i) dx_i = \frac{L_i}{2}
\end{aligned}
\end{equation}
and similar for the integrals over $y_i$. Therefore,
\begin{equation}
\frac{1}{V^2} \int_{V} \int_{V} |\hat{p}|^2 d^3 \boldsymbol x d^3 \boldsymbol y = \frac{c_0^4 |\hat{Q}|^2}{V^2} \sum_{n} \frac{1}{\left| \omega_n^2 + 2j \zeta_n \omega_n -\omega^2  \right|^2}.
\end{equation}

Since the modal density is high, we can replace the sum over modes with eigenfrequencies $\omega_n$ with an integral over continuous angular frequency $\omega$. The oblique modes have the continuous modal density from eq.~(\ref{eq:modes_rect_room_fn_density}),
\begin{equation}
\frac{dN_{\omega_n<\omega}}{d\omega} = \frac{1}{2\pi} \frac{dN_{f_n<f}}{df} \approx \frac{V \omega^2}{2 \pi^2 c_0^3}.
\end{equation}
Accordingly,
\begin{equation}\label{modes_superposition_to_diffuse_field}
\begin{aligned}
\frac{1}{V^2} \int_{V} \int_{V} |\hat{p}|^2 d^3 \boldsymbol x d^3 \boldsymbol y &= \frac{c_0 |\hat{Q}|^2}{2\pi^2 V} \int_{\omega_c - \Delta \omega/2}^{\omega_c + \Delta \omega/2} \frac{\omega^2}{\left| \omega^2 + 2j \zeta \omega -\omega_c^2  \right|^2} d\omega \\
&= \frac{c_0 |\hat{Q}|^2}{2\pi^2 V} \int_{\omega_c - \Delta \omega/2}^{\omega_c + \Delta \omega/2} \frac{\omega^2}{ (\omega^2 - \omega_c^2)^2 + 4 \zeta^2 \omega^2 } d\omega.
\end{aligned}
\end{equation}
Rather then integrating over all frequencies, we split the frequency range into finite frequency bands. A frequency band of width $\Delta \omega$ is centred at $\omega_c$. It is small enough that the supposed low damping constant is frequency-independent in it, $\zeta_n = \zeta \ll \omega$, but large enough and including many modes for energy summation to be valid. Low damping also implies that most of the energy of the mode $n$ is at frequencies $\omega \approx \omega_n$, so we replaced $2j\zeta_n \omega_n$ with $2j\zeta_n \omega = 2j\zeta \omega$, as already anticipated in the context of eq.~(\ref{eq:solution_tailored_Green_Helmholtz_mode}). Moreover, only the modes with frequencies $\omega_n$ within the frequency band $(\omega_c-\Delta \omega/2, \omega_c+\Delta \omega/2)$ have a significant contribution to the total energy in it and practically their entire energy is contained in it. This suggests that $\Delta \omega$ is much larger than the half-width of the modes, $2\zeta$, and we can formally integrate over an infinitely large frequency range, as long as only the modes with eigenfrequencies inside the frequency band are considered.

The last integral can be solved by introducing the auxiliary variable $u = (\omega^2-\omega_c^2)/(2\omega)$. Since $\omega \in (\omega_c-\Delta \omega/2, \omega_c+\Delta \omega/2)$ and $\Delta \omega \ll \omega$, $u \sim \Delta \omega \ll \omega$, $du = (1-u/\omega)d\omega \approx d\omega$, and $u \approx \Delta \omega/2$ for $\omega = \omega_c+\Delta \omega/2$ and $u \approx -\Delta \omega/2$ for $\omega = \omega_c-\Delta \omega/2$. The integral becomes
\begin{equation}
\begin{aligned}
\frac{1}{4} \int_{-\Delta \omega/2}^{\Delta \omega/2} \frac{1}{ u^2 + \zeta^2 } du &= \frac{1}{4 \zeta} \left( \arctan \left( \frac{\Delta \omega}{2 \zeta} \right) - \arctan \left( -\frac{\Delta \omega}{ 2\zeta} \right) \right) \\
&= \frac{1}{4 \zeta} \left( \frac{\pi}{2} - \left( -\frac{\pi}{2} \right) \right) = \frac{\pi}{4 \zeta},
\end{aligned}
\end{equation}
for $\Delta \omega \gg \zeta$. Substituting this in eq.~(\ref{modes_superposition_to_diffuse_field}), we obtain
\begin{equation}
\begin{aligned}
\frac{1}{V^2} \int_{V} \int_{V} |\hat{p}|^2 d^3 \boldsymbol x d^3 \boldsymbol y = \frac{c_0 |\hat{Q}|^2}{8\pi V \zeta},
\end{aligned}
\end{equation}
which matches $|\hat{p}_{diff}|^2$ in eq.~(\ref{mean_square_pressure_diffuse_field}). 

As expected, the expression for stationary acoustic energy in a diffuse field can be derived without switching back to time domain. It is equal to the spatially (for both source and receiver locations and over the entire room volume) and frequency (over a finite frequency band) averaged expression for energy obtained using the modal analysis. The key additional assumption is that the considered frequencies are high enough (the Helmholtz numbers are much larger than one) so that the modal density is high and continuous and dominated by many overlapping oblique modes, whose energies (rather than complex amplitudes) sum. The damping is also low and does not vary significantly within the frequency band, which is nevertheless much larger than the half-width of the modes.

Although the derivation here is more specific than the one in section~\ref{ch:stationary_state}, because it assumed a rectangular room, the same conclusion actually holds for acoustically large rooms with arbitrary shapes (like equations~(\ref{eq:modes_rect_room_Nfn})-(\ref{eq:Schroeder_freq})). The statistical theory is thus a high Helmholtz number approximation of the modal theory. Still, a certain frequency range between $kL \sim 1$ and $kL \gg 1$ remains, where the former theory is inaccurate and the latter is inefficient. This problematic range is in the focus of many new numerical techniques being developed, often as extensions of the techniques suitable for low or high frequencies. Rather than going into details of those, we discuss next another high-frequency theory, which does not require spatial averaging and is therefore suitable for non-diffuse fields.

\subsection{Ray tracing in a rectangular room}\label{ch:ray-tracing_rectangular_room}

Although the energy-based statistical theory is very simple and sufficiently accurate in many situations in practice, a more general theory is desirable, especially concerning the assumption of a diffuse field, the main constraint of the statistical approach. With regard to this, we shall go back and rather than considering only overall sound energy in a room (as in eq.~(\ref{eq:conservation_of_energy_integral}) and further), we shall examine distinct sound rays (which in general do not carry equal energies), their propagation through the room, and reflections at the surfaces in time. This is known as \textbf{geometrical (ray) acoustics} and we study its applications in room acoustics.

Similarly as with modal analysis, complexity of calculations based on the geometrical acoustics depends mainly on the complexity of the room geometry and boundary conditions. Ray tracing simulations are typically performed on computers and they will be described in more details in section~\ref{ch:ray_tracing}. In order to come up with an analytical solution, here we restrict ourselves again to a rectangular room as well as a diffuse field. Moreover, we assume that all six surfaces of the room are acoustically large. In terms of the Helmholtz number from eq.~(\ref{eq:Helmholtz_number}), this means that $kL_i \gg 1$, for $i = 1,2,3$, where $L_1, L_2$ and $L_3$ are, as before, dimensions of the room. Sound rays are thus assumed to specularly reflect from all surfaces of the room\footnote{This already introduces a certain error, because waves do not reflect specularly in the areas close to the edges or corners of the room. Two or three surfaces meeting at the edges or corners are not acoustically far from each other and simple plane waves do not provide an accurate model for sound propagation there. However, we neglect these areas in favour of much larger regions in which specular reflections do take place.}. In fact, like section~\ref{diffuse_field_modal_analysis}, the following analysis complements the one in section~\ref{ch:rectangular_room_wave_theory} for high Helmholtz numbers, when the latter becomes too demanding, and diffuse fields.

If sound dissipation in air is negligible, a sound ray loses energy only when it reflects from absorbing (or transmitting) surfaces, according to eq.~(\ref{eq:energy_loss_ray}). The remaining power, which is reflected from the surface element back into the room, equals
\begin{equation}\label{eq:reflected_power_ray}
d\langle P_{ray,refl}(\Omega) \rangle_T = (1-\alpha_{s,\Omega}(\Omega)) d\langle P_{s,ray}(\Omega) \rangle_T.
\end{equation}
For further simplicity, we assume a \textbf{diffuse} field and replace the angularly dependent absorption coefficient $\alpha_{s,\Omega}(\Omega)$ with its diffuse-field value averaged over all surfaces of the room, $\langle \alpha_{s,diff} \rangle_S$, for each surface and each sound ray in the room, regardless of its angle of incidence\footnote{In reality, certain degree of diffuseness and stochasticity of the angles of incidence would apply even for a single ray, owing to the scattering at non-flat surfaces, as well as successive specular reflections from non-parallel surfaces (as in an empty reverberation chamber). However, we cannot use this argument for the rectangular room discussed here, so we assume that the diffuseness is achieved by many rays, which originate from essentially omnidirectional sources and hit the room surfaces at different angles of incidence (indicated in eq.~(\ref{eq:reflected_power_ray_diffuse}) with $\Omega$ as the argument, even though the rays of a diffuse field carry equal energies regardless of their angles). Hence, they also reach all receiver locations with uniform energy from all angles.}. Therefore, the average value of absorption coefficient, $\langle \alpha_{s,diff} \rangle_S$, applies here not only to the room as a whole, but to each separate surface as well, which appears to be more restrictive than in the statistical theory. However, the diffuse field implies that the energy and absorption are equally distributed over surfaces (recall eq.~(\ref{eq:irradiation_strength_diffuse})), so the difference between the two constraints vanishes. Power of the reflected ray becomes
\begin{equation}\label{eq:reflected_power_ray_diffuse}
\begin{split}
d\langle P_{ray,diff,refl}(\Omega) \rangle_T = (1-\langle \alpha_{s,diff} \rangle_S) d\langle P_{s,ray,diff}(\Omega) \rangle_T.
\end{split}
\end{equation}
After $N_{ray}(\Omega)$ reflections, the remaining power of the ray is
\begin{equation}\label{eq:reflected_power_ray_alphaConst_Nreflections}
d\langle P_{ray,diff,N refl}(\Omega) \rangle_T = (1-\langle \alpha_{s,diff} \rangle_S)^{N_{ray}(\Omega)} d\langle P_{s,ray,diff}(\Omega) \rangle_T.
\end{equation}

For a rectangular room, we can calculate $N_{ray}(\Omega)$ for all rays explicitly. First we notice that
\begin{equation}\label{eq:number_reflections_ray_2surfaces}
N_{ray,i}(\Omega) = N_{ray,i}(\theta_i) =\frac{\Delta t c_0 \cos(\theta_i)}{L_i}
\end{equation}
is the number of reflections of a ray during the time interval $\Delta t$ from the two surfaces which are normal to the axis $x_i$ (at the distance $L_i$ from each other) and $\theta_i$ is the angle of incidence to those surfaces. Since the surfaces are parallel to each other and all reflections are specular, $\theta_i$ remains constant for any single ray. The ray travels the distance $L_i/\cos(\theta)$ between two successive reflections from the two surfaces, which leads to eq.~(\ref{eq:number_reflections_ray_2surfaces}).

In a diffuse field we can estimate the average number of reflections of all rays arriving uniformly from all incident angles $\theta_i$ to the two surfaces:
\begin{equation}\label{eq:number_reflections_diffuse_2surfaces}
\begin{aligned}
\langle N_i(\Omega) \rangle_\Omega &= \frac{1}{2 \pi} \int_{0}^{2 \pi} N_{ray,i}(\Omega) d\Omega = \frac{1}{2 \pi} \int_{0}^{2 \pi} \frac{\Delta t c_0 \cos(\theta_i)}{L_i} d\Omega &\\
&= \frac{\Delta t c_0}{2\pi L_i} 2\pi \int_{0}^{\pi /2} \cos(\theta_i) \sin(\theta_i) d\theta_i = \frac{\Delta t c_0}{2 L_i}.&
\end{aligned}
\end{equation}
The half-space in front of the surfaces is covered by the solid angle $2\pi$, which cancels out in front of the integral in the third equality, because the integration over the azimuthal angle $\phi$ gives the same factor. Therefore, the average number of reflections of all rays from all six surfaces equals
\begin{equation}\label{eq:N_reflections_ray}
\langle N(\Omega) \rangle_\Omega = \frac{\Delta t c_0}{2} \sum_{i=1}^{3} \frac{1}{L_i} = \frac{\Delta t c_0}{2} \frac{L_1 L_2 + L_1 L_3 + L_2 L_3}{L_1 L_2 L_3} = \frac{\Delta t c_0}{2} \frac{S/2}{V} = \frac{\Delta t c_0 S}{4 V}.
\end{equation}
As before, $S$ denotes total area of the surfaces and $V$ is volume of the room. The ratio $S/V$ in a rectangular room is shown in Fig.~\ref{fig:L_S_V_ratios_rect_room} (right). Although we do not prove it, the last equality too holds for rooms with arbitrary shapes, if the sound field is diffuse. Therefore, the results which follow can be generalized to non-rectangular rooms. In fact, eq.~(\ref{eq:N_reflections_ray}) can be used as an indicator of a diffuse field. If $\langle N(\Omega) \rangle_\Omega$ (calculated for example using ray tracing simulations) matches approximately the value $\Delta t c_0 S/(4V)$, the field in the room is likely to be diffuse. Nevertheless, eq.~(\ref{eq:N_reflections_ray}) refers only to the values averaged over all rays. Details of the distributions of ray path lengths (or time intervals) between successive reflections or $N(\Omega)$ around their average values do depend on particular shapes of rooms.

From equations (\ref{eq:reflected_power_ray_alphaConst_Nreflections}) and (\ref{eq:N_reflections_ray}), it follows that total sound energy of all rays in a diffuse field decays proportionally to $(1-\langle \alpha_{s,diff} \rangle_S)^{\langle N(\Omega) \rangle_\Omega} = (1-\langle \alpha_{s,diff} \rangle_S)^{\Delta t c_0 S / (4 V)}$. We can also include the dissipation in air from equations~(\ref{eq:p_energy_exp_decay}) and (\ref{eq:attenuation_constant}) by multiplying this factor with $e^{-2 \zeta_{air} \Delta t}$. Switching off the source at the beginning of the time interval $\Delta t$, reverberation time $T_{60}$ can be expressed by substituting $\Delta t = T_{60}$:
\begin{align*}
\begin{aligned}
10 \log_{10} & \left[ (1-\langle \alpha_{s,diff} \rangle_S)^{\frac{T_{60} c_0 S}{4 V}} e^{-2 \zeta_{air} T_{60}} \right] &\\
&= 10 \log_{10} \left[ e^{\frac{T_{60} c_0 S}{4 V} \ln(1-\langle \alpha_{s,diff} \rangle_S)-2 \zeta_{air} T_{60}} \right] = -60 \text{\,dB}.&
\end{aligned}
\end{align*}
It equals
\begin{equation}
T_{60} = \frac{\ln(10^{-6})}{\frac{c_0 S \ln(1-\langle \alpha_{s,diff} \rangle_S)}{4 V} - 2 \zeta_{air}} = \frac{-24 V \ln(10)}{c_0 (S \ln(1-\langle \alpha_{s,diff} \rangle_S) - 4 V m_{air})}
\end{equation}
or
\begin{equation}\label{eq:T60_Eyring}
\boxed{ T_{60} \approx 0.16 \frac{V \cdot 1\text{\,s/m}}{4 m_{air} V - S \ln(1-\langle \alpha_{s,diff} \rangle_S)} }.
\end{equation}
This is \textbf{Eyring's reverberation formula}. If we expand the natural logarithm into series,
\begin{align*}
\ln(1-\langle \alpha_{s,diff} \rangle_S) = -\langle \alpha_{s,diff} \rangle_S -\frac{1}{2}\langle \alpha_{s,diff} \rangle_S^2 -\frac{1}{3}\langle \alpha_{s,diff} \rangle_S^3 - ...,
\end{align*}
we see that for $\langle \alpha_{s,diff} \rangle_S \ll 1$, $\ln(1-\langle \alpha_{s,diff} \rangle_S) \approx -\langle \alpha_{s,diff} \rangle_S$ and Eyring's formula simplifies to eq. (\ref{eq:T60_Sabine_air}) or, for negligible dissipation in air, to Sabine's formula~(\ref{eq:T60_Sabine}). Analogously to eq.~(\ref{eq:SPL_diffuse_field}), sound pressure level in the room can be estimated with
\begin{equation}
\boxed{ L \approx L_W - 10 \log_{10}\left( \frac{- S \ln(1-\langle \alpha_{s,diff} \rangle_S)}{1 \text{\,m}^2} \right) + 6 \text{\,dB} },
\end{equation}
for negligible dissipation in air.

Since it simplifies to Sabine's formula for small $\langle \alpha_{s,diff} \rangle_S$, Eyring's formula appears to be more general and more accurate when the room is not very weakly damped, say for $\langle \alpha_{s,diff} \rangle_S > 0.15$. However, this is not necessarily true, since derivations of the two equations included somewhat different assumptions and the extent to which these assumptions are satisfied in a particular problem determines ultimately accuracy of the used equation. In particular, we derived Eyring's equation for a rectangular room (although, as mentioned above, eq.~(\ref{eq:N_reflections_ray}) and the results which follow from it are valid for arbitrary shapes), diffuse field (as Sabine's equation), all specular reflections in the room (which is irrelevant for the statistical theory), and equal absorption coefficient $\alpha_{s,diff} = \langle \alpha_{s,diff} \rangle_S$ of all surfaces (although the diffuse field already implies that the absorption is distributed uniformly). The last two are additional requirements in comparison to the derivation of Sabine's formula, which might affect the accuracy of Eyring's formula in certain cases, for example, when relatively small surfaces in the room have different values of the absorption coefficient or scatter the sound. In any case, an unevenly distributed absorption of the surfaces can lead to a non-diffuse sound field, which is a prerequisite for both theories, besides energy summation in the acoustic far field. The introduced assumptions for both formulas are summarized in Table~\ref{tab:Sabine_Eyring_assumptions}.

\begin{table}[h]
	\caption{Assumptions in the derivations of Sabine's and Eyring's formulas for reverberation time of a room.}
	\label{tab:Sabine_Eyring_assumptions}
	\begin{tabular}{ | p{3.3cm} | p{11.5cm} |}
		\hline
		\textbf{formula} & \textbf{assumptions} \\
		\hline
		Sabine, eq.~(\ref{eq:T60_Sabine}) and eq.~(\ref{eq:T60_Sabine_air}) & energy summation (broadband and incoherent reflections), acoustic far field, diffuse field, average absorption coefficient of the surfaces \\
		\hline
		Eyring, eq. (\ref{eq:T60_Eyring}) & energy summation (broadband and incoherent reflections), acoustic far field, diffuse field, specular reflections, equal absorption coefficients of the surfaces \\
		\hline
	\end{tabular}
\end{table}

Unlike the analysis of overall sound energy in the room from section~\ref{ch:statistical_theory}, the ray tracing analysis from this section reveals the important relation between energy decay (reverberation) and rate of reflections (number of reflections per second), in addition to the absorption at the surfaces. By definition, \textbf{rate of reflections} is
\begin{equation}\label{eq:reflections_rate}
\dot{N} = \frac{N}{\Delta t}.
\end{equation}
From eq.~(\ref{eq:N_reflections_ray}), average rate of reflections in a room with arbitrary shape with a diffuse field is
\begin{equation}\label{eq:reflections_rate_average}
\langle \dot{N} \rangle_\Omega = \frac{\langle N(\Omega) \rangle_\Omega}{\Delta t} = \frac{c_0 S}{4 V}.
\end{equation}
Accordingly, the mean \textbf{free path length} of the rays between two successive reflections is
\begin{equation}\label{eq:ray_path_length_average}
\boxed{ l = \frac{c_0}{\langle \dot{N} \rangle_\Omega} = \frac{4 V}{S} }.
\end{equation}
As already anticipated, either one of the two quantities, rate of reflection or free path length, and its distribution from the sample of all rays in the room can be used to assess diffuseness of the field, when available (typically from ray tracing simulations). If the distribution peaks at the mean value in eq.~(\ref{eq:reflections_rate_average}) or eq.~(\ref{eq:ray_path_length_average}), the field is largely diffuse and higher accuracy of the statistical theory and Eyring's formula can be expected. On the other hand, occurrence of other peaks, for example at shorter free path lengths may indicate partly decoupled spaces in the room, such as niches, in which more frequent local reflections take place. These reflections can cause irregularities in the exponential decay of energy, making the reverberation time formulas, as well as other formulas based on the diffuse field assumption, less accurate.

In general, reverberation time and distribution of sound energy in a room do depend on the geometry of the room and its surfaces, as well as the distribution of absorption, even in the far field of omnidirectional sources. Irregular room shapes or non-uniform placement of absorption may cause large differences in free path lengths or energy of different rays, resulting in a \textbf{non-diffuse} field. In such cases, certain absorbing surfaces may become less irradiated than other surfaces in the room and therefore less efficient (absorbing less energy) than in a diffuse field with uniform energy distribution (eq.~(\ref{eq:irradiation_strength_diffuse})). In other words, their effective absorption coefficient in the particular room can be lower than the diffuse-field value, which is measured in a reverberation chamber and usually used in the calculations. Depending on the area of the absorbing surface compared to the total area of the room, the reverberation time is then underestimated by equations~(\ref{eq:T60_Sabine_air}) and (\ref{eq:T60_Eyring}), even if all surfaces have similar absorption coefficient values.

In real rooms, sound absorption is often unevenly distributed. Important examples are auditorium, which is often the only substantially absorbing surface of the room, such as in concert halls or opera houses, and absorbing suspended ceilings or walls, which are frequent in many rooms, such as offices, lecture halls and classrooms, public places, etc. If all other surfaces are much less absorbing, sound fields in these rooms are not diffuse in general. Sound waves (or rays) reflected from the absorbing surfaces carry much less energy than other reflected waves and the sound energy at particular locations in the room is angularly dependent. Consequently, all room surfaces are not irradiated equally or uniformly over different angles of incidence. The absorbed power (eq.~(\ref{eq:energy_loss_surface_element}) or contributions to the integral in eq.~(\ref{eq:conservation_of_energy_integral}) involving $d\langle P_{s,ray}(\Omega) \rangle_T$ of each surface) depends on the locations of particular surfaces with respect to the highly absorbing surface(s) and the averaging over angles of incidence, as introduced in equations~(\ref{eq:conservation_of_energy_integral_diffuse}) and (\ref{eq:reflected_power_ray_diffuse}), and the substitution of $\alpha_{s,\Omega}(\Omega)$ with $\alpha_{s,diff}$ are not justified. Sound energy in the room is not uniform and the statistical theory becomes inaccurate.

More particularly, reverberation time can be overestimated by Eyring's and Sabine's formulas when area of the absorbing floor (auditorium) is large compared to the areas of non-absorbing walls (the room's length or width is significantly larger than the height). The reason is that the reflections reaching the floor from all other, more reflecting surfaces and from all directions are largely absorbed. This leaves the other surfaces less irradiated by the reflections from the floor, which results in a comparatively larger effective absorption of the floor than if the field was ideally diffuse. If the reflecting ceiling is high, reverberation time can be longer in the upper, more reflecting and diffuse part of the room, than in the lower, absorbing part. This is sometimes even perceivable in the auditorium, as additional longer reverberation of the upper part coupled to the shorter local reverberation in the auditorium. The opposite case is when an absorbing surface is located in a part of the room with lower sound energy, for example, at one of the two shorter walls of a long rectangular room. The surface is less irradiated and absorbs less efficiently than in a diffuse field and longer reverberation time should be expected than predicted based on the diffuse field assumption.

Of course, the exact deviations of the actual values of acoustic quantities from the values obtained using the diffuse field theory are case-dependent and accurate corrections are not easy to introduce. Overall (non-)uniformity of sound energy and distribution of absorption can at least indicate if positive or negative corrections of the simple diffuse-field formulas should be expected. More details of non-diffuse sound fields can be obtained by means of \textbf{ray tracing simulations}, which also rely on the assumption of energy summation and acoustic far field, as well as predominantly specular reflections. Further details on numerical calculations using ray tracing will be discussed in section \ref{ch:ray_tracing}. 

Having covered the three main theories of acoustics of closed spaces, the modal, geometrical, and statistical theory, we can continue with other related and more particular topics, such as measurement, assessment, modelling of room acoustics, and its optimization in rooms with different purposes and requirements (discussed in section~\ref{ch:introduction}) using various acoustic elements. These are the subjects of the remaining sections.

\section{Measurement and descriptors of room acoustics}\label{ch:measurements_and_descriptors}

In section~\ref{LTI_systems_and_impulse_response} we concluded that, like tailored Green's function, impulse response of a room observed as an LTI system contains all information about its acoustics (apart from the propagation time of the direct sound), for the given source and receiver locations. Therefore, it is no surprise that room acoustics is typically assessed from measured or estimated impulse responses at the locations of interest and compared against subjective and objective criteria discussed in section~\ref{ch:introduction}. Nevertheless, two things should be immediately emphasized. First, impulse responses are objective, while the most relevant judgment of room acoustics is usually subjective. This means that certain relations between different subjective criteria and impulse responses as objective signals have to be established, which is associated with many difficulties, as will be discussed below. 

Second, certain important properties of room acoustics or sound fields can be determined also without acquisition of impulse responses. That is, impulse responses carry too much information for some applications. For instance, spatial distribution of \textbf{sound level} in a room can be measured using a source of stationary noise (usually pink noise), broadband or in octave or third-octave frequency bands, for example, for the assessment of sound coverage in the auditorium or before obtaining spatially averaged values. Similarly, \textbf{reverberation time} can be measured according to the definition (so-called interrupted noise technique), by switching off the source of stationary sound, after the stationary regime is achieved (as in section \ref{ch:stationary_state} for the case of a diffuse field; it usually suffices to leave the source on for at least the time interval equal to the expected reverberation time of the room). From the recording of the decaying sound, the reverberation time can be estimated, for example, with the least squares linear fitting of the sound level decay curve, with a relatively short time interval of averaging (recall eq.~(\ref{eq:RMS})), say $\sim \mathcal{O} (10\text{\,ms})$. Of course, the recorded signal can be filtered using octave or third-octave filters prior to the fitting, in order to obtain frequency-dependent reverberation time values\footnote{If the achieved signal-to-noise ratio is smaller than the required 60\,dB, reverberation time can be estimated on a smaller range, similarly as $T_{30}$, $T_{20}$, and $T_{10}$ in equations~(\ref{eq:T30_Schroeder_curve})-(\ref{eq:T10_Schroeder_curve}).}.

Such simplified procedures are often sufficient for noise control (see Fig.~\ref{fig:room_acoustics_object_applications} (right) and Tables~\ref{tab:room_acoustics_applicatons} and \ref{tab:rooms_subjective_properties}). \textbf{Modal analysis} of frequency responses from section~\ref{ch:modal_analysis} is appropriate when resonance behaviour of the room should be investigated, especially in small rooms at relatively low Helmholtz numbers. If, however, more details of the (broadband) room acoustics are required, which is normally the case with the rooms in which auditory information is produced, impulse responses can be obtained quite routinely by implementing standard procedures. In this section we explain how impulse responses can be measured in existing rooms and which descriptors of room acoustics can be calculated from them, regardless of whether they have been acquired by measurements or some other means. Their prediction using different modelling techniques will be studied in section~\ref{ch:modelling}.

\subsection{Measurement of a room impulse response}\label{ch:acquisition_room_impulse_response}

Like tailored Green's function $G_{tail}(\boldsymbol x, t | \boldsymbol y, \tau)$, room impulse response does not depend only on the room geometry and boundary conditions, but also on the particular locations of the source and receiver. On the other hand, it should not depend on any other properties of the source or receiver (such as directivity and other properties listed in section~\ref{ch:characteristics_of_sources}). In certain problems, these can also be specified, relevant, and taken into account. However, impulse responses imply by definition \textbf{point omnidirectional} sources and receivers. Table~\ref{tab:imp_response_source_directivity} shows typical tolerance of directional radiation patterns in octave bands. Notice that sound pressure levels are not specified and their mean values can differ between bands. The tolerance is higher at higher frequencies, due to the fact that directivity of real sources usually increases with frequency.

\begin{table}[h]
	\caption{Typical maximum allowed variations of the far-field sound pressure level of a source used for measurements of room impulse responses over the polar and azimuthal angles of the spherical coordinate system centred at the source location and for a fixed distance from the source $r \gg 1/k$.}
	\label{tab:imp_response_source_directivity}
	\begin{tabular}{ | l | l | l | l | l | l | l |}
		\hline
		\textbf{octave} & \textbf{125\,Hz} & \textbf{250\,Hz} & \textbf{500\,Hz} & \textbf{1\,kHz} & \textbf{2\,kHz} & \textbf{4\,kHz} \\
		\hline
		deviation & $\pm 1$\,dB & $\pm 1$\,dB & $\pm 1$\,dB & $\pm 3$\,dB & $\pm 5$\,dB & $\pm 6$\,dB \\
		\hline
	\end{tabular}
\end{table}

Apart from being omnidirectional, the source should be \textbf{broadband} enough, that is, provide sufficient sound power over the entire frequency range of interest. Ideally, the sound power should not vary strongly with frequency, especially within a single frequency band (as by the sources of tonal sound). However, as we will see in section~\ref{ch:descriptors_of_room_acoustics}, large majority of the descriptors of room acoustics, which are commonly calculated from impulse responses, do not depend on absolute sound pressure levels and therefore sound power of the source. Hence, it is not strictly necessary that the source radiates with equal power in different frequency bands (recall, for example, the impulse of a balloon from Figures~\ref{fig:imp_response} and \ref{fig:freq_response}), as long as sufficient \textbf{signal-to-noise ratio} (typically above 30\,dB) is achieved in each band of interest. Since the background noise most often has energy distribution similar to pink noise, which is indeed constant over octave and third-octave bands, the same frequency distribution is preferred for sound power of the sources, because it minimizes the required total sound power of the source. However, a true source-independent response to an ideal (delta) impulse is not necessary to obtain for most of the applications and a response to a non-ideal impulse suffices. In contrast to this, angle- and frequency-independent sensitivity of microphones as well as high dynamic range and low self-noise are nowadays not difficult to achieve in practically entire audible frequency range, which makes these criteria less problematic for receivers.

The source can produce a short and strong \textbf{sound impulse}, thus approximating the ideal delta function in time. The room impulse response is acquired simply by recording the emitted impulse followed by the decay of sound at a receiver location, with sufficient sampling frequency and dynamic range (like the signal $p(t) \approx g(t)$ in Fig.~\ref{fig:imp_response}). Commonly used are starting pistols, firecrackers, balloons, clappers, and electric spark generators. However, their radiation patterns must not vary over different angles more than several decibels (Table \ref{tab:imp_response_source_directivity}) and they have to provide sufficient signal-to-noise ratio in the entire frequency range of interest, which unfortunately often makes them unsuitable for measurements according to the standards. While the directivity of starting pistols, popped balloons, and clappers can be an issue (especially with the person performing the measurements in their immediate proximity), sources of too short or weak impulses, such as small firecrackers and spark generators, may not provide sufficient sound energy, especially at low frequencies and in large rooms. If the measurements have to be conducted according to the standards, these properties must be tested for each source prior to the measurements. In return, many of the listed sources are lightweight and compact and allow quick and simple measurements of impulse responses, even on remote locations, although some of them, like pistols, firecrackers, and spark generators, require careful handling.

As an alternative to the listed sources of impulse sound, loudspeakers (together with signal generators and amplifiers) can be used as more controllable sources of sound. Omnidirectional radiation in a broad frequency range is difficult to achieve with a single loudspeaker, so multiple loudspeakers are commonly arranged in the form of a dodecahedron or sphere with twelve equal drivers, or less commonly a cube with six drivers on its sides. Radiating in phase, the loudspeakers emit an approximately spherical wave, thus imitating a point monopole. However, insufficient signal-to-noise ratio can still be a problem in large spaces, as well as non-linearity of loudspeakers' response at high sound power levels. In addition to this, the measurement setup is much more complicated and cumbersome compared to the sources of sound impulses mentioned above and power supply is usually required on site.

Although possible, radiation of short impulses from loudspeakers is usually impractical, due to the limited dynamic range and non-linear distortions. In fact, the main advantage of loudspeakers over the conventional sources of impulse sound is that they can emit other well defined \textbf{deterministic signals}, with certain properties which are suitable for acquisition of impulse responses, $g(t)$, after additional post-processing of their recordings, $p(t)$. These are maximum length sequence (MLS) and swept-sine (also called sweep or chirp). These deterministic sequences have the property of a random noise signal that their autocorrelation approaches ideal delta function for a sufficiently long duration. This is expressed as
\begin{equation}\label{eq:autocorrelation}
\lim\limits_{T \rightarrow \infty} \frac{1}{T} \int_{-T/2}^{T/2} s(t) s(t+\tau) dt = C \delta(\tau),
\end{equation}
where $C$ is a real constant proportional to the mean energy of the real signal $s(t)$ representing the sequence. Of course, duration $T$ of real signals is finite, which turns the equality in eq.~(\ref{eq:autocorrelation}) into approximation.

From eq.~(\ref{eq:convolution_causal}) it is clear that the impulse response $g(t)$ can be obtained after the deconvolution of signal $p(t)$, recorded with an omnidirectional microphone, with the known input signal $s(t)$ (recall that the emission time is implicit and the information on the difference between emission and reception time is lost). Alternatively, a convolution of $p(t)$ (or $s(t)$) with time-inverted sequence $s(-t)$ (or $p(-t)$) can be performed with the same outcome.
This is then essentially equivalent to the \textbf{cross-correlation} of $p(t)$ and $s(t)$ (compare the integrals in eq.~(\ref{eq:convolution}) with $p$ instead of $g$ and $-$ replaced with $+$ with the integral in eq.~(\ref{eq:autocorrelation}) with one $s$ replaced with $p$ and $t$ and $\tau$ switched). Since the deterministic sequences discussed here have autocorrelation approximating an ideal impulse, the deconvolution of $p(t)$ (which is essentially a superposition of delayed and scaled versions of $s(t)$ due to the reflections) with them indeed gives an impulse response. This is also the reason why an arbitrary sequence cannot be used as $s(t)$, because its autocorrelation will in general give additional impulses after the deconvolution, which are not due to the actual reflections in the room.

In particular, cross correlation with MLS can be efficiently performed using the fast Hadamard transformation (which we do not present here). Deconvolution with swept-sines is much simpler in frequency domain, as it comes down to the division of two spectra, according to eq.~(\ref{eq:convolution_freq}):
\begin{equation}\label{eq:deconvolution_freq}
 \hat{g}(\omega) = \frac{\hat{p}(\omega)}{\hat{s}(\omega)},
\end{equation}
where $\hat p(\omega)$ is the (fast) Fourier transform of $p(t)$, as in eq.~(\ref{eq:Fourier_p}), and similarly for $\hat g(\omega)$ and $\hat s(\omega)$. The impulse response $g(t)$ is then obtained as the inverse Fourier transform of $\hat g(\omega)$ (eq.~(\ref{eq:inv_Fourier_p})). Notice that in either case it is not necessary to synchronize the source and receiver, as long as the propagation time of the direct signal is irrelevant.

The recorded signal $p(t)$ at the receiver location, however, must include the complete sequence emitted by the loudspeaker as well as the entire decaying sound after the emission is over. The extraneous noise outside this time interval can be cut out, especially if the recorded signal is weak or the silent period is long. In addition to this, signal-to-noise ratio can be increased further either by \textbf{repeating} the sequence (followed by averaging over the repetitions) or by increasing its \textbf{duration} (length of the MLS or time the swept-sine takes to cover the entire frequency range). In both cases total energy of the sequence is increased by the factor of four per doubling the duration or number of repetitions, because the amplitude at each frequency is doubled. On the other hand, energy of a supposedly random background noise is only doubled, because the noise signals in the recordings of two repetitions or the two halves of the doubled time interval are mutually incoherent, so they sum energetically. This results in around 3\,dB higher signal-to-noise ratio per each doubling of the duration or number of repetitions. In general, increasing the duration is preferred whenever the conditions in the room are stable, since averaging of discrete signals can be sensitive to jitter, which decreases the signal-to-noise ratio. If the sequence is repeated, the duration of each successive repetition should also be longer than the expected reverberation time of the room, in order to avoid significant overlapping of impulse responses acquired by different repetitions.

\textbf{Swept-sines} are generally preferred over MLS. The input signal has the form
\begin{equation}\label{eq:swept-sine}
s(t) = A \sin(2 \pi f(t) t),
\end{equation}
where $A$ is amplitude of the swept-sine, the frequency of which, $f(t)$, is a function of time. For a linear swept-sine, it is a linear function,
\begin{equation}\label{eq:linear_swept_sine}
f(t) = f_1 + (f_2-f_1) \frac{t}{T},
\end{equation}
with $f_1$ the lowest (start) frequency, $f_2$ the highest (end) frequency, and $T$, as above, duration of the sequence. Energy of the linear swept-sine is distributed uniformly over all frequencies between $f_1$ and $f_2$ and the amplitude spectrum of the sequence approaches that of white noise. In contrast to this, amplitude spectrum of background noise usually decays with frequency similarly to pink noise, which means that the signal-to-noise ratio achieved with \textbf{linear} swept-sines decreases at low frequencies.

One of the key advantages of swept-sine over MLS is that its spectrum can be quite easily adjusted to imitate shape of the spectrum of background noise by modifying the function $f(t)$. In this way, signal-to-noise ratio can be kept more or less constant over the entire frequency range, which minimizes the required power of the used loudspeakers. Of course, the same can be achieved with a frequency- (and thus time-)dependent amplitude $A$, but this is suboptimal since keeping $A$ constant (and low enough to avoid non-linear distortions) allows using the entire dynamic range of the loudspeaker at all frequencies. Accordingly, \textbf{logarithmic} swept-sine is most frequently used, since it approximates the amplitude spectrum of pink noise. Its frequency increases faster than linearly (exponentially) with time, resulting in more energy at lower than at higher frequencies and fairly equal energy in octave or third-octave bands. The exponential increase of frequency from $f_1$ to $f_2$ over time $T$ is given by the function
\begin{equation}\label{eq:logarithmic_swept_sine}
f(t) = f_1 \left( \frac{f_2}{f_1} \right)^{t/T}.
\end{equation}

Another important advantage of swept-sine over MLS is that additional tonal components due to \textbf{non-linear distortions} can very often be detected and removed from the acquired impulse responses. Since frequency of the swept-sine increases monotonically, the unphysical tonal components appear in $p(t)$ before or after the true tone of the emitted sequence at those frequencies reaches the receiver. Consequently, the obtained impulse response contains spurious reflections before or after the direct sound. If they precede the direct sound or if they are strong and very late, in the weak reverberation tail of the response, they can be easily identified and windowed out. This makes the swept-sine technique very robust to (moderate) non-linear distortions, which often occur when loudspeakers are used as sources. Nevertheless, non-linear distortions should be avoided, since they generally increase the noise level.

Regardless of which technique is used for acquiring impulse responses, the sources and receivers should be placed at appropriate locations in the room, where the actual sources and receivers are expected (see Table~\ref{tab:room_types}). In the case of large stages, three or more evenly distributed source locations can be used, while for small stages and podiums a single location may suffice. If the actual receivers are distributed in space (as in an auditorium), the microphone locations should cover the entire area (at the height of the listeners' ears) with preferably at least one microphone location per 100-200 seats. In order to reduce the number of measurements, an advantage can be taken if the room has a vertical plane of symmetry which splits the stage and auditorium into two identical halves. The measurements can then be performed only in one half of the auditorium or one half of the stage.

When appropriate (usually in larger acoustic projects), room impulse responses should be calculated even before the room is built, during the design phase. This allows estimation of the room's acoustics, values of the descriptors introduced below, as well as prevention of large failures in the earliest stage. It can be done numerically, using computer simulations, or in scale models. Both techniques will be discussed in section~\ref{ch:modelling}. When the room acoustics is of major importance, such as in concert halls and opera houses, impulse response measurements on site can be performed at different stages of the construction of the hall, in order to control the building process and avoid mistakes. As a rule, correction of mistakes in the acoustic design and during construction is more difficult and expensive at later stages and potential mistakes should therefore be identified as early as possible. Finally, measurements of impulse responses in finished rooms allow assessment of the room and comparison with the specified requirements and predictions, or calibration of numerical simulations and adjustments of scale models, when these are used for testing future acoustic treatment.

\subsection{Descriptors of room acoustics}\label{ch:descriptors_of_room_acoustics}

Next we introduce numerical descriptors which are commonly used for an assessment of room acoustics. They can all be calculated from impulse responses of a room, which, as already discussed, contain all the information about the room acoustics for the specified source and receiver locations at which they are acquired. The frequency-dependent descriptors are usually calculated in \textbf{octave bands}. In such cases, the impulse response $g(t)$ in the following expressions should be first filtered with appropriate octave filters. The filtered responses resemble the one in Fig.~\ref{fig:imp_response}. However, the signal-to-noise ratio can vary substantially between octaves, depending on the source power and background noise spectra, and must be checked for each band separately, for example, by determining the peak-to-noise level as in Fig.~\ref{fig:imp_response} (right). Neglecting this may lead to high inaccuracy of the descriptor values, especially if they are calculated automatically in a software, from broadband responses, and noise in the late part of a response is misinterpreted as part of the response (late reflections). Even if only broadband values should be estimated, it is advisable to remove low-frequency noise below the frequency range of interest with a low-pass (or wide band-pass) filter. At very low frequencies, the responses are also stretched in time and one should make sure that no energy of the response, especially of the direct sound and early reflections, is cut out by the octave filters. For example, zero-samples can be added before and after the actual response before filtering.

Calculated from impulse responses, the descriptors of room acoustics are objective quantities. On the other hand, impulse responses are not readily related to the subjective criteria considered in section~\ref{ch:subjective_criteria}. Therefore, the main purpose of various descriptors is to isolate and quantify different subjectively relevant aspects of room acoustics from the general, redundant information contained in an impulse response. The important questions are then how much of the considered perceivable information can be extracted with the descriptors and how accurately (how well do they \textbf{correlate} with subjective assessments)? Moreover, quantifying many subjectively relevant components of room acoustics (Table~\ref{tab:rooms_subjective_properties}) necessarily requires a set of such descriptors. The additional questions of interest are then -- do different descriptors relate to the same components, that is, are they mutually correlated and thus partly redundant, and what is the \textbf{minimum set} of descriptors needed for a sufficient assessment of room acoustics in a particular application? These questions do not have definite answers. In the search for a comprehensive set of mutually uncorrelated relevant descriptors, new descriptors are continuously being proposed and studied, commonly with the methods of psychoacoustics. We will give a quite exhaustive list of the most important ones, which are well established in literature and norms regarding room acoustics and, accordingly, commonly used in practice.

Reverberation time is the most frequently measured, estimated, and assessed parameter of room acoustics. By definition, it is the time interval during which sound pressure level in a room drops 60\,dB after a source of stationary sound (usually broadband random noise) in the room is switched off (section~\ref{ch:damping_and_reverberation_time}). Note that this is not the decay of sound level of an impulse response and reverberation time cannot be assessed directly from the curve $10 \log_{10} (|g(t)|^2)$, such as the one in Fig.~\ref{fig:imp_response} (right), even though it too is approximately linear. In fact, if the source emits a noise signal $s(t)$ and the room impulse response is $g(t)$, the signal at the microphone is given by the convolution in eq.~(\ref{eq:convolution_causal}). If the decay of energy starts at the reception time $t=0$, $s(t)=0$ for $t > 0$, because the source is off, and
\begin{equation}\label{eq:convolution_causal_noise}
p(t > 0) = \int_{t}^{\infty} g(\tau) s(t-\tau) d\tau,
\end{equation}
since $s(t-\tau) = 0$ for $\tau < t$.

Reverberation time is a room property and should not depend on any single member $s(t)$ of the entire ensemble of possible random noise signals. This is the reason why measurements of reverberation time with the interrupted noise technique (described briefly in the introductory part of section~\ref{ch:measurements_and_descriptors}) require averaging over multiple measurements, with different noise sequences $s(t)$, for the same source and receiver locations. Alternatively, we can use the fact that ensemble averaging of $N$ random signals, denoted with $\langle \text{ } \rangle_N$, is equivalent to their time averaging, $\langle \text{ } \rangle_T$. Ensemble-averaged energy of $p(t>0)$ is proportional to (we are interested in the rate of decay of energy, not its absolute values, so we can observe simply the squared sound pressure)
\begin{equation}\label{eq:convolution_causal_noise_decay}
\begin{aligned}
\langle p^2(t > 0) \rangle_N & = \Big\langle \int_{t}^{\infty} g(\tau) s(t-\tau) d\tau \int_{t}^{\infty} g(\tau') s(t-\tau') d\tau' \Big\rangle_N &\\
&= \Big\langle \int_{t}^{\infty} \int_{t}^{\infty} g(\tau) g(\tau') s(t-\tau) s(t-\tau') d\tau d\tau' \Big\rangle_N &\\
&= \int_{t}^{\infty} \int_{t}^{\infty} g(\tau) g(\tau') \langle s(t-\tau) s(t-\tau') \rangle_N d\tau d\tau'.
\end{aligned}
\end{equation}
The impulse responses do not change between measurements with different sequences, so the ensemble averaging acts only on $s$.

Averaging the term in the angle brackets over an infinite time interval $T \rightarrow \infty$ rather than over large $N$ gives
\begin{equation}
\begin{aligned}
\langle s(t-\tau) s(t-\tau') \rangle_T &=\lim\limits_{T \rightarrow \infty} \frac{1}{T} \int_{-T/2}^{T/2} s(t-\tau) s(t-\tau') dt &\\
&= \lim\limits_{T \rightarrow \infty} \frac{1}{T} \int_{-T/2}^{T/2} s(t) s(t+\tau-\tau') dt.&
\end{aligned}
\end{equation}
Since $T$ is infinite, we can shift $t$ by $\tau$ without affecting the result of integration. On the other hand, the last integral is equal to the autocorrelation in eq.~(\ref{eq:autocorrelation}) with the shift $\tau-\tau'$ and $s(t)$ is a random signal, so
\begin{equation}
\begin{aligned}
\langle s(t-\tau) s(t-\tau') \rangle_T = C \delta(\tau-\tau').&
\end{aligned}
\end{equation}
Hence, the dependence on the random sequences comes down to the constant $C$, regardless of their time forms, and
\begin{equation}\label{eq:convolution_causal_noise_decay_autocorr}
\begin{aligned}
\langle p^2(t > 0) \rangle_N & = C \int_{t}^{\infty} \int_{t}^{\infty} g(\tau) g(\tau') \delta(\tau - \tau') d\tau d\tau' &\\
&= C \int_{t}^{\infty} g(\tau) \left( \int_{t}^{\infty} g(\tau') \delta(\tau - \tau') d\tau' \right) d\tau = C \int_{t}^{\infty} g^2(\tau) d\tau &\\
&=C \left( \int_{0}^{\infty} g^2(\tau) d\tau - \int_{0}^{t} g^2(\tau) d\tau \right),&
\end{aligned}
\end{equation}
according to the selectivity of delta function, eq.~(\ref{eq:Dirac_sampling}). The first of the two resulting terms does not depend on time. It is proportional to the total energy of the impulse response. Therefore, the ensemble-averaged energy decays over time $t > 0$ with the same rate at which the cumulative energy of the impulse response increases (the second term, which is subtracted from the first constant term). This is exactly opposite to the energy decay which is evident in Fig.~\ref{fig:imp_response} (right). The initial (maximum) energy at the time $t=0$ is proportional to the total energy of the impulse response and the result in eq.~(\ref{eq:convolution_causal_noise_decay_autocorr}) is often referred to as \textbf{backward integration}. To underline again, sound energy in the room after the source of stationary noise is switched off does not decay proportionally to $g^2(t)$. It does so only if the source indeed emitted an impulse, so $p(t) = g(t)$, but this is not how reverberation time is defined and $s(t)$ is not a random sequence.

We can finally obtain the sound pressure level drop by calculating the logarithm of eq.~(\ref{eq:convolution_causal_noise_decay_autocorr}). Moreover, the energy can be divided with the maximum value at $t=0$, which removes the constant $C$ and gives a normalized function of time depending only on the impulse response and expressed in dB:
\begin{equation}\label{eq:Schroeder_curve_norm}
\begin{aligned}
SC(t) = 10 \log_{10} \left( \frac{\int_{t}^{\infty} g^2(\tau) d\tau}{\int_{0}^{\infty} g^2(\tau) d\tau} \right) = 10 \log_{10} \left( 1 - \frac{\int_{0}^{t} g^2(\tau) d\tau}{\int_{0}^{\infty} g^2(\tau) d\tau} \right).
\end{aligned}
\end{equation}
This monotonically decreasing function of time is called \textbf{Schroeder curve}. It decays starting with the value 0\,dB at $t = 0$. Its minimum value at $t \rightarrow \infty$ approaches theoretically $-\infty$, but in reality its value at the end of the response is determined by the background noise (assuming that the integration interval, which is only formally infinite, does include the entire impulse response), which prevents the recorded impulse response from decaying to zero. Still, the function $SC(t)$ continues to decay monotonically after the actual impulse response, due to the backward integration, even if the background noise is stationary. It is therefore critical to note that this decay is not related to the decay of sound energy in the room and should be omitted from the evaluation of reverberation time. Especially critical are usually octave bands at low frequencies, when the signal-to-noise ratio is low or even insufficient for accurate estimations. When calculated from the Schroeder curve automatically, reverberation time can be substantially overestimated if the decay of the Schroeder curve due to noise is included, because by moderate levels of noise in impulse responses, this decay is much slower than the decay of reflections.

\begin{figure}[h]
	\centering
	\includegraphics[width=0.7\linewidth]{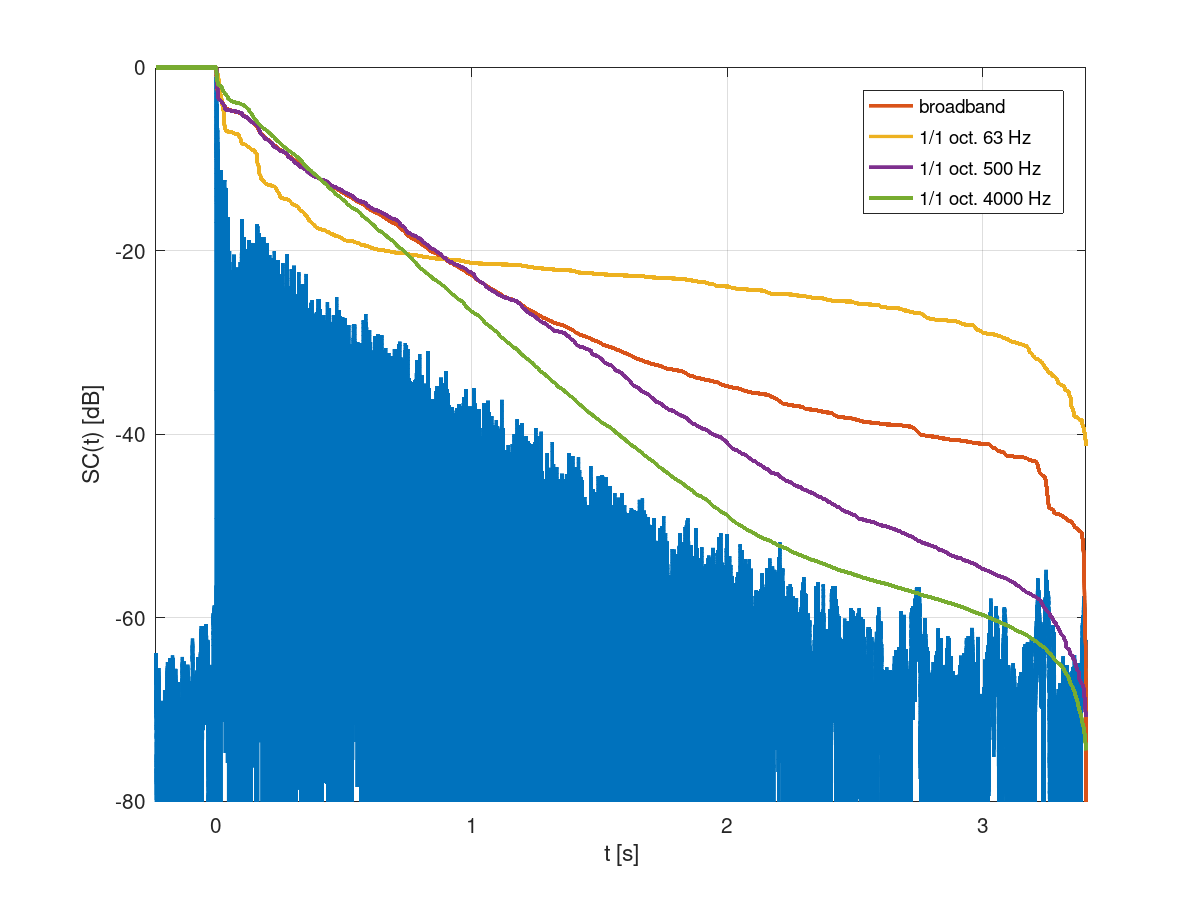}
	\caption{Schroeder curves obtained from the impulse response in Fig.~\ref{fig:imp_response}.}
	\label{fig:Schroed_curve}
\end{figure}

As an example, Fig.~\ref{fig:Schroed_curve} shows Schroeder curves calculated from the impulse response shown in Fig.~\ref{fig:imp_response}, broadband and in three octave bands, plotted on the top of Fig.~\ref{fig:imp_response} (right). As discussed above, even the broadband Schroeder curve does not match the decaying logarithm of the squared impulse response. Starting at 0\,dB, all curves exhibit a sharp drop of several decibels, after the energy of the direct sound and strong early reflections is subtracted by the backward integration. This is followed by a relatively smooth linear decay of the curves corresponding to an exponential decay of sound energy of a hypothetical source of stationary noise switched off at $t=0$. However, this decay extends, after a more or less pronounced knee of the Schroeder curve, into a slower drop which is due to the background noise, not reflections. The transition is especially evident in the octave 63\,Hz, where the signal-to-noise ratio is below 20\,dB, and part of the curve after around $t=0.3$\,s is useless for calculations. This is not obvious from the broadband response. The broadband signal-to-noise ratio is around 30\,dB (the knee is not very pronounced), possibly dictated by the strong low-frequency noise, and around 50\,dB at 500\,Hz, where the used source of impulse has sound power maximum (recall Fig.~\ref{fig:freq_response}), and at 4000\,Hz. Eventually, all curves drop to a very low value at the end of the recording of the response.

Commonly, when Schroeder curves have approximately linear decay within a sufficient dynamic range, \textbf{reverberation time} can be estimated fairly easily. From the definition it follows directly
\begin{equation}\label{eq:T60_Schroeder_curve}
T_{60} = t(SC = -60\text{\,dB}),
\end{equation}
because the decay starts at $t=0$ and equals 60\,dB when $SC(t)=-60$\,dB. However, signal-to-noise ratio of 60\,dB is quite difficult to achieve in practice, especially at low frequencies. Moreover, strong early reflections often cause an abrupt or irregular, non-linear drop in the initial part of $SC$, as in Fig.~\ref{fig:Schroed_curve}. For these reasons, reverberation time is usually estimated on a smaller dynamic range, such as
\begin{equation}\label{eq:T30_Schroeder_curve}
T_{30} = 2 \left[ t(SC = -35\text{\,dB})-t(SC = -5\text{\,dB}) \right].
\end{equation}
Multiplication with 2 is necessary, since $T_{30}$ is estimated on the time interval of the decay of 30\,dB (as indicated by the subscript), which is two times shorter than the decay of 60\,dB, which is according to the definition of reverberation time. As usual, it is assumed that the sound pressure level decays linearly also after the time the Schroeder curve reaches -35\,dB. If the signal-to-noise ratio of 35\,dB is still not achievable, $T_{20}$ can be used,
\begin{equation}\label{eq:T20_Schroeder_curve}
T_{20} = 3 \left[ t(SC = -25\text{\,dB})-t(SC = -5\text{\,dB}) \right],
\end{equation}
or even $T_{10}$,
\begin{equation}\label{eq:T10_Schroeder_curve}
T_{10} = 6 \left[ t(SC = -15\text{\,dB})-t(SC = -5\text{\,dB}) \right].
\end{equation}

Starting point of the interval at $SC=-5$\,dB is used for all three descriptors of reverberation, $T_{10}$, $T_{20}$, and $T_{30}$, in order to exclude the initial decay governed by the early reflections, which does not match the rest of the decay process. In contrast to this, \textbf{early decay time} is defined as
\begin{equation}\label{eq:EDT_Schroeder_curve}
EDT = 6 t(SC = -10\text{\,dB}).
\end{equation}
Unlike reverberation time, it should indeed quantify decay of the early reflected sound energy, which is considered to be tightly related to the subjective perception of reverberation (reverberance) in a room. This is because usual sounds produced in rooms, such as music or speech, rarely allow large drops of sound energy to take place. Apart from occasional longer breaks, time intervals between syllables, words, or tones are too short for the entire decay of the sounds to be heard. Consequently, reverberation is perceived mostly based only on the early decay. Nevertheless, multiplication with 6 is introduced to match the definition of reverberation time and allow comparison of $EDT$ with $T_{60}$ and other associated quantities. For the same reasons, reverberation time is sometimes estimated based on the initial 160\,ms of the impulse response (the usual index $160$ refers here to the time interval in milliseconds, not the drop in decibels):
\begin{equation}
T_{160} = \frac{60\text{\,dB}}{-SC(t=0.16\text{\,s})} 0.16\text{\,s}.
\end{equation}
The value is again scaled such that the obtained time interval corresponds to the linearly extrapolated drop of 60\,dB.

\textbf{Bass ratio} is usually calculated in rooms for music performances, in which low frequencies carry more auditory information than in speech. It is defined as the ratio of reverberation times at low frequencies (in octaves 125\,Hz and 250\,Hz) and middle frequencies (in octaves 500\,Hz and 1000\,Hz),
\begin{equation}\label{eq:BR}
BR = \frac{T_{60,125\text{Hz}}+T_{60,250\text{Hz}}}{T_{60,500\text{Hz}}+T_{60,1000\text{Hz}}}.
\end{equation}
This descriptor should quantify frequency balance of the room's response. In practice, $T_{60}$ is usually replaced with $T_{30}$. While $BR \approx 1$ indicates balanced low and middle frequencies in a room's response, somewhat larger values are common. This can be desirable for music listening, but undesirable in room for speech, since the low-frequency sound energy in them is mostly due to the ambient noise, which should be suppressed. 

While the descriptors based on reverberation time quantify rate of energy decay, other descriptors involve total energy within certain intervals of time. \textbf{Definition} $D_{50}$ is one such parameter, which is associated with speech intelligibility. It is the fraction of energy of the impulse response contained in the first 50\,ms after the direct sound. Since $g(t=0)$ represents the direct sound,
\begin{equation}\label{eq:D_50}
D_{50} = \frac{\int_{0}^{50\text{ms}} g^2(\tau) d\tau}{\int_{0}^{\infty} g^2(\tau) d\tau} \leq 1.
\end{equation}
It is sometimes given as a percentage, after multiplication with 100\%. A very similar quantity which is used for rating transparency of music is \textbf{clarity} $C_{80}$. It is the ratio of energy in the first 80\,ms and the rest of the impulse response, expressed in decibels:
\begin{equation}\label{eq:C_80}
C_{80} = 10 \log_{10} \left( \frac{\int_{0}^{80\text{ms}} g^2(\tau) d\tau}{\int_{80\text{ms}}^{\infty} g^2(\tau) d\tau} \right).
\end{equation}
Unlike $D_{50}$, it can take both positive and negative values. By changing the corresponding time intervals, variations of these parameters are possible, for example, $C_{50} = 10 \log_{10}(D_{50}/(1-D_{50}))$ or $D_{80} = 1/(1+10^{-C_{80}/10})$. However, the interval of 50\,ms is always used for assessing speech intelligibility and 80\,ms for clarity of music. While the reflected energy after already around 50\,ms is detrimental for perception of speech, music allows somewhat longer delays.

Conceptually similar is a parameter which can be used for assessing room acoustics from the perspective of performers -- \textbf{stage support factor}. It is defined as
\begin{equation}\label{eq:ST}
ST = 10 \log_{10} \left( \frac{\int_{t_{min}}^{t_{max}} g^2(\tau) d\tau}{\int_{0\text{ms}}^{10\text{ms}} g^2(\tau) d\tau} \right).
\end{equation}
Several pairs of values $(t_{min},t_{max})$ are used in practice: (20\,ms, 100\,ms) for $ST1 = ST_{early}$, (20\,ms, 200\,ms) for $ST2$, (100\,ms, 1000\,ms) for $ST_{late}$, and (20\,ms, 1000\,ms) for $ST_{total}$. However, the impulse response $g(t)$ is in this case measured at the locations of the performers (on the stage or in an orchestra pit), not in the auditorium, at the distance 1\,m from an omnidirectional source. Energy of the impulse response within the interval $(t_{min},t_{max})$ is scaled with the energy in the first 10\,ms. The parameter $ST_{early}$ is believed to quantify well audibility between performers (ease of ensemble) and $ST2$ their perception of reverberance and feedback of the room (support). According to some authors, only $ST_{early}$ describes well both support and ease of ensemble. There is still no strong consensus on the optimal values of these parameters. In rooms in which they are usually measured, concert halls and opera houses, they often take negative values between -17\,dB and -10\,dB and some authors recommend the values of $ST_{early}$ between $-15\,\text{dB}$ and $-10\,\text{dB}$.

Another approach to estimate the ratio of early and late energy in an auditorium is with \textbf{centre time}
\begin{equation}\label{eq:t_s}
t_s = \frac{\int_{0}^{\infty} \tau g^2(\tau) d\tau}{\int_{0}^{\infty} g^2(\tau) d\tau}.
\end{equation}
In contrast to reverberation time, it is usually expressed in milliseconds and represents the first moment of energy of the impulse response (analogously to the centre of mass, for example, when $g^2$ is replaced with mass per unit length and the integration is over a one-dimensional body's length). The index $_s$ has origin in the German term ``Schwerpunktzeit''. Lower values of centre time are usually preferred, since they indicate that more energy is contained in the early part of the impulse response.

Table~\ref{tab:imp_response_source_descriptors} gives the values of most of the descriptors of room acoustics introduced so far, for the impulse response from Fig.~\ref{fig:imp_response}. The values are given broadband (BB) and in octave bands and rounded to two decimal places (one is commonly sufficient for all listed descriptors except $D_{50}$). Broadband values of reverberation time, as well as at $63$\,Hz, $500$\,Hz, and $4000$\,Hz, can be compared against the Schroeder curves in Fig.~\ref{fig:Schroed_curve}. Notice the similarity of $T_{10}$, $T_{20}$, and $T_{30}$ due to fairly linear decays of the Schroeder curves, when the signal-to-noise ratio is well above 35\,dB, which is not the case in the octaves below 500\,Hz. As the signal-to-noise ratio drops at low frequencies, the accuracy of $T_{30}$ decreases (if the parameter can be calculated at all), followed by $T_{20}$ and $T_{10}$. As usual in rooms, reverberation time decreases at high frequencies due to higher damping. The values of $EDT$ and $T_{160}$ are lower than the (late) reverberation time, which is often the case in relatively large acoustically untreated rooms. Especially sensitive to low signal-to-noise ratio are descriptors which separate (high) early and (low) late energy, such as $T_{160}$, $t_s$, $D_{50}$, and $C_{80}$. Fortunately, the values at middle frequencies are usually the most relevant. Another important difference between the descriptors which depend strongly on early reflections, which are most of the descriptors presented here apart from reverberation time and bass ratio, and those which depend mostly on late reflections is that values of the former ones vary more significantly over the room. This is because the distribution of strong early reflections in time is much more sensitive to the source and receiver locations in the room (with regard to the reflecting surfaces, absorption, etc.) than the late decay of many weaker reflections.

\begin{table}[h]
	\caption{Values of some descriptors of room acoustics for the impulse response in Fig.~\ref{fig:imp_response}.}
	\label{tab:imp_response_source_descriptors}
	\begin{tabular}{ | l | l | l | l | l | l | l |}
		\hline
		\textbf{octave [Hz]} & \textbf{BB} & \textbf{31.5} & \textbf{63} & \textbf{125} & \textbf{250} & \textbf{500} \\
		\hline
		$T_{10}$ [s] & 2.87 & 1.34 & 1.12 & 1.90 & 2.13 & 2.93 \\
		\hline
		$T_{20}$ [s] & 2.95 & 1.65 & 1.35 & 2.11 & 2.64 & 2.99 \\
		\hline
		$T_{30}$ [s] & 2.82 & / & 1.21 & 2.23 & 2.40 & 2.84 \\
		\hline
		$EDT$ [s] & 1.80 & 0.98 & 0.97 & 1.02 & 1.09 & 1.80 \\
		\hline
		$T_{160}$ [s] & 1.48 & 1.02 & 0.98 & 1.06 & 1.14 & 1.55 \\
		\hline
		$t_{s}$ [ms] & 97.45 & 83.10 & 48.51 & 70.53 & 55.94 & 99.98 \\
		\hline
		$D_{50}$ [/] & 0.65 & 0.87 & 0.82 & 0.70 & 0.80 & 0.65 \\
		\hline
		$C_{80}$ [dB] & -1.73 & 0.07 & -0.83 & -1.34 & -0.94 & -1.79 \\
		\hline
		\hline
		\textbf{octave [Hz]} & \textbf{1000} & \textbf{2000} & \textbf{4000} & \textbf{8000} & \textbf{16000} & \\
		\hline
		$T_{10}$ [s] & 3.29 & 2.69 & 2.26 & 1.34 & 0.66 & \\
		\hline
		$T_{20}$ [s] & 3.20 & 2.83 & 2.34 & 1.38 & 0.73 & \\
		\hline
		$T_{30}$ [s] & 2.93 & 2.68 & 2.20 & 1.39 & 0.75 & \\
		\hline
		$EDT$ [s] & 2.19 & 2.17 & 1.95 & 0.95 & 0.26 & \\
		\hline
		$T_{160}$ [s] & 1.72 & 1.77 & 1.61 & 0.96 & 0.51 & \\
		\hline
		$t_{s}$ [ms] & 122.43 & 131.85 & 112.90 & 54.43 & 16.88 & \\
		\hline
		$D_{50}$ [/] & 0.62 & 0.51 & 0.56 & 0.71 & 0.92 & \\
		\hline
		$C_{80}$ [dB] & -1.94 & -2.62 & -2.28 & -1.08 & -0.19 & \\
		\hline
	\end{tabular}
\end{table}

A very simple and useful broadband quantity is \textbf{initial time delay gap} $\Delta t_{init}$. It is the time interval between arrival of the direct sound and the first reflection which reaches the receiver. Energy of the direct sound which is scattered from the body of the source or receiver and thus immediately follows the direct sound is not taken into account. The quantity is correlated with the subjective perception of distance from the source, that is, intimacy: the larger its value (and, thus, delay of the reflections with regard to the direct sound) is, the closer the source appears to be and the effect of the room is perceived as weaker. The short delays normally vary between a few milliseconds in small rooms to $\mathcal{O}(10\text{\,ms})$ in large halls. Hence, too large values of $\Delta t_{init}$ are undesirable in rooms for sound production, when the room critically affects the judgment of sound fields, such as in concert halls.

\textbf{Strength factor} $G$ quantifies the energy gain (hence the symbol $G$) in a room compared to an unbounded (free) space. As such, it is well correlated to the subjective perception of loudness in the room, for a given power of the source of sound. By definition, it is the difference between sound pressure level measured at some location in the room, generated by an omnidirectional source in it, and sound pressure level which would be produced by the same source in free space at the distance 10\,m. This distance should represent an average distance of listeners in larger rooms, such as concert halls. Therefore, the strength factor captures typical gain of the room with respect to free space, even though the chosen receiver locations in the room do not have to be placed at the same distance from the source. Importantly, strength factor can also be estimated directly from an impulse response recorded with an omnidirectional source:
\begin{equation}\label{eq:G}
G = 10 \log_{10} \left( \frac{\int_{0}^{\infty} g^2(\tau) d\tau}{\int_{0}^{t_{dir}} g_{10\text{m}}^2(\tau) d\tau} \right),
\end{equation}
where $g_{10\text{m}}(t)$ is a reference impulse response of the room measured at the distance 10\,m from the source and $(0,t_{dir})$ is the time interval in which only the direct sound occurs, as in free space. The interval should be long enough to include the entire energy of the direct impulse, but short enough (typically a few milliseconds and shorter than $\Delta t_{init}$) to exclude all reflections and, as much as possible, scattered and diffracted energy of the direct sound.

Strength factor is practically the only descriptor which is not normalized with the same impulse response. Consequently, if $g(t)$ and $g_{10\text{m}}(t)$ are recorded simultaneously, the two microphones should have equal sensitivity and the recorded signals must be equally amplified. If a single microphone is used for both recordings (by replacing the microphone), absolute sound power of the source must also remain equal, that is, the emitted impulse (or a deterministic sequence) has to be repeatable, which is not critical for other normalized descriptors. Theoretically, strength factor should (like reverberation time in eq.~(\ref{eq:T60_Sabine}) and sound pressure level in eq.~(\ref{eq:SPL_diffuse_field})) be spatially independent in a diffuse field, at distances larger than the critical distance. In real rooms, however, it usually decays gradually with the distance from the source, due to the limited diffuseness, before it increases again close to reflecting surfaces.

Spaciousness is an important quality of sound fields in rooms for music performances and it is often quantified by \textbf{lateral energy fraction}
\begin{equation}\label{eq:LEF}
LEF = \frac{\int_{5\text{ms}}^{80\text{ms}} g_{fig8}^2(\tau) d\tau}{\int_{0}^{80\text{ms}} g^2(\tau) d\tau} = \frac{\int_{5\text{ms}}^{80\text{ms}} g^2(\tau, \theta) \cos^2(\theta) d\tau}{\int_{0}^{80\text{ms}} g^2(\tau,\theta) d\tau}.
\end{equation}
Here, $g_{fig8}(t)$ denotes an impulse response recorded with a bi-directional microphone (with the figure-of-eight pattern as in Fig.~\ref{fig:directivity_monopole_dipole}), oriented such that the minimum of sensitivity (for $\theta = \pi/2$) points towards the source and both maxima (for $\theta = 0$ and $\theta = \pi$) lie in the horizontal plane, thus pointing to the sides. This and the integration interval starting at $t = 5$\,ms ensure that the direct sound is excluded in the numerator in eq.~(\ref{eq:LEF}), which is associated with lateral reflections and normalized with usual ``omnidirectional'' impulse response. In general, spaciousness cannot be captured with a single omnidirectional microphone and in the case of $LEF$ it refers only to relatively early energy, in the first 80\,ms of the impulse responses. The second equality in eq.~(\ref{eq:LEF}) (and Fig.~\ref{fig:directivity_monopole_dipole}) indicates that $0 \leq LEF \leq 1$ and higher values imply more spaciousness (listener envelopment). Information on the angular dependence of $g(t,\theta)$ is lost when the impulse response is obtained with a single omnidirectional microphone ($g(t)$).

Another approach for quantifying spaciousness is based on the fact that the spatial information is obtained owing to the binaural hearing and differences between the two, to a certain extent mutually uncorrelated received sounds. A descriptor which takes this into account is interaural cross correlation coefficient. It is a measure of similarity between sounds received simultaneously by two ears of a listener. Normalized interaural cross correlation function is defined as (compare the numerator with the autocorrelation on the left-hand side of eq.~(\ref{eq:autocorrelation}))
\begin{equation}\label{eq:IACF}
IACF(t) = \frac{\int_{t_{min}}^{t_{max}} g_r(\tau) g_l(\tau + t) d\tau}{\sqrt{ \int_{t_{min}}^{t_{max}} g_l^2(\tau) d\tau \int_{t_{min}}^{t_{max}} g_r^2(\tau) d\tau}}.
\end{equation}
It is a function of time within the specified interval $-1\text{\,ms} < t < 1\text{\,ms}$, where 1\,ms is approximately the time in which sound waves propagate over the distance between the two ears (width of the head). Hence, the range (-1\,ms, 1\,ms) covers all delays of sound between the two ears, for all angles of incidence. The impulse responses $g_l(t)$ and $g_r(t)$ should be recorded with the left and right microphones placed in the corresponding ears of an artificial (dummy) head or left and right in-ear binaural microphones placed in the ears of a human listener, respectively. The function $IACF(t)$ is normalized by the denominator, so that its values are within the interval $[-1,1]$.

\textbf{Interaural cross correlation coefficient} is defined as the maximum absolute value of the interaural cross correlation function,
\begin{equation}\label{eq:IACC}
IACC = \max \{ |IACF(t)| \}.
\end{equation}
Therefore, it describes the maximum similarity between the two signals with a single-number value in the interval $[0,1]$. Lower values of $IACC$ indicate lower correlation between the sounds at the two ears and, consequently, more spaciousness. Alternatively, \textbf{binaural quality index} is also used, which is simply
\begin{equation}\label{eq:binaural_quality_index}
BQI = 1-IACC,
\end{equation}
so that its higher values (towards 1) indicate more spaciousness.

The integration interval $(t_{min},t_{max})$ is commonly $(0\text{\,s},1\text{\,s})$ for $IACC_A$, $(0\text{\,ms},80\text{\,ms})$ for early interaural cross correlation coefficient $IACC_E$, or $(80\text{\,ms},1\text{\,s})$ for late interaural cross correlation coefficient $IACC_L$. Early interaural cross correlation coefficient, $IACC_E$, is associated with apparent source width and late $IACC_L$ with listener envelopment. As frequency-dependent quantities, both $LEF$ and $IACC$ can be calculated in a certain limited frequency range. For example, $IACC_{E,3oct}$ can denote early interaural cross correlation coefficient calculated in the three middle octaves only, 500\,Hz, 1000\,Hz, and 2000\,Hz. In this way, poor spaciousness and high correlation which can always be expected at low frequencies (when listener's head presents an acoustically small obstacle) are excluded from the assessment.

In contrast to rooms for music, spaciousness is far less important in rooms for speech. The essential quality, speech intelligibility, is largely reduced when the spoken syllables are masked either by the ambient noise or by the decaying sound energy of the preceding syllables. In order to derive a descriptor which captures the effect of masking, (slow) variations of the time-averaged sound energy of speech, $\langle E \rangle_T$, can be roughly modelled as simple oscillations around a constant value $\langle E_0 \rangle_T$:
\begin{equation}\label{speech_energy_modulation}
\langle E \rangle_T = \langle E_0 \rangle_T [1+m(\Omega)\cos(\Omega (t - t_0))],
\end{equation}
with $\Omega$ the modulation frequency, which approximates the frequency of syllables in the speech (not the sound waves), and $t_0$ is a delay introduced for generality. In this expression $m(\Omega)$ satisfies $0 \le m(\Omega) \le 1$, because energy is non-negative and negative sign of $m(\Omega)$ can be absorbed in the phase shift $\Omega t_0$. Thus, it can be seen as the real amplitude of a complex modulation transfer function $\hat{m}(\Omega) e^{j \Omega t}$, that is, $m(\Omega) = |\hat{m}(\Omega)|$.

It can be shown that if the modulated signal (speech) has a flat spectrum in the frequency range of interest (for example, an octave band) and the ambient noise is negligible, the complex modulation transfer function is equal to the normalized Fourier transform of the squared impulse response, with modulation frequency (rather than $\omega$) as the argument of the transform:
\begin{equation}\label{eq:MTF}
\hat{m}(\Omega) = \frac{\int_{0}^{\infty} g^2(\tau) e^{-j\Omega \tau} d\tau}{\int_{0}^{\infty} g^2(\tau) d\tau}.
\end{equation}
The dynamics of $\langle E \rangle_T$ is theoretically maximal (with $m=1$) when $g(t)$ contains only the direct sound and minimal ($m = 0$) when the impulse response does not decay at all. Larger dynamics provides higher speech intelligibility. If the ambient noise is not negligible, $m$ can still be estimated as $m(\Omega) = |\hat{m}(\Omega)|/(1+10^{-SNR/10})$, where $SNR$ is the signal-to-noise ratio. Lower signal-to-noise ratio results in a lower dynamics and intelligibility of speech. As above, the definitions can be applied both to a broadband sound and in frequency bands. Therefore, the modulation transfer function in an octave band can be calculated using eq.~(\ref{eq:MTF}) after filtering a broadband impulse response with an appropriate octave filter.

Apparent signal-to-noise ratio for the energy given in eq.~(\ref{speech_energy_modulation}) equals
\begin{equation}\label{eq:SNR_apparent}
SNR_a(\Omega) = 10 \log_{10} \left( \frac{m(\Omega)}{1-m(\Omega)} \right).
\end{equation}
It interprets as noise everything which decreases the dynamics of speech, such as energy of the overlapping preceding syllables, and approaches $+\infty$ for $m=1$ and $-\infty$ for $m=0$. Moreover, it provides a basis for the definition of descriptors associated with speech intelligibility in rooms. For example, in the case of \textbf{rapid speech transmission index} ($RASTI$), it is calculated in two octave bands, at 500\,Hz for the modulation frequencies 1\,Hz, 2\,Hz, 4\,Hz, and 8\,Hz and at 2\,kHz for the modulation frequencies 0.7\,Hz, 1.4\,Hz, 2.8\,Hz, 5.6\,Hz, and 11.2\,Hz. The value of $RASTI$ is then defined as
\begin{equation}\label{eq:RASTI}
RASTI = \frac{avg\{SNR_a(\Omega)\} + 15\text{\,dB}}{30\text{\,dB}},
\end{equation}
where averaging $avg\{\text{ }\}$ is performed over all (maximum nine, for nine modulation frequencies in the two octaves) apparent signal-to-noise ratios which lie in the range $[-15,15]$\,dB. The less likely values outside this range are ignored. Consequently, $RASTI$ takes the values in the range $[0,1]$. Speech intelligibility is assesses as bad for its values below 0.3, poor between 0.3 and 0.45, fair between 0.45 and 0.6, good between 0.6 and 0.75, and excellent above 0.75.

$RASTI$ is often used instead of the basic \textbf{speech transmission index} ($STI$), which involves more modulation frequencies and all octaves from 125\,Hz to 8\,kHz. In addition to this, apparent signal-to-noise ratios in the seven octave bands are weighted with the factors 0.13, 0.14, 0.11, 0.12, 0.19, 0.17, and 0.14, respectively. However, it has been shown that the difference between the values of $RASTI$ and $STI$ is normally not large, so $RASTI$ can be used as a simpler alternative to $STI$\footnote{Another alternative, $STIPA$ (speech transmission index for public address systems), uses the same set of octaves as $STI$ with only two modulation frequencies per octave.}.

A different descriptor which is used for assessment of speech intelligibility (also for public address systems) is \textbf{articulation loss of consonants},
\begin{equation}\label{eq:articulation_loss_of_consonants}
	AL_{cons} = 0.65\text{\,s}^{-1} \left( \frac{r}{r_c} \right)^2 T_{60} \cdot 1\%,
\end{equation}
where $r_c$ is critical distance in the room, estimated, for example, using eq.~(\ref{eq:critical_distance_diffuse_directional_source}), and $r$ is distance between the source and receiver. This expression is given for a signal-to-noise ratio of minimum 30\,dB. Higher percentage of the lost (not perceived) consonants and lower intelligibility are, of course, expected if the signal-to-noise ratio is lower. The value of $AL_{cons}$ should be below 15\%, preferably below 7\%, and ideally below 3\% for a good speech intelligibility. The descriptor should not be used for very large distances from the source $r$, when its value can become even larger than 100\%. If the value of speech transmission index is known, the value of $AL_{cons}$ can also be estimated using the formula
\begin{equation}\label{eq:articulation_loss_of_consonants_from_STI}
	AL_{cons} = 170.5405\% \cdot e^{-5.419 \cdot STI}.
\end{equation}
The two parameters are thus mutually correlated.

Recalling that the descriptors of room acoustics should correlate well with different subjectively relevant properties of room acoustics, Table~\ref{tab:subj_and_objective_parameters} relates the introduced descriptors with the subjective criteria discussed in section~\ref{ch:subjective_criteria}. Although it is not clear from the table how high the correlation actually is, it is evident that some descriptors point to the same criteria, particularly intelligibility/clarity and reverberance. It is then possible to calculate and compare values of several parameters and obtain a more reliable assessment. On the other hand, still sought for are appropriate, reliable, and efficient descriptors of the effects of strong distinct (usually early) reflections, such as echo, coloration, and apparent source shift (see section~\ref{ch:subjective_criteria}), as well as the texture of an impulse response (time distribution of reflections and diffuseness of the field). Phase-dependent superposition of waves and complicated relationships between physical sound fields and human perception make these more difficult to define. Impulse responses evidently contain more information than can be extracted by the descriptors above. A usual procedure includes additional visual inspection of impulse responses in search of strong, typically first- or second-order reflections. The corresponding reflecting surfaces can then often be identified using the simple model of specular reflections, as in section~\ref{ch:early_reflections_strategy}.

\begin{table}[h]
	\caption{Subjective criteria and related objective descriptors of room acoustics. Note: $IACC$ can be replaced with $BQI = 1-IACC$, according to eq.~(\ref{eq:binaural_quality_index}).}
	\label{tab:subj_and_objective_parameters}
	\begin{tabular}{ | l | l | }
		\hline
		\textbf{subjective property} & \textbf{descriptor} \\
		\hline
		loudness & $G$ \\
		\hline
		intelligibility/clarity & $D_{50}$, $C_{80}$, $t_s$, $STI$, $RASTI$, $AL_{cons}$ \\
		\hline
		reverberance & $T_{10}$, $T_{20}$, $T_{30}$, $T_{160}$, $EDT$ \\
		\hline
		listener envelopment & $LEF$, $IACC_L$ \\
		\hline
		apparent source width & $IACC_E$ \\
		\hline
		intimacy & $\Delta t_{init}$ \\
		\hline
		frequency balance & $BR$ \\
		\hline
		ease of ensemble & $ST1$ ($ST_{early}$) \\
		\hline
		support & $ST2$, $ST_{late}$ \\
		\hline
	\end{tabular}
\end{table}

Once the descriptors are calculated, their values should be compared against the \textbf{optimal values} (or, more commonly, ranges of values), depending on the room purpose. If single-number values are specified for frequency-dependent descriptors, they are often compared with the average descriptor values in the most critical mid-frequency range. Larger deviations from the optimal values are usually expected and tolerated at low and high frequencies. For a general overview, Table~\ref{tab:imp_response_param_values} lists typical appropriate values of the descriptors in rooms for music production (including singing) and speech. The values are given for usual conditions in the room, for example, half occupied auditorium\footnote{Impulse responses are most often measured in empty rooms. For this reason, if empty seats in an auditorium have a very low absorption, they can be covered with thin sheets of porous fabric, in order to compensate for the absorption of the absent audience.}. Still, they should be taken only as a rough indication of the true optimal values. For example, reverberation time value should be specified with regard to the room volume. Further details for particular types of rooms will be considered in section~\ref{ch:acoustic_design} and more accurate application-dependent values of the descriptors can be found in literature and relevant norms and guidelines. Table~\ref{tab:imp_response_param_values} also indicates types of the reflections which most significantly affect values of the descriptors and thus should be in the focus when improvements of the parameter values are necessary. 

\begin{table}[h]
	\caption{Suggested values of descriptors of acoustics in rooms for speech and music production and the most relevant types of reflections. Descriptor values which are less relevant for a particular type of the room are denoted with n/a.}
	\label{tab:imp_response_param_values}
	\begin{tabular}{ | l | l | l | p{3.3cm} | }
		\hline
		\textbf{descriptor} & \textbf{reflections} & \textbf{speech} & \textbf{music} \\
		\hline
		reverberation time ($T_{30}$) & late & 0.7-1.2\,s & 1.5-2.2\,s (concert halls) 1.2-1.8\,s (opera houses) \\
		\hline
		early decay time (EDT) & early & n/a & 2-2.3\,s \\
		\hline
		centre time ($t_s$) & early & $<100$\,ms & 100-150\,ms \\
		\hline
		initial time delay gap ($\Delta t_{init}$)& the earliest & n/a & 12-20\,ms \\
		\hline
		strength factor ($G$) & early & n/a & 4-6\,dB \\
		\hline
		definition ($D_{50}$) & early & $>50$\% & n/a \\
		\hline
		clarity ($C_{80}$) & early & n/a & -2\,dB to 3\,dB \\
		\hline
		speech transmission index ($STI$) & early, late & $>60$\% & n/a \\
		\hline
		articulation loss of consonants ($AL_{cons}$) & early, late & $<7$\% & n/a \\
		\hline
		lateral energy fraction ($LEF$) & lateral & n/a & $> 0.15$ \\
		\hline
		interaural cross corr. ($IACC_{E,3oct}$) & lateral & n/a & $<0.4$ \\
		\hline
		bass ratio (BR) & late & n/a & 1.1-1.4 \\
		\hline
	\end{tabular}
\end{table}

As already mentioned in the beginning of this subsection, relevance of different descriptors and their correlation with subjective criteria for room acoustics are a permanent object of research. There is no clear consensus on which set of (ideally mutually uncorrelated) descriptors should be used for a sufficient assessment for each application of room acoustics. New descriptors are continuously being defined and proposed, which should provide a more accurate or complete assessment. One example is surface diffusivity index (SDI), which is based on a visual rating of the scattering capacity of large surfaces, mainly ceiling and side walls in rooms for music performances. Properties of the surfaces, such as roughness and geometric irregularity, are compared with the textual descriptions given in a look-up table, from which the appropriate numerical rates are read out for each surface. High average rate of all surfaces in the room (high value of SDI) indicates a diffuse field in it and high level of spaciousness. Diffuseness is also associated with the texture of an impulse response. However, there is no generally adopted descriptor for it.

If calculated values of descriptors of room acoustics are unsatisfactory, different measures can be taken to improve them. Some basic methods to estimate the effects of such measures and common approaches to the optimization of acoustics in various rooms are considered in the next section.
\section{Estimation and optimization of sound fields in rooms}\label{ch:basic_strategies}

In this section we suppose that acoustics of a room should be optimized, either in the design phase of a room, which is yet to be built or renovated, or because the acoustic measurements and assessment in an existing room delivered unsatisfactory results. Choice of the optimization measures is tightly related to the estimation or prediction of relevant properties of sound fields in rooms, which is usually done by an acoustic consultant. This can be achieved relatively accurately using the modelling techniques which will be considered in section~\ref{ch:modelling}. Here we discuss some very elementary, simple, and approximate methods based on the theoretical models introduced in the earlier sections. These methods can be implemented in rooms with very different types and strategies of acoustic optimization. Further aspects of the acoustic design for particular types of rooms will be the subject of section~\ref{ch:acoustic_design}.

Suitable strategy for achieving good acoustics in a room depends primarily on its purpose. In section~\ref{ch:introduction} we distinguished between three general goals of room acoustics -- suppression of noise, accurate sound reproduction, and appropriate enhancement of the sound field generated by a source of sound in the room. More concretely, the goal is \textbf{optimization} of the direct sound and reflections inside the room. With regard to that, Table~\ref{tab:basic_strategies} provides an overview of very general strategies and means for achieving such a goal. The distinction is made between early (fewer and stronger) reflections and late reverberation (many weaker reflections during the decay process), which are usually recognizable in room impulse responses (see Fig.~\ref{fig:imp_response}, for example). The table can be seen as an extension of Table~\ref{tab:rooms_subjective_properties} relating the subjective criteria with certain acoustic phenomena which affect the assessment. Associated with it is the treatment of ambient noise, which does not directly belongs to room acoustics (noise control usually refers to some specific sources of noise in the room, not the ambient noise in general), but must be taken into consideration.

\begin{table}[h]
	\caption{General acoustic treatment in the applications of room acoustics from Table~\ref{tab:rooms_subjective_properties}.}
	\label{tab:basic_strategies}
	\begin{tabular}{ | l | p{3.5cm} | p{3.5cm} | p{3.5cm} |}
		\hline
		\textbf{application} & \textbf{direct sound} & \textbf{early reflections}& \textbf{late reflections} \\
		\hline
		noise control & / & absorption & absorption \\
		\hline
		sound reproduction & appropriate strength and propagation time (source-receiver distance) & absorption or scattering/reflection away from the receiver & absorption \\
		\hline
		sound production & appropriate strength and propagation path length (source-receiver distance) & appropriate number, distribution over time and angle of incidence, strength, and spectral content & scattering or absorption, appropriate reverberation \\
		\hline
	\end{tabular}
\end{table}

As always, the direct and reflected sound components and acoustic measures refer to particular locations of sources and receivers in the room. For example, the \textbf{direct sound} is practically irrelevant for noise control whenever the receiver is located outside the zone of the direct sound dominance. In a diffuse field, this means that its distance from the source of noise is larger than the critical distance from eq.~(\ref{eq:critical_distance_diffuse_T}) or eq.~(\ref{eq:critical_distance_diffuse_T_directional_source}). If this is not the case or the direct sound dominates over the reflected sound, appreciable decrease of noise level cannot be achieved merely by changing room acoustics. It is then often necessary to modify the source or add sound insulation between the source and receiver, such as by placing the source in a closed cavity (enclosure) or a separate room.

In contrast to this, listener in a room for sound reproduction, such as a control room, should be close enough to the source (loudspeaker), in order to receive the information carried by the direct sound without distorting effects of the room. The direct sound should dominate over the reflected sound, which is easier to achieve (at least for a certain range of frequencies) if the source is directional and the receiver is localized. It is much more complicated to achieve appropriate strength and propagation time of the direct sound when the sources and listeners are distributed, such as in cinemas with multichannel reproduction, where obviously not all listeners can be in the zone of the direct sound of each source. Apart from uneven coverage, different times of arrival of the direct sound emitted from different loudspeakers can substantially affect the spatial information contained in a multichannel recording, which is thus not equally well reproduced in every part of the auditorium.

Direct sound is also important in rooms for sound production. Its propagation time from the source to a listener should not be too long to disturb natural synchronization between the visual and auditory components. This usually limits the source-receiver distance to maximum 20-30\,m. Moreover, direct sound of conventional sources, such as musical instruments and human voice, cannot be amplified as in the case of sound reinforcement. Its strength at the receiver location can be increased only by orienting a directional source towards the receiver or decreasing the distance between them. Any additional attenuation of the direct sound (typically due to obstacles along the propagation path or grazing propagation over an absorbing surface, such as auditorium) should be avoided.

Usual treatment of rooms for sound reproduction and rooms in which noise should be suppressed involves absorption of both early and late \textbf{reflected sound}. However, excessive absorption is not advisable, especially if humans are expected to stay in the room for a longer time. Lack of reflections appears unnatural and anechoic conditions are often associated with a feeling of unease. This is one of the reasons why musicians cannot perform very well in overdamped recording studios. Furthermore, strong suppression of late reflected sound allows higher intelligibility of speech and clarity of background sounds, which can be disturbing or spoil privacy in shared spaces, when these sounds are unwanted, such as in open plan offices. Numerous indistinct weak late reflections can partially mask the unwanted sounds and at the same time make the environment more natural. Finally, large amount of absorbing materials can be economically unfeasible or occupy considerable part of the room. Fully anechoic conditions should be approached normally only in anechoic chambers, such as those used for acoustic tests.

As an alternative to absorption, undesired reflected sound can be scattered or reflected away from the listener. This is particularly recommendable in rooms for sound production, since power of natural sources of sound is limited. Any damping in such rooms, besides the unavoidable absorption by the auditorium, can lead to an insufficient gain of the room and poor coverage with useful sound energy. Rather than absorbed, reflected sound energy can be distributed in a controlled way, evenly over the room. In fact, after a proper value of reverberation time has been achieved by means of absorption (including the auditorium), carefully steered early reflections and scattered late reflections lead to far better room impulse responses than additional damping -- well distributed energy of the reflections, both in time and over different angles of incidence. Reflectors can also be used as a simple and efficient replacement for thick absorbers of early reflections in control rooms. With appropriate placement, the reflections can be directed towards a diffuser or absorber in the rear part of the room. Thus, energy of the strong early reflections can be distributed over many less intrusive weak late reflections, without an increase of the overall damping in the room. 

Diverse subjective criteria (Table~\ref{tab:rooms_subjective_properties}) make the acoustic treatment in rooms for sound production the most challenging. As indicated by Table~\ref{tab:imp_response_param_values}, particularly critical are early reflections. Indeed, room acoustics of concert halls, opera houses, theatres, lecture halls, and similar rooms is to a large extent management of early reflections. Although their desired properties (strength, time and direction of arrival, spectral content) depend on the particular type of the room, as will be discussed later, several universal remarks on their optimization can be given regardless of the application. In general, acoustic treatment indicated in Table~\ref{tab:basic_strategies} can be achieved with an appropriate choice of the following \textbf{features} of the room:
\begin{itemize}
	\item room geometry (macro-geometry)
	\begin{itemize}
		\setlength\itemsep{-5pt}
		\item volume
		\item shape
	\end{itemize}
	\item large elements (large surfaces, typically walls, ceiling, and floor) and additional (small) elements (reflectors, diffusers, and absorbers)
	\begin{itemize}
		\setlength\itemsep{-5pt}
		\item size and shape
		\item position and orientation
		\item material and micro-geometry (for example, roughness)
	\end{itemize}
	\item source and receiver (complementing the room acoustics)
	\begin{itemize}
		\setlength\itemsep{-5pt}
		\item location and size (spatial distribution)
		\item directivity.
	\end{itemize}
\end{itemize}
The difference between large and small elements is rather formal. Basically, it distinguishes between boundary surfaces of the room and additional elements which can be used as acoustic measures.

Unfortunately, not everything listed above can always be applied or optimized in practice and the freedom of choosing an appropriate acoustic treatment is limited. If the room has already been built, its size and shape, as well as the size, shape, position, and orientation of the boundary surfaces are usually fixed. Number of seats (capacity) and area of the auditorium in concert halls, opera houses, and theatres are often determined by an economic analysis. Locations of musical instruments on the stage and their radiation patterns are also defined, at least for classical music and standard symphonic orchestra. Even the use of common acoustic elements -- reflectors, absorbers, and diffusers -- can be restricted by the aesthetic criteria, architectural design, limited space, or (especially in the case of rooms with historical value) need to preserve the visual identity of the existing space. Of course, the available budget usually sets further constraints to the acoustic treatment.

Acoustic consultant proposes optimization measures or an acoustic design based on the purpose of the room and other available data, such as technical drawings of the room (at least the horizontal and vertical cross sections are usually provided) and relevant constraints. The suggestions should be made bearing in mind that later corrections are most often associated with higher costs, especially if they involve macro-geometry of the room and large surfaces. This is even more pronounced in rooms of greater cultural value (concert halls, opera houses, and theatres) and large projects in general. The suggestions also have to be discussed with and eventually approved by the other persons in charge of the project, architect or civil engineer (see Table \ref{tab:room_acoustics_applicatons}), and the investor.

Acoustic design has to rely on the \textbf{predictions} of the sound field, which are also made by the acoustic consultant. Different calculation tools can be used for this purpose:
\begin{itemize}
	\setlength\itemsep{-5pt}
	\item analytical or empirical formulas for quick and rough estimations of basic properties,
	\item numerical simulations (typically ray tracing in combination with image source technique and less frequently finite element analysis or boundary element analysis) for more detailed calculations, and
	\item experimental methods (laboratory tests, \textit{in-situ} measurements, and measurements on scale models) as closest to reality.
\end{itemize}
Some of the most frequently used formulas, which we derived earlier, will be recapitulated next in the context of sound field predictions. Ray tracing, image source technique, and scale models will be covered in section~\ref{ch:modelling}. \textit{In-situ} measurements of impulse responses and the associated descriptors of room acoustics were discussed in section~\ref{ch:measurements_and_descriptors}. In larger projects, the predictions should be made at various stages of the project, parallel to the measurements of impulse responses, particularly when new acoustically relevant data are available. As more details of the room, elements, and materials become known, the predictions should become more detailed and accurate, but there is also less room for substantial modifications of the design.

According to Table~\ref{tab:basic_strategies}, the acoustic analyses should generally result in the estimation and optimization of
\begin{itemize}
	\setlength\itemsep{-5pt}
	\item ambient (background) noise level,
	\item energy and propagation time of the direct sound and visibility of the source,
	\item reverberation time (energy of late reflections) and sound level distribution, and
	\item temporal and angular distribution of early reflections and their energy.
\end{itemize}
Estimation of background noise in a room depends most often on sound insulation\footnote{Which is often part of the same project.} (for exterior sources of noise) and particular mechanisms of sound generation and thus will not be discussed further here. Still, too high levels of ambient noise can ultimately affect the acoustic rating of a room and the estimation of noise is always necessary. If it is generated by the sources in the room, such as various installations, ventilation, light, etc., the treatment is very similar as in the application of noise control considered here. In any case, special care should be taken to identify the most significant sources of noise, even (or especially) when they are not under direct control of the acoustic consultant. In the following we consider the remaining three points. While the first two of them can be treated fairly accurately, even with simple analytical or semi-empirical models, important early reflections can be estimated only roughly, which inevitably leaves certain amount of inaccuracy and risk of the predictions. Engineering experience is especially helpful in such considerations.

\subsection{Direct sound and visibility}\label{ch:direct_sound_and_visibility}

In the absence of obstacles, a line of sight is established between a source and listener and the \textbf{propagation time} of the direct sound can be estimated simply as $r/c_0$, where $r$ is the source-receiver distance. According to equations~(\ref{eq:solution_tailored_Green_wave_eq_free_space_compact_source}) and (\ref{eq:solution_tailored_Green_wave_eq_free_space_compact_directed_source_emission_time}), amplitude of the direct sound of an acoustically compact source decays proportionally to $1/r$, or around 6\,dB per doubling the distance from the source. For a known sound power level of a directional point source, the direct \textbf{sound pressure level} in the acoustic far field can be calculated using eq.~(\ref{eq:acoustic_power_level_complex_time_average_plane_wave_directional_source_pressure}). However, if the wave propagates just above an absorbing surface, for example an auditorium, and parallel to it (at a very low, grazing angle), somewhat faster decay with the distance should be expected, because part of the energy is refracted and absorbed by the surface. As a result, remote listeners in the back rows of the auditorium may receive appreciably weaker direct sound than in free space, besides a poor visibility of the source and possibly asynchronous auditory and visual information. This problem can be handled in two ways -- by decreasing the distance between the source and receiver or by increasing the visibility angle (angle between the line of sight and auditorium surface at the listener location) to avoid additional attenuation.

In terms of the average \textbf{distance} between listeners and a source on the stage, for a fixed area of the auditorium, round auditoriums (in the form of a full circle or its segment with the stage in the centre) outperform all other shapes. They are followed by horseshoe-shaped, fan-shaped (or trapezoidal), and diamond-shaped (rhomboidal) auditoriums, with the stages placed next to the shortest plane walls of the rooms. The least favoured rooms according to this criterion are narrow rectangular rooms with the stage in front of one of the two shorter walls. However, the criterion, which promotes stronger direct sound and better visibility of the source (or larger auditorium with more seats), turns out to be in contrast to the requirement for strong lateral reflections (early reflections will be discussed in more details in subsection~\ref{ch:early_reflections_strategy}) which are essential for the perception of spaciousness. Namely, relatively narrow rectangular rooms provide more lateral sound energy (and therefore more spaciousness) in a larger part of the auditorium than rooms with distant side walls or larger angles between the front wall (behind the source) and side walls, as in fan-shaped halls. This explains why rectangular shape is still preferred for concert halls, in which spaciousness presents a very important subjective criterion, while horseshoe and fan shapes are more common in opera houses and theatres, where good visibility and strong direct sound bring more significant information. Spaciousness is very poor in simple round auditoriums, due to the lack of lateral reflections as well as useful early reflections in general, and they are usually avoided in rooms for sound production (without sound reinforcement).

\textbf{Visibility angle} can be increased either with an ascent (gradual increase of height) of the auditorium in the form of a staircase, or by lifting the source above the plane of the auditorium (for example, by hanging a loudspeaker on the ceiling, or placing the sources on a stage with a descent towards the auditorium). This avoids grazing propagation of both direct sound and important first-order lateral and front-wall reflections, which improves spaciousness and increases the early-to-late energy ratio. At large distances from the source, lines of sight of the listeners located one behind the other are approximately parallel, so the distance between two adjacent lines is sometimes used instead of the visibility angle. This distance should generally be at least 6\,cm and preferably around 10\,cm or larger. The visibility angle should ideally be at least 15$^\circ$.

In order to achieve a constant visibility angle, independent of the distance from the source, slope of the auditorium should gradually increase with the distance from the stage. On the other hand, linear ascent is more convenient for realization and easier to adjust to the stairs leading to different rows of the auditorium. Accordingly, a continuous increase of the heigh of an auditorium can be approximated by dividing it into several sections with different but constant slope angles, which increase with the distance from the stage.

One way to increase capacity of an auditorium but keep acceptable distances of the listeners from the source is with \textbf{balconies} or \textbf{galleries} at one or more levels above the main part of the auditorium. However, deep balconies with many rows of seats can result in poor coverage with the reflected sound in the rear parts of the room, especially below them. In contrast to this, relatively short galleries (with only a few rows of seats) introduce large geometric irregularities in the room and can improve diffuseness of the field even at low frequencies, which would otherwise be difficult to achieve with micro-geometry of large room surfaces. At higher frequencies they also provide valuable early reflections, such as from the bottom of side galleries. As a rule of thumb, ratio of the height (of the opening) and depth of balconies should not be smaller than one.

\subsection{Reverberation time and sound level}\label{ch:reverberation_time_sound_level}

\textbf{Reverberation time} is most commonly estimated using Sabine's formula~(\ref{eq:T60_Sabine_air}). Unless the room is very large and weakly damped, dissipation in air can be neglected in most of the cases. Eyring's equation~(\ref{eq:T60_Eyring}) can also be used, especially if the room is not very weakly damped. On the other hand, very damped rooms do not comply with the assumption of a diffuse field and both formulas are likely inaccurate. Alternatively, reverberation time can be estimated numerically, from the echograms obtained using ray tracing simulations (section~\ref{ch:ray_tracing}). In any case, accuracy of the estimation is limited by the availability and validity of the absorption coefficient values of various materials and surfaces in the room, especially large surfaces. The values are usually collected from data sheets of the materials and other literature or, when necessary, measured in a laboratory in a diffuse field (reverberation chamber) or at normal incidence (Kundt's tube). Deviation of the actual sound field in the room from the diffuse field or plane waves at normal incidence can also increase the error of the estimations (recall the discussion at the end of section~\ref{ch:geometrical_statistical_theory}). The values in octave bands are typically presented in tables or diagrams. 

Rooms for sound production are usually weakly damped and the main absorbing surface is the auditorium. For simple (initial) estimations of reverberation time, \textbf{absorption coefficient} of a fully occupied auditorium can be set to 0.8 at middle and high frequencies, while a small positive value, say 0.05, can be adopted for all other reflecting surfaces in the room. In addition to this, the effective seating area of the auditorium (which is required for an estimation of its equivalent absorption area) is often approximated as the area of the auditorium extended by 0.5\,m in all directions without crossing room boundaries. Such extension takes into account additional absorption at the edges of the auditorium, for example, its side surfaces. When the auditorium is divided into several separate sections, the extensions should be added to each section.

Absorption coefficient of unoccupied seats is relatively difficult to estimate in general, since it depends on the material of the seats and size of the upholstery. If the specific data or measured values of the absorption coefficient are not available, they can be estimated from the data for similar types of seats in literature. As a rule, it is advisable that (moderately upholstered) unoccupied seats provide similar absorption coefficient values as when they are occupied by the listeners, which are around 0.8-0.85 at middle frequencies, so that the total absorption and therefore reverberation time in the room do not vary significantly with the occupancy.

When present, other possibly significant contributors to the total absorption, such as porous or membrane absorbers, wall linings, carpets, curtains, large openings, windows and doors, organs (in churches and concert halls), diffusers with non-negligible absorption, as well as dissipation in air in large reverberant spaces (for example, churches and sport halls), can also be included in the estimation of reverberation time. It is up to the acoustic engineer to assess how relevant these surfaces are and the reliability and applicability of the available data on their absorption, which depend on the type of the materials, location in the room, orientation, mounting, etc. If especially high risk of an inaccurate estimation of reverberation time is associated with certain (large) elements, for example seats of the auditorium, their absorption should be measured in a laboratory.

Estimation of reverberation time is more complicated if the main room (where the listeners are located) is coupled to another adjacent room (or several rooms) via an open surface (interface) which is much smaller than the total surface of both main and adjacent room. Typical examples are stage houses in theatres and opera houses, which are coupled with the main hall via the proscenium behaving as the interface, as well as boxes (when they are mostly closed on the sides) and sometimes aisles and naves in churches. Different sound energy decays in the \textbf{coupled rooms} can in general result in a non-linear sound pressure level decay, in contrast to the diffuse field and definition of reverberation time. In such cases reverberation times of the coupled rooms can be estimated independently by treating the interface as a boundary surface of the rooms which it couples, with a certain absorption coefficient value assigned to it. If reverberation time in the coupled room is much shorter than in the main room with the listeners, it does not significantly affect the perceived reverberance. The interface can then be treated as an absorbing boundary surface of the main room, with absorption coefficient above 0.5. In the opposite case, when reverberation time in the coupled room is longer than in the main room, the coupled reverberation can be noticeable in the main room, especially if the source or listener is close to (or even inside) the coupled room. This is usually undesired and the coupled room should be additionally damped, in order to decrease its reverberation time relative to the main room.

As indicated in Table~\ref{tab:imp_response_param_values}, reverberation time should be around 1\,s in rooms for speech and around 2\,s in rooms for music. Opera performances involve both types of sound information, so the optimal values are around 1.5\,s. These values hold at middle frequencies, in octaves 500-2000\,Hz. It is common that reverberation time increases at low frequencies and decreases at high frequencies, although the increase at low frequencies should not be too high, in order to preserve the intelligibility of speech and clarity of music (avoid masking by the low-frequency background noise). Usually acceptable are an increase at low frequencies up to 20\% at 63\,Hz and drop at high frequencies up to 40\% at 8\,kHz with respect to the values at the middle frequencies. The tolerance is somewhat higher for music than for speech. Optimal reverberation time values depend also on the volume of the room. For the volume of 1000\,m$^3$, reverberation time in rooms for speech should be around 1\,s and for music around 1.5\,s. The optimal values increase roughly 0.1-0.15\,s per doubling the room volume. Since concert halls are in average larger than rooms for speech, optimal reverberation time in them is typically around 2\,s. Baroque music and large volumes allow even longer reverberation times in churches.

In rooms with appropriate size and shape, suggested reverberation time values can be reached by ensuring adequate amount of absorption in the room. If the room is intended for different types of sources of both speech and music (a multipurpose room), \textbf{adaptive} reverberation time is sometimes achieved with movable walls (effective volume of the room is decreased by closing one part of the room) or variable absorbing surfaces, for example, by opening or closing curtains or exposing absorbing or reflecting sides of adaptable absorbing elements. The achieved range of values of reverberation time depends, according to eq.~(\ref{eq:T60_Sabine_air}), on the absorption coefficient of the variable elements, fraction of the total area of the room which they occupy, and the change of room volume.

Reverberation time in a room is related to the \textbf{sound pressure level} produced by a source in it. For example, constant sound pressure level in a diffuse field of an omnidirectional point source at distances larger than the critical distance can be estimated using eq.~(\ref{eq:SPL_diffuse_field}) and the equivalent absorption area is related to the reverberation time via Sabine's formula. In reality, lack of diffusion causes decay of sound level with the distance from the source (except close to reflecting surfaces, where it increases). As already discussed, this is more pronounced in rooms with irregular shapes, such as long or flat rooms, as well as when absorption is not evenly distributed inside the room.

Still, the model of a diffuse field and the statistical theory are most commonly used for simple predictions of the effects of noise control measures in rooms. In order to appreciably decrease sound pressure level due to a source of noise by means of room acoustics, sufficiently large area of the room should be covered with materials with sufficiently high values of absorption coefficient (much higher than the covered surface). When feasible, even several decibels lower noise level can significantly improve acoustic comfort in the room, unless the room is overdamped.

\subsection{Early reflections}\label{ch:early_reflections_strategy}

Accurate calculation of early reflections in a room is in general more difficult than estimation of the overall decay of late reflections (reverberation time) or time-averaged energy (sound pressure level). While late reflections are numerous and suitable for the statistical treatment, early reflections are often distinct and separable. Their number, rate of occurrence at the receiver location, strength, spectral content, time of arrival after the direct sound, path lengths, and directions or arrival depend not only on the (possibly angle-dependent) absorption coefficients of the involved reflecting surfaces, but also their geometry, size, shape, position in the room, orientation, roughness, as well as positions of the sources and receivers relative to them and their directivity.

It takes some practice to \textbf{identify} the most relevant early reflections in a room, determine their propagation paths from the source to the listener, and estimate the listed acoustically relevant properties. Besides technical drawings of the room, ray tracing and image-source simulations and (partial) scale models can make this task easier. When strong reflections are recognized in a room impulse response, their propagation paths can often be determined from the technical drawings, based on the delay with respect to the direct sound and assuming that the reflections are specular, especially in the case of the earliest (low-order) reflections. Of particular interest are strong distinct reflections (or groups of reflections, such as those coming from concave reflecting surfaces), which may cause a detrimental jump of relatively late sound energy at particular locations in the room. The resulting lack of diffusion and non-uniformity of the sound field as well as possible occurrence of echo are usually undesired. Nevertheless, spectral content and temporal and angular distribution of useful early reflections are also important for the assessment of room acoustics.

Difference between sound pressure levels of the direct sound and an $N^{\text{th}}$-order ($N$ times reflected) \textbf{specular} reflection of an omnidirectional point source can be estimated at the receiver location as
\begin{equation}\label{eq:SPL_dir-refl}
\begin{split}
L_d - L_r \approx 20 \log_{10} \left( \frac{r_r}{r_d} \right) - \sum_{n=1}^{N} 10 \log_{10}[1-\alpha_{s,n}(\theta_{i,n})] - \sum_{n=1}^{N} 10 \log_{10} [1-s_{s,n}(\theta_{i,n})].
\end{split}
\end{equation}
The first term follows from eq.~(\ref{eq:acoustic_power_level_complex_time_average_plane_wave_omni_source_pressure}) and involves ratio of the total path length of the reflection, $r_r$, and path of the direct sound, $r_d$. It captures the decay of sound pressure level with distance. Besides the absorption coefficient $\alpha_s(\theta_{i})$ of each reflecting surface on the path, the scattering coefficient $s_s(\theta_{i})$ has been introduced, which represents the ratio of non-specularly reflected (scattered) energy and total reflected energy, regardless of the absorption. It approaches zero for large flat surfaces. In general, both quantities can depend on the angle of incidence $\theta_i$ (we neglect possible dependence on the azimuthal angle) as well as frequency.

Similarly as the direct sound, reflections which propagate over large absorbing surfaces, such as auditorium, can be additionally attenuated. As discussed in section~\ref{ch:direct_sound_and_visibility}, this can be avoided by lifting the source or the receiver and thus increasing the propagation angle to the surface. Dissipation in air is most often negligible for relatively short propagation paths of low-order reflections. Time difference of arrival of the reflection and direct sound can also be estimated as $(r_r - r_d)/c_0$. This simple ``specular'' model follows directly from the image source model which will be treated in section~\ref{ch:image_sources}. It is applicable for acoustically large surfaces and its accuracy decreases with the reflection order, largely due to the uncertainty of the values of the two coefficients.  In particular, actual values of the scattering coefficients are rarely well known, which affects accuracy of the estimation using eq.~(\ref{eq:SPL_dir-refl}).

It is usually reasonable to expect that the lowest-order reflections with $N=1$ and $N=2$ are dominant and the most relevant for room acoustics. Moreover, their delay compared to the direct sound is short and thus they provide a useful sound energy in rooms for sound production. For this reason, for example, a performer or a speaker on a stage should not be located too close to the front edge of the stage, close to the auditorium, if they should benefit from the immediate reflection from the stage floor. This sound energy can be redirected towards the auditorium as a second-order reflection from the ceiling or a hanging reflector above the stage or as higher-order lateral reflections. When identified, undesired early reflections can be suppressed by increasing absorption or scattering of the associated reflecting surfaces, or redirected by changing orientation of the surfaces, for instance, by placing acoustic reflectors. In general, favourable properties of early reflections and their \textbf{optimization} are application-dependent. Therefore, they will be considered in more details in the rest of the section for particular types of rooms.

\subsection{Acoustic design}\label{ch:acoustic_design}

In the rest of this section we cover briefly the main issues and basic strategies for optimization of room acoustics in some of the typical room types. With the exception of noise control in rooms, they predominantly pertain to the early reflected energy. Further details on each room type, acoustic requirements, and rules of best practice can be found in more specific literature.

\subsubsection{Noise control in rooms}

Room acoustics is relevant for noise control when more efficient acoustic measures on the sources of noise or sound propagation path, such as enclosing the sources with casings, separating them from the receivers by means of building partitions with appropriate sound insulation, or affecting the mechanisms of noise generation, are not sufficient or feasible. This is often the case with large, distributed or moving sources, humans as sources of unwanted sounds, equipment which has to be cooled or ventilated, and similar. In addition to this, receivers are also often distributed or moving. Since the main task is \textbf{overall noise} suppression, without more sophisticated aesthetic criteria or distinction between particular reflections, predictions of the effects of acoustic measures are normally made using simple models and statistical theory, at least for broadband random noises.

As in section~\ref{ch:reverberation_time_sound_level}, diffuse field equation~(\ref{eq:SPL_diffuse_field}) indicates that sound level in the room can be decreased by increasing the \textbf{equivalent absorption area} in it. If total surface area of the room is fixed, the average absorption coefficient should be increased. However, this holds only for the locations in an essentially diffuse field, outside the zone of the direct sound dominance, that is for $r > r_c$, where $r$ is distance from the source of noise and $r_c$ is the critical distance (see Fig.~\ref{fig:critical_distance} (left) and equations~(\ref{eq:critical_distance_diffuse_T}) and~(\ref{eq:critical_distance_diffuse_T_directional_source})). Equation~(\ref{eq:SPL_diffuse_field}) can then be used for calculations of noise level in octave bands. A single-number value can be obtained using NR-curves (Fig.~\ref{fig:NR_curves}) or A-weighting, as in the case of background noise considered in section~\ref{ch:sources_dynamics}, and compared against the maximum allowed values. 

It should be noticed that the amount of noise attenuation (which rarely surpasses several decibels in practice) depends on the relative increase of the average absorption coefficient of the room, compared to its initial state. Consequently, such acoustic measures are efficient mostly in very weakly damped rooms, in which the average absorption coefficient can be appreciably increased with a reasonable amount of the introduced absorber, for example, with an absorbing suspended ceiling. Especially appropriate for covering them with absorbing materials are large reflecting surfaces -- hard walls and ceiling. If the room is already damped, adding more absorption can be impractical and the measures based on sound insulation are usually much more appropriate. As a side effect, increase of absorption results also in a larger zone of the direct sound, which may then include the listeners located relatively close to the sources of noise and thus additionally limit the efficiency of noise control. 

\subsubsection{Rooms for sound reproduction}

As indicated in Table~\ref{tab:rooms_subjective_properties}, rooms for sound reproduction should have neutral acoustics with minimum influence on the reproduced sound. Acoustic properties of rooms for critical listening and monitoring, such as control rooms in music studios, are optimized for particular \textbf{locations} of the sources (loudspeakers) and listeners (typically at the height around 1.2\,m from the floor). These are often dictated by the audio format, such as stereo or 5.1 surround sound, and optimal response of the room must be achieved only at those locations. All other parts of the room can benefit from the optimization, but do not have to meet some rigorous criteria, which gives more freedom for both acoustic and architectural design of the interior. The situation is more complicated in cinemas, due to multiple sources and distributed listeners.

At the listener's locations in control rooms it is recommendable that all reflections are at least 20\,dB weaker than the direct sound from the loudspeakers, in order to avoid colouration or change of the apparent location of the source by the \textbf{early reflections} (recall section~\ref{ch:subjective_criteria}). This value is sometimes reduced for the earliest reflections to 10\,dB below the direct sound, which is considerably easier to achieve. A common measure is building the monitor loudspeakers into the walls rather than immediately in front of them, in order to avoid the very early first-order reflections from the walls. The first-order reflections from the mixing desk should also be avoided by placing the loudspeakers either low enough or further behind the desk.

In general, parts of the walls and ceiling which are closest to the loudspeakers should either absorb or reflect the sound away from the listener's location. The latter is achieved most efficiently by appropriate orientation of  large reflecting surfaces. Strong reflections from the rear wall (behind the listener) are usually avoided by mounting a diffuser or absorber on it. The combination of diffusively reflecting rear part of the room and highly absorbing front part is also known as ``Live End, Dead End'' (LEDE). If necessary, additional absorption can be provided by the ceiling or side walls. In cinemas, broadband absorbers are usually placed on both front (behind the screen) and rear walls, as well as on the floor (carpet) and the seats, due to their larger volumes and thus inherently longer reverberation times. Exceptionally, front parts of the side walls can be left reflecting and used to achieve a more uniform coverage in the back rows.

\textbf{Late reflections} should be as diffuse as possible and reverberation time should be low, only about 0.3\,s for the room volume 100\,m$^3$, but not much shorter. An approximate relation between the optimal reverberation time and the room volume is $T_{60} = 0.3(V/100\text{\,m}^3)^{1/3}$\,s (or sometimes $T_{60} = 0.25(V/100\text{\,m}^3)^{1/3}$\,s) for $V>35\text{\,m}^3$. Typical tolerance is $\pm 20\%$ and somewhat larger below 250\,Hz and above 2\,kHz, up to around +50\% at 63\,Hz and -50\% at 8\,kHz. Volumes below 100\,m$^3$ should be avoided because of probable issues with room resonances at low frequencies. Optimal volume of a cinema can be estimated with around 4\,m$^3$ per seat.

Specific problems appear in control rooms due to their relatively small common sizes compared to most of the other rooms considered below. The associated Schroeder frequency from eq.~(\ref{eq:Schroeder_freq}) can be high, well in the audible range, in spite of high damping and distinct \textbf{room modes} from eq.~(\ref{eq:solution_tailored_Green_Helmholtz_modes}) can dominate in the frequency responses at low frequencies, regardless of the room shape. As a result, frequency responses are not neutrally flat in this part of the frequency range, as they should ideally be. The remedy requires high damping even at low frequencies. On the other hand, lack of space limits the possibilities for application of thick porous absorbers, which would be efficient at low frequencies. Low-frequency membrane absorbers, bass traps, and careful choice of the source and listener locations (for example, avoiding the central zone of the room) can help dealing with this difficult problem. As already mentioned, room volumes below 100\,m$^3$ should be avoided and proportions should not be simple (integer values), in order to avoid strong overlapping of the lowest modes, which intensifies the resonance behaviour. Simple room shapes such as rectangular should not be used, although left-right symmetry with respect to the listener is often desirable, since it gives approximately equal frequency responses of the rooms for symmetrically placed loudspeakers. Room shapes with non-parallel walls also prevent flutter echo at higher frequencies.

A specific issue in cinemas is appropriate \textbf{coverage} of the auditorium with the direct sounds of multiple sources. In order to reproduce the created sound image accurately, not only their strengths should be balanced, but the times of arrival at a listener's location, as well. Although directional surround loudspeakers provide more control than omnidirectional sources, uniform coverage with all components of the multichannel sound can only be approximated. This is a much simpler task in control rooms with the receivers localized in the ``sweet spot''.

\subsubsection{Recording studios}

Here belong primarily studios for recording music and speech, but broadcasting (radio and TV) studios can also be included. Unlike in most of the rooms in which sound is produced, the receivers in recording studios are not human listeners, but \textbf{microphones}. Moreover, similarly as in rooms for sound reproduction, acoustic effects of these rooms should usually be suppressed and optimal impulse responses with the dominant direct sound should be achieved for specific locations of sources and receivers, if they are defined. Otherwise, optimal locations in the room can be determined by an inspection, if sources and receivers are movable. 

Characteristic for these rooms is that proximity and directivity of the microphones can be utilized to neutralize defects of room acoustics. By placing a microphone in the zone of the direct sound dominance and directing the maximum of its sensitivity towards the source, a response free of strong reflections from most of the directions can be achieved without excessive damping. This is even easer if the source is directional, too. More pleasant ambient for sound production is obtained with less damping in the room and, possibly, more scattering surfaces, which further avoid strong specular reflections. The use of microphones thus allows much greater flexibility in the design of recording studios than in the case of control rooms, shifting the focus to other non-acoustic qualities.

If the room is acoustically small (occasionally even smaller than 30\,m$^3$, with reverberation time sometimes below 0.3\,s), similar remarks regarding the effects of \textbf{room modes} at low frequencies apply as for control rooms. However, if the room is used only for voice recordings (as in radio studios), its frequency response below 100\,Hz is less critical than in music studios. Furthermore, locations in the room where resonant behaviour is least pronounced can be chosen for recording. In average, TV studios have larger volumes, which make this issue less critical, but variable studio set design can sometimes cause detrimental reflections. Still, these can be handled with appropriate microphone setups, such as using lavalier microphones.

\subsubsection{Lecture halls}

In comparison with classrooms, lecture halls are relatively large rooms for speech. According to section~\ref{ch:reverberation_time_sound_level} and Table~\ref{tab:imp_response_param_values}, appropriate \textbf{reverberation time} in them at middle frequencies is around 1\,s ($\pm 20\%$), for the room volume around 1000\,m$^3$. The dependence on volume can be expressed roughly with $T_{60} \approx 1\text{\,s}+0.1\text{\,s} \cdot \log_2(V/1000\text{\,m}^3)$ or similarly $T_{60} \approx 0.37\text{\,s} \cdot \log_{10}(V/1\text{\,m}^3) - 0.14$\,s. The allowed tolerance increases linearly in octaves below 250\,Hz and above 2\,kHz up to around (-50\%, +70\%) at 63\,Hz and (-50\%, +20\%) at 8\,kHz. For comparison, in classrooms with volumes below 500\,m$^3$, recommended reverberation time is about 0.2\,s shorter for equal volume (more precisely, $T_{60} \approx 0.26\text{s} \cdot \log_{10}(V/1\text{m}^3) - 0.14$s). Ideally, reverberation time in rooms for speech should decrease at the lowest frequencies, outside the frequency range of speech (below the octave 125Hz), as a measure against low-frequency ambient noise. In some rooms this is achieved simply owing to the transmission of low-frequency sound through windows.

A parameter which is frequently used for determining optimal volume of a room is ratio of the room volume and number of seats. In larger lecture halls with volumes up to 5000\,m$^3$ it should be 4-6\,m$^3/$seat and in smaller classrooms 3-5\,m$^3/$seat. Very large lecture halls, with more than 500 seats, usually require the use of sound reinforcement and accordingly somewhat lower reverberation time than specified above.

Besides an appropriate reverberation time, room for speech have to ensure good speech intelligibility, especially in the absence of sound reinforcement. Sufficient \textbf{early sound energy}, within the first 50\,ms after the direct sound or so, is crucial for that. As indicated in Table~\ref{tab:imp_response_param_values}, value of the definition $D_{50}$ should be at least 0.5 and speech transmission index above 0.6 or ideally above 0.75. Accordingly, the listeners should be provided with many early reflections. Since source area on the podium is normally less distributed than the auditorium, this is most efficiently achieved by careful placement and orientation of the reflecting surfaces (walls, ceiling, or reflectors) close to the source. 

In order to increase the number of useful early reflections in the auditorium, the ceiling should not be higher than 8-10\,m and its front parts above the podium can be inclined such to distribute the energy of first-order reflections over the auditorium. All areas of the ceiling which provide useful reflections should be left acoustically hard and reflecting, in particular the front and central parts. Since short times of arrival of the reflections are critical, especially in larger halls, the reflecting surfaces should not scatter the sound (which might be preferable in rooms for music performances) and thus extend the propagation paths to the auditorium. If the ceiling is too high, hanging reflectors can decrease the propagation path lengths. Similar remarks hold for the surfaces surrounding the podium, left and right from the speaker, which also play a major role in providing listeners with early energy. Very wide halls with large podiums should be avoided or additional reflecting elements on the podium may be necessary.

In contrast to the surfaces close to the source, rear walls and remote parts of the ceiling in larger rooms are often diffuse or inclined such to prevent too late and strong low-order reflections from reaching the front parts of the auditorium, as well as to avoid flutter echo between the front and back wall of the room. The redirected reflected energy is in such a way delivered mostly to the rear parts of the auditorium, where the delay with respect to the direct sound is shorter. If the reverberation time should be decreased with additional absorption besides the auditorium, these remote surfaces can also be treated with absorbing materials. 

In general, diffuse reflections and direction of arrival of sound energy to the listeners have much less significance than in rooms for music performances. The associated perception of spaciousness is less relevant and the emphasis is rather on decreasing the average \textbf{distance} from the source and improving visibility and the direct sound coverage. Accordingly, fan-shaped and round auditoriums are common, which can accommodate a relatively large number of listeners at acceptable distances from the speaker. Galleries and balconies can be introduced to further increase the number of seats. As discussed in section~\ref{ch:direct_sound_and_visibility}, slope of the auditorium (both on the floor and the galleries) avoids attenuation of the direct sound and some reflections due to grazing propagation over the auditorium and increases the intelligibility. When sound reinforcement is used, the loudspeakers should be placed well above the stage and the front rows of the auditorium and directed towards the auditorium.

\subsubsection{Theatres}

Here we consider primarily drama theatres, in which spoken information is produced by the actors on the stage. As such, theatres share many similarities, including requirements for optimal room acoustics, with lecture halls. For example, fan-shaped auditoriums are quite common, largely for visual reasons (gesticulation and facial expressions of the actors should be accessible to the audience) as well as higher speech intelligibility due to stronger \textbf{direct sound}. Galleries and balconies often increase the hall capacity. Although the room volumes larger than in lecture halls are common (with up to around 1000 seats even without sound reinforcement, owing to the trained voices of the actors), ratio of the volume and number of seats should be similar, 4-7\,m$^3/$seat, as well as the reverberation time values. Apart from the ascent of the auditorium, slope of the stage (when the stage design allows it) can further improve visibility and the direct sound. If sound reinforcement is present, it should be placed well above the stage floor, often above the proscenium.

In terms of room acoustic, the most important difference between large lecture halls and drama theatres is the stage. In theatres it is usually significantly larger than podiums in lecture halls. Moreover, \textbf{stage houses} in theatres are typically much higher and wider than the proscenium and can thus present coupled spaces to the main rooms, causing additional late reflections in them, for example, from the back wall and ceiling of the stage. The undesired late energy is usually suppressed with absorbers on the surfaces of the stage house. Proscenium then acts essentially as a highly absorbing boundary surface of the main room with the auditorium. In rough predictions of the sound field, it is often approximated as a fully absorbing surface, or a high value of absorption coefficient (above 0.6) is assigned to it. Together with the auditorium it constitutes a large equivalent absorption area, which is why most of the other room surfaces should be reflecting and the reflected energy optimally distributed. To a certain extent, acoustic behaviour of the stage house depends also on a particular scenery on the stage.

As in lecture halls, speech intelligibility in theatres should be supported with a sufficient number of strong \textbf{early reflections} reaching the listeners not later than approximately 50\,ms after the direct sound. Their directions of arrival and subjective impression of spaciousness are not critical. Since the auditorium is distributed, it is reasonable to have efficiently reflecting surfaces close to the stage. However, in contrast to lecture halls in which they can often be placed closer to the sources with a greater flexibility, additional reflectors on the stage or just above it conflict with the stage design (unless they happen to be part of the scenery). This and the large size of the stage house limit the possibilities for acoustic treatment of early reflections and make the fewer reflections from parts of the side walls and ceiling closest to the stage crucial for early sound energy. Ideally, these surfaces should also reflect part of the sound energy back to the performers, giving them an acoustic feedback (support). With regard to that, the ceiling should not be higher than around 10\,m. This is somewhat lower than in rooms for music performances (with heights up to around 15\,m), which can be directly related to the difference of the early-energy time intervals for definition $D_{50}$ and clarity $C_{80}$ (the difference of 30\,ms corresponds to the path length difference of around 10\,m, twice the height difference).

Mostly closed \textbf{boxes} (separated with side walls) should be avoided in general. When strongly damped, they can cause an excessive damping in the room, in addition to the unavoidable absorption of the auditorium. If they are weakly damped, they can act as small rooms coupled to the main part of the hall, as already discussed. Another potential cause of undesired absorption are empty slits in the ceiling above the auditorium, which are often used for \textbf{stage lighting}. If this is the case, the slits should be closed. On the other hand, the inclined surfaces of the ceiling below the lighting can be conveniently used for improving the coverage of the auditorium with the sound energy reflected from the ceiling.

\subsubsection{Opera houses}

Opera houses are rooms for both speech and music (including singing) and their optimal acoustics is a compromise of their dual nature (a combination of the last two columns of Table~\ref{tab:imp_response_param_values}). For instance, optimal \textbf{reverberation time} is somewhat higher than in rooms for speech and somewhat lower than in rooms for music performances, usually between 1.2\,s and 1.8\,s (roughly 1.2\,s for the room volume 1000\,m$^3$ and increasing 0.12\,s per doubling the volume). A small increase of the value at low frequencies is often acceptable to support the music. The volume of 5-8\,m$^3$ per seat is considered as appropriate and the total volume should not exceed 15000\,m$^3$, corresponding to the maximum of around 2000 seats.

Surface area and shape of the auditorium (typically horseshoe) are determined primarily by the visual criteria. Similarly as in theatres, actions on the stage, including the details such as facial expressions of the performers, should be visible over the auditorium. This limits considerably the maximum \textbf{distance} between the listeners and the stage and practically forces the use of galleries, in order to increase the capacity of the hall. As usual, ascent of the auditorium on the floor and galleries improves the visibility as well as strength of the direct sound.

Another similarly with theatres is presence of the \textbf{stage house}. Coupled reverberation and additional strong late reflections in the auditorium are most often avoided by means of absorbing materials inside the stage house. However, horseshoe shape of the hall and the large size of the stage house leave even smaller reflecting surfaces of the side walls and ceiling close to the stage, which could contribute useful early reflections to the listeners and performers, both on the stage and in the orchestra pit.

\textbf{Orchestra pit} present a peculiarity of an opera house, from which the orchestra follows actions on the stage without introducing a visual obstacle between the auditorium and the stage. Height of the pit is usually adjustable by lifting and lowering its floor. During the performance it should not be too shallow (less than 2-2.5\,m deep) and its floor area should be around 1.5\,m$^2$ per member of the orchestra. Such a geometry allows strong early reflections inside the pit, which help the musicians hear each other well and improves their common performance (ease of ensemble). On the other hand, some low-frequency absorption in the pit is often desirable to dampen the resonances of the small partly closed space, which could otherwise affect the frequency balance in the entire hall. Apart from this, most of the acoustic optimization aspects related to the \textbf{early reflections}, speech intelligibility, and design of \textbf{boxes} are essentially the same as in lecture halls and theatres. The remaining aspects related to the music component are discussed next in the context of concert halls.

\subsubsection{Concert halls}

Compared to the rooms for speech, in rooms for music performances much more attention should be given to the lateral reflections, \textbf{diffuseness} of the sound field, and therefore scattering of the reflected sound energy. In general, large flat specularly reflecting surfaces should be avoided. Diffuse field approximated with many reflections reaching the listeners from various directions is rated positively by the listeners and associated with the listener envelopment. In terms of the descriptors of room acoustics, lateral energy fraction and binaural quality index should be high (Table~\ref{tab:imp_response_param_values}). Time interval of the early sound energy which contributes to the clarity of music is longer than for speech intelligibility, around 80\,ms after the direct sound. Therefore, even the earliest reflections from the surfaces close to the stage should be scattered. In extreme cases, focused specular reflections can even distort the perceived reverberation at particular places in the room.

Conventional rectangular shape of the hall with the stage in front of a shorter wall provides most easily sufficient lateral sound energy even in the rear parts of the auditorium, especially closer to the side walls. This is much more difficult to achieve in round or fan-shaped halls. On the other hand, rectangular shape results in a larger average distance of the listeners from the stage (meaning weaker direct sound and less visibility) for the same total number of seats, which is not as essential as in theatres and opera houses. Larger number of seats can nevertheless be achieved in vineyard-shaped rooms, which are essentially round, with the auditorium distributed around the stage. However, the auditorium is divided into sections with floors at different heights. This introduces additional side walls for lower parts of the auditorium and therefore valuable lateral reflections there. Still, box-shaped concert halls acoustically outperform the vineyard-shaped rooms in average.

As in theatres and opera houses, the number of seats can also be increased with \textbf{galleries}. They should not be too deep, preferably with up to several rows of seats only (section~\ref{ch:direct_sound_and_visibility}). Together with the closest wall, bottom of a gallery can give additional strong second-order reflections and bring more lateral sound energy which is not attenuated by the grazing propagation over the auditorium. Slopes of the auditorium on both the floor and the galleries and/or the stage improve further the visibility, direct sound strength, and the strength of early lateral reflections. Low-frequency diffusion is another benefit of galleries.

Late reverberation influences the perception of space in which music is performed and therefore the overall aesthetic evaluation. For this reason, \textbf{late reflections} are much more significant and useful than in rooms for speech and somewhat longer reverberation times are desirable. Optimal reverberation time for symphonic music is around 1.4\,s for the room volume 1000\,m$^3$ and increases around 0.15\,s per doubling the volume. However, optimal reverberation time depends also on the type of music. For organ music it is about 0.3\,s longer for the same room volume. A slight increase at the lower frequencies is common and acceptable. If necessary, additional absorption at low frequencies can be achieved with membrane absorbers on the walls or ceiling and less commonly with Helmholtz resonators.

Small concert halls (for small orchestra and chamber music) with volumes below 10000\,m$^3$ should provide 6-10\,m$^3$ per seat, while large halls for symphonic orchestra, with the volumes usually up to 25000\,m$^3$, should have a larger ratio of the volume and number of seats, 8-12\,m$^3/$seat. Even larger volumes and the ratio values up to 14\,m$^3/$seat can be appropriate for organ music. The total number of seats should be between 1500 and 2000 for symphonic music, in order to keep a good coverage with sound of the entire auditorium, and somewhat smaller, less than around 1000, for small orchestra and chamber music.

Surfaces which are far from the stage are a potential cause of strong late reflections, especially in the front part of the auditorium. In extreme cases, these can be perceived as echo, which can be avoided by placing absorbing materials on the remote surfaces. However, this often leaves a too short reverberation time of the room (in combination with the absorption of the auditorium). Therefore, redirecting the reflections with flat inclined reflecting surfaces or scattering with non-flat surfaces is usually preferred.

As usual, side walls and ceiling are essential for the optimization of \textbf{early sound energy}, particularly their position and orientation with respect to the stage and auditorium. The initial time delay gap should be below around 20\,ms for appropriate intimacy, which can be challenging in very large halls with high ceilings. The ceiling should optimally be 5-10\,m high. If it is much higher, above 15\,m, reflectors can be mounted on the side walls or hanged from the ceiling to shorten the initial time delay gap. Ceilings in rooms for music performances are even more critical than in rooms for speech, because some important musical instruments, such as violin and piano, radiate large fraction of sound energy in vertical direction, particularly at high frequencies. This energy can be captured and directed to the auditorium by the parts of the ceiling above and in front of the stage, or, alternatively, with hanging reflectors. Simple flat reflectors are usually not very large (with characteristic dimensions $\sim 1$\,m; compare with section~\ref{ch:rigid_motionless_plate}), in order to avoid specular reflections at middle frequencies.

Area of the \textbf{stage} should be around 1.8-2\,m$^2$ per musician, which is in practice around 200\,m$^2$ for large symphonic orchestra and around 4 times less in halls for chamber music. The stage floor should be massive enough and should not resonate noticeably. Apart from supplying the audience with enough early and diffuse sound energy, walls and ceiling, and other surfaces close to the sources should give appropriate feedback to the performers on the stage, as already mentioned in section~\ref{ch:subjective_criteria}. Ease of ensemble is particularly critical for large orchestras. The performers distributed over large area of the stage should hear each other well to keep the intonation and synchronisation. To some extent this can be achieved by decreasing distance between the musicians. Still, strong local early reflections are necessary for a successful common performance of the musicians on the opposite sides of the stage. If the side walls and ceiling of the room cannot provide them (which is often the case with large halls and outdoor stages), the reflections can be ensured with a stage enclosure. Alternatively, stage is sometimes built in a recess of the hall, surrounded by the reflecting local walls and ceiling. In such cases, these surfaces take over part of the role of the side walls and ceiling in the rest of the room. The key geometric properties are still their distances from the performers (which should not be too short either) and orientations with regard to the stage. Parallel surfaces are not recommended, since they can lead to a flutter echo.

As in other rooms for sound production, an important factor of room acoustics are the \textbf{seats} of the auditorium. They should be lightly to moderately upholstered with porous absorbing materials. The materials are placed on the parts of the seats which are mostly covered by the audience when the seats are occupied. In this way the damping and reverberation time in the room do not depend largely on the occupancy of the hall.

\bigskip

After discussing the optimal acoustic properties of rooms and general strategies how they can be realized in practice, we shall study in more details the most important acoustic elements which are used for various acoustic measures. Knowledge of the physical mechanisms of their work and the associated quantities are necessary for a proper prediction of their effects in rooms and estimation of sound fields. They are in the focus of the next two sections.
\section{Basic acoustic elements}\label{ch:basic_elements}

In this section, we study acoustic behaviour of large surfaces in a room, such as walls or ceiling, either as simple flat rigid surfaces or with more complicated micro-geometry or acoustic treatment. Although bounded in reality, their finite sizes (surface areas) are not critical for describing the basic mechanisms of their work, which are thus often treated assuming infinite surface areas. Surfaces for which finite dimensions have a more important role will be considered separately, in section~\ref{ch:small_elements}.

For all types of surfaces we consider some simple theoretical models which provide important insights into the physical mechanisms. The point of these simple models is obviously not achieving a very high accuracy of calculations of the acoustic quantities. For this purpose, more detailed analytical and empirical models can be found in literature or (when available) measurement data should be used in practice. The main purpose of the models considered here is to explain the acoustic behaviour of different surfaces and express it in terms of the theory developed in previous sections.

\subsection{Infinite uniform plane surface}\label{ch:infinite_plane_surface}

First we observe a \textbf{flat} surface normal to the $x_1$-axis of a Cartesian coordinate system, which extends to infinity in all directions around the axis. In reality, its dimensions should be \textbf{acoustically large}, much larger than the wavelength. The surface is assumed to be \textbf{uniform} and in the \textbf{far field} of any source of sound (or other reflecting surfaces, which could be observed as secondary sources). Therefore, an incident sound wave represented by the sound pressure $p_i$ can be modelled as a plane wave. According to eq.~(\ref{eq:v_solution_tailored_Green_wave_eq_free_space_compact_source_emission_time_far_field_amlitude}),
\begin{equation}\label{eq:incident_plane_wave}
\hat{p}_i(\boldsymbol x) = \hat{p}_Q e^{- j kr} = \rho_0 c_0 \hat{\boldsymbol v}_i(\boldsymbol x) \cdot \boldsymbol e_i,
\end{equation}
where $\boldsymbol e_i$ is the unit vector pointing in the direction of wave propagation towards the surface and the source- and distance-dependent factor $\hat{Q}(\boldsymbol y)/(4 \pi r)$ is replaced by the complex amplitude $\hat{p}_Q$, which also includes information on the initial phase of the wave. Since we consider the field only in the vicinity of the surface, $\hat{p}_Q$ does not decay significantly with the propagation path length ($r$) and thus we can treat it as constant in both space and time, as for a simple plane wave in free space. The entire spatial dependence of the incoming wave is contained in the phase term, $e^{-jkr}$.

Due to the axial symmetry of the geometry (the surface is infinite, flat, and uniform), we can treat the problem as two-dimensional, in the $x_1 x_2$-plane, without a loss of generality. The surface is placed at $x_1 = 0$ and the incident sound wave reaches it at the origin $(x_1,x_2) = (0,0)$ from the quadrant $x_1,x_2<0$. If $0 \le \theta \le \pi/2$ denotes the angle of incidence (which is equal to the angle between $\boldsymbol e_i$ and the unit vector $\boldsymbol e_1$ of the $x_1$-axis), then the two components of $\boldsymbol e_i$ are $(\cos(\theta), \sin(\theta))$. Since we work with the Cartesian coordinates $x_1$ and $x_2$, we replace $kr$ (with the radial coordinate $r$) in the exponent with the scalar product $\boldsymbol k_i \cdot \boldsymbol x$, where $\boldsymbol k_i = k \boldsymbol e_i = \omega \boldsymbol e_i / c_0$ is wave vector of the incident wave. Thus we can write
\begin{equation}\label{eq:incident_plane_wave_surface}
\hat{p}_i(\boldsymbol x) = \hat{p}_Q e^{- j \boldsymbol k_i \cdot \boldsymbol x} = \hat{p}_Q e^{- j (k_{i1} x_1 + k_{i2} x_2)} = \hat{p}_Q e^{- j [k x_1 \cos(\theta) + k x_2 \sin(\theta)]}.
\end{equation}

The surface is acoustically large, flat, and uniform, so we expect only specular reflections from it, at the same angle to the surface normal as the incident sound wave ($\theta < \pi/2$) that is, $\boldsymbol e_r = (-\cos(\theta), \sin(\theta))$ and $\boldsymbol k_r = k \boldsymbol e_r$ are the direction and wave vector of the reflected plane wave, respectively. Direction of the wave with respect to the $x_1$-axis is reversed after the reflection, while it remains unchanged with respect to the $x_2$-axis. The reflected wave can be written most generally as
\begin{equation}\label{eq:reflected_plane_wave_surface}
\begin{aligned}
\hat{p}_r(\boldsymbol x) &= \hat{p}_Q \hat{R}_s(\theta) e^{- j \boldsymbol k_r \cdot \boldsymbol x} = \hat{p}_Q \hat{R}_s(\theta) e^{- j (k_{r1} x_1 + k_{r2} x_2)} \\
&= \hat{p}_Q \hat{R}_s(\theta) e^{- j [-k x_1 \cos(\theta) + k x_2 \sin(\theta)]},
\end{aligned}
\end{equation}
where $\hat R_s(\theta) = \hat{p}_r(x_1=0)/\hat{p}_i(x_1=0)$ is by definition \textbf{reflection coefficient} of the surface (since the surface is uniform and axisymmetric, it is not a function of $x_2$ or the azimuthal angle, but it can depend on the angle of incidence). It is a complex parameter which quantifies both amplitude and phase change of the sound wave after the reflection from the surface\footnote{Similarly as with absorption coefficient, change of amplitude may be due to the transmission of sound through the surface, not only true energy losses.}. Complex sound pressure amplitude in front of the surface (for $x_1 < 0$) equals
\begin{equation}\label{eq:p_plane_surface}
\hat{p}(\boldsymbol x) = \hat{p}_i(\boldsymbol x) + \hat{p}_r(\boldsymbol x) = \hat{p}_Q e^{-j k x_2 \sin(\theta)} \left( e^{- j k x_1 \cos(\theta)} + \hat{R}_s(\theta) e^{ j k x_1 \cos(\theta)} \right).
\end{equation}

It should be noted that the same result is obtained if the surface is replaced with a copy of the incident plane wave scaled with $\hat R_s(\theta)$ and reaching the surface from ``behind'' (the side opposite to the incident sound, as in a mirror), at the angle $\pi-\theta$ to the vector $\boldsymbol e_1$. In fact, an acoustically large, flat, uniform, and reflecting wall can be represented by an \textbf{image source} located behind its surface, at the straight line perpendicular to it, which also includes the location of the original source, and at the same distance from the surface as the original source. The sound field in front of the wall is equal in the two cases (as long as no obstacles are present on the propagation path of the incident wave) and the strength of the image source depends on the reflection coefficient of the wall. This is the basis of the image source technique for estimation of sound fields in rooms, which is often used in combination with ray tracing and which will be discussed in section~\ref{ch:image_sources}. The image source approximation holds even if the incident wave is not plane, for example, when the wall is in the near field of the source of sound. The original source can also be a secondary source, for example, a lower-order image source of another reflecting surface of the room. If the source is directional, far-field radiation pattern of its image should be oriented such that it mirrors the radiation pattern of the original source.

In the special case of motionless and fully reflecting rigid wall (\textbf{hard wall} as in section~\ref{ch:rectangular_room_wave_theory}, reflecting both specularly and without a change of sound amplitude or phase), $\hat{R_s} = 1$ and the pressure amplitude from eq.~(\ref{eq:p_plane_surface}) becomes
\begin{equation}\label{eq:p_rigid_wall}
\hat{p}(\boldsymbol x) = \hat{p}_Q e^{-j k x_2 \sin(\theta)} \left( e^{- j k x_1 \cos(\theta)} + e^{ j k x_1 \cos(\theta)} \right) = 2 \hat{p}_Q \cos(k x_1 \cos(\theta)) e^{-j k x_2 \sin(\theta)}
\end{equation}
with the modulus
\begin{equation}\label{eq:p_amplitude_rigid_wall}
|\hat{p}(\boldsymbol x)| = 2 |\hat{p}_Q \cos(k x_1 \cos(\theta))|.
\end{equation}
For a given frequency, the cosine factor represents a \textbf{standing wave} in front of the wall (the ``wave'' does not propagate over $x_1$) and the modulus has the maximum $2 |\hat{p}_Q|$ at the fixed planes parallel to the surface of the hard wall given by $x_1 = -n\pi/(k \cos(\theta)) = -n \lambda/(2 \cos(\theta))$, with $n=0,1,2...$
From eq.~(\ref{eq:intensity_energy_complex_time_average_plane_wave}), time-averaged energy equals
\begin{equation}\label{eq:energy_complex_time_average_plane_wave_rigid_wall}
\langle E \rangle_T = \frac{|\hat{p}|^2}{2 \rho_0 c_0^2} = \frac{2 |\hat{p}_Q|^2}{\rho_0 c_0^2} \cos^2(k x_1 \cos(\theta)) = \frac{2 |\hat{p}_Q|^2}{\rho_0 c_0^2} \frac{1+\cos(2 k x_1 \cos(\theta))}{2}.
\end{equation}

If the sound field in front of the wall is \textbf{diffuse}, $|\hat{p}_{diff}|^2 = 4\pi |\hat{p}_Q|^2$ (the factor $4\pi$ is due to integration over the full solid angle as in the ray approximation in section~\ref{ch:statistical_theory} and eq.~(\ref{eq:intensity_energy_diff})). Therefore, total sound energy in the half space in front of the wall is
\begin{equation}\label{eq:energy_complex_time_average_plane_wave_rigid_wall_diffuse}
\begin{aligned}
\langle E_{diff} \rangle_T &= \frac{|\hat{p}_{diff}|^2}{2 \pi \rho_0 c_0^2} \int_{0}^{2\pi} \frac{1+\cos(2 k x_1 \cos(\theta))}{2} d\Omega &\\
&= \frac{|\hat{p}_{diff}|^2}{2 \pi \rho_0 c_0^2} \frac{1}{2} \int_{0}^{\pi/2} \int_{0}^{2 \pi} \left[ 1 + \cos(2 k x_1 \cos(\theta)) \right] \sin(\theta) d\phi d\theta &\\
&= \frac{|\hat{p}_{diff}|^2}{2 \rho_0 c_0^2} \left( \int_{0}^{\pi/2} \sin(\theta) d\theta + \int_{0}^{\pi/2} \cos(2 k x_1 \cos(\theta)) \sin(\theta) d\theta \right) &\\
&= \frac{|\hat{p}_{diff}|^2}{2 \rho_0 c_0^2} \left( 1 - \frac{\sin(2 k x_1 \cos(\pi/2)) - \sin(2 k x_1 \cos(0))}{2 k x_1} \right) &\\
&= \frac{|\hat{p}_{diff}|^2}{2 \rho_0 c_0^2} \left( 1 + \frac{\sin(2 k x_1)}{2 k x_1} \right).&
\end{aligned}
\end{equation}
The second term expresses the difference compared to the diffuse-field value in the absence of the wall (eq.~(\ref{eq:energy_complex_time_average_plane_wave_diffuse})). It involves the sinc function (compare with Fig.~\ref{fig:energy_sum_additional_terms} (right)), which equals 1 for the zero argument. Hence, at the surface of the wall ($x_1=0$) $\langle E_{diff} \rangle_T = |\hat{p}_{diff}|^2 / (\rho_0 c_0^2)$, which is double the value far from the wall (for $kx_1 \gg 1$, with $x_1$ as the characteristic length in the Helmholtz number), when the sinc function vanishes. As the distance from the wall increases, the energy oscillates according to the sinc function and converges to the value from eq.~(\ref{eq:energy_complex_time_average_plane_wave_diffuse}), $\langle E_{diff} \rangle_T = |\hat{p}_{diff}|^2 / (2\rho_0 c_0^2)$. This explains the increase of sound pressure level close to the reflecting boundaries of a room, even if the field in bulk of the room is fairly diffuse and uniform. Since it follows from the superposition of waves (summation of the incident and reflected sound amplitudes), this effect is not captured by the energy-based statistical theory.

According to eq.~(\ref{eq:impedance}), boundary condition at a locally reacting surface (not necessarily a hard wall) can be expressed in terms of \textbf{impedance}. Moreover, we can specify the impedance for an imaginary surface, as well, such as any plane control surface parallel to the actual surface considered above. It equals
\begin{equation}\label{eq:impedance_in_front_of_plane_surface}
Z(\boldsymbol x) = \frac{\hat{p}(\boldsymbol x)}{\hat{\boldsymbol v}(\boldsymbol x) \cdot \boldsymbol e_1} = \frac{\hat{p}(\boldsymbol x)}{\hat{v}_1(\boldsymbol x)},
\end{equation}
where $x_1 \neq 0$ in general. This is very useful if we are primarily interested in the sound field in front of the imaginary surface. Sufficient information on the relevant acoustic phenomena behind the surface is then contained in the impedance of the imaginary surface, which behaves as an apparent boundary. From the conservation of momentum in eq.~(\ref{eq:Euler_momentum_sine_wave}) multiplied with $\boldsymbol e_1$ and divided with $e^{j \omega t}$ and eq.~(\ref{eq:p_plane_surface}), we obtain
\begin{equation}\label{eq:Euler_momentum_sine_wave_porous}
\begin{aligned}
j \omega \rho_0 & \hat{\boldsymbol v} \cdot \boldsymbol e_1 = j \omega \rho_0 \hat{v}_1 = - (\nabla_x \hat{p}) \cdot \boldsymbol e_1 = - \frac{\partial \hat{p}}{\partial x_1} &\\
&= -\hat{p}_Q e^{-j k x_2 \sin(\theta)} \left(- j k \cos(\theta) e^{- j k x_1 \cos(\theta)} + j k \cos(\theta) \hat{R}_s(\theta) e^{ j k x_1 \cos(\theta)} \right) &\\
&= j k \cos(\theta) \hat{p}_Q e^{-j k x_2 \sin(\theta)} \left(e^{- j k x_1 \cos(\theta)} - \hat{R}_s(\theta) e^{j k x_1 \cos(\theta)} \right).&
\end{aligned}
\end{equation}
Therefore, the impedance for arbitrary $\boldsymbol x$ with $x_1 \leq 0$ is
\begin{equation}\label{eq:impedance_in_front_of_plane_surface_R}
\begin{split}
Z(\boldsymbol x) = \frac{\hat{p}(\boldsymbol x)}{\hat{v}_1(\boldsymbol x)}
= \frac{Z_0 \left( e^{- j k x_1 \cos(\theta)} + \hat{R}_s(\theta) e^{ j k x_1 \cos(\theta)} \right)}{\cos(\theta) \left(e^{- j k x_1 \cos(\theta)} - \hat{R}_s(\theta) e^{j k x_1 \cos(\theta)} \right)},
\end{split}
\end{equation}
where $Z_0$ is characteristic impedance of the fluid from eq.~(\ref{eq:impedance_plane_wave}). Notice that the reflection coefficient $\hat R_s$ is specified at the actual surface, while sound propagation between the imaginary and actual surface is captured by the exponential terms. The impedance at the actual uniform locally reacting surface ($x_1 = 0$) equals
\begin{equation}\label{eq:impedance_plane_surface_R}
\begin{split}
Z_s(\theta) = \frac{\hat{p}(x_1 = 0)}{\hat{v}_1(x_1 = 0)} = \frac{Z_0 \left( 1 + \hat{R}_s(\theta) \right)}{\cos(\theta) \left(1 - \hat{R}_s(\theta) \right)}.
\end{split}
\end{equation}
This is a simple relation between complex impedance and reflection coefficient of the surface.

For a hard wall with $\hat{R}_s = 1$,
\begin{equation}\label{eq:impedance_in_front_of_plane_reflecting_surface_R}
\boxed{ Z(\boldsymbol x) = \frac{Z_0 \left( e^{- j k x_1 \cos(\theta)} + e^{ j k x_1 \cos(\theta)} \right)}{\cos(\theta) \left(e^{- j k x_1 \cos(\theta)} - e^{j k x_1 \cos(\theta)} \right)} 
= \frac{j Z_0 \cot(k x_1 \cos(\theta))}{\cos(\theta)} },
\end{equation}
while the surface impedance $Z_s$ is infinite for any $\theta < \pi/2$ (velocity component normal to the wall is zero for any pressure). In contrast to it, acoustically \textbf{soft wall} has $\hat{R}_s = -1$ and $Z_s = 0$ for any $\theta < \pi/2$ and $j \cot(k x_1 \cos(\theta))$ in eq.~(\ref{eq:impedance_in_front_of_plane_reflecting_surface_R}) becomes $-j \tan(k x_1 \cos(\theta))$. 
From eq.~(\ref{eq:impedance_in_front_of_plane_reflecting_surface_R}), we also see that for $\theta < \pi/2$ the impedance is zero at the planes $x_1 = -(2n+1) \pi/(2 k \cos(\theta)) = -(2n+1) \lambda/(4 \cos(\theta))$ in front of the hard wall. This implies that the particle velocity component $v_1$ at these planes parallel to the wall is very high even for moderate values of sound pressure, unlike at the surface of the wall or in general the planes $x_1 = -n \pi/(k \cos(\theta)) = -n \lambda/(2 \cos(\theta))$, where it vanishes.

For the important special case of \textbf{normal incidence} ($\theta = 0$) and acoustically \textbf{small distance} from the wall ($|kx_1| = -kx_1 \ll 1 \Rightarrow \cot(kx_1) \approx 1/(kx_1)$, after expanding the cotangent into series around $kx_1 = 0$),
\begin{equation}\label{eq:impedance_in_front_of_plane_reflecting_surface_R_lumped}
\boxed{ Z(\boldsymbol x) = \frac{j \rho_0 c_0 }{k x_1} = \frac{j \rho_0 c_0^2 }{\omega x_1} }.
\end{equation}
Comparing this with the mechanical impedance in eq.~(\ref{eq:oscillator_impedance}), we see that acoustically thin layer of fluid (with the thickness $d = -x_1$ and small value of the associated Helmholtz number $kd \ll 1$) in front of an acoustically large and flat hard wall behaves analogously to a spring with stiffness $S = \rho_0 c_0^2/d$, for normal incidence of the incoming plane sound wave. It should be underlined that the mechanical analogy with the lumped elements (in this case spring) holds only for normal incidence (one-dimensional geometry, since the dependence on $x_2$ disappears) and acoustically small length scales of the elements (thickness $d$ in this case). In the electro-acoustic analogy the thin layer of fluid acts as a lumped capacitor with the capacitance $C = d/(\rho_0 c_0^2)$.

For energy-based calculations, \textbf{absorption coefficient} of the surface, which was introduced in eq.~(\ref{eq:energy_loss_ray}), can be related to the reflection coefficient and impedance from above. The first relation is
\begin{equation}\label{eq:absorption_coeff_plane_surface}
\alpha_s(\theta) = \frac{d\langle P_{s,ray,loss}(\theta,t) \rangle_T}{d\langle P_{s,ray}(\theta,t) \rangle_T} = \frac{|\hat{p}_i(\theta,x_1 = 0)|^2 - |\hat{p}_r(\theta,x_1 = 0)|^2}{|\hat{p}_i(\theta,x_1 = 0)|^2} = 1 - |\hat{R}_s(\theta)|^2.
\end{equation}
As before, the absorption coefficient treats the sound energy transmitted behind the surface as absorbed. It equals 0 for fully reflecting surfaces, such as acoustically hard and soft walls, and 1 for fully absorbing surfaces. The second relation follows from eq.~(\ref{eq:impedance_plane_surface_R}), which can be written as
\begin{equation}\label{eq:R_plane_surface_impedance}
\begin{split}
\hat{R}_s(\theta) = \frac{ Z_s(\theta) \cos(\theta) - Z_0}{ Z_s(\theta) \cos(\theta) + Z_0}.
\end{split}
\end{equation}
Since $Z_0$ is real,
\begin{align*}
\begin{split}
|\hat{R}_s(\theta)|^2 = \frac{ |Z_s(\theta) \cos(\theta) - Z_0|^2}{ |Z_s(\theta) \cos(\theta) + Z_0|^2} = \frac{(\mathcal{R}_e(Z_s(\theta)) \cos(\theta) - Z_0)^2 + (\mathcal{I}_m(Z_s(\theta)) \cos(\theta))^2}{(\mathcal{R}_e(Z_s(\theta)) \cos(\theta) + Z_0)^2 + (\mathcal{I}_m(Z_s(\theta)) \cos(\theta))^2}
\end{split}
\end{align*}
and we can express absorption coefficient in terms of impedance with
\begin{equation}\label{eq:absorption_coeff_plane_surface_Z}
\begin{aligned}
\alpha_s(\theta) &= 1 - \frac{(\mathcal{R}_e(Z_s(\theta)) \cos(\theta) - Z_0)^2 + (\mathcal{I}_m(Z_s(\theta)) \cos(\theta))^2}{(\mathcal{R}_e(Z_s(\theta)) \cos(\theta) + Z_0)^2 + (\mathcal{I}_m(Z_s(\theta)) \cos(\theta))^2} &\\
&= \frac{4 \mathcal{R}_e(Z_s(\theta)) \cos(\theta) Z_0}{(\mathcal{R}_e(Z_s(\theta)) \cos(\theta) + Z_0)^2 + (\mathcal{I}_m(Z_s(\theta)) \cos(\theta))^2} &\\
&= \frac{4 \mathcal{R}_e(Z_s(\theta)) \cos(\theta) Z_0}{\mathcal{R}_e(Z_s(\theta))^2 \cos^2(\theta) + 2 \mathcal{R}_e(Z_s(\theta)) \cos(\theta) Z_0 + Z_0^2 + \mathcal{I}_m(Z_s(\theta))^2 \cos(\theta)^2} &\\
&= \frac{4 \mathcal{R}_e(Z_s(\theta)) \cos(\theta) Z_0}{(|Z_s(\theta)| \cos(\theta))^2 + 2 \mathcal{R}_e(Z_s(\theta)) \cos(\theta) Z_0 + Z_0^2}.&
\end{aligned}
\end{equation}

\begin{figure}[h]
	\centering
	\begin{subfigure}{.5\textwidth}
		\centering
		\includegraphics[width=1\linewidth]{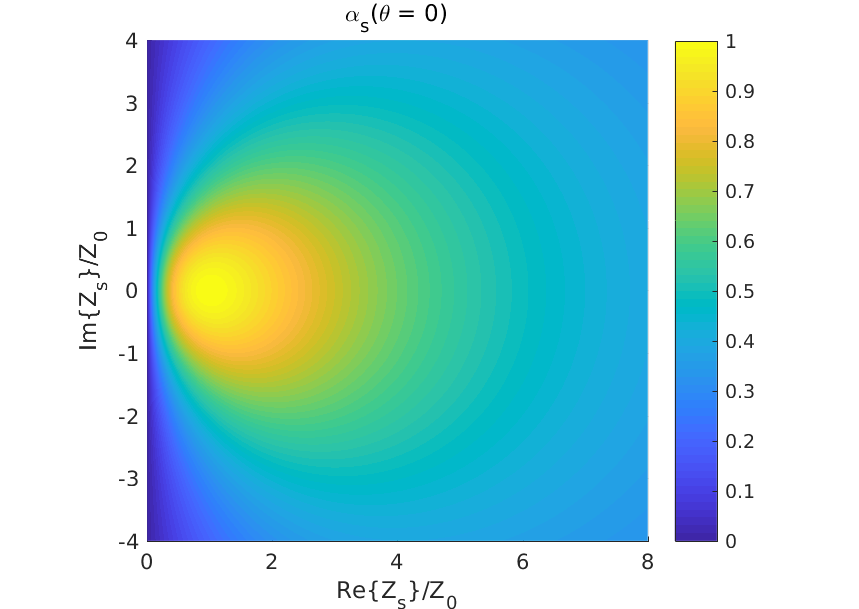}
		\label{fig:alpha_Z_s_norm_normal_incidence}
	\end{subfigure}%
	\begin{subfigure}{.5\textwidth}
		\centering
		\includegraphics[width=1\linewidth]{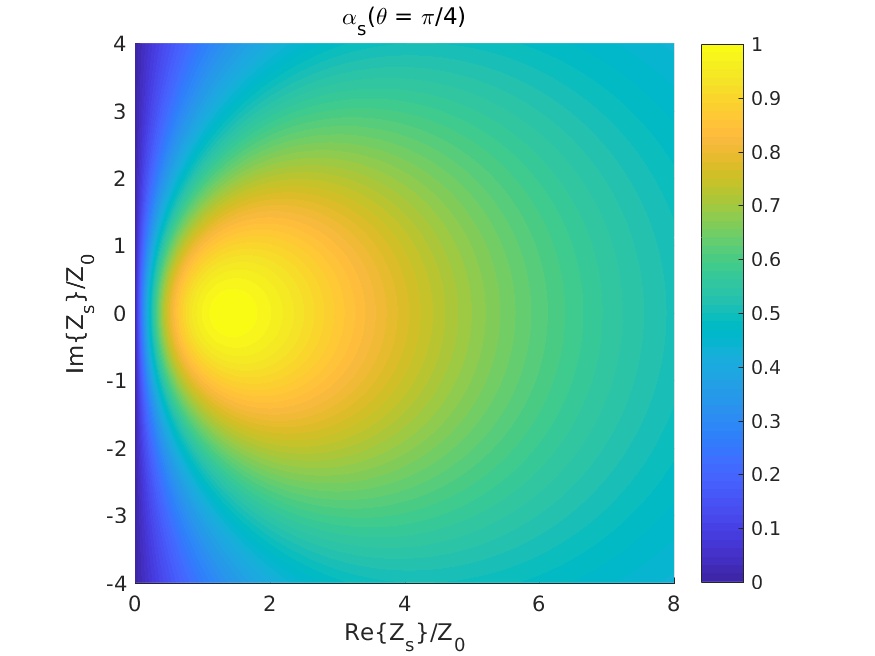}
		\label{fig:alpha_Z_s_norm_theta_pi_4}
	\end{subfigure}
	\caption{Absorption coefficient according to eq.~(\ref{eq:absorption_coeff_plane_surface_Z_Zconst}) as a function of real and imaginary parts of specific impedance for (left) normal incidence and (right) angle of incidence $\theta = \pi/4$.}
	\label{fig:alpha_Z_s_norm}
\end{figure}

In many cases of practical importance, impedance does not essentially depend on the angle of incidence. Equations~(\ref{eq:R_plane_surface_impedance}) and (\ref{eq:absorption_coeff_plane_surface_Z}) read then
\begin{equation}\label{eq:R_plane_surface_impedance_Zconst}
\begin{split}
\hat{R}_s(\theta) = \frac{ Z_s \cos(\theta) - Z_0}{ Z_s \cos(\theta) + Z_0}
\end{split}
\end{equation}
and
\begin{equation}\label{eq:absorption_coeff_plane_surface_Z_Zconst}
\begin{split}
\boxed{ \alpha_s(\theta) = \frac{4 \mathcal{R}_e(Z_s) \cos(\theta) Z_0}{(|Z_s| \cos(\theta))^2 + 2 \mathcal{R}_e(Z_s) \cos(\theta) Z_0 + Z_0^2} },
\end{split}
\end{equation}
respectively. Note that the reflection coefficient and absorption coefficient are still functions of the angle of incidence. Figure \ref{fig:alpha_Z_s_norm} shows value of the absorption coefficient as a function of real and imaginary part of the normalized (specific) impedance $Z_s/Z_0$ for two angles of incidence, $\theta = 0$ and $\theta = \pi/4$. Only passive surfaces with $\mathcal{R}_e(Z_s) \geq 0$ (recall eq.~(\ref{eq:acoustic_power_complex_time_average1})) are considered. According to the figure, full absorption of sound energy with $\alpha_s = 1$ is achievable for $Z_s = Z_0$ and normal incidence ($\theta = 0$). The surface impedance matches the characteristic impedance of the fluid in front of it. However, for a different angle of incidence, surface impedance which provides maximum absorption is different from $Z_0$ (around $1.4 Z_0$ for $\theta = \pi/4$). Consequently, the impedance boundary condition $Z_s = \rho_0 c_0$ does not represent a fully absorbing surface (with $\alpha_s = 1$) for arbitrary angles of incidence. This is important, for example, in acoustic simulations of free space radiation. Simple condition $Z = Z_0$ at the boundaries of the computational domain can cause unphysical reflections, if the angle of incidence is not $\theta = 0$. In fact, no single value of the impedance can be used for arbitrary angles of incidence and other techniques have to be used, such as perfectly matched layers (PML) and infinite elements. Surfaces with $\mathcal{R}_e(Z_s) = 0$ are fully reflecting, as evident from eq.~(\ref{eq:absorption_coeff_plane_surface_Z_Zconst}).

At \textbf{grazing incidence} ($\theta = \pi/2$) no reflection or absorption take place and $\hat p_r(\boldsymbol x) = 0$. According to the model considered here, the surface does not affect the incident wave, regardless of its acoustic properties. In reality, however, if the surface is absorbing, sound waves refract into it (which is not taken into account by the model) and certain fraction of their energy is absorbed. An example of this is already mentioned additional attenuation due to sound propagation over auditorium. In such situations, amplitude of the direct sound of a point source or certain reflections can decay faster than 6\,dB per doubling the distance. In the opposite case, when the angle of incidence approaches zero, the surface is expected to react more locally and angularly independent, which gives more justification for  equations~(\ref{eq:R_plane_surface_impedance_Zconst}) and (\ref{eq:absorption_coeff_plane_surface_Z_Zconst}), even for surfaces which might not be strictly locally reacting, such as plates and membranes.

\subsection{Diffuse surface}

If a surface is not flat or uniform (for example, it has irregular micro-geometry or spatially dependant acoustic properties), \textbf{scattering} of sound in non-specular directions is likely to occur, especially at wavelengths which are comparable to the characteristic length scale of the non-uniformities\footnote{If the characteristic length scale of geometric deviations from a flat surface (that is of roughness) is much larger than the wavelength, the surface may behave much like multiple inclined specularly reflecting surfaces, which can be treated as above with adjusted coordinated systems. Similarly, if its acoustic properties vary only slightly over acoustically large areas, the total surface can be divided into several uniform surfaces. On the other hand, if the acoustic properties or micro-geometry of the surface vary with the length scales much shorter than the wavelength, the former can be averaged and the latter can be neglected, so the surface can be treated again as uniform and flat.}. Fraction of energy which is reflected non-specularly is given by the scattering coefficient $s_{s}(\theta_{i})$ introduced in eq.~(\ref{eq:SPL_dir-refl}), which in general depends on the angle of incidence $\theta_i$. Energy of the specularly reflected sound, $E_{spec}$, can be estimated accordingly by multiplying the total reflected energy, $E_r$, with the factor $1-s_{s}(\theta_{i})$,
\begin{equation}
E_{spec} = E_r - E_{scatt} = E_r (1-s_{s}(\theta_{i})),
\end{equation}
where $E_{scatt} = s_{s}(\theta_{i}) E_r$ is total energy scattered in all non-specular directions.

\begin{figure}[h]
	\centering
	\includegraphics[width=0.6\textwidth]{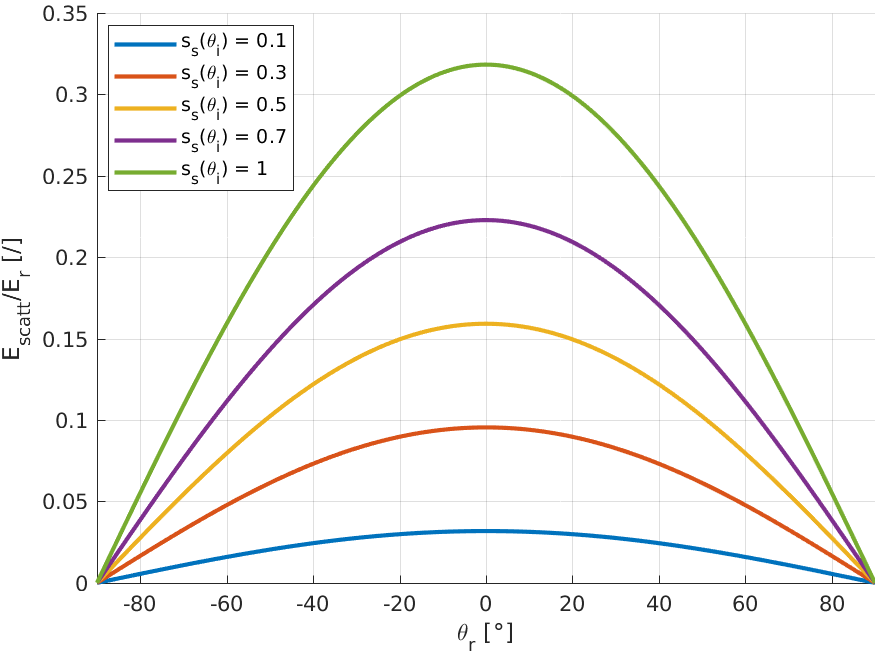}
	\caption{Ratio of scattered and specularly reflected energy according to Lambert's cosine law as a function of the reflection angle and for several values of the scattering coefficient.}
	\label{fig:Lambert_cosine_law}
\end{figure}

A simple model which is frequently used for angular dependence of the scattered energy is \textbf{Lambert's cosine law}:
\begin{equation}\label{eq:Lambert}
\boxed{ E_{scatt}(\theta_r) = s_{s}(\theta_{i}) E_r \frac{\cos(\theta_r)}{\pi} },
\end{equation}
where $\theta_r$ is angle of the scattered sound waves to the normal of the surface. The factor of $1/\pi$ is necessary for the total scattered energy to satisfy
\begin{equation}\label{eq:Lambert_integral}
\begin{aligned}
\int_{0}^{2\pi} E_{scatt}(\theta_r) d\Omega_r &= \int_{0}^{\pi/2} \int_{0}^{2 \pi} s_{s}(\theta_{i}) E_r \frac{\cos(\theta_r)}{\pi} \sin(\theta_r) d\phi_r d\theta_r &\\
&= 2 s_{s}(\theta_{i}) E_r \int_{0}^{\pi/2} \cos(\theta_r) \sin(\theta_r) d\theta_r = s_{s}(\theta_{i}) E_r.&
\end{aligned}
\end{equation}
Value of the ratio $E_{scatt}(\theta_r)/E_r$ as a function of the angle $\theta_r$ and scattering coefficient $s_s(\theta_i)$ is shown\footnote{\label{ftn:spherical_to_polar_angle}Merely for the graphical representation we allow negative values of $\theta_r$ and let $-\pi/2 \leq \theta_r \leq \pi/2$. In fact, $E_{scatt}(\theta_r)/E_r$ is a symmetric function of $\theta_r$ and its values for the negative angles are equal to the values for the corresponding positive angles. The graph can be rotated around the axis $\theta_r = 0$ in order to obtain a three-dimensional scattering pattern in front of the surface, since $E_{scatt}$ does not depend on $\phi_r$ according to the model.} in Fig.~\ref{fig:Lambert_cosine_law}. The scattered energy increases with $s_s$ (less energy is reflected specularly) and it has a maximum at $\theta_r = 0$ and minimum at $\theta_r = \pi/2$ (for any given angle of incidence $\theta_i$), which is reasonable for many real diffuse surfaces. Moreover, the scattered energy includes certain amount of energy reflected in the specular direction, for $\theta_r = \theta_i$.

Of course, Lambert's cosine law is only approximate. Like scattering coefficients, scattering patterns can vary significantly between surfaces. If higher accuracy is necessary (for example, if reflections from a particular surface are of special interest), these as well as first-order acoustic quantities can be measured in laboratory or estimated analytically (see sections~\ref{ch:rigid_motionless_plate} and \ref{ch:Schroeder_diffuser}) or numerically (for example, using boundary element method). Still, simple models like Lambert's cosine law are often satisfactory for room acoustic calculations, such as ray tracing simulations, especially when many diffuse reflections take place at many surfaces in the room and at many different angles. Exact scattering coefficients and patterns of surfaces become thus less relevant for numerous late reflections. Unfortunately, the same does not hold for low-order reflections, which largely depend on few surfaces and angles. This presents a major difficulty and a source of inaccuracies of ray tracing simulations.

\subsection{A layer of porous absorber}\label{ch:layer_poros_abosrber}

In section~\ref{ch:infinite_plane_surface} we calculated the field in front of an infinitely large flat surface characterized with impedance, reflection coefficient, or absorption coefficient. Values of these parameters can be determined simply for several idealized elements, such as acoustically hard and soft walls. On the other hand, very important porous materials have a complicated structure which makes the exact analytical (and numerical) treatment difficult, even though they can normally be treated as homogeneous at the scale of sound wavelengths and surfaces of large flat layers of such materials as uniform. If measured values of the parameters are not available, relatively simple empirical or analytical models are often used in practice. One analytical approach is based on eq.~(\ref{eq:complex_wave_number}). Exponential decay of sound pressure amplitude during wave propagation inside the material is modelled by adding the imaginary part $- j \zeta / c$ to the real wave number $k$ of a forward propagating wave (assuming that no reflection occurs in the porous material), where $\zeta$ is real and non-negative damping constant and $c$ is speed of sound in the porous material, which is in general different from the speed of sound in the fluid, $c_0$.

The required quantities $\zeta$ and $c$ should be estimated from the physical properties of the material, such as its geometry, inner structure, density, etc. A very simple but crude is the \textbf{Rayleigh model}. It treats a porous material as consisting of many straight narrow channels filled with the fluid (air), parallel to each other and separated by rigid solid walls (solid phase of the material). The model considers only viscous and not thermal losses and the only required parameter of the material is its specific flow resistance (flow resistance per unit length, also called \textbf{flow resistivity}), $\Xi'$. It represents the ratio of pressure drop and velocity of the fluid flow through the narrow channels\footnote{The value of flow resistivity is practically equal for mean (time-independent) and unsteady flow (such as due to sound propagation), as long as the fluid dynamics inside the channels is governed by the viscous effects. This means that width of the channels formed by the solid skeleton of the material should not be larger than the viscous boundary layer thickness from eq.~(\ref{eq:viscous_thermal_boundary_layer_thickness}).},
\begin{equation}\label{flow_resistivity}
\Xi' = -\frac{\partial p/ \partial x_1}{v_1},
\end{equation}
where we suppose the flow in the direction of the $x_1$-axis (specified by a single component $v_1$) and all channels are parallel to the same axis. Due to the minus sign in the definition, pressure drop with increasing $x_1$ gives  positive value of the flow resistivity. The unit is kg/(m$^3$s) = \mbox{Pa s/m$^2$} = Rayl/m.

Flow resistivity is thus defined based on the flow in the fluid phase of the material, regardless of the solid phase and its properties, such as thickness of walls or volume. However, it is normally measured on a sample of a material with certain porosity $\sigma$, which is ratio of the volume of channels (air inside the material) and the total volume of the material. Flow resistivity is accordingly
\begin{equation}\label{flow_resistivity_porosity}
\Xi' = \Xi \sigma,
\end{equation}
where $\Xi$ denotes the measured flow resistivity, including the solid phase. Table~\ref{tab:flow_resistivity_porous} indicates typical ranges of values of density and flow resistivity of various porous materials which are often used as sound absorbers. Common porous materials, such as mineral wool, have high porosity $\sigma \approx 1$ (therefore, $\Xi' \approx \Xi$) and flow resistivity which increases with density, in the range 5000-100000\,kg/(m$^3$s). Actual values can depend on the direction of the flow in the material (angle of incidence in the case of an incoming sound wave). However, we assume that the inner structure of the material is stochastic enough so that it is essentially \textbf{isotropic}.

\begin{table}[h]
	\caption{Density and flow resistivity of some porous materials.}
	\label{tab:flow_resistivity_porous}
	\begin{tabular}{ | l | c | l | }
		\hline
		\textbf{material} & \textbf{$\rho$ [kg/m$^3$]} & \textbf{$\Xi$ [$\cdot$10$^3$\,kg/(m$^3$s)]} \\
		\hline
		mineral/glass wool & 15-50 & 5-15 \\
		\hline
		mineral/glass wool & 50-100 & 15-40 \\
		\hline
		mineral/glass wool & 100-200 & 40-80 \\
		\hline
		wood wool & 350-500 & 0.5-2 \\
		\hline
		cotton & 25-100 & 10-150 \\
		\hline
	\end{tabular}
\end{table}

If flow resistivity is measured on a layer of the material with thickness $d$ as $\Xi = \Delta p/(v_1 d)$, where the pressure drop $\Delta p$ divided with $d$ approximates the negative gradient from eq.~(\ref{flow_resistivity}), \textbf{flow resistance} in kg/(m$^2$s) can be calculated as $\Xi' d = \sigma \Xi d = \sigma \Delta p/v_1$. For normal sound incidence at the surface of the layer and acoustically thin layer ($kd \ll 1$), flow resistance quantifies pressure drop of the sound wave due to propagation through the absorbing material. The thin layer acts like a lumped resistor with the resistance  $R = \Delta p/ v_1 = \Xi' d/\sigma$. Notice the electro-acoustic analogy between electric potential and current and sound pressure and particle velocity, respectively. Similarly as in eq.~(\ref{eq:impedance_in_front_of_plane_reflecting_surface_R_lumped}), the analogy holds only for normal incidence and acoustically thin layers.

Next we proceed with calculation of the sound field, for which we refer to the conservation laws. As discussed in section~\ref{ch:basic_equations}, linearized momentum equation~\ref{eq:Euler_momentum} holds for an inviscid fluid and has to be modified in order to include absorption due to viscosity in porous materials. By equalizing it with the pressure gradient from eq.~(\ref{flow_resistivity}) on the right-hand side, we can obtain a one-dimensional form (assuming an isotropic material) of the conservation of momentum inside the narrow channels of the material:
\begin{equation}\label{eq:Euler_momentum_porous}
\rho_0 \frac{\partial v_1}{\partial t} + \frac{\partial p}{\partial x_1} = -\Xi' v_1.
\end{equation}
The equation is still linear and in frequency domain reads
\begin{equation}\label{eq:Euler_momentum_porous_freq}
j \omega \rho_0 \hat{v}_1 + \frac{\partial \hat{p}}{\partial x_1} + \Xi' \hat{v}_1 = (j \omega \rho_0 + \Xi') \hat{v}_1 + \frac{\partial \hat{p}}{\partial x_1} = 0.
\end{equation}
In this way we include the attenuation inside the viscous boundary layer at the supposedly rigid and motionless solid inner surfaces the porous material, while neglecting energy losses due to heat exchange between the fluid and the structure.

We can now observe a forward propagating plane wave inside the porous material with the \textbf{complex wave number}
\begin{equation}\label{eq:wave_number_porous}
k_{porous} = \frac{\omega}{c} - j \frac{\zeta}{c},
\end{equation}
which should be expressed in terms of $\Xi'$, angular frequency, and properties of the fluid, for which we suppose air with density $\rho_0$ and speed of sound $c_0$. Similarly as in eq.~(\ref{eq:incident_plane_wave_surface}), complex amplitude of the wave can be written as a function of the only spatial coordinate,
\begin{equation}\label{eq:incident_plane_wave1}
\hat{p}(x_1) = \hat{p}_Q e^{- j k_{porous} x_1} = \hat{p}_Q e^{-\zeta x_1/c} e^{- j \omega x_1 / c}.
\end{equation}
It decays exponentially with increasing $x_1$. After inserting
\begin{align*}
\frac{\partial \hat{p}}{\partial x_1} = -j k_{porous} \hat{p}.
\end{align*}
into eq.~(\ref{eq:Euler_momentum_porous_freq}), we obtain
\begin{equation}\label{eq:Euler_momentum_porous_freq2}
(j \omega \rho_0 + \Xi') \hat{v}_1 -j k_{porous} \hat{p} = 0.
\end{equation}

The second relation between $\hat{p}$ and $\hat{v}_1$ follows from the conservation of mass, eq.~(\ref{eq:Euler_mass}), and the equation of state~(\ref{eq:p_rho}), which have the same form as before. In one-dimensional form and frequency domain they give
\begin{equation}\label{eq:Euler_mass_freq_1D}
\frac{j \omega}{c_0^2} \hat{p} + \rho_0 \frac{\partial \hat{v}_1}{\partial x_1} = \frac{j \omega}{c_0^2} \hat{p} - j k_{porous} \rho_0 \hat{v}_1 = 0.
\end{equation}
We can multiply this equation with $k_{porous}c_0^2/\omega$ and use the result in eq.~(\ref{eq:Euler_momentum_porous_freq2}) to obtain
\begin{equation}\label{eq:velocity_porous}
(j \omega \rho_0 + \Xi') \hat{v}_1 - \frac{j k_{porous}^2 \rho_0 c_0^2}{\omega} \hat{v_1} = 0
\end{equation}
and therefore
\begin{equation}\label{eq:k_squared_porous_flow_resistivity}
k_{porous}^2 = \frac{\omega^2}{c_0^2} \left(1 -j \frac{\Xi'}{\rho_0 \omega} \right).
\end{equation}
Since $\mathcal{R}_e(k_{porous}) = k = \omega/c_0>0$, the only valid solution is
\begin{equation}\label{eq:k_porous_flow_resistivity}
\boxed{ k_{porous} = \frac{\omega}{c_0} \sqrt{1 -j \frac{\Xi'}{\rho_0 \omega} } }.
\end{equation}
This is the required relation between the complex wave number and flow resistivity as a physical property of the material, which can be used for calculations of damping inside the (unbounded) porous material. In general, damping associated with $\mathcal{I}_m(k_{porous})$ increases with the flow resistivity, while for vanishing resistivity $k_{porous}$ takes real value of the wave number in air and no viscous damping occurs.

It should be emphasized that the Rayleigh model is very crude. Real porous materials do not consist of straight narrow channels, but rather a network of small interconnected cavities with irregular shapes and various sizes or mostly fluid with only coarsely intertwined fibres of the solid structure. The materials can be non-isotropic. The solid parts are not perfectly rigid and motionless and they do conduct heat, which adds thermal to the viscous losses. In order to improve accuracy, more sophisticated models (involving more parameters) have been suggested in literature (by Delany and Bazley, Miki...) and many of them are empirical, based on the average results of a large number of measurements of various absorbing materials. In practice, absorption coefficients and other acoustic parameters of porous absorbers (usually placed in front of hard surfaces, as in the next subsection) are often provided by the manufacturers, measured in a reverberation chamber (in a diffuse field) or Kundt's tube (at normal incidence), or estimated from the available data in literature for similar types of materials and elements. In addition to this, true properties of a porous absorber in a room depend on the specific arrangement, distribution of the material in the room (see also the comments at the end of section~\ref{ch:ray-tracing_rectangular_room}), and mounting, especially when its shape deviates from a simple flat layer with constant thickness. One example are absorbing wedges in anechoic chambers. Their geometry provides gradual change of the surface impedance, by gradually increasing the absorbing area of the porous material. This increases efficiency of the absorber by avoiding reflections due to an abrupt change of impedance, at least for wavelengths which are not much larger than the depth of the wedges.

\subsection{Porous absorber in front of a hard wall}\label{ch:porous_absorber_rigid_wall}

Porous materials alone are not very efficient sound absorbers. Thickness of a single layer of porous material should be at least comparable to the sound wavelength for appreciable absorption, which often leads to a large amount of the material required for a desired damping, especially at low frequencies. In order to increase their efficiency, porous materials are usually placed close to highly reflecting hard surfaces. A reason for this can be found in eq.~(\ref{flow_resistivity}), which indicates that for a fixed flow resistivity of the material, pressure drop is proportional to the particle velocity. Reflections from hard surfaces can substantially increase particle velocity in the fluid phase of the material and thus viscous losses, as will be demonstrated next.

First we analyse absorption at the surface of a layer of porous material with thickness $d$, which was considered in the previous subsection, mounted directly on an acoustically hard wall from section~\ref{ch:infinite_plane_surface}. Using the same coordinate system as in section~\ref{ch:infinite_plane_surface}, the free surface of the porous layer is located at $x_1 = -d$ and the wall at $x_1 = 0$. From eq.~(\ref{eq:impedance_in_front_of_plane_reflecting_surface_R}), impedance of the surface equals
\begin{equation}\label{eq:impedance_porous_rigid_wall}
\begin{split}
Z_{porous,wall}(\boldsymbol x) = \frac{\hat{p}}{\sigma \hat{v}_1}= -\frac{j Z_0 \cot(k_{porous} d \cos(\theta))}{\sigma \cos(\theta)}.
\end{split}
\end{equation}
A division with \textbf{perforation ratio} $\sigma$ in necessary to take into account the material's rigid solid structure (recall the discussion with regard to eq.~(\ref{flow_resistivity_porosity})), which has a very high (theoretically infinite) impedance. The perforation ratio is ratio of the surface area of all small openings of the material at its free surface and the total free surface area of the material. Assuming that the material is isotropic and statistically uniform (homogeneous), this ratio is constant over the entire surface of the material. Moreover, according to the Rayleigh model, the material consists of straight parallel channels with constant cross sections, which are normal to the free surface and extend over the entire thickness of the layer. Therefore, porosity, which we defined in section~\ref{ch:layer_poros_abosrber}, and perforation ratio are equal, which is why we use the same symbol ($\sigma$). Including $\sigma$ in eq.~(\ref{eq:impedance_porous_rigid_wall}) follows also from the conservation of mass. Since $v_1=0$ at the solid parts of the material at $x_1 = -d$ and the mass is conserved, velocity component parallel to the $x_1$-axis immediately in front of the material, which is in the definition of impedance at $x_1 = -d$, is equal to $v_1$ at the openings of the material multiplied with the factor $\sigma$. The fluid flow outside the material is contracted into the openings, which increases the velocity (for the conserved mass) by the same factor as the surface area contraction. Commonly, however, $\sigma \approx 1$ and this correction is not critical.

After inserting $k_{porous}$ from eq.~(\ref{eq:k_porous_flow_resistivity}),
\begin{equation}\label{eq:impedance_porous_rigid_wall1}
\begin{split}
Z_{porous,wall}(\boldsymbol x) = -\frac{j Z_0 }{\sigma \cos(\theta)} \cot \left( \frac{\omega}{c_0} \sqrt{1 -j \frac{\Xi'}{\rho_0 \omega} } d \cos(\theta) \right)
\end{split}
\end{equation}
and for normal incidence ($\theta = 0$)
\begin{equation}\label{eq:impedance_porous_rigid_wall_normal_incidence}
\begin{split}
Z_{porous,wall}(\boldsymbol x) = -\frac{j Z_0 }{\sigma} \cot(k_{porous} d) = -\frac{j Z_0 }{\sigma} \cot \left( \frac{\omega}{c_0} \sqrt{1 -j \frac{\Xi'}{\rho_0 \omega} } d \right).
\end{split}
\end{equation}
If the Helmholz number $|k_{porous}|d \ll 1$ (acoustically thin layer), $\cot(k_{porous} d \cos(\theta)) \approx 1/(k_{porous} d \cos(\theta))$ and $Z_{porous,wall}(\boldsymbol x) \rightarrow \infty$. The impedance approaches that of a hard wall and absorption is very low. Consequently, thickness of the porous layer at the hard wall should not be much smaller than the wavelength, if significant absorption should be achieved.

Hence, thin layers of porous materials placed directly on a large acoustically hard surface are not very efficient absorbers, because normal component of the particle velocity is low close to the hard surface. Viscous losses can be much higher at the locations in the sound field where particle velocity is high. For example, eq.~(\ref{eq:impedance_in_front_of_plane_reflecting_surface_R}) showed that such locations are at the distances from a flat hard wall which are odd multiples of $\lambda/(4 \cos(\theta))$. Conversely, normal velocity component is zero at hard surfaces (as well as even multiples of $\lambda/(4 \cos(\theta))$). This means that absorption of a relatively thin layer of porous absorber (say, $d \approx \lambda /10$ or larger) can still be high if it is displaced from the wall, at distance $d_0$, leaving a \textbf{gap} between the layer and the wall with the thickness $d_0$. The gap does not have to be filled with the absorbing material, so higher absorption can be achieved with the same amount of the material or equal absorption can be achieved with less material. If very low angles of incidence dominate, the highest absorption is achieved at the distance around $\lambda/4 = c_0/(4 f)$ (its odd multiples are rarely considered in practice for saving space in the room and because higher angles of incidence become more likely further from the wall). However, the optimal distance depends both on the angles of incidence and frequency of the incoming sound. The angle-dependent absorption coefficient can be estimated by inserting the impedance from eq.~(\ref{eq:impedance_porous_rigid_wall1}) into eq.~(\ref{eq:absorption_coeff_plane_surface_Z}) and its diffuse field value afterwards from eq.~(\ref{eq:absorption_coeff_diffuse}). Reflection coefficient can also be found using eq.~(\ref{eq:impedance_porous_rigid_wall1}) in eq.~(\ref{eq:R_plane_surface_impedance}) and the sound field in front of the absorber, a superposition of the incident plane wave and its specular reflection from the surface, using eq.~(\ref{eq:p_plane_surface}).

For simplicity, we restrict ourselves to normal incidence and assume that the absorber is acoustically thin, that is $kd \ll 1$ and we can treat it as a lumped resistance $\Xi' d/\sigma$ (see section \ref{ch:layer_poros_abosrber}). The total impedance analogous to eq.~(\ref{eq:oscillator_impedance}) is then sum\footnote{The two analogue electric elements are connected in series, because their currents are equal (both amplitudes and phases of the particle velocity at the two sides of the thin porous layer are equal). Notice, however, that the second element is not lumped, because the gap is not necessarily much smaller than wavelength.} of this resistance and the impedance of the gap with thickness $d_0$ in front of the rigid wall and behind the porous layer, which was expressed in eq.~(\ref{eq:impedance_in_front_of_plane_reflecting_surface_R}). After inserting $\theta = 0$ and $d_0 = -x_1$, we obtain
\begin{equation}\label{eq:porous_gap_wall_impedance}
Z_{porous,wall} = \Xi' d/\sigma - j Z_0 \cot(k d_0).
\end{equation}
We can also express absorption coefficient from eq.~(\ref{eq:absorption_coeff_plane_surface_Z_Zconst}) with $\theta = 0$:
\begin{equation}\label{eq:absorption_coeff_plane_surface_Z_Zconst_porous_gap_wall}
\begin{aligned}
\alpha_{s,porous,wall} &= \frac{4 \Xi' d Z_0/\sigma}{\{ (\Xi' d/\sigma)^2 + [Z_0 \cot(k d_0)]^2 \} + 2 (\Xi' d/\sigma) Z_0 + Z_0^2} &\\
&= \frac{4 \Xi' d Z_0/\sigma}{ (\Xi' d/\sigma+Z_0)^2 + [Z_0 \cot(k d_0)]^2}.&
\end{aligned}
\end{equation}
As expected, it approaches zero when $\cot(k d_0) \rightarrow \pm\infty$, which is for $k d_0 = n\pi$ and $n = 0,1,2,...$, and reaches the maximum value for $k d_0 = (2n+1)\pi/2$, when $\cot(k d_0) = 0$. The maximum value depends essentially on flow resistance of the porous layer and it is $\alpha_{s,porous,wall} = 1$ only when $\Xi' d/\sigma = Z_0$. When $\Xi' d/\sigma$ is larger (smaller) than $Z_0$, the absorption coefficient drops with further increasing (decreasing) value of $\Xi' d/\sigma$. While these are fixed parameters of the porous layer, the cotangent term dictates frequency dependence of the absorption coefficient, which includes periodic occurrence of zeros and maxima over frequency. In practice, the differences between minima and maxima are less pronounced when averaging over frequency is performed, such as in octave bands. They are even smaller in a diffuse field, after averaging over different angles of incidence, because the frequencies of minima and maxima due to more general $\cot(k d_0 \cos(\theta))$ vary with $\theta$ for any fixed $d_0$.

If the gap is acoustically thin, as well, $kd_0 \ll 1$ and the gap behaves as a lumped stiffness $\rho_0 c_0^2/d_0$, since $\cot(kd_0) \approx 1/(kd_0)$, as shown in eq.~(\ref{eq:impedance_in_front_of_plane_reflecting_surface_R_lumped}). The impedance and absorption coefficient become
\begin{equation}\label{eq:porous_gap_wall_impedance_small_distance}
\boxed{ Z_{porous,wall} = \Xi' d/\sigma - j\frac{\rho_0 c_0^2}{\omega d_0} }
\end{equation}
and
\begin{equation}\label{eq:absorption_coeff_plane_surface_Z_Zconst_porous_gap_wall_small_distance}
\begin{aligned}
\alpha_{s,porous,wall} = \frac{4 \Xi' d Z_0/\sigma}{ (\Xi' d/\sigma+Z_0)^2 + [\rho_0 c_0^2/(\omega d_0)]^2},
\end{aligned}
\end{equation}
respectively. The gap is too thin for propagation of sound waves, so the unsteady acoustic quantities in it are uniform and in phase. Nevertheless, the absorption is not necessarily negligible, as in the case of a thin layer placed directly on the hard wall. At sufficiently high frequencies, when $\rho_0 c_0^2/(\omega d_0)$ is negligible compared to $\Xi' d/\sigma+Z_0$, maximum absorption is achieved, up to $\alpha_{s,porous,wall} = 1$ when $\Xi' d/\sigma = Z_0$. Evidently, ``sufficiently high" frequency will be lower for larger distances from the wall $d_0$, which should be preferred if low-frequency absorption is required. Indeed, increasing the thin gap leads to higher absorption coefficient values at low frequencies, without compromising absorption at higher frequencies. Still, $k d_0 \ll 1$ must be satisfied for this model to hold or additional effects (as in eq.~(\ref{eq:absorption_coeff_plane_surface_Z_Zconst_porous_gap_wall})) should be expected.

\begin{figure}[h]
	\centering
	\begin{subfigure}{.5\textwidth}
		\centering
		\includegraphics[width=1\linewidth]{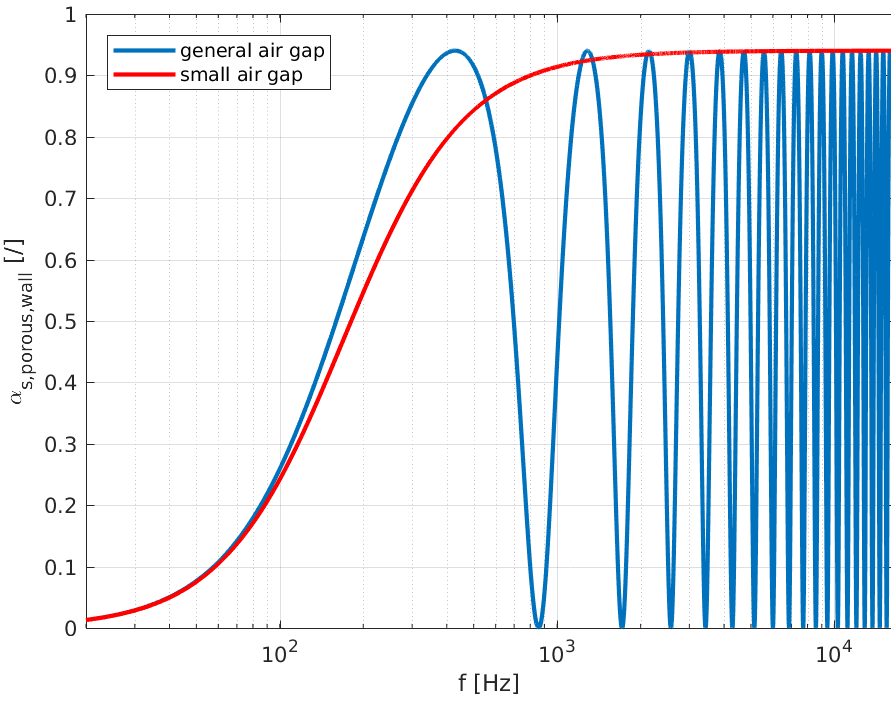}
		\label{fig:abs_coeff_porous_wall_normal_incidence}
	\end{subfigure}%
	\begin{subfigure}{.5\textwidth}
		\centering
		\includegraphics[width=1\linewidth]{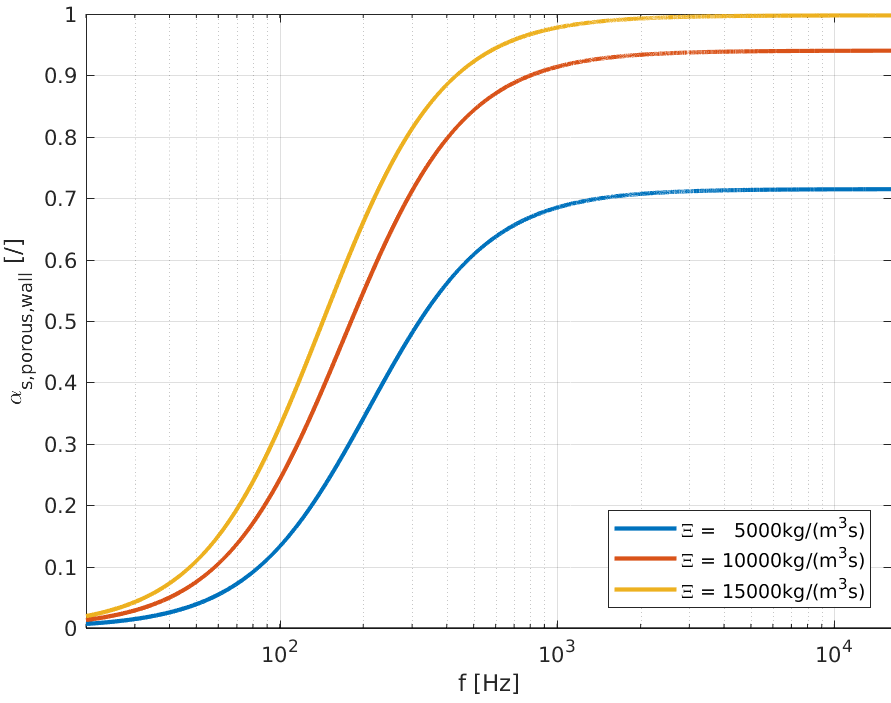}
		\label{fig:abs_coeff_porous_wall_normal_incidence_resistivity}
	\end{subfigure}
	\begin{subfigure}{.5\textwidth}
		\centering
		\includegraphics[width=1\linewidth]{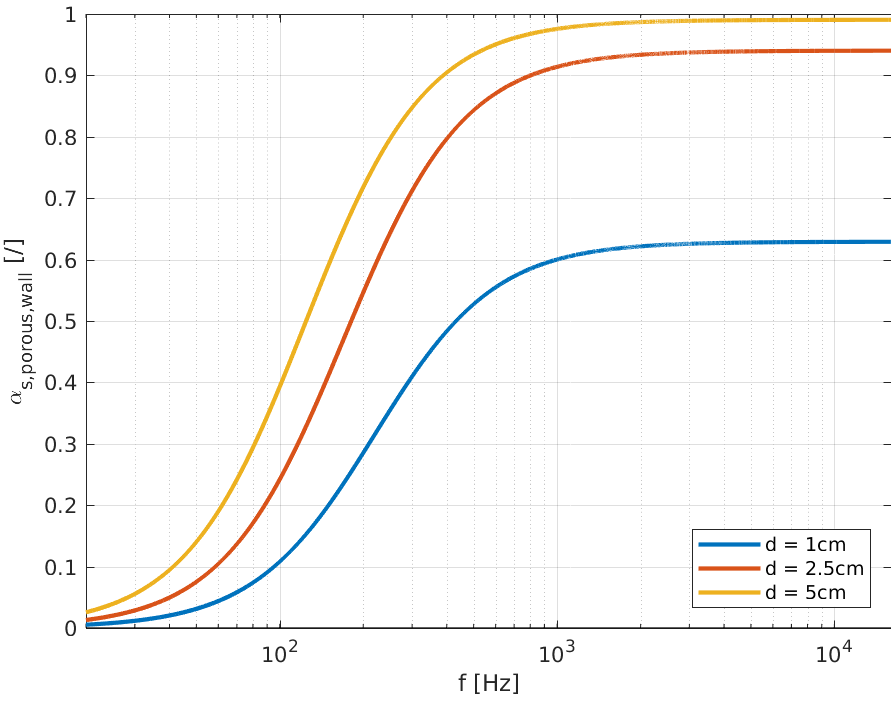}
		\label{fig:abs_coeff_porous_wall_normal_incidence_mat_thickness}
	\end{subfigure}%
	\begin{subfigure}{.5\textwidth}
		\centering
		\includegraphics[width=1\linewidth]{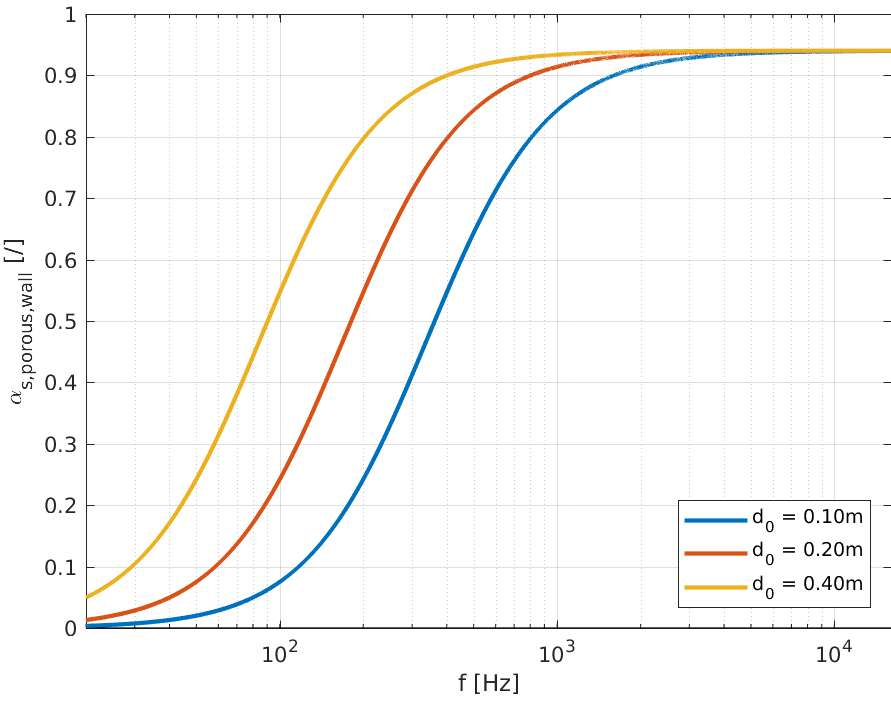}
		\label{fig:abs_coeff_porous_wall_normal_incidence_gap_thickness}
	\end{subfigure}
	\caption{Absorption coefficient of a thin layer of porous absorber with porosity $\sigma=1$ ($\Xi = \Xi'$) in front of a hard wall for normal incidence: (above left) calculated using eq.~(\ref{eq:absorption_coeff_plane_surface_Z_Zconst_porous_gap_wall}) and eq.~(\ref{eq:absorption_coeff_plane_surface_Z_Zconst_porous_gap_wall_small_distance}) for $\Xi =  10000$\,kg/(m$^3$s), $d = 2.5$\,cm, $d_0 = 0.20$\,m; calculated using eq.~(\ref{eq:absorption_coeff_plane_surface_Z_Zconst_porous_gap_wall_small_distance}) for (above right) $d = 2.5$\,cm, $d_0 = 0.20$\,m, (below left) $\Xi =  10000$\,kg/(m$^3$s), $d_0 = 0.20$\,m, (below right) $\Xi =  10000$\,kg/(m$^3$s), $d = 2.5$\,cm.}
	\label{fig:abs_coeff_porous_wall}
\end{figure}

Figure~\ref{fig:abs_coeff_porous_wall} shows absorption coefficient values obtained with equations~(\ref{eq:absorption_coeff_plane_surface_Z_Zconst_porous_gap_wall}) and  (\ref{eq:absorption_coeff_plane_surface_Z_Zconst_porous_gap_wall_small_distance}) and demonstrates the effects of varying flow resistivity, thickness of the porous layer, and thickness of the air gap between the layer and the wall, with usual values of $c_0$ and $\rho_0$. Periodic maxima an minima are not captured if acoustically thin gap is assumed, which is in this case justified for $f \ll c_0/(2\pi d_0) \approx 273$\,Hz with $d_0 = 0.2$\,m. Nevertheless, even this simple model captures well behaviour of the absorption coefficient curves, for example increase of absorption at low frequencies for larger $d_0$ and increase of broadband absorption for larger $d$. As already discussed, certain value of flow resistivity (in this case slightly above 15000\,kg/(m$^3$s)) provides maximal broadband absorption, above which the absorption coefficient values drop, which is not visible in Fig.~\ref{fig:abs_coeff_porous_wall}.

The analysis above indicates that low frequencies and associated large wavelengths require thicker porous materials placed at relatively large distances from the hard walls, which may occupy considerable volume of the room for the acoustic treatment. With regard to that, distances larger than 20-30\,cm are rarely used in practice. In order to achieve a more compact absorber, local particle velocity in the porous material can be increased by means of acoustic resonators, at least for a certain range of frequencies close to their resonances. Particularly compact are Helmholtz resonators, which are covered next.

\subsection{Helmholtz resonator and perforated panels}

Helmholtz resonator is basically an acoustically compact \textbf{cavity} with volume $V_c$, acoustically hard boundaries, and arbitrary shape. The only opening of the cavity extends into a narrow \textbf{neck} with the length $l_n$, which is also much smaller than the wavelength ($k l_n \ll 1$) and comparable to or smaller than the characteristic length scale of the cavity. For convenience, we position the straight neck along the horizontal right-oriented $x$-axis (instead of the $x_1$-axis above), with the cavity located at the left end of the neck. Since the entire resonator is compact, sound waves cannot propagate inside it and both cavity and neck can be treated as lumped elements. Limiting ourselves to plane incoming sound waves parallel to the $x$-axis and accordingly one-dimensional sound field, we can consider only two distinct velocities in the resonator which are essentially functions of time only -- inside the cavity, $v_c(t)$, and inside the neck, $v_n(t)$.

Resonance of a Helmholtz resonator is due to its geometry, in particular due to the more or less abrupt change from the very narrow neck\footnote{Still, the neck is assumed to be wider than the boundary layer thickness. We neglect viscous and thermal effects on the walls of the neck and the cavity.} with cross section area $S_n$ to the wider cavity with cross section area $S_c$. If these are not constant, average values can be used, since details of the geometry do not play a significant role in the mechanism of resonance. Since the two essential parts of the geometry are finite and compact, we can write (linearized and inviscid) mass and momentum equations~(\ref{eq:Euler_mass}) and (\ref{eq:Euler_momentum}) in the integral form. With $S_c \gg S_n$, conservation of mass at the interface between the neck and cavity\footnote{\label{fn:mass_conserv_control_volume}We centre a very thin control volume at the interface with its curved boundaries coinciding with the hard surfaces of the resonator's neck and cavity and two remaining flat surfaces (on each side of the interface) normal to the $x$-axis. Since mass of the fluid is conserved and walls of the resonator are rigid and motionless, the volume flux at the left (cavity) side of the control volume, $S_c v_c$, has to be equal to the flux at the right (neck) side, $S_n v_n$, so $S_c \gg S_n$ implies $|v_c| \ll |v_n|$.} implies $|v_c| \ll |v_n|$ and in the linear approximation
\begin{equation}\label{Helmholtz_resonator_conservation_of_mass}
\abs*{ \left( \frac{\partial v}{\partial t} \right)_c } \ll \abs*{ \left( \frac{\partial v}{\partial t} \right)_n } \text{ and } \abs*{ \left( \frac{\partial v}{\partial x} \right)_c } \ll \abs*{ \left( \frac{\partial v}{\partial x} \right)_n },
\end{equation}
where $(\text{ })_c$ means that the expression in the brackets is evaluated inside the cavity and $(\text{ })_n$ inside the neck. Together with the momentum equation~(\ref{eq:Euler_momentum}) in the integral form, the first inequality in eq.~(\ref{Helmholtz_resonator_conservation_of_mass}) gives
\begin{equation}
\abs*{ \left( \frac{\partial v}{\partial t} \right)_c } \ll \abs*{ \left( \frac{\partial v}{\partial t} \right)_n } \Rightarrow \abs*{ \left( \frac{\partial p}{\partial x} \right)_c } \ll \abs*{ \left( \frac{\partial p}{\partial x} \right)_n }.
\end{equation}
Hence, pressure distribution in the bulk of the cavity can be considered to be uniform in comparison to the neck. On the other hand, the acoustically compact volumes $V_n \ll V_c$ imply that the much smaller neck behaves like a plug of incompressible fluid (air) relative to the cavity, so the density perturbation satisfies $|\rho_n| \ll |\rho_c|$ and therefore $|(\partial \rho / \partial t)_n| \ll |(\partial \rho / \partial t)_c|$.

With these insights we can write the conservation of mass for the entire volume of the resonator $V = V_c + V_n$,
\begin{equation}\label{eq:Euler_mass_Helmholtz_res}
\begin{aligned}
\int_{V} \frac{\partial \rho}{\partial t} d^3 \boldsymbol x + \rho_0 \int_{V} \frac{\partial v}{\partial x} d^3 \boldsymbol x &\approx \int_{V_c} \frac{\partial \rho}{\partial t} d^3 \boldsymbol x + \rho_0 \int_{l_n} \frac{\partial v}{\partial x} S_n d x &\\
&= V_c \frac{\partial \rho_c}{\partial t} + \rho_0 S_n (v_n-v_c) \approx V_c \frac{\partial \rho_c}{\partial t} + \rho_0 S_n v_n = 0,
\end{aligned}
\end{equation}
and the conservation of momentum,
\begin{equation}\label{eq:Euler_momentum_Helmholtz_res}
\begin{aligned}
\rho_0 \int_{V} \frac{\partial v}{\partial t} d^3 \boldsymbol x + \int_{V} \frac{\partial p}{\partial x} d^3 \boldsymbol x &\approx \rho_0 \int_{V_n} \frac{\partial v}{\partial t} d^3 \boldsymbol x + \int_{l_n} \frac{\partial p}{\partial x} S_n d x &\\
&= \rho_0 V_n \frac{\partial v_n}{\partial t} + S_n (p_n - p_c) = 0,
\end{aligned}
\end{equation}
where $v_n$ and $p_n$ are also velocity and pressure at the opening of the neck. Compressibility of practically constant volume of air in the cavity is balanced by the motion of essentially incompressible plug of air in the neck (eq.~(\ref{eq:Euler_mass_Helmholtz_res})). The latter is determined by the difference of pressure at the opening of the neck and inside the cavity (eq.~(\ref{eq:Euler_momentum_Helmholtz_res})).

After using $\rho_c = p_c/c_0^2$ from the equation of state~(\ref{eq:p_rho}) as well as the ansatz $p = \hat{p} e^{j\omega t}$ and $v = \hat{v} e^{j\omega t}$, we obtain two equations for the complex amplitudes,
\begin{equation}\label{eq:Euler_mass_Helmholtz_res_freq_domain}
j \omega V_c \frac{\hat{p}_c}{c_0^2} + \rho_0 S_n \hat{v}_n = 0
\end{equation}
and
\begin{equation}\label{eq:Euler_momentum_Helmholtz_res_freq_domain}
j \omega \rho_0 V_n \hat{v}_n + S_n (\hat{p}_n - \hat{p}_c) = 0.
\end{equation}
We can express $\hat{v}_n$ from eq.~(\ref{eq:Euler_mass_Helmholtz_res_freq_domain}) in terms of $\hat p_c$ and insert it in eq.~(\ref{eq:Euler_momentum_Helmholtz_res_freq_domain}), which gives
\begin{equation}
\frac{\hat{p}_n}{\hat{p}_c} = 1 - \frac{\omega^2 V_n V_c}{c_0^2 S_n^2}.
\end{equation}
At angular frequencies around
\begin{equation}
\omega_0 = c_0 S_n \sqrt{\frac{1}{V_n V_c}},
\end{equation}
sound pressure inside the cavity becomes much larger than pressure at the opening of the resonator and the amplitude of oscillations of the neck is very high. This is the \textbf{resonance frequency} (a single frequency, without harmonics). Theoretically infinite, the maximum amplitude is always limited, for example by viscous losses in the neck or conversion of sound energy into kinetic energy of vorticity generated due to non-linear effects at high sound levels. Both effects are neglected here.

Geometric volume of the neck equals $S_n l_n$. However, effective length of the plug of air in the neck is somewhat larger than $l_n$, owing to the inertia of the fluid at both ends of the neck (part of the fluid outside the neck moves with similar velocity as fluid inside the neck). Taking this into account, we write the effective length of the neck as
\begin{equation}\label{eq:end_correction}
l_{n,eff} = \frac{V_n}{S_n} = l_n+2\delta,
\end{equation}
where $\delta$ is \textbf{end correction} of the length (at one end of the neck, hence multiplication with 2 in eq.~(\ref{eq:end_correction})) and $V_n > S_n l_n$. The end correction depends on the geometry of the neck and for a circular opening (perforation) in a rigid baffle it is approximately $0.85 a$, where $a$ is radius of the perforation. It is of the same order of magnitude as $a$, which makes the end correction critical for thin perforated screens, the thickness of which ($l_n$) is often much smaller than $a$. The resonance frequency becomes
\begin{equation}\label{eq:Helmholtz_resonator_resonance_frequency}
\boxed{ f_0 = \frac{c_0}{2\pi} \sqrt{\frac{S_n}{l_{n,eff} V_c}} = \frac{c_0}{2\pi} \sqrt{\frac{\sigma S_c}{l_{n,eff} V_c}} = \frac{c_0}{2\pi} \sqrt{\frac{\sigma}{l_{n,eff} d}} },
\end{equation}
where $d = V_c/S_c$ is characteristic length of the cavity and $\sigma = S_n/S_c$.

Helmholtz resonators are occasionally used as separate acoustic elements, usually damped, as low-frequency absorbers. Another very common implementation is in sound absorbing \textbf{perforated panels}. A perforated screen backed by a hard wall can be modelled as a raster of Helmholtz resonators. Each perforation is opening of the neck of one resonator. For simplicity, we assume that all perforations are equal and equidistantly distributed over the screen. In such a case, $\sigma$ represents the \textbf{perforation ratio} of the panel (as in section~\ref{ch:porous_absorber_rigid_wall}), which should be much smaller than 1, according to the assumption $S_c \gg S_n$. $d$ is distance between the screen and wall.

Impedance at the outer surface of the perforated panel can be estimated in the following way. First, we insert $\hat{p}_c$ from eq.~(\ref{eq:Euler_mass_Helmholtz_res_freq_domain}) into eq.~(\ref{eq:Euler_momentum_Helmholtz_res_freq_domain}), which gives
\begin{equation}
\frac{\hat{p}_n}{\hat{v}_n} = -\frac{j \omega \rho_0 V_n}{S_n} - \frac{\rho_0 S_n c_0^2}{j \omega V_c} = -j \omega \rho_0 l_{n,eff} + \frac{j \rho_0 \omega_0^2 l_{n,eff}}{\omega} = j \rho_0 l_{n,eff} \left( \frac{\omega_0^2}{\omega} - \omega \right).
\end{equation}
In the definition of impedance, eq.~(\ref{eq:impedance}), the unit vector $\boldsymbol n$ points into the surface. Therefore, $\hat{\boldsymbol v}_i \cdot \boldsymbol n$ equals $-\hat{v}_i$ here (the cavity is left from the neck) and $\hat{\boldsymbol v}_i$ is particle velocity due to the incoming plane wave at normal incidence. Furthermore, conservation of mass at the outer surface of the screen (interface between the perforation and the room) gives $v_n = v_i/\sigma$ (compare with footnote \footref{fn:mass_conserv_control_volume} as well as eq.~(\ref{flow_resistivity_porosity})), while $\hat p_n = \hat p_i$. Hence, the impedance equals
\begin{equation}
Z_{perf} = -\frac{\hat{p}_i}{\hat{v}_i} = -\frac{\hat{p}_n}{\sigma \hat{v}_n} = \frac{j \rho_0 l_{n,eff}}{\sigma} \left( \omega - \frac{\omega_0^2}{\omega} \right).
\end{equation}
By comparing this with the mechanical impedance in eq.~(\ref{eq:oscillator_impedance}), we see that the perforated screen backed by a hard wall behaves as two lumped elements in series -- mass $\rho_0 l_{n,eff}/\sigma$, which is due to the fluid in the neck (the plug), and stiffness $\rho_0 l_{n,eff} \omega_0^2/\sigma = \rho_0 c_0^2 S_n/(\sigma V_c) = \rho_0 c_0^2 S_c/V_c = \rho_0 c_0^2/d$, which is due to the cavity. This stiffness is equal to the stiffness of an acoustically thin layer of air for normal sound incidence, which was derived in section~\ref{ch:infinite_plane_surface} (eq.~(\ref{eq:impedance_in_front_of_plane_reflecting_surface_R_lumped})). The thin air gap behind the screen thus plays the same role of a spring (or lumped electric capacitor), while the perforated screen adds the mass (lumped inductor) to it. The two together form a resonator.

In order to introduce absorption and build a damped resonator, it is the most efficient to place a thin layer of porous material inside the neck or close to it, where the particle velocity is high (section~\ref{ch:porous_absorber_rigid_wall}). Felt or fabric is usually glued on the back side of the screen, towards the wall. In section~\ref{ch:layer_poros_abosrber} (eq.~(\ref{flow_resistivity_porosity}) and the discussion below it) we saw that a thin layer of porous material acts as a lumped resistance $\Xi' d_{porous}/\sigma_{porous}$ connected in series, with $\Xi'$ flow resistivity, $\sigma_{porous}$ porosity, and $d_{porous} = d-d_0$ thickness of the porous material ($d_0$ is thickness of the remaining air gap inside the cavity). As a result, impedance of the entire assembly equals
\begin{equation}
\boxed{ Z_{perf,absorb} = \frac{j \rho_0 l_{n,eff}}{\sigma} \left( \omega - \frac{\omega_0^2}{\omega} \right) + \frac{\Xi' d_{porous}}{\sigma_{porous}} }.
\end{equation}
Absorption coefficient can now be calculated from eq.~(\ref{eq:absorption_coeff_plane_surface_Z_Zconst}) with $\theta = 0$:
\begin{equation}\label{eq:absorption_coeff_plane_surface_Z_Zconst_perf_panel}
\begin{aligned}
\alpha_{s,perf,absorb} &= \frac{4 Z_0 \Xi' d_{porous}/\sigma_{porous}}{ (\Xi' d_{porous}/\sigma_{porous}+Z_0)^2 + [\rho_0 l_{n,eff} (\omega^2 - \omega_0^2)/(\omega \sigma)]^2} &\\
&= \frac{4 Z_0 \Xi' d_{porous}/\sigma_{porous}}{ (\Xi' d_{porous}/\sigma_{porous}+Z_0)^2 + [\rho_0 c_0^2 (\omega^2/\omega_0^2 - 1)/(\omega d)]^2}.&
\end{aligned}
\end{equation}
It reaches the maximum value at $\omega = \omega_0$, when it equals
\begin{equation}\label{eq:max_absorption_coeff_plane_surface_Z_Zconst_perf_panel}
\begin{aligned}
\alpha_{s,perf,absorb} &= \frac{4 \Xi' d_{porous}/(\sigma_{porous}Z_0)}{ [\Xi' d_{porous}/(\sigma_{porous}Z_0)+1]^2}.&
\end{aligned}
\end{equation}
As before, maximum absorption $\alpha_{s,perf,absorb} = 1$ is achieved at the resonance frequency only if $\Xi' d_{porous}/\sigma_{porous} = Z_0$.

\begin{figure}[h]
	\centering
	\begin{subfigure}{.45\textwidth}
		\centering
		\includegraphics[width=1\linewidth]{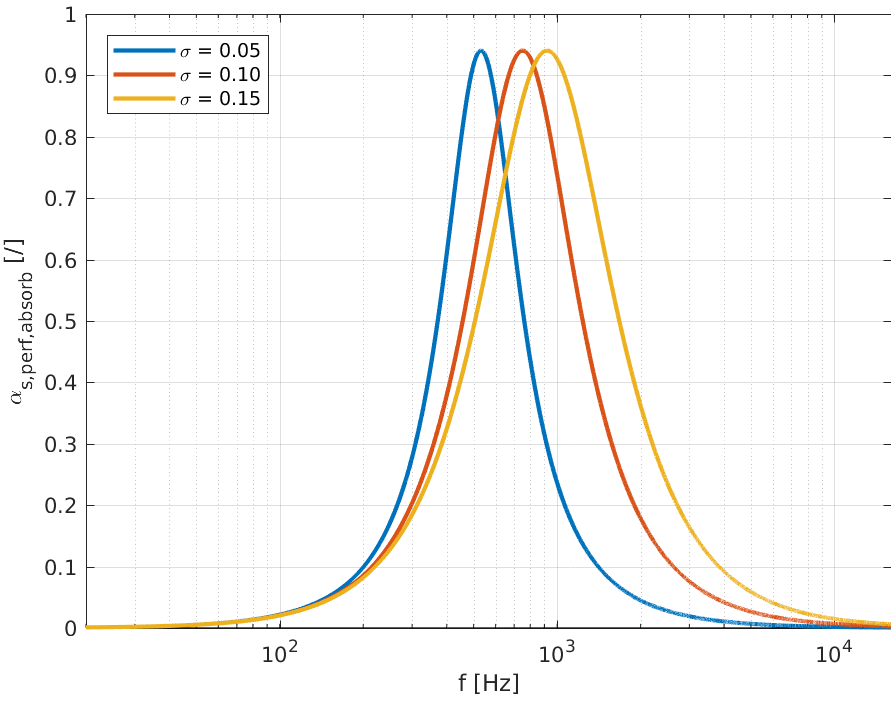}
		\label{fig:abs_coeff_perforated_panel_normal_incidence_perforation_ratio}
	\end{subfigure}%
	\begin{subfigure}{.45\textwidth}
		\centering
		\includegraphics[width=1\linewidth]{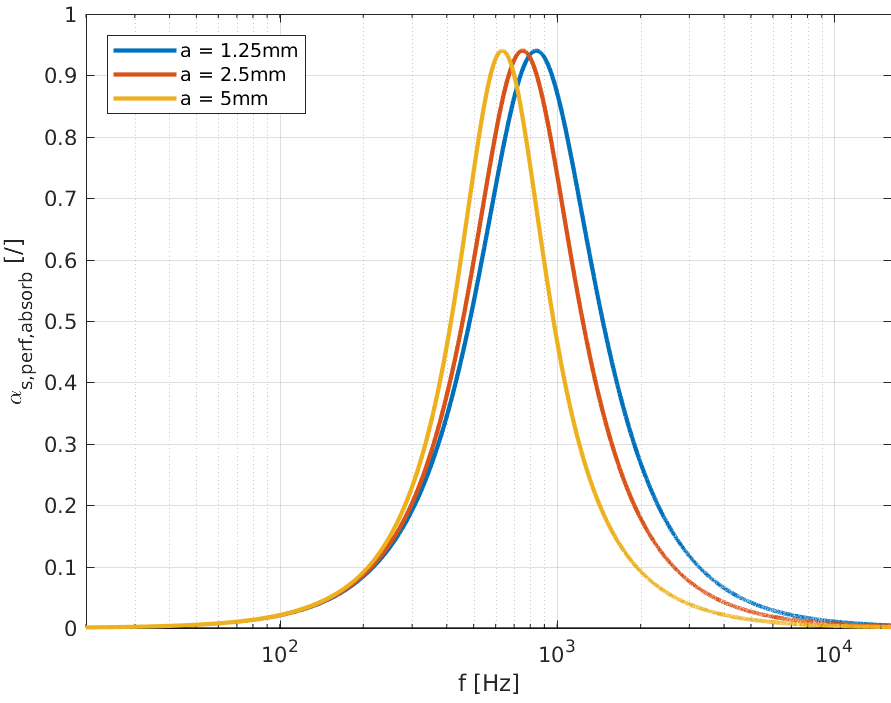}
		\label{fig:abs_coeff_perforated_panel_normal_incidence_perforation_radius}
	\end{subfigure}
	\begin{subfigure}{.45\textwidth}
		\centering
		\includegraphics[width=1\linewidth]{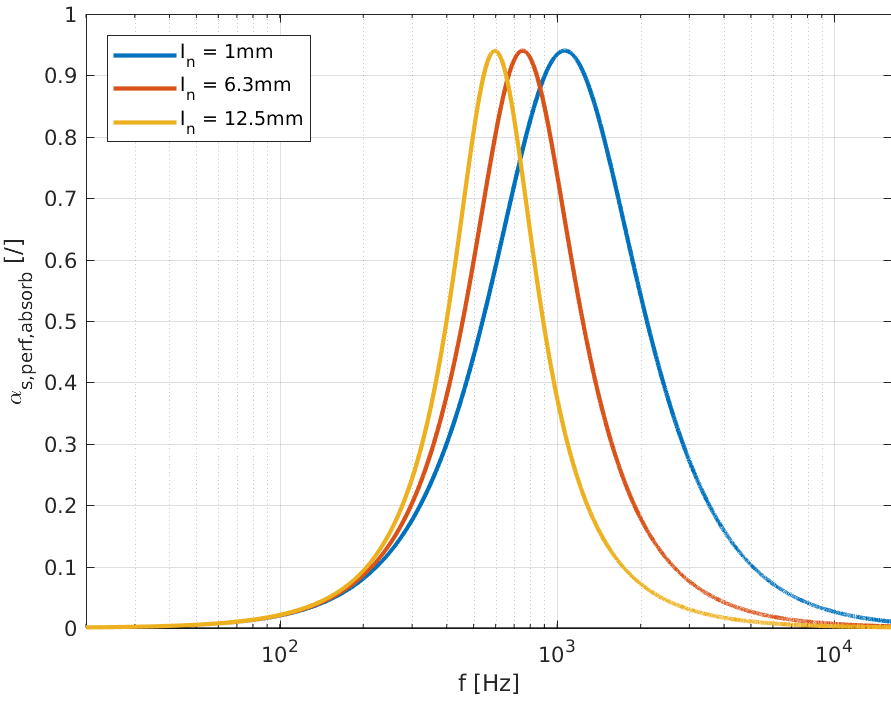}
		\label{fig:abs_coeff_perforated_panel_normal_incidence_thickness_panel}
	\end{subfigure}%
	\begin{subfigure}{.45\textwidth}
		\centering
		\includegraphics[width=1\linewidth]{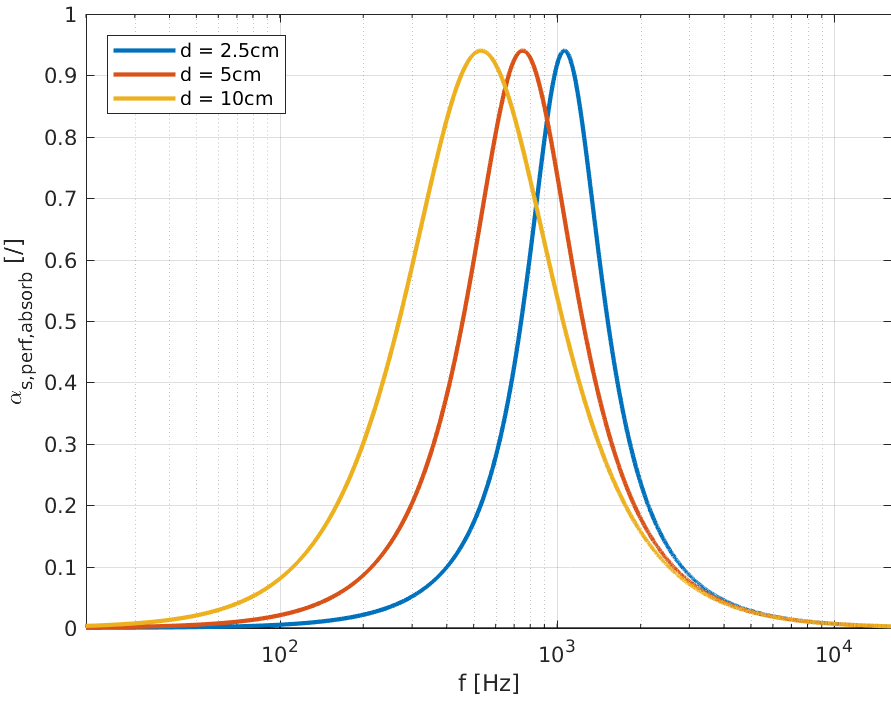}
		\label{fig:abs_coeff_perforated_panel_normal_incidence_cavity_length}
	\end{subfigure}
	\begin{subfigure}{.45\textwidth}
			\centering
			\includegraphics[width=1\linewidth]{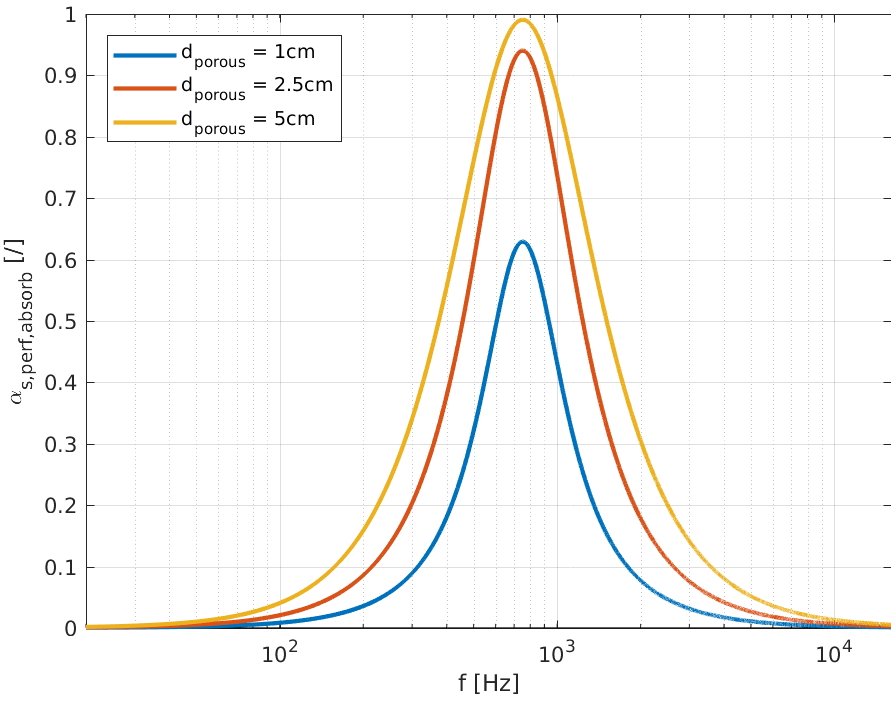}
			\label{fig:abs_coeff_perforated_panel_normal_incidence_thickness_porous_material}
	\end{subfigure}%
	\begin{subfigure}{.45\textwidth}
			\centering
			\includegraphics[width=1\linewidth]{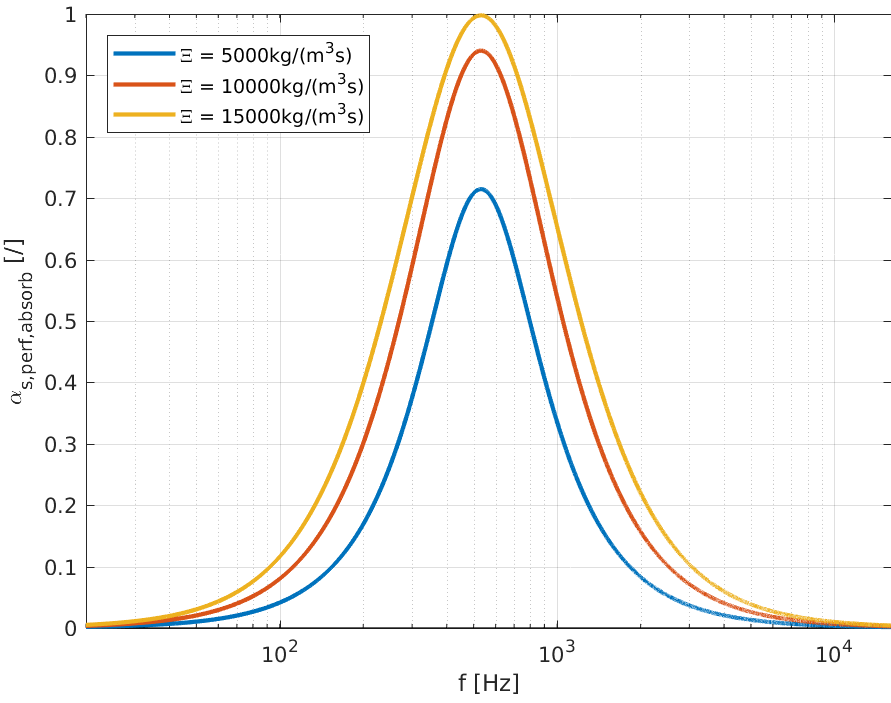}
			\label{fig:abs_coeff_perforated_panel_normal_incidence_flow_resistivity}
	\end{subfigure}
	\caption{Absorption coefficient of a perforated panel with circular perforations backed by a layer of porous absorber ($\sigma_{porous}=1 \Rightarrow \Xi = \Xi'$) in front of a hard wall for normal incidence: (above left) $a = 2.5$\,mm, $l_n = 6.3$\,mm, $d = 5$\,cm, $d_{porous} = 2.5$\,cm, $\Xi = 10000$\,kg/(m$^3$s), (above right) $\sigma = 0.1$, $l_n = 6.3$\,mm, $d = 5$\,cm, $d_{porous} = 2.5$\,cm, $\Xi = 10000$\,kg/(m$^3$s), (middle left) $\sigma = 0.1$, $a = 2.5$\,mm, $d = 5$\,cm, $d_{porous} = 2.5$\,cm, $\Xi = 10000$\,kg/(m$^3$s), (middle right) $\sigma = 0.1$, $a = 2.5$\,mm, $l_n = 6.3$\,mm, $d_{porous} = 2.5$\,cm, $\Xi = 10000$\,kg/(m$^3$s), (below left) $\sigma = 0.1$, $a = 2.5$\,mm, $l_n = 6.3$\,mm, $d = 5$\,cm, $\Xi = 10000$\,kg/(m$^3$s), (below right) $\sigma = 0.1$, $a = 2.5$\,mm, $l_n = 6.3$\,mm, $d = 5$\,cm, $d_{porous} = 2.5$\,cm.}
	\label{fig:abs_coeff_perforated_panel_normal_incidence}
\end{figure}

Figure~\ref{fig:abs_coeff_perforated_panel_normal_incidence} gives estimation of the absorption coefficient using eq.~(\ref{eq:absorption_coeff_plane_surface_Z_Zconst_perf_panel}) for a panel with circular perforations and different values of the relevant parameters. It clearly demonstrates that resonance-based absorbers are efficient only at frequencies around the resonance frequency. The overall absorption increases with flow resistivity and thickness of the porous material. However, values of $\Xi' d_{porous}/\sigma_{porous}$ in the figure are all smaller than $Z_0$. For $\Xi' d_{porous}/\sigma_{porous} > Z_0$, absorption drops with further increase of $\Xi'$ or $d_{porous}$ (see for comparison the last graph in Fig.~\ref{fig:abs_coeff_membrane_resonator_normal_incidence} below, in which the maximum absorption drops with the increase of damping $D$, which is larger than $Z_0$). Besides that, resonance frequency is lower for a larger cavity (the gap thickness $d$), longer neck, larger radius of perforation (which leads to both larger effective length of the neck and width of the cavity, if perforation ratio is fixed), or lower perforation ratio (which leads to a larger width of the cavity, when radius of the perforation is fixed). This is in agreement with eq.~(\ref{eq:Helmholtz_resonator_resonance_frequency}).

In the analysis above, we considered only normal incidence of the incoming sound. This is a reasonable assumption if cavities of the resonators behind the perforations are physically separated with rigid walls, so that sound waves behind the screen cannot propagate parallel to the wall. This is, however, rarely the case in practice. Nevertheless, if absorption behind the screen is high, incident sound at higher angles of incidence will tend to refract normal to the wall. Indeed, measurements show that the impedance of perforated panels changes only slightly with the angle of incidence, at least for relatively low angles (up to around 60$^\circ$) and not too low frequencies (above 300\,Hz for common geometries of the perforated panels). Prediction of resonance frequency and equivalent absorption area of Helmholtz resonators with the simple model considered here is still rather inaccurate and additional adjustments of the geometry and laboratory measurements are often necessary, especially for narrow-band (weakly damped) absorbers. On the other hand, large Helmholtz resonators are suitable for treatment of low-frequency sound, which is difficult with conventional porous absorbers and perforated panels.

\subsection{Membrane resonator}

In the previous section we saw how efficiency of a porous absorber backed by a hard wall can be further increased with a perforated screen in front of it. Mass of the air plugs inside the perforations (necks of the Helmholtz resonators) ads to the stiffness of the air gap between the screen and wall and builds a damped resonator. At the resonance frequency, particle velocity is particularly high inside the perforations, as well as the porous material in their vicinity, which increases the energy losses. Physical mechanism of membrane absorbers is essentially the same, except that the added mass is due to a thin solid material (a membrane or thin plate), which is again placed in front of a hard surface and close to an absorbing material. Moreover, damping of membrane resonators can be caused by viscous losses inside the material of the membrane itself or its connections at the edges, which sometimes makes use of additional absorbers unnecessary. With regard to that, we use here a general symbol $D$ to denote total damping of the resonator, regardless of the mechanism of dissipation. If additional porous absorber is placed close to the membrane to increase the damping, the two should not be fixed together (for example, by gluing the porous material on the membrane), which would cause them to move synchronously. This would decrease the effective particle velocity inside the material and therefore its efficiency.

Pressure drop on the two sides of the membrane positioned normal to the $x_1$-axis can be estimated from the conservation of momentum as
\begin{equation}\label{eq:conservation_momentum_membrane}
\Delta p = M_m \frac{d v_1}{d t},
\end{equation}
where $M_m$ is surface mass of the membrane in kg/m$^2$. We again assumed normal incidence of a plane sound wave propagating in the direction of the $x_1$-axis and one-dimensional geometry of the problem. After switching to frequency domain, $\Delta \hat p / \hat v_1 = j \omega M_m $ can be compared with eq.~(\ref{eq:oscillator_impedance}), confirming that an acoustically thin membrane acts as a lumped mass (or electric inductor) for normal sound incidence. Impedance of the whole assembly is thus
\begin{equation}\label{eq:membrane_absorber_impedance}
\boxed{ Z_{memb,wall} = j\omega M_m + D + \frac{\rho_0 c_0^2}{j\omega d} },
\end{equation}
where stiffness of the gap between the membrane and wall with acoustically small thickness $d$ is, as before, $S = \rho_0 c_0^2/d$. Since the velocity is equal on the two sides of the membrane, all three lumped elements are connected in series.

\begin{figure}[h]
	\centering
	\begin{subfigure}{.5\textwidth}
		\centering
		\includegraphics[width=1\linewidth]{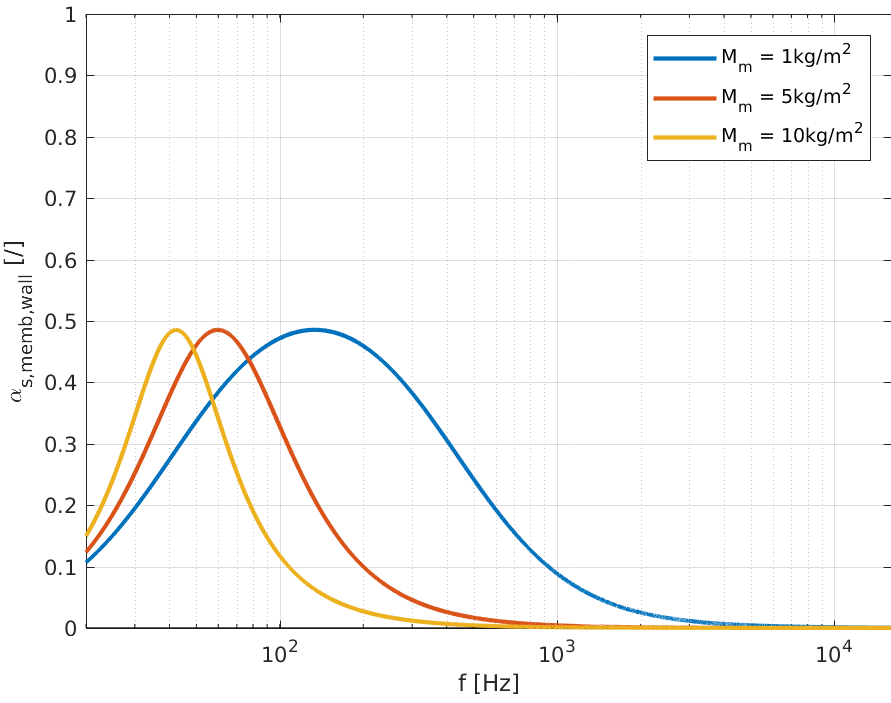}
		\label{fig:abs_coeff_membrane_resonator_normal_incidence_surface_mass}
	\end{subfigure}%
	\begin{subfigure}{.5\textwidth}
		\centering
		\includegraphics[width=1\linewidth]{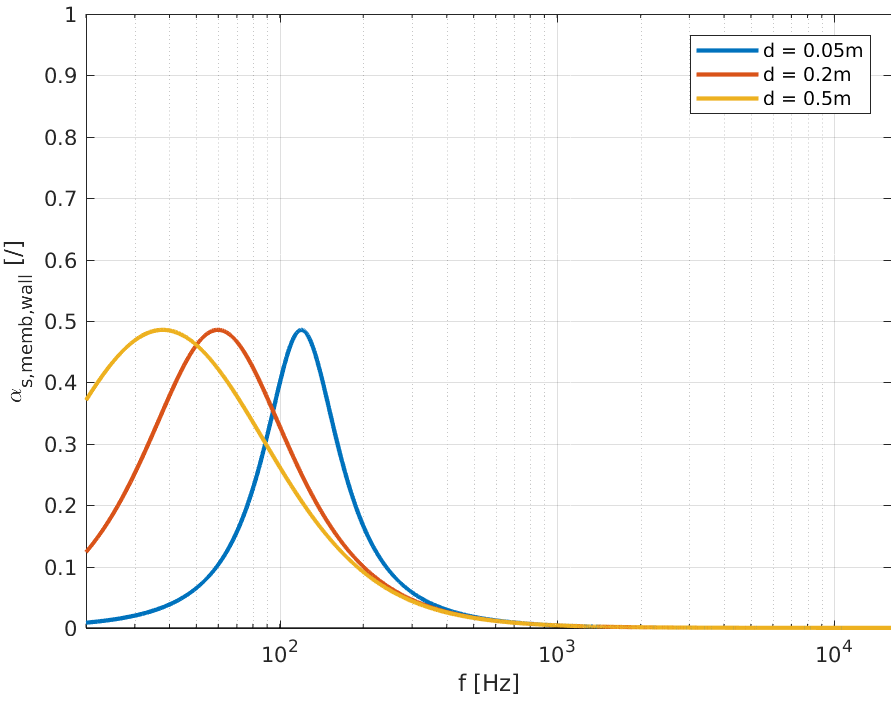}
		\label{fig:abs_coeff_membrane_resonator_normal_incidence_air_gap}
	\end{subfigure}
	\begin{subfigure}{.5\textwidth}
		\centering
		\includegraphics[width=1\linewidth]{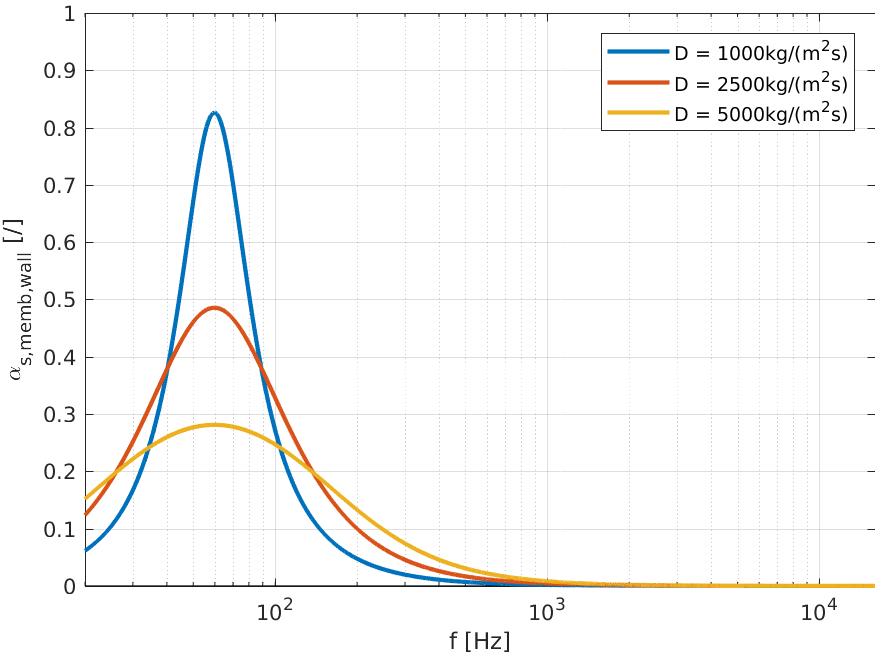}
		\label{fig:abs_coeff_membrane_resonator_normal_incidence_damping}
	\end{subfigure}%
	\caption{Absorption coefficient of a damped membrane in front of a hard wall for normal incidence: (above left) $d = 0.2$\,m, $D = 2500$\,kg/(m$^2$s), (above right) $M_m = 5$\,kg/m$^2$, $D = 2500$\,kg/(m$^2$s), (below) $M_m = 5$\,kg/m$^2$, $d = 0.2$\,m.}
	\label{fig:abs_coeff_membrane_resonator_normal_incidence}
\end{figure}

Neglecting the damping, resonance frequency is as in the case of a linear oscillator discussed in section~\ref{ch:room_eigenmodes},
\begin{equation}\label{membrane_resonator_resonance_frequency}
\boxed{ f_0 = \frac{\omega_0}{2 \pi} = \frac{\sqrt{S/M_m}}{2 \pi} = \frac{c_0}{2\pi} \sqrt{\frac{\rho_0}{d M_m}} }.
\end{equation}
Absorption coefficient can also be calculated from eq.~(\ref{eq:absorption_coeff_plane_surface_Z_Zconst}) with $\theta = 0$:
\begin{equation}\label{eq:absorption_coeff_plane_surface_Z_Zconst_membrane}
\begin{aligned}
\alpha_{s,memb,wall} &= \frac{4 D Z_0}{\{ D^2 + [\omega M_m - \rho_0 c_0^2/(\omega d)]^2 \} + 2 D Z_0 + Z_0^2} &\\
&= \frac{4 D Z_0}{ (D+Z_0)^2 + [\omega M_m - \omega_0^2 M_m/\omega]^2} &\\
&= \frac{4 D Z_0}{ (D+Z_0)^2 + [(\omega^2 - \omega_0^2)M_m/\omega]^2}.&
\end{aligned}
\end{equation}
It has the maximum value
\begin{equation}\label{eq:max_absorption_coeff_plane_surface_Z_Zconst_membrane}
\begin{aligned}
\alpha_{s,memb,wall} &= \frac{4 D/ Z_0}{ [D/Z_0+1]^2}&
\end{aligned}
\end{equation}
at the resonance frequency ($\omega = \omega_0$). For $D = Z_0$, the maximum value is $\alpha_{s,memb,wall} = 1$, otherwise it is lower. Values of the absorption coefficient from eq.~(\ref{eq:absorption_coeff_plane_surface_Z_Zconst_membrane}) are shown in Fig.~\ref{fig:abs_coeff_membrane_resonator_normal_incidence}. Due to the similar mechanisms of work of the resonators, the absorption coefficient curves have similar forms as in Fig.~\ref{fig:abs_coeff_perforated_panel_normal_incidence} for perforated panels. However, relatively large masses of membranes and plates result normally in lower resonance frequencies and the mismatch between $D$ and $Z_0$ makes membrane resonators less efficient as sound absorbers in comparison to Helmholtz resonators. Increasing mass of the membrane or its distance from the wall lowers the resonance frequency, according to eq.~(\ref{membrane_resonator_resonance_frequency}). For $D>Z_0$, absorption coefficient decays with further increase of damping.

Using the results from above, we can also write impedance of a thin \textbf{membrane} or plate (without internal damping) which is not backed by a hard surface. At normal incidence it equals
\begin{equation}\label{eq:free_membrane_impedance}
\boxed{ Z_{memb} = j\omega M_m + Z_0 },
\end{equation}
where $Z_0 = \rho_0 c_0$ is simply characteristic impedance of air behind the membrane, which has to be taken into account when evaluating impedance at the surface of the membrane. Absorption coefficient is then (eq.~(\ref{eq:absorption_coeff_plane_surface_Z_Zconst}))
\begin{equation}\label{eq:absorption_coeff_plane_surface_Z_Zconst_free_membrane}
\begin{split}
\alpha_{s,memb} = \frac{4 Z_0^2}{4Z_0^2 + \omega^2 M_m^2}.
\end{split}
\end{equation}
It decreases with the increase of the membrane's surface mass or frequency and approaches 1 for a very light membrane or low frequencies. Since $D=0$, the non-reflected energy is actually transmitted through the membrane, and the absorption coefficient quantifies its fraction.

Unfortunately, modelling a membrane absorber with lumped elements gives very often unsatisfactory accuracy of the predicted values of both resonance frequency and equivalent absorption area (even more often than in the case of Helmholtz resonators and perforated panels). One of the main reasons for this is that the membrane is not a locally reacting surface. It supports wave propagation over its surface (bending waves) and has its own eigenmodes. These effects are neglected by the assumption of normal sound incidence. Furthermore, sound waves behind the membrane can propagate parallel to the wall, which is also ignored in the one-dimensional approximation. In addition to this, connections at the edges of the membrane necessarily hinder its motion, while eq.~(\ref{eq:conservation_momentum_membrane}) holds for a freely moving membrane. Consequently, the connections can decrease effective absorbing surface of a membrane absorber. Laboratory measurements and tuning of membrane resonators are often necessary in practice to avoid unexpected acoustic effects. In general, more damped resonators with broader absorption coefficient characteristics carry less risk of erroneous predictions.

\bigskip

By comparing Figures~\ref{fig:abs_coeff_porous_wall}, \ref{fig:abs_coeff_perforated_panel_normal_incidence}, and \ref{fig:abs_coeff_membrane_resonator_normal_incidence}, we can conclude quite generally that absorbers involving only layers of porous materials in front of hard surfaces provide high absorption at middle to high frequencies. To some extent, their frequency range can be expended towards low frequencies by increasing the distance from the wall. In contrast to this, absorbers based on resonators are efficient in a limited frequency range around the resonance frequency, which is usually higher for perforated panels than for membrane absorbers or a large Helmholtz resonator. These absorbers are in general more expensive and difficult for predicting their acoustic performances and, particularly in the case of membrane absorbers, often provide moderate maximum absorption coefficient. However, they are useful for the parts of the frequency range in which porous absorbers are inefficient, at low to middle frequencies.

\subsection{Quarter-wavelength resonator and seat dip effect}

Another type of resonators which are occasionally used as sound absorbers is quarter-wavelength resonator. As the name suggests, length of the resonator is equal to one quarter of the wavelength ($\lambda/4$) corresponding to the resonance frequency and it is open at one end and closed with a hard surface at the other. Distance between the hard side walls is in contrast much smaller, which supports normal incidence of incoming plane sound waves. The mechanism of resonance can be understood from eq.~(\ref{eq:impedance_in_front_of_plane_reflecting_surface_R}) with $\theta = 0$. According to the discussion there, the highest values of particle velocity are reached at the opening of the resonator, which is at the distance $\lambda/4$ from the rigid end, and introducing damping there results in an efficient sound absorption, similarly as in the case of a layer of porous absorber in front of a hard wall. Moreover, the resonance frequency can be estimated fairly accurately for a given geometry of the resonator. However, a practical disadvantage of quarter-wavelength resonators is that the length of $\lambda/4$ implies much larger dimensions of a resonator's body than in the case of acoustically compact Helmholtz absorbers, which can be critical for low-frequency absorption.

An curious phenomenon associated with quarter-wavelength resonance, which is occasionally encountered in rooms, is the so-called \textbf{seat dip effect}. It is an occurrence of excessive attenuation of sound by empty seats of an auditorium, usually in a narrow frequency range between 100\,Hz and 200\,Hz (and to less extent at the odd multiples of the lowest frequency). The effect can be explained by the formation of vertical quarter-wavelength resonators between the adjacent rows of seats of the auditorium. The unoccupied seats act as side walls of the resonators and damping is introduced by the upholstery of the seats. Less commonly, the effect can also be observed when the seats are occupied, but it is less pronounced, because interior volume and length of the ``resonators'' decrease.

\section{Small elements}\label{ch:small_elements}

Room acoustics is often optimized by means of relatively small elements, with surface areas much smaller than the room's boundary surfaces and not necessarily much larger than sound wavelengths of interest. Finiteness of their sizes and micro-geometry can essentially affect their acoustic behaviour. Two of such elements, reflectors and diffusers, are considered in the following. Similarly as in section~\ref{ch:basic_elements}, we estimate some relevant parameters and sound fields based on simple geometries of the elements. The focus is again on physical mechanisms which lead to (diffuse) reflections, rather than accuracy of the estimations.

\subsection{Rectangular surface with finite dimensions}\label{ch:rectangular_surface}

In contrast to section~\ref{ch:infinite_plane_surface}, reflecting surfaces in room often have dimensions which are not much larger than the sound wavelength, at least in certain (low) frequency ranges. As a consequence, less energy of the incident sound will be reflected specularly and more will diffract around or scatter at different angles to the surface, even if it is flat and uniform. In fact, simple rigid plates or more generally reflecting flat surfaces with finite dimensions are often used both as (specular) reflectors and diffusers. For geometric simplicity, we estimate first the sound reflected from a rectangular surface.

As in section~\ref{ch:basic_elements} and eq.~(\ref{eq:incident_plane_wave}), we assume that the surface is placed in \textbf{acoustic far field} and that the incident sound wave given by the pressure $p_i$ is plane. However, we cannot suppose only specular reflections and the form of the reflected sound from eq.~(\ref{eq:reflected_plane_wave_surface}). A more general solution for the reflected sound is necessary. Recall that the general solution of the time-independent Helmholtz equation was given by eq.~(\ref{eq:solution_Green_Helmholtz}). It captures effects of all reflecting surfaces in frequency domain (for stationary sound and stable conditions) and its integral form is suitable for the finite dimensions of the observed surface. Since we analyse reflections from a single solid body, we can treat it as placed in free space so the second integral in eq.~(\ref{eq:solution_Green_Helmholtz}) is over its surface only and the Green's function is the free-space Green's function. Furthermore, we replace $\hat{p} (\boldsymbol y)$ at the surface with the total (incident plus reflected sound) pressure amplitude $(1+\hat{R}_s(\theta_i)) \hat{p}_i(\boldsymbol y)$ and $\nabla_y \hat{p}(\boldsymbol y)$ with $-j\omega \rho_0 \hat{\boldsymbol v}(\boldsymbol y)$ from the conservation of momentum, eq.~(\ref{eq:Euler_momentum_sine_wave}). Here, $\theta_i$ is the angle of incidence and $\hat R_s$ is reflection coefficient of the surface. The volume integral represents the free-space component which is independent of the surface, that is, the incident sound $\hat{p}_i(\boldsymbol x)$. Hence, the equation becomes
\begin{equation}\label{eq:solution_Green_Helmholtz_from_inc_sound}
\hat{p} (\boldsymbol x) = \hat{p}_i(\boldsymbol x) - \oint_S \{ j\omega \rho_0 \hat{G} \hat{\boldsymbol v}(\boldsymbol y) \cdot \boldsymbol n + (1+\hat{R}_s(\theta_i)) \hat{p}_i(\boldsymbol y) (\nabla_y \hat{G}) \cdot \boldsymbol n \} d^2 \boldsymbol y.
\end{equation}

For a \textbf{rectangular} surface, we can define the Cartesian coordinates such that the surface lies in the plane $x_3 = 0$, centred at the origin, and extending from $x_1 = -L_1$ to $x_1 = L_1$ and from $x_2 = -L_2$ to $x_2 = L_2$. Moreover, due to the geometrical symmetry, we can suppose without a loss of generality that the incident sound reaches the surface from the octant\footnote{\label{ftn:incident_wave_direction}Notice that the direction of arrival of the incident sound is in a way opposite to the one in section~\ref{ch:infinite_plane_surface}. The incident wave travels ``backwards'' with the wave vector components $k_1,k_2,k_3 \le 0$ (the wave number remains positive nevertheless). The inclination and azimuthal angles of the wave vector are actually $\pi-\theta_i$ (for $0 \leq \theta_i \leq \pi/2$ defined as the usual angle of incidence to the normal of the surface) and $\phi_i + \pi$ (for $0 \leq \phi_i \leq \pi/2$), respectively, since the angles of the spherical coordinate system are defined with respect to the directions of the $x_3$-axis and $x_1$-axis. This explains the apparently opposite signs of the exponents in the last equality of eq.~(\ref{eq:incident_plane_wave_plate}) and eq.~(\ref{eq:incident_plane_wave_surface}), since $\cos(\pi - \theta_i) = -\cos(\theta_i)$, $\sin(\pi - \theta_i) = \sin(\theta_i)$, $\sin(\phi_i+\pi) = -\sin(\phi_i)$, and $\cos(\phi_i+\pi) = -\cos(\phi_i)$. Another difference from the geometry in section~\ref{ch:infinite_plane_surface} is that the surface lies in the plane $x_3 = 0$, not $x_1 = 0$.} $x_1,x_2,x_3 \ge 0$. The angle of incidence $0 \le \theta_i \le \pi/2$ is angle to the $x_3$-axis (corresponding to the inclination angle) and the angle of the projection of the plane wave's trajectory in the $x_1x_2$-plane to the $x_1$-axis, $0 \le \phi_i \le \pi/2$, corresponds to the azimuthal angle. The rectangular surface is not axisymmetric, so both angles must be taken into account and the problem has to be solved in all three spatial dimensions.

The incident plane sound wave can be written similarly as in eq.~(\ref{eq:incident_plane_wave_surface}):
\begin{equation}\label{eq:incident_plane_wave_plate}
\begin{aligned}
\hat{p}_i(\boldsymbol x) &= \hat{p}_Q e^{- j \boldsymbol k_i \cdot \boldsymbol x} = \hat{p}_Q e^{- j (k_{i1} x_1 + k_{i2} x_2 + k_{i3} x_3)} &\\
&= \hat{p}_Q e^{-j k [x_1 \sin(\pi-\theta_i) \cos(\phi_i+\pi) + x_2 \sin(\pi-\theta_i) \sin(\phi_i+\pi) + x_3 \cos(\pi-\theta_i)]}&\\
&= \hat{p}_Q e^{j k [x_1 \sin(\theta_i) \cos(\phi_i) + x_2 \sin(\theta_i) \sin(\phi_i) + x_3 \cos(\theta_i)]},&
\end{aligned}
\end{equation}
from which its value on the irradiated part of the surface equals
\begin{equation}\label{eq:incident_plane_wave_at_plate}
\begin{aligned}
\hat{p}_i(\boldsymbol y) &= \hat{p}_i(-L_1 \leq y_1 \leq L_1, -L_2 \leq y_2 \leq L_2, y_3 = 0) \\
&= \hat{p}_Q e^{j k [y_1 \sin(\theta_i) \cos(\phi_i) + y_2 \sin(\theta_i) \sin(\phi_i)]}.
\end{aligned}
\end{equation}
The fact that this pressure distribution at the surface is well defined makes eq.~(\ref{eq:solution_Green_Helmholtz}) useful for solving the scattering problem in free space, even though the field appears on its right-hand side. If additional surfaces were present, the distribution would depend on reflections from them, which would complicate the calculation, especially if multiple reflections are possible, as in closed spaces. 

Only the irradiated side of the surface reflects the sound back into the half space $x_3 \ge 0$, where we calculate the field. We shall not calculate the field behind the surface, which is due to diffraction. Therefore, the integral in eq.~(\ref{eq:solution_Green_Helmholtz_from_inc_sound}) is actually over the open irradiated flat surface $S$ with area $(2L_1)(2L_2)=4 L_1 L_2$ and with $\boldsymbol n = \boldsymbol e_3$ pointing outwards from it (into the field). Total sound pressure in front of the surface equals
\begin{equation}\label{eq:solution_Green_Helmholtz_from_inc_sound_z3}
\hat{p} (\boldsymbol x) = \hat{p}_i(\boldsymbol x) - \int_S \{ j\omega \rho_0 \hat{G} \hat{v}_3(\boldsymbol y) + (1+\hat{R}_s(\theta_i)) \hat{p}_i(\boldsymbol y) \frac{\partial \hat{G}}{\partial y_3} \} d^2 \boldsymbol y.
\end{equation}
From the definition of the surface impedance $Z_s$ in eq.~(\ref{eq:impedance}),
\begin{equation}
\begin{aligned}
\hat{v}_3(\boldsymbol y) &= -\frac{\hat{p}(\boldsymbol y)}{Z_s(\boldsymbol y)} = -\frac{(1+\hat{R}_s(\theta_i)) \hat{p}_i(\boldsymbol y)}{Z_s(\boldsymbol y)} &\\& = - \frac{(1+\hat{R}_s(\theta_i)) \hat{p}_i(\boldsymbol y) \cos(\theta_i) (1-\hat{R}_s(\theta_i))}{Z_0 (1+\hat{R}_s(\theta_i))} &\\
&= - \frac{\hat{p}_i(\boldsymbol y) \cos(\theta_i) (1-\hat{R}_s(\theta_i))}{\rho_0 c_0},&
\end{aligned}
\end{equation}
where eq.~(\ref{eq:impedance_plane_surface_R}) was also used. The minus sign appears in the first equality because $\boldsymbol e_3$ (in $\hat v_3 = \hat {\boldsymbol v} \cdot \boldsymbol e_3$) points outwards from the surface, while the impedance is defined for a unit normal vector pointing into the surface. Using this in eq.~(\ref{eq:solution_Green_Helmholtz_from_inc_sound_z3}) to remove the particle velocity,
\begin{equation}\label{eq:solution_Green_Helmholtz_from_inc_sound_impedance_y3}
\hat{p} (\boldsymbol x) = \hat{p}_i(\boldsymbol x) + \int_S \{ jk \hat{G} \hat{p}_i(\boldsymbol y) \cos(\theta_i) (1-\hat{R}_s(\theta_i)) - (1+\hat{R}_s(\theta_i)) \hat{p}_i(\boldsymbol y) \frac{\partial \hat G}{\partial y_3} \} d^2 \boldsymbol y.
\end{equation}

Free space Green's function is given in eq.~(\ref{eq:Green_Helmholtz_free_space}),
\begin{equation}\label{eq:Green_Helmholtz_half_free_space}
\hat{G}(\boldsymbol{x}|\boldsymbol{y}) = \frac{e^{-j k |\boldsymbol x-\boldsymbol y|}}{4 \pi |\boldsymbol x-\boldsymbol y|} = \frac{e^{-j k r}}{4 \pi r},
\end{equation}
where $r = |\boldsymbol x-\boldsymbol y|$. Although bulk of the surface reflects (or radiates as a secondary source) effectively into half space, we should calculate with the usual free-space Green's function, because contribution of the reflection from the surface is already accounted for by the factor $(1+\hat{R_s})$ in eq.~(\ref{eq:solution_Green_Helmholtz_from_inc_sound}). The Green's function depends on the location at the surface, $\boldsymbol y$, which makes the integration over $S$ non-trivial when the surface is not acoustically compact. However, if we are interested in the sound field far from the surface, at distances much larger than both $L_1$ and $L_2$, the denominator can be approximated as $4 \pi |\boldsymbol x-\boldsymbol y| \approx 4 \pi |\boldsymbol x|$. This is \textbf{geometric far field} ($|\boldsymbol x| \gg |\boldsymbol y|$), in which sound attenuation due to the decay of 6\,dB per doubling the distance does not depend significantly on the exact location in the (secondary) source region centred at $\boldsymbol y = 0$. Unlike the acoustic far field, in which $|\boldsymbol x - \boldsymbol y| \approx |\boldsymbol x| \gg 1/k = \lambda/(2\pi)$ (and which will also be assumed below in eq.~(\ref{eq:Green_Helmholtz_half_free_space_derivative})), it is defined with respect to the dimensions of the source region (or the reflecting body), not the sound wavelength.

The approximation $|\boldsymbol x-\boldsymbol y| \approx |\boldsymbol x|$ is justified for the negligible decay of amplitude with $\boldsymbol y$ (compared to the decay with $\boldsymbol x$). However, it cannot be readily applied in the exponent of the Green's function, $-jk|\boldsymbol x - \boldsymbol y|$, since phase of the wave can still vary significantly with location at the surface $\boldsymbol y$, depending on its size and the frequency (that is, the associated Helmholtz number), unless $k |\boldsymbol y| \ll 1$. We can use the low of cosines instead to write
\begin{equation}\label{eq:r_far_field_approx}
r = |\boldsymbol x - \boldsymbol y| = \sqrt{|\boldsymbol x|^2 - 2 \boldsymbol x \cdot \boldsymbol y + |\boldsymbol y|^2} = |\boldsymbol x| \sqrt{1-\frac{2 \boldsymbol x \cdot \boldsymbol y}{|\boldsymbol x|^2}+\frac{|\boldsymbol y|^2}{|\boldsymbol x|^2}}
\end{equation}
for $|\boldsymbol x| \neq 0$. The Taylor expansion $\sqrt{1-a} = 1 - a/2 + \mathcal{O}(a^2) \approx 1-a/2$ for $a \ll 1$ gives
\begin{equation}
\begin{aligned}
|\boldsymbol x - \boldsymbol y| &\approx |\boldsymbol x| \left[ 1-\frac{\boldsymbol x \cdot \boldsymbol y}{|\boldsymbol x|^2}+\frac{|\boldsymbol y|^2}{2|\boldsymbol x|^2} + \mathcal{O} \left( \left(\frac{2 \boldsymbol x \cdot \boldsymbol y}{|\boldsymbol x|^2}-\frac{|\boldsymbol y|^2}{|\boldsymbol x|^2}\right)^2 \right) \right] &\\
&= |\boldsymbol x| \left[ 1-\frac{\boldsymbol x \cdot \boldsymbol y}{|\boldsymbol x|^2} + \mathcal{O} \left(\frac{|\boldsymbol y|^2}{|\boldsymbol x|^2} \right) \right],&
\end{aligned}
\end{equation}
where we used $\mathcal{O}(|\boldsymbol x \cdot \boldsymbol y|) = \mathcal{O}(|\boldsymbol x| |\boldsymbol y|)$ and $\mathcal{O}(a^2)$ denotes order of $a^2$ or higher. Since $|\boldsymbol y| \ll |\boldsymbol x|$,
\begin{equation}\label{eq:r_far_field_approx_Taylor}
r = |\boldsymbol x - \boldsymbol y| \approx |\boldsymbol x|-\frac{\boldsymbol x \cdot \boldsymbol y}{|\boldsymbol x|}.
\end{equation}
If we express $\boldsymbol x$ and $\boldsymbol y$ with the spherical coordinates $(r,\theta,\phi)$ and $(r_y,\theta_y,\phi_y)$, respectively, with $0 \leq \theta \leq \pi$, $\theta_y = \pi/2$, and $0 \leq \phi,\phi_y < 2\pi$, their scalar product reads
\begin{equation}
\boldsymbol x \cdot \boldsymbol y = |\boldsymbol x||\boldsymbol y| \cos(\theta - \pi/2) \cos(\phi - \phi_y) = |\boldsymbol x||\boldsymbol y| \sin(\theta) \cos(\phi - \phi_y)
\end{equation}
and therefore
\begin{equation}
\begin{aligned}
r &= |\boldsymbol x - \boldsymbol y| \approx |\boldsymbol x|-|\boldsymbol y| \sin(\theta) \cos(\phi - \phi_y) &\\
&= |\boldsymbol x|-|\boldsymbol y| \sin(\theta) \left[ \cos(\phi) \cos(\phi_y) + \sin(\phi) \sin(\phi_y) \right] &\\
&= |\boldsymbol x|-\sin(\theta) \left[ y_1 \cos(\phi) + y_2 \sin(\phi) \right].&
\end{aligned}
\end{equation}
We can insert this into eq.~(\ref{eq:Green_Helmholtz_half_free_space}) to obtain an approximation of the free-space Green's function in the geometric far field,
\begin{equation}\label{eq:Green_Helmholtz_half_free_space1}
\hat{G}(\boldsymbol{x}|\boldsymbol{y}) = \frac{1}{4 \pi |\boldsymbol x|} e^{-j k \left( |\boldsymbol x|- y_1 \sin(\theta) \cos(\phi) - y_2 \sin(\theta) \sin(\phi) \right) }.
\end{equation}

In order to calculate the spatial derivative of the Green's function which appears in eq.~(\ref{eq:solution_Green_Helmholtz_from_inc_sound_impedance_y3}), we first notice that
\begin{equation}\label{free_space_G_geometric_far_field_gradient}
\frac{\partial \hat{G}}{\partial y_3} = \frac{\partial \hat{G}}{\partial r}\frac{\partial r}{\partial y_3},
\end{equation}
where (compare with eq.~(\ref{eq:p_solution_tailored_Green_wave_eq_free_space_compact_source_emission_time}))
\begin{equation}\label{eq:dG_dr_derivation}
\frac{\partial \hat{G}}{\partial r} = \frac{\partial}{\partial r}\frac{e^{-j k r}}{4 \pi r} = \frac{1}{4 \pi r} (-jk) e^{-j k r} + \frac{e^{-j k r}}{4 \pi} \left( -\frac{1}{r^2} \right) = \frac{e^{-j k r}}{4 \pi r} \left( -jk-\frac{1}{r} \right).
\end{equation}
On the other hand, $r = |\boldsymbol x - \boldsymbol y|$, $\boldsymbol n = \boldsymbol e_3$, and
\begin{equation}
\begin{aligned}
\frac{\partial r}{\partial y_3} &= (\nabla_y r) \cdot \boldsymbol n = -(\nabla_x r) \cdot \boldsymbol n = -\frac{\partial r}{\partial x_3} &\\
&= -\frac{\partial}{\partial x_3} (|\boldsymbol x|-\sin(\theta) \left[ y_1 \cos(\phi) + y_2 \sin(\phi) \right]) &\\
&= -\frac{\partial}{\partial x_3} (x_1 \sin(\theta) \cos(\phi) + x_2 \sin(\theta) \sin(\phi) + x_3 \cos(\theta) &\\
&-\sin(\theta) \left[ y_1 \cos(\phi) + y_2 \sin(\phi) \right])
= -\cos(\theta).
\end{aligned}
\end{equation}
In the \textbf{acoustic far field} of the surface, $r \gg 1/k$ and we can neglect the second term in the last equality in eq.~(\ref{eq:dG_dr_derivation}) in favour of the first term (as in eq.~(\ref{eq:v_solution_tailored_Green_wave_eq_free_space_compact_source_emission_time_far_field})). Equation~(\ref{free_space_G_geometric_far_field_gradient}) becomes
\begin{equation}\label{eq:Green_Helmholtz_half_free_space_derivative}
\begin{aligned}
\frac{\partial \hat{G}}{\partial y_3} &= \frac{\partial \hat{G}}{\partial r}\frac{\partial r}{\partial y_3} = jk \cos(\theta) \frac{e^{-j k r}}{4 \pi r} = jk \cos(\theta) \hat{G}.
\end{aligned}
\end{equation}
We can now insert the last equality into eq.~(\ref{eq:solution_Green_Helmholtz_from_inc_sound_impedance_y3}) to obtain
\begin{equation}\label{eq:solution_Green_Helmholtz_from_inc_sound_impedance_y3_G}
\begin{aligned}
\hat{p}& (\boldsymbol x) = \hat{p}_i(\boldsymbol x) + j k \int_S \hat{G} \hat{p}_i(\boldsymbol y) \{ \cos(\theta_i) (1-\hat{R}_s(\theta_i)) - (1+\hat{R}_s(\theta_i)) \cos(\theta) \} d^2 \boldsymbol y &\\
&= \hat{p}_i(\boldsymbol x) - j k \int_S \hat{G} \hat{p}_i(\boldsymbol y) \{ \hat{R}_s(\theta_i) (\cos(\theta) + \cos(\theta_i)) + (\cos(\theta) - \cos(\theta_i)) \} d^2 \boldsymbol y.&
\end{aligned}
\end{equation}

From eq.~(\ref{eq:solution_Green_Helmholtz_from_inc_sound_impedance_y3_G}) it is evident that angular distribution of the scattered sound amplitude (the second term) depends on the reflection coefficient values at the surface $S$, independent from the dimensions of the surface and regardless of its flat shape. Additional scattering (and diffraction) can be expected at its edges. If $\hat R_s$ does not depend on the angle of incidence, the reflected sound pressure amplitude $\hat p - \hat p_i$ equals from equations (\ref{eq:solution_Green_Helmholtz_from_inc_sound_impedance_y3_G}), (\ref{eq:incident_plane_wave_at_plate}), and (\ref{eq:Green_Helmholtz_half_free_space1})
\begin{equation}\label{eq:solution_Green_Helmholtz_from_inc_sound_impedance_y3_G1_Ry1}
\begin{aligned}
\hat{p}_r & (\boldsymbol x) = -j k \int_S \hat{G} \hat{p}_i(\boldsymbol y) \{ \hat{R}_s(\boldsymbol y) (\cos(\theta) + \cos(\theta_i)) + (\cos(\theta) - \cos(\theta_i)) \} d^2 \boldsymbol y &\\
&= -j k \int_{-L_1}^{L_1} \int_{-L_2}^{L_2} \hat{G} \hat{p}_i(y_3 = 0) \{ \hat{R}_s(y_1, y_2) (\cos(\theta) + \cos(\theta_i)) &\\& + (\cos(\theta) - \cos(\theta_i)) \} dy_2 dy_1
= -\frac{j k \hat{p}_Q}{4 \pi |\boldsymbol x|} e^{-j k |\boldsymbol x|} &\\& \int_{-L_1}^{L_1} \{ \hat{R}_s(y_1) (\cos(\theta) + \cos(\theta_i)) + (\cos(\theta) - \cos(\theta_i)) \} e^{j A y_1 } dy_1 \int_{-L_2}^{L_2} e^{j B y_2} dy_2.&
\end{aligned}
\end{equation}
For simplicity, we further restricted $\hat{R}_s$ in the last equality to be a function of $y_1$ only. We also introduced the parameters $A = k [\sin(\theta_i) \cos(\phi_i) + \sin(\theta) \cos(\phi)]$ and $B = k [\sin(\theta_i) \sin(\phi_i) + \sin(\theta) \sin(\phi)]$ for brevity. Furthermore, the last two integrals represent two (or a single two-dimensional) inverse \textbf{spatial Fourier transforms} analogue to eq.~(\ref{eq:inv_Fourier_p}), namely $\mathcal{F}_{y_1}^{-1} \{ \hat{R}_s(y_1) (\cos(\theta) + \cos(\theta_i)) + (\cos(\theta) - \cos(\theta_i)) \}$ and $\mathcal{F}_{y_2}^{-1} \{ 1 \}$, respectively, with the angle-dependent parameters $A$ and $B$ instead of time, spatial coordinates $y_1$ and $y_2$ instead of angular frequency, and over the finite intervals $(-L_1,L_1)$ and $(-L_2,L_2)$. In the following two subsections we inspect different forms of the function $\hat R_s(y_1)$ and the resulting reflected sound field.

\subsubsection{Hard plate as a reflector}\label{ch:rigid_motionless_plate}

If the rectangular surface is acoustically \textbf{hard} (a rigid, motionless, fully reflecting plate), the reflection coefficient is constant, $\hat{R}_s = 1$. The first term in the curly brackets in the first integral in eq.~(\ref{eq:solution_Green_Helmholtz_from_inc_sound_impedance_y3_G}) vanishes, so the reflected sound equals
\begin{equation}\label{eq:solution_Green_Helmholtz_plate}
\hat{p}_r (\boldsymbol x) = -2 j k \cos(\theta) \int_S \hat{G} \hat{p}_i(\boldsymbol y) d^2 \boldsymbol y = -2 jk \cos(\theta) \int_{-L_1}^{L_1} \int_{-L_2}^{L_2} \hat{G} \hat{p}_i(y_3 = 0) d y_1 d y_2.
\end{equation}
Notice the similarity with eq.~(\ref{eq:solution_tailored_Green_Helmholtz}) for $\hat G_{tail} = \hat G_{free} = \hat G$. Indeed, the surface acts as a secondary source in free space, with the source function which depends on the incident sound. Moreover, $\cos(\theta)$ indicates a dipole mechanism of radiation (recall Fig.~\ref{fig:directivity_monopole_dipole}), which is typical for acoustically hard surfaces, as well as the far-field ``gradient factor'' $-jk$ (the first term in eq.~(\ref{eq:p_solution_tailored_Green_wave_eq_free_space_compact_source_emission_time})), which limits the efficiency of radiation at low frequencies. Nevertheless, additional effects due to the finite dimensions of the surface (the two integrals) will further determine the angular dependence of $\hat p_r$. Finally, the multiplication with 2 is due to $\hat R_s = 1$ and superposition of the incident and reflected sound at the surface. This factor can also be absorbed into $\hat G$ to give the half-space Green's function with $2\pi r$ in the denominator.

\begin{figure}[h]
	\centering
	\begin{subfigure}{.33\textwidth}
		\centering
		\includegraphics[width=1\linewidth]{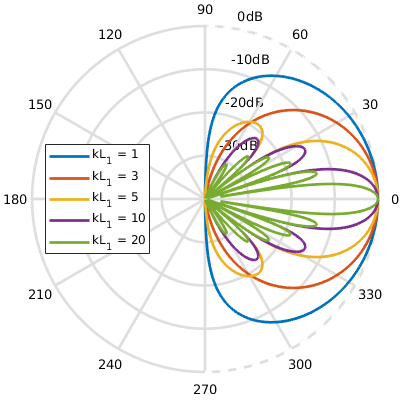}
		\label{fig:rect_plate_thetai_0}
	\end{subfigure}%
	\begin{subfigure}{.33\textwidth}
		\centering
		\includegraphics[width=1\linewidth]{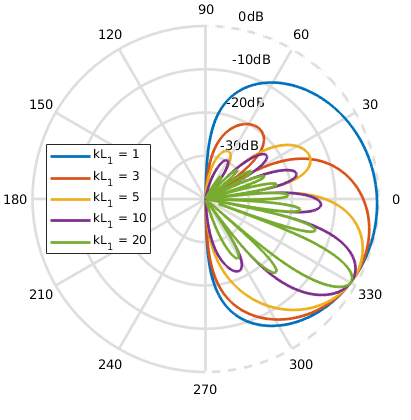}
		\label{fig:rect_plate_thetai_pi_6}
	\end{subfigure}
	\begin{subfigure}{.33\textwidth}
		\centering
		\includegraphics[width=1\linewidth]{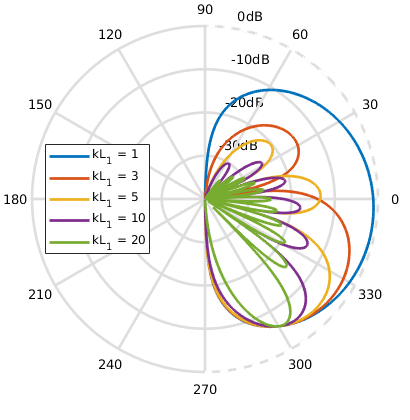}
		\label{fig:rect_plate_thetai_pi_3}
	\end{subfigure}
	\caption{Scattering pattern of an acoustically hard rectangular plate, the function $20\log_{10}(|\hat p_r| \pi |\boldsymbol x| /(2 k L_1 L_2 |\hat p_Q| ))$ with $\hat p_r$ given in eq.~(\ref{eq:solution_Green_Helmholtz_plate3}), calculated in the $x_1x_3$-plane ($\phi = 0$ and $-\pi/2 \leq \theta \leq \pi/2$ in the shown polar coordinates, where $\theta < 0$ corresponds to $-\theta$ and $\phi=\pi$ for the spherical coordinates used in the text), with $kL_2 = 5$, $\phi_i = 0$, and (left) $\theta_i = 0$, (middle) $\theta_i = \pi/6$, (right) $\theta_i = \pi/3$.}
	\label{fig:rect_plate_scattering}
\end{figure}

From eq.~(\ref{eq:solution_Green_Helmholtz_from_inc_sound_impedance_y3_G1_Ry1}),
\begin{equation}\label{eq:solution_Green_Helmholtz_plate1}
\begin{split}
\hat{p}_r (\boldsymbol x) = -\frac{jk \cos(\theta) \hat{p}_Q}{2 \pi |\boldsymbol x|} e^{-jk|\boldsymbol x|} \int_{-L_1}^{L_1} e^{j A y_1} dy_1 \int_{-L_2}^{L_2} e^{j B y_2} dy_2.
\end{split}
\end{equation}
Both Fourier integrals in eq.~(\ref{eq:solution_Green_Helmholtz_plate1}) can be solved analytically to give
\begin{equation}\label{eq:solution_Green_Helmholtz_plate2}
\begin{aligned}
\hat{p}_r (\boldsymbol x) &= -\frac{jk \cos(\theta) \hat{p}_Q}{2 \pi |\boldsymbol x|} e^{-jk|\boldsymbol x|} \frac{e^{j A L_1} - e^{-j A L_1}}{jA}
\frac{e^{j B L_2} - e^{-j B L_2}}{jB} &\\
&= \frac{jk \cos(\theta) \hat{p}_Q}{2 \pi |\boldsymbol x|} e^{-jk|\boldsymbol x|} \frac{2 j \sin(A L_1)}{A}
\frac{2 j \sin(B L_2)}{B} &\\
&= -\frac{2jk L_1 L_2 \cos(\theta) \hat{p}_Q}{\pi |\boldsymbol x|} e^{-jk|\boldsymbol x|} \frac{\sin(A L_1)}{A L_1}
\frac{\sin(B L_2)}{B L_2}.&
\end{aligned}
\end{equation}
The last two fractions are sinc functions\footnote{Sinc function is indeed Fourier transform of a rectangular pulse. In the limiting case when length of the pulse approaches zero as for a delta impulse (here for $L_1 \rightarrow 0$ or $L_2 \rightarrow 0$), the sinc function is approximately constant, $\lim_{x \rightarrow 0} \text{sinc} (x) = 1$.} (or spherical Bessel functions of the first kind and zero order, $j_0$), $\text{sinc}(x) = j_0(x) = \sin(x)/x$. Therefore, we can write
\begin{equation}\label{eq:solution_Green_Helmholtz_plate3}
\begin{aligned}
\hat{p}_r (\boldsymbol x) &= -\frac{2jk L_1 L_2 \cos(\theta) \hat{p}_Q}{\pi |\boldsymbol x|} e^{-jk|\boldsymbol x|} \text{sinc}(A L_1) \text{sinc}(B L_2) &\\
&= -\frac{2jk L_1 L_2 \cos(\theta) \hat{p}_Q}{\pi |\boldsymbol x|} e^{-jk|\boldsymbol x|} \text{sinc}(k L_1 [\sin(\theta_i) \cos(\phi_i) + \sin(\theta) \cos(\phi)]) &\\
&\text{sinc}(k L_2 [\sin(\theta_i) \sin(\phi_i) + \sin(\theta) \sin(\phi)]).&
\end{aligned}
\end{equation}
Normalized value of $20 \log_{10}(|\hat p_r|)$, the function $20 \log_{10} (|\cos(\theta) \text{sinc}(AL_1) \text{sinc}(BL_2)|)$, is shown in Fig.~\ref{fig:rect_plate_scattering} for three different angles of incidence and five sizes of the rectangular plate (values of the Helmholtz number associated with the dimension $L_1$, $kL_1$). It represents \textbf{scattering pattern} of the surface.

The scattering pattern depends primarily on the Helmholtz numbers $kL_1$ and $kL_2$, defined with regard to the two dimensions of the plate, and its orientation on the angle of incidence. Lower values of the Helmholtz numbers imply fewer side lobes of less directional scattering patterns (with less dominant specular reflections). This behaviour is dictated by the sinc functions. When $kL_1$ and $kL_2$ approach zero, the scattering pattern differs from an omnidirectional only by the factor $\cos(\theta)$ (the blue curve in Fig.~\ref{fig:rect_plate_scattering}~(middle) approaches the red curve in Fig.~\ref{fig:directivity_monopole_dipole}). However, this does not mean that low Helmholtz number values should be preferred for diffuse surfaces, because the lower values also imply less efficiency of the reflector (lower total reflected energy). This is determined by the factor $k L_1 L_2$ in eq.~(\ref{eq:solution_Green_Helmholtz_plate3}), which is not captured by the scattering patterns due to the normalization. An acoustically compact plate is transparent for the incoming sound which diffracts around it. As values of the Helmholtz numbers increase, less energy of the incoming sound is diffracted and more energy is reflected specularly. Acoustically large hard plates behave essentially like hard walls and reflect mostly specularly with zero absorption.

Still, it should be mentioned that $\hat{p}_r$ in eq.~(\ref{eq:solution_Green_Helmholtz_plate3}) does not converge to a single specular reflection from an infinite hard wall when $kL_1,kL_2 \rightarrow \infty$. From eq.~(\ref{eq:incident_plane_wave_plate}) with $k_{r3} = -k_{i3}$ (because the surface is normal to the $x_3$-axis), the specular reflection is given by \begin{equation}\label{eq:incident_plane_wave_plate1}
\begin{split}
\hat{p}_r(\boldsymbol x) = \hat{p}_Q e^{j k [x_1 \sin(\theta_i) \cos(\phi_i) + x_2 \sin(\theta_i) \sin(\phi_i) - x_3 \cos(\theta_i)]},
\end{split}
\end{equation}
which is similar to eq.~(\ref{eq:reflected_plane_wave_surface}) with $\hat R_s = 1$. The reason for this mismatch is the assumed geometric far field, $|\boldsymbol x| \gg |\boldsymbol y|$, in eq.~(\ref{eq:r_far_field_approx_Taylor}). This condition cannot be satisfied for every $\boldsymbol y$ on the plate if it is infinitely large (while the wave number is finite). Hence, the model is inadequate for very large plates, because contributions of the reflections at the points which do not satisfy $|\boldsymbol y| \ll |\boldsymbol x|$ are not calculated accurately. It should also be noted that the reflected energy depends on the amplitude of the incident sound and distance from the surface. These contributions are also excluded from the normalized scattering patterns.

Figure~\ref{fig:rect_plate_scattering} demonstrates that even flat acoustically hard surfaces can exhibit complicated acoustic behaviour. Depending on their finite sizes and frequency of the sound, they can act as (specular) reflectors, diffuse, or acoustically transparent surfaces. Such multifaceted function can be advantageous. For example, reflecting plates of appropriate sizes can be used as a simple and inexpensive alternative to more sophisticated diffusers (which will be discussed shortly). In concert halls and opera houses, they can scatter the mid-frequency sound of a singer's voice (singer's formant) and remain transparent for the low-frequency sound of other, more powerful sources on the stage. As both reflectors and diffusers, they can substitute absorbers in rooms, when damping should be avoided, but strong (typically first-order) reflections from particular surfaces (with more or less well-defined direction of arrival of the incident sound) should be suppressed. Placed in front of such surfaces and properly oriented, they can redirect or scatter the reflected energy away from the listener to other parts of the room.

Reflectors as separate acoustic elements are useful not only for avoiding reflections, but also for shortening their propagation paths, when room boundaries are too remote to provide desired early reflections. Typical examples are reflectors hanged on high ceilings of large halls. Apart from being made of rigid materials (acoustically hard), curved shapes are often preferred over flat geometries, especially when the locations of sources and listeners are distributed or changeable. Coverage of the auditorium with the sound energy reflected from such surfaces can be estimated using ray tracing simulations. Pronounced resonances of the reflectors in the frequency range of interest should be avoided in general, since they may cause a prolonged reverberation time or, if damped, unexpected increase of absorption around the resonance frequency.

\subsubsection{Schroeder diffuser}\label{ch:Schroeder_diffuser}

In section~\ref{ch:rectangular_surface} we saw that angular distribution of sound energy reflected from a flat rectangular surface depend on the spatial distribution of reflection coefficient at the surface. While the distribution in section~\ref{ch:rigid_motionless_plate} was uniform, some non-uniform surfaces can be treated analytically, as well. If we limit ourselves to relatively \textbf{low angles} of incidence and reflections to the surface normal, say $\theta, \theta_i < \pi /4$, then $|\cos(\theta)-\cos(\theta_i)| \ll |\cos(\theta)+\cos(\theta_i)|$ in equations~(\ref{eq:solution_Green_Helmholtz_from_inc_sound_impedance_y3_G}) and (\ref{eq:solution_Green_Helmholtz_from_inc_sound_impedance_y3_G1_Ry1}). Moreover, if $|\hat{R}_s| \rightarrow 1$ (which is more general than $\hat R_s = 1$; we allow phase changes at the \textbf{fully reflecting}, not necessarily hard rectangular surface with zero absorption), eq.~(\ref{eq:solution_Green_Helmholtz_from_inc_sound_impedance_y3_G}) simplifies to
\begin{equation}\label{eq:solution_Green_Helmholtz_from_inc_sound_impedance_y3_G_low_angles}
\begin{aligned}
\hat{p}_r& (\boldsymbol x) = - j k \int_S \hat{G} \hat{p}_i(\boldsymbol y) \hat{R}_s(\theta_i) (\cos(\theta) + \cos(\theta_i)) d^2 \boldsymbol y.&
\end{aligned}
\end{equation}
Strength of the secondary source and its radiation pattern depend on both incident sound (including the angle of incidence) and reflection coefficient. For $\hat R_s = 1$ and specular reflections ($\theta = \theta_i < \pi/4$), the equation reduces to eq.~(\ref{eq:solution_Green_Helmholtz_plate}).

Equation~(\ref{eq:solution_Green_Helmholtz_from_inc_sound_impedance_y3_G1_Ry1}) simplifies to
\begin{equation}\label{eq:solution_Green_Helmholtz_from_inc_sound_impedance_y3_G1_Ry2}
\begin{aligned}
\hat{p}_r (\boldsymbol x) &= -\frac{j k \hat{p}_Q}{4 \pi |\boldsymbol x|} e^{-j k |\boldsymbol x|} (\cos(\theta) + \cos(\theta_i)) \int_{-L_1}^{L_1} \hat{R}_s(y_1) e^{j A y_1} dy_1 \int_{-L_2}^{L_2} e^{j B y_2} dy_2 &\\
&= -\frac{j k L_2 \hat{p}_Q}{2 \pi |\boldsymbol x|} e^{-j k |\boldsymbol x|} \text{sinc}(BL_2)(\cos(\theta) + \cos(\theta_i)) \int_{-L_1}^{L_1} \hat{R}_s(y_1) e^{j A y_1} dy_1.&
\end{aligned}
\end{equation}
Ignoring the multiplication factor in front of the integral\footnote{According to the assumption $0 \leq \theta, \theta_i < \pi /4$, so the factor $\sqrt{2} < (\cos(\theta) + \cos(\theta_i)) \leq 2$ does not substantially affect the reflected energy or scattering pattern. For the remaining terms, the same remarks apply as in eq.~(\ref{eq:solution_Green_Helmholtz_plate3}).}, the angularly dependent reflected sound is determined by the inverse spatial Fourier transform of the reflection coefficient, $\mathcal{F}_{y_1}^{-1} \{ \hat{R}_s(y_1) \}$, with the parameter $A = k [\sin(\theta_i) \cos(\phi_i) + \sin(\theta) \cos(\phi)]$. If we observe the field only in the $x_1 x_3$-plane\footnote{In general, less diffusion can be expected in other planes, which are not considered here, and efficiency of a diffuser depends on the angle of incidence. For example, since $\hat R_s(y_1)$ does not change along the $x_2$-axis, the surface acts as a homogeneous surface in the $x_2x_3$-plane.} and switch to polar coordinates (letting $-\pi/4 < \theta, \theta_i < \pi/4$, where negative values represent $\phi = \pi$ and $\phi_i = \pi$, respectively, so $\phi = \phi_i = 0$ is forced, as in Fig.~\ref{fig:rect_plate_scattering}), the parameter $A$ becomes $k [\sin(\theta_i) + \sin(\theta)]$ and the reflected sound is proportional to
\begin{equation}\label{eq:solution_Green_Helmholtz_from_inc_sound_impedance_y3_G1_Ry2_proportional}
\begin{split}
\hat{p}_r (\boldsymbol x) \propto \int_{-L_1}^{L_1} \hat{R}_s(y_1) e^{j k [\sin(\theta_i) + \sin(\theta)] y_1} dy_1.
\end{split}
\end{equation}
As in eq.~(\ref{eq:solution_Green_Helmholtz_plate3}), the total reflected energy depends also on the Helmholtz numbers $kL_1$ and $kL_2$, amplitude of the incident sound, and distance from the surface. However, we focus on the normalized scattering patterns in the following. We also ignore the ``flat hard plate'' contribution, $\text{sinc}(BL_2)$, which amounts to the assumption $kL_2 \gg 1$ (scattering from the edges of the surface at $x_2 = -L_2$ and $x_2 = L_2$ is neglected and the surface reflects specularly in the $x_2 x_3$-plane). 

Equation~\ref{eq:solution_Green_Helmholtz_from_inc_sound_impedance_y3_G1_Ry2_proportional} points to the key idea behind (one-dimensional, since $\hat R_s$ depends only on $y_1$) \textbf{Schroeder diffusers}. It is to select such surface distribution of the reflection coefficient which maximizes diffusion, that is, gives approximately constant value (independent of $\theta$ and $\theta_i$) of the Fourier integral, at least in a certain frequency range. Of course, $kL_1 \ll 1$ must not hold, which would make the diffuser acoustically compact and diminish the reflection altogether. In fact, angular distribution of the reflected sound amplitude cannot be uniform, because parameter of the transform in the exponent in eq.~(\ref{eq:solution_Green_Helmholtz_from_inc_sound_impedance_y3_G1_Ry2_proportional}) depends not only on $\theta$, but also on $\theta_i$ and $k$. The scattering patterns contain peaks and lobes separated by dips. The total number and widths of the lobes depend on frequency and size of the diffuser in a similar manner as for a rectangular plate (which is, indeed, a special case for $\hat R_s=1$). Still, much more omnidirectional scattering patterns can be achieved than those in Fig.~\ref{fig:rect_plate_scattering}, even for relatively large Helmholtz number values.

In practical realizations, one-dimensional Schroeder diffusers are divided (here over the length $2L_1$) into $N$ equally wide homogeneous sections, each with the width $l_1 = 2L_1/N$. In reality, $N$ rarely exceeds 20. Different values (corresponding to different sections) of reflection coefficient at the rectangular (imaginary) interface surface of a diffuser are given with a discrete function $\hat{R}_s(y_1) = \hat{R}_n$, for $n = 1, 2,... N$. Accordingly, the integral in eq.~(\ref{eq:solution_Green_Helmholtz_from_inc_sound_impedance_y3_G1_Ry2_proportional}) can be expressed as a sum over $n$. If $k l_1 \ll 1$ (\textbf{acoustically narrow sections}), we can replace $y_1$ with $n l_1 - L_1$ (the phase does not change significantly over one section width) and write
\begin{equation}\label{eq:Schroeder_scattering}
\begin{split}
\hat{p}_r (\boldsymbol x) \propto \sum_{n = 1}^{N} \hat{R}_n e^{j k [\sin(\theta_i) + \sin(\theta)] (n l_1 - L_1)}.
\end{split}
\end{equation}
Next, we assume that the sequence of reflection coefficients $\hat R_n$ \textbf{repeats} along the $x_1$-axis with the period $2L_1$. As a consequence, the scattering pattern (inverse Fourier transform) exhibits angular periodicity. Scattering maxima\footnote{These peaks and lobes are due to the repetitive pattern of the inhomogeneous surface and should be distinguished from the lobes discussed above, in the context of eq.~(\ref{eq:solution_Green_Helmholtz_from_inc_sound_impedance_y3_G1_Ry2_proportional}), which occur even for a hard plate.} are at the angles $\theta$ which satisfy
\begin{equation}\label{eq:Schroeder_lobes_angles}
\sin(\theta) + \sin(\theta_i) = \frac{m \pi}{k L_1} = \frac{m \lambda}{2 L_1},
\end{equation}
where $m$ is an integer called diffraction order. For $m=0$, the lobe is specular, with the angle $\theta = -\theta_i$ in polar coordinates. The sum becomes
\begin{equation}\label{eq:Schroeder_scattering2}
\begin{aligned}
\hat{p}_r (\boldsymbol x) &\propto \sum_{n = 1}^{N} \hat{R}_n e^{j k m \lambda (n l_1 - L_1) / (2 L_1)} = \sum_{n = 1}^{N} \hat{R}_n e^{j 2 \pi m (n l_1 - N l_1/2) / (N l_1)} &\\
&= e^{-j \pi m} \sum_{n = 1}^{N} \hat{R}_n e^{j 2 \pi m n / N}.&
\end{aligned}
\end{equation}
Since $m$ is integer, $e^{-j\pi m} = \pm 1$ determines only the sign of the expression, which can be absorbed into the omitted multiplication factor from eq.~(\ref{eq:solution_Green_Helmholtz_from_inc_sound_impedance_y3_G1_Ry2}).

Different reflection coefficient values are due to different phase shifts $\Phi$ of the reflected waves in the sections, that is, $\hat{R}_n = |\hat{R}| e^{j \Phi_n}$, with $|\hat R| \rightarrow 1$. In the most common Schroeder diffusers, the phase shifts are achieved with different depths $d_n$ of the sections (\textbf{wells}) with equal widths $l_1$, separated from each other by thin rigid plates. Free edges of the plates are parallel to each other and lie in the plane of the rectangular interface surface of the diffuser, $x_3 = 0$. At this outer surface of the diffuser\footnote{Modelling diffuser as a plane rectangular surface from eq.~(\ref{eq:solution_Green_Helmholtz_from_inc_sound}) does not capture sound propagation behind the surface. Consequently, all considered effects at the surface and behind it are contained in the reflection coefficients $R_n$, which are assumed to be constant for each well. Hence, details of sound propagation inside the wells, such as reflections from the side walls of the wells, cannot be captured with this model, although they can affect performances of the diffusers, for example, by increasing damping in the wells.} (its interface to the room), phase shift of a plane wave propagating normal to the surface and reflecting back from the supposedly acoustically hard bottom of the well $n$ is $\Phi_n = 2 \pi (2 d_n/ \lambda) = 2kd_n$. Modulus of the complex amplitude of the reflected sound is thus proportional to
\begin{equation}\label{eq:Schroeder_scattering_amplitude}
\begin{aligned}
|\hat{p}_r (\boldsymbol x)| &\propto \abs*{ \sum_{n = 1}^{N} e^{j (2 \pi m n / N + \Phi_n)} } = \abs*{ \sum_{n = 1}^{N} e^{j (2 \pi m n / N + 2kd_n)} } & \\
&= \abs*{ \sum_{n = 1}^{N} e^{(j/N) (2 k L_1 n (\sin(\theta)+\sin(\theta_i)) + 2 \pi s_n k/k_d)} }.&
\end{aligned}
\end{equation}
The phase shift depends on frequency (wave number), so the diffuser is expected to be efficient within a certain frequency range around the \textbf{design frequency} $f_d$ (and generally its multiples, if the scattering mechanism is not obstructed by other unconsidered phenomena at higher frequencies). With regard to that, depths of the wells can be universally expressed normalized with the design wavelength $\lambda_d = c_0/f_d = 2\pi/k_d$, as integer numbers $s_n = 2 N d_n/ \lambda_d = N k_d d_n/ \pi$, regardless of the design frequency. Therefore, $\Phi_{d,n} = 2k_d d_n = 2\pi s_n/N$. Finally, the scattering pattern equals
\begin{equation}\label{eq:Schroeder_scattering_pattern}
\begin{aligned}
\boxed{ SP_{Schroed} = 20 \log_{10} \left( \abs*{ \sum_{n = 1}^{N} e^{(j/N) (2 k L_1 n (\sin(\theta)+\sin(\theta_i)) + 2 \pi s_n k/k_d)} } \right) },
\end{aligned}
\end{equation}
down to a constant value in dB which depends on the disregarded multiplication factor from eq.~(\ref{eq:solution_Green_Helmholtz_from_inc_sound_impedance_y3_G1_Ry2}). A normalization can be performed simply by subtracting maximum value of the expression in eq.~(\ref{eq:Schroeder_scattering_pattern}): $SP_{Schroed} - \max\{SP_{Schroed} \}$.

As already indicated, the design wavelength should be significantly larger than the well width $l_1$ (say, $k_d l_1 < 1$). A rule of thumb $f_d < c_0/(2 l_1)$ is often used in practice. In any case, some scattering should be expected at higher frequencies, as well, even though the model presented here might not be valid any more. On the other hand, the diffuser is inefficient if the design wavelength is much larger than the width of one whole repetition of the diffuser, the length $2L_1 = N l_1$. The condition $k_d N l_1 > 1$ leads to a rule of thumb for the lower limit of the design frequency: $f_d > c_0/(2 N l_1)$. In reality, certain diffusion is achieved even one or two octaves below this frequency. The two established limits for the design frequency can be expressed simultaneously as $c_0/(2 N l_1) < f_d < c_0/(2 l_1)$ or more compactly as $\pi < 2k_dL_1 < N\pi$.

\begin{figure}[h]
	\centering
	\begin{subfigure}{.4\textwidth}
		\centering
		\includegraphics[width=1\linewidth]{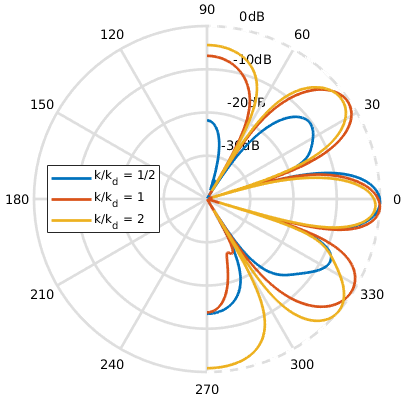}
		\label{fig:Schroed_diffuser_k_kd_ratio}
	\end{subfigure}%
	\begin{subfigure}{.4\textwidth}
		\centering
		\includegraphics[width=1\linewidth]{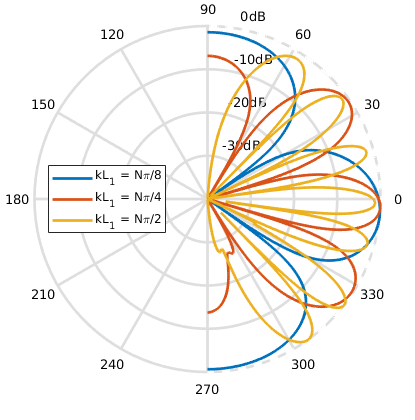}
		\label{fig:Schroed_diffuser_k_L1}
	\end{subfigure}
	\begin{subfigure}{.4\textwidth}
		\centering
		\includegraphics[width=1\linewidth]{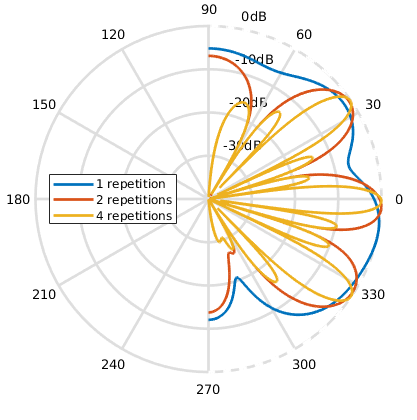}
		\label{fig:Schroed_diffuser_repetitions}
	\end{subfigure}%
	\begin{subfigure}{.4\textwidth}
		\centering
		\includegraphics[width=1\linewidth]{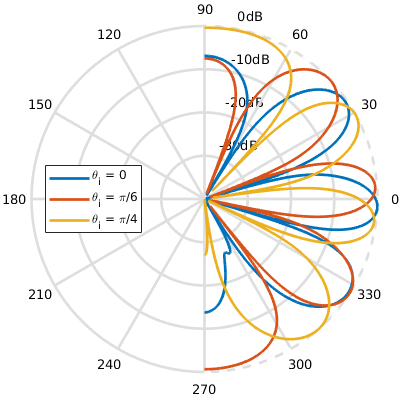}
		\label{fig:Schroed_diffuser_theta_i}
	\end{subfigure}
	\caption{Normalized scattering pattern ($SP_{Schroed} - \max\{SP_{Schroed}\}$) of a one-dimensional Schroeder diffuser based on the quadratic residue sequence with $N=7$ ($s_n = 0, 1, 4, 2, 2, 4, 1$) and: (above left) $kL_1 = N\pi/4$, 2 repetitions, $\theta_i = 0$, (above right) $k/k_d = 1$, 2 repetitions, $\theta_i = 0$, (below left) $k/k_d = 1$, $kL_1 = N\pi/4$, $\theta_i = 0$, (below right) $k/k_d = 1$, $kL_1 = N\pi/4$, 2 repetitions.}
	\label{fig:Schroed_diffuser}
\end{figure}

For the given design frequency and appropriate well width, scattering pattern of a Schroeder diffuser is essentially determined by the \textbf{sequence} $s_n$. Several types of sequences are used which give especially efficient diffusers. For a quadratic residue sequence, $s_n = (n-1)^2 \text{ mod } N$, where $N$ is a prime number, and mod gives the least non-negative remainder after division with $N$. In principle, any multiple of $N$ can be added to any value of $s_n$, because it adds a multiple of $2\pi$ to the corresponding phase shift $\Phi_{d,n}$, which has no effect on the reflection coefficient $\hat R_n$. The least remainder is used for practical reasons, to avoid unnecessarily deep wells, which provide the phase shifts larger than $2\pi$ when $s_n > N$. A quadratic residue sequence gives equal peaks of the lobes described by eq.~(\ref{eq:Schroeder_lobes_angles}). Figure~\ref{fig:Schroed_diffuser} shows normalized scattering patterns estimated using eq.~(\ref{eq:Schroeder_scattering_pattern}) for a Schroeder diffuser based on the quadratic residue sequence with $N = 7$ and different values of some of the relevant parameters. Other possibilities for the sequences include:
\begin{itemize}
	\item maximum length sequences (binary sequences which give only two depths of the wells, 0 and $d$); such diffusers are efficient in a very limited frequency range of around one octave, but have advantage of the small depth
	\item primitive root sequences ($s_n = r_{N+1}^n \text{ mod } (N+1)$, with $N+1$ a prime number and $r_{N+1}$ its primitive root, which results in all unique values of $s_n$ for $n = 1, 2,... N$); the specular reflection is more suppressed than with a quadratic residue sequence (see the lower right diagram in Fig.~\ref{fig:Schroed_diffuser}), especially for large $N$, which can be advantageous for some applications
	\item Legendre sequences; these diffusers perform similarly as the diffusers based on primitive root sequences; however, some of the wells are filled with absorbers, which leads to higher damping
\end{itemize}

For simplicity of the mathematical treatment, we have considered only one-dimensional Schroeder diffusers. Openings of the wells of these diffusers were treated as acoustically narrow rectangular imaginary surfaces with different phase parts of the reflection coefficients. Their depths are given by the sequences $s_n$, which are repeated periodically along a single axis. \textbf{Two-dimensional} diffusers can be constructed by introducing additional (often the same) sequence along a different (typically perpendicular) axis. If the axis are normal to each other, the reflection coefficient in eq.~(\ref{eq:solution_Green_Helmholtz_from_inc_sound_impedance_y3_G1_Ry1}) remains a function of $y_1$ and $y_2$. Two-dimensional diffusers provide more flexibility if the direction of arrival of the incident sound is not well known or receiver locations are distributed.

As already discussed, Schroeder diffusers are efficient in a limited frequency range around the design frequency. One way to expand the range is by means of so-called \textbf{fractal diffusers}. In such diffusers, the bottom of each relatively wide well is not flat and acoustically hard, but constitutes the interface surface of another smaller, high-frequency diffuser. Thus, smaller diffusers are placed inside a larger diffuser. Another disadvantage of Schroeder diffusers is their relatively large size dictated by the depths of the deepest wells, especially for low design frequencies. The overall depth of a diffuser can sometimes be decreased by bending the deepest wells behind the shorter ones, while keeping their depths unchanged. However, this requires more complicated and mechanically less robust designs.

In addition to the analysis above, it should be mentioned that actual performances of Schroeder diffusers are affected in reality by many factors which are not captured well by the used model, such as their finite extension (finite number of repetitions with period $2L_1$, as well as the extension along the $x_2$-axis), additional damping mechanisms, and three-dimensional sound propagation inside the wells at high frequencies. These phenomena can decrease scattering as well as lower the accuracy of its estimation. The latter can be increased with the aid of numerical simulations.

Of course, Schroeder diffusers are not the only acoustic elements which can substantially contribute to the diffuseness of the sound field in a room. Any non-planar reflecting surface is expected to scatter the sound, at least within a certain range of Helmholtz number values around $kL \approx 1$, where $L$ is the characteristic length scale (such as depth or width) of the geometric irregularities. Diffusers are also occasionally used in combination with absorbing materials to form hybrid surfaces. Many variations and products exist on the market. In general, scattering patterns and scattering coefficients are difficult to estimate, especially taking into account the large variety of diffuse surfaces which occur in practice. Compared, for example, with absorption coefficient values of porous materials, there are less measurement data and models available in literature which can be used for estimations of sound fields. Choice of appropriate values of scattering coefficients is, indeed, one of the major challenges in ray tracing simulations. 

\section{Modelling and prediction of room acoustics}\label{ch:modelling}

In this section we consider three major techniques for prediction of sound fields in rooms -- ray tracing, image source technique, and scale models. Compared to the simple analytical methods discussed in section~\ref{ch:basic_strategies}, they provide higher accuracy and more details of the calculated sound fields. This makes them more appropriate for larger projects and complex real scenarios, which can deviate significantly from the theoretical assumptions. As mentioned in section~\ref{ch:basic_strategies}, the fields and associated acoustic parameters can be estimated analytically, empirically, numerically, or be measured. In this section we focus on the last two possibilities. While on-site measurements, such as those treated in section~\ref{ch:measurements_and_descriptors}, generally give the most reliable results, measurements in scale models offer greater flexibility for testing various possible scenarios and modifications on a physical model of the actual room and can be used for predictions of room acoustics even before the room is built. The focus of this section is on how room acoustics can be predicted, assessed, and optimized without access to the actual room. The most common techniques for computational room acoustics (of relatively large spaces) are based on ray tracing and image sources. For certain applications scale models are a useful complement to these techniques.

\subsection{Ray tracing}\label{ch:ray_tracing}

Ray tracing technique relies on the high-frequency approximation of plane sound waves with rays, as described in section~\ref{ch:sound_rays}, and time-averaged energy summation discussed in section~\ref{ch:energy_summation}. There we concluded that energy summation of plane waves (in the acoustic far field of relevant sources or reflecting bodies, according to eq.~(\ref{eq:far_field_condition})) is valid for sufficiently broadband incoherent waves. Accordingly, ray tracing technique should be used only for acoustically large rooms, with the characteristic dimension $L$ much larger than the wavelengths of interest ($kL \gg 1$), and only for receiver locations in the far field. Another frequently cited condition is that the analysed frequency range lies above the Schroeder frequency given in eq.~(\ref{eq:Schroeder_freq}), so that distinct modes are unlikely to dominate in any of the finite frequency bands in which the energy is summed. If we suppose that the room volume is $V = L^2 H$, where $L$ is characteristic length/width of the room and $H$ its height, we can estimate the Schroeder frequency as
\begin{equation}\label{eq:Schroeder_freq2}
f_{Schroed} \approx 2100 \sqrt{\frac{T_{60,\Delta f_n} \cdot 1\text{m}^3/\text{s}^3}{V}} = \frac{2100}{L}\sqrt{\frac{T_{60,\Delta f_n} \cdot 1\text{m}^3/\text{s}^3}{H}}.
\end{equation}
For common values of height $H$ (up to 10-20\,m) and reverberation time $T_{60}$ (down to around 0.5\,s), the condition $f > f_{Schroed}$ appears to be somewhat more rigorous than $f \gg c_0/(2\pi L) \approx 55\text{\,m}\cdot\text{s}^{-1}/L$. However, the condition $kH \gg 1$ should also be checked if $H<L$. The superposed waves should also be mutually incoherent at the receiver locations. While this is usually satisfied by numerous late reflections, it might not hold for the earliest low-order reflections and direct sound. This already implies a limited accuracy of ray tracing analyses for estimation of early reflected sound.

Next we describe how ray tracing can be performed numerically. While section~\ref{ch:ray-tracing_rectangular_room} provided an analytical solution for a rectangular room and diffuse sound field, ray tracing is usually performed in the form of computer simulations, for more general room geometries and fields. Although the following discussion outlines a hypothetical ray tracing algorithm, practical realizations can differ significantly. The main reason is a demand for high computational efficiency of the algorithms, which is not considered here. The goal is to introduce key concepts and components of ray tracing simulations, as well as to demonstrate their geometrical nature.

\subsubsection{Numerical procedure}

Energy-based ray tracing cannot be used for direct calculations of impulse responses. For specified source and receiver locations, it can output an energy impulse response of the room (also called echogram or reflectogram) to a short excitation by the source. For this purpose, sound rays emitted by the source are represented by \textbf{sound particles} defined in section~\ref{ch:sound_rays}, which transport sound energy along the ray paths at the speed of sound $c_0$. The particles are followed from the source location (or the point of last reflection from a surface), which we can place at the origin of a coordinate system, on their straight paths to a receiver or surface. The direction of particle propagation (and the associated ray) is thus defined with polar and azimuthal angles ($\theta, \phi$) in usual spherical coordinates. Calculations in each frequency band of interest (typically in octave bands) are performed in parallel, so each particle's energy is given with an array of non-negative real numbers, with the number of elements of the array equal to the number of frequency bands. However, since calculations in different bands are completely equal, we can observe a particle with a single-number value of energy $E_p$, bearing in mind that the value can represent energy in one of several frequency bands.

Continuous sound field is represented by a finite number of particles travelling at certain discrete angles ($\theta, \phi$), samples of the continuous sets of all possible angles, $\theta \in [0,\pi]$ and $\phi \in [0,2\pi)$. The traced particles should provide a good representation of the actual field. Without \emph{a priori} knowledge of the field, it is reasonable to perform random sampling of discretized quantities. The \textbf{stochasticity} is achieved through generation of random numbers and the entire method is sometimes called stochastic ray tracing. In the following we use $z$ to denote a real random number taking values in the range between 0 and 1 ($0 \le z \le 1$). Occasionally we need more than one random number in a single expression, so we add indices for distinction: $z_1$, $z_2$, $z_3$, etc. As will be discussed below (see footnote \footref{ft:number_of_particles}), the total number of particles should be large enough to ensure statistical validity of the results obtained with the used finite sample, but small enough to keep the computational costs (in terms of memory and processing power) low enough. 

Particles are generated by sources of sound. An omnidirectional point \textbf{source} in the origin of its local coordinate system radiates particles with equal energies at random angles $\theta = \text{round}(z_1) \arccos(z_2) + (1-\text{round}(z_1)) (\pi - \arccos(z_3))$ and $\phi = 2\pi z_4$. The function round($z_1$) outputs 0 for $z_1<0.5$ or 1 for $z_1\geq 0.5$. In the expression for $\theta$ it randomly selects one of the two hemispheres around the source (for any fixed distance from the origin $r$), which correspond to the subsets $\theta \in [0,\pi/2]$ and $\theta \in [\pi/2,\pi]$. Notice that the distribution of $\theta$ is not uniform as in the case of $\phi$. This is because all particles carry the same energy. Since energy radiated from the omnidirectional source should be distributed uniformly over a spherical control surface with the source in its centre, distribution of $\cos(\theta)$ (or $\sin(\theta)$, but not $\theta$) should be uniform, which follows from the last equality in eq.~(\ref{eq:surface_integral}) ($d\Omega =  \sin(\theta) d\theta d\phi$ and the energy is uniform over the full solid angle).

From eq.~(\ref{eq:intensity_energy_complex_time_average_ray}), energy of a single particle radiated by the omnidirectional source equals
\begin{equation}\label{eq:ray_energy_omni_source2}
E_{p} = \frac{|\langle \boldsymbol I_p\rangle_T|}{c_0} = \frac{\langle P_{q}\rangle_T/ 1\text{m}^2 }{N_p c_0},
\end{equation}
where $\langle P_{q}\rangle_T$ is time-averaged sound power of the source. The full solid angle is covered with $N_p$ particles, so $4\pi$ (which is in the denominator for infinitesimal $d\Omega$) is replaced with $N_p$. On the other hand, absolute sound pressure level values are often irrelevant, for example, if ray tracing simulations are used for estimating energy impulse responses and descriptors of room acoustics from section~\ref{ch:descriptors_of_room_acoustics}). In such cases, multiplication of $E_p$ with a positive constant equal for each particle does not affect the results. For example, $E_p = 1$\,J/m$^3$ can be set for each particle in each frequency band. As we will see throughout the rest of this section, $E_p$ can often be kept constant during the entire ray tracing analysis, which is computationally much more efficient than re-evaluating it repeatedly during the particle propagation. Decay of sound energy is then accounted for by omission (\textbf{annihilation}) of some particles, rather than scaling their energies. Even if the values of $\langle P_{q}\rangle_T$ are specified in each frequency band, this approach can still be used and the particle energies can be scaled accordingly only at the end of the calculation. 

If necessary, directivity of the source can be included simply by multiplying $E_p$ of each radiated particle with $D_i^2(\theta, \phi)$. However, as just mentioned, it is computationally favourable to keep the particle energies unchanged. Rather then scaling them, the angularly dependent energy distribution proportional to $D_i^2(\theta, \phi)$ can be achieved by annihilating (removing) particles radiated at angles for which radiation pattern of the source drops. This decreases the overall energy emitted in these directions and emulates directivity of the source. The annihilation is performed stochastically when the condition $z > D_i^2(\theta, \phi) / D_{i,max}^2$ is met. The remaining particles preserve their energies regardless of their directions and an appropriate scaling which recovers the specified power of the source is done at the end of the analysis. For a very large number of particles, $N_p$, energy scaling and particle annihilation lead to the same results. While the latter technique is less accurate for smaller numbers of particles (since fewer particles negatively affect statistical validity of the results), it is computationally cheaper, as some particles are left out from further calculations and multiplication with the scaling factor $D_i^2(\theta, \phi)$ is rendered unnecessary.

Energy impulse is generated by a point source when all particles start from the source location, say $\boldsymbol x_0$, at the same time $t_0$. Particles travel along the ray paths at the speed $c_0$. Starting from $t_0$, each particle is followed (\textbf{traced}) and checks are made whether it hits a surface or receiver in the room. The particles are traced until (a) they are annihilated, (b) their energy drops below some predefined value $E_{p,min}$ (set sufficiently low for an accurate extraction of the relevant acoustic quantities; it plays a role very similar to the energy of ambient noise in measurements), if energy scaling is used, or (c) some specified time $t_{end} > t_0$ is reached (which terminates the calculation even if neither of the two previous conditions is met). Similarly as with the number of particles, suitable values of $E_{p,min}$ and $t_{end}$ are a compromise between (a sufficient) accuracy and computational load. For example, $E_{p,min}$ can be set to $E_p/10^{4}$ to achieve a dynamic range of 40\,dB and $t_{end}$ to $t_0 + 2 T_{60}/3$, where $T_{60}$ can be estimated using Sabine's formula.

For easier geometrical treatment, all \textbf{surfaces} are assumed to consist of plane polygons. Accordingly, curved surfaces have to be approximated with sets of non-overlapping plane surfaces. Furthermore, we suppose that all surfaces have a triangular shape (any plane polygon can be divided into triangles). Vertices of a triangle, the points $\boldsymbol x_A$, $\boldsymbol x_B$, and $\boldsymbol x_C$ in global Cartesian coordinates, determine its supporting plane. Assigned to it, we define a unit vector normal to the supporting plane according to the right-hand rule, that is, if a right-handed screw is placed normal to the plane and rotated in the direction from $\boldsymbol x_A$ over $\boldsymbol x_B$ to $\boldsymbol x_C$, it translates in the direction of the normal. The unit vector is thus
\begin{equation}\label{eq:supporting_plane_normal}
\begin{aligned}
\boldsymbol n &= \frac{(\boldsymbol x_B - \boldsymbol x_A) \times (\boldsymbol x_C - \boldsymbol x_A)}{|(\boldsymbol x_B - \boldsymbol x_A) \times (\boldsymbol x_C - \boldsymbol x_A)|} = \frac{(\boldsymbol x_C - \boldsymbol x_B) \times (\boldsymbol x_A - \boldsymbol x_B)}{|(\boldsymbol x_C - \boldsymbol x_B) \times (\boldsymbol x_A - \boldsymbol x_B)|} &\\
&= \frac{(\boldsymbol x_A - \boldsymbol x_C) \times (\boldsymbol x_B - \boldsymbol x_C)}{|(\boldsymbol x_A - \boldsymbol x_C) \times (\boldsymbol x_B - \boldsymbol x_C)|}.&
\end{aligned}
\end{equation}
For any point $\boldsymbol y$ on the supporting plane the following equality holds:
\begin{equation}\label{eq:supporting_plane}
\boldsymbol n \cdot \boldsymbol y = P_p,
\end{equation}
regardless of the location of origin of the coordinate system. The constant $P_p$ can be found using any vertex of a triangle in the supporting plane, for example, $P_p = \boldsymbol n \cdot \boldsymbol x_A$.

In order to check whether a particle hits a triangular surface, first we need to determine the intersection point $\boldsymbol x_s$ on the supporting plane of the surface given by eq.~(\ref{eq:supporting_plane}). The particle is assumed to be emitted (or reflected from or transmitted through a surface) from the point $\boldsymbol x_0$ in the direction given by the unit vector $\boldsymbol e_r$. Cartesian components of the vector can be calculated from the angles $\theta$ and $\phi$ of the local spherical coordinate system with the origin at $\boldsymbol x_0$: $e_{r1} = \sin(\theta) \cos(\phi)$, $e_{r2} = \sin(\theta) \sin(\phi)$, and $e_{r3} = \cos(\theta)$. Locus of the particle's path (as well as the sound ray) consists only of the points $\boldsymbol x$ which satisfy
\begin{equation}\label{eq:ray_line}
\boldsymbol x = \boldsymbol x_0 + R \boldsymbol e_r,
\end{equation}
where $R$ is any real number, but only $R>0$ implies a forward travelling particle, in the direction of the ray. If $\boldsymbol n \cdot \boldsymbol e_r = 0$, the particle travels parallel to the supporting plane, there is no intersection point, and the particle cannot hit the surface. In all other cases, the value of $R$ for the intersection point is found after replacing $\boldsymbol y$ from eq.~(\ref{eq:supporting_plane}) with $\boldsymbol x$ from eq.~(\ref{eq:ray_line}),
\begin{equation}
\boldsymbol n \cdot (\boldsymbol x_0 + R \boldsymbol e_r) = P_p,
\end{equation}
from which
\begin{equation}
R = \frac{P_p - \boldsymbol n \cdot \boldsymbol x_0}{\boldsymbol n \cdot \boldsymbol e_r}.
\end{equation}
If $R<0$, the particle travels away from the surface and cannot hit it. If $R>0$, the intersection point is
\begin{equation}\label{eq:intersection_point}
\boldsymbol x_s = \boldsymbol x_0 + \frac{P_p - \boldsymbol n \cdot \boldsymbol x_0}{\boldsymbol n \cdot \boldsymbol e_r} \boldsymbol e_r.
\end{equation}

Next it needs to be checked whether the intersection point $\boldsymbol x_s$ lies inside the triangle with the vertices $\boldsymbol x_A$, $\boldsymbol x_B$, and $\boldsymbol x_C$. If it is outside, at least one cross product $(\boldsymbol x_B - \boldsymbol x_A) \times (\boldsymbol x_s - \boldsymbol x_A)$, $(\boldsymbol x_C - \boldsymbol x_B) \times (\boldsymbol x_s - \boldsymbol x_B)$, or $(\boldsymbol x_A - \boldsymbol x_C) \times (\boldsymbol x_s - \boldsymbol x_C)$ must point into the direction opposite from $\boldsymbol n$. Therefore, an intersection point which lies inside the triangle or at its edges has to satisfy all of the following conditions:
\begin{equation}\label{eq:intersection_point_inside_triangle}
\begin{split}
[(\boldsymbol x_B - \boldsymbol x_A) \times (\boldsymbol x_s - \boldsymbol x_A)] \cdot \boldsymbol n \ge 0, \\
[(\boldsymbol x_C - \boldsymbol x_B) \times (\boldsymbol x_s - \boldsymbol x_B)] \cdot \boldsymbol n \ge 0, \\
[(\boldsymbol x_A - \boldsymbol x_C) \times (\boldsymbol x_s - \boldsymbol x_C)] \cdot \boldsymbol n \ge 0.
\end{split}
\end{equation}
The equalities hold when the intersection point is at the corresponding edge (the first term of the cross product) or vertex (in the second term of the cross product) appearing on the left-hand side. For faster computations, the algorithm can stop the check as soon as one of the conditions is not satisfied. If the intersection point does belong to the triangular surface, length of the particle's path from $\boldsymbol x_0$ to the surface is
\begin{equation}\label{path_part_length}
r = |\boldsymbol x_s - \boldsymbol x_0|
\end{equation}
and the elapsed travel time is $t - t_0 = r/c_0$. If multiple surfaces contain intersection points, the one closest to $\boldsymbol x_0$ (with the shortest path length $r$) should be observed. In rare occasions when \textbf{dissipation} in air cannot be neglected, it can be accounted for by scaling the particle energy as
\begin{equation}\label{eq:ray_tracing_air_dissipation}
E_p e^{-m_{air} r},
\end{equation}
with $m_{air}$ attenuation constant in air (and $4.34 m_{air}$ attenuation expressed as usual in dB/m), according to equations~(\ref{eq:p_energy_exp_decay}) and (\ref{eq:attenuation_constant}).

Once a surface hit by a ray is identified, energy losses due to \textbf{absorption} are included either by scaling energy of every reflected particle, $(1-\alpha_s)E_p$, or by random annihilation of particles for $z < \alpha_s$, where $\alpha_s$ is absorption coefficient of the surface (the diffuse-field value is normally used). As already pointed out, sound \textbf{transmission} through boundary surfaces is treated as absorption ($\alpha_s = 1 - |\hat{R}_s|^2$) and the entire non-reflected energy is captured by the absorption coefficient. If the surface is not a boundary surface (for example, a hanging reflector, curtain, lining panel, etc.), its transmission coefficient can be defined independently as the (angle-independent) ratio of complex transmitted and incident sound amplitudes, $\hat T_s = \hat p_{trans}/\hat p_{inc}$. For energy-based ray tracing simulations, only its squared value $0 \leq |\hat T_s|^2 \leq 1$ is relevant, which is sometimes referred to as transparency coefficient. In such cases, absorption coefficient should not include the transmitted energy (eq.~(\ref{eq:absorption_coeff_plane_surface}) becomes $\alpha_s = 1 - |\hat{R}_s|^2 - |\hat{T}_s|^2$). The two expressions for reflected particles from above become $(1-\alpha_s - |\hat T_s|^2)E_p$ and $z < \alpha_s + |\hat T_s|^2$, respectively. Similarly, energy of the transmitted particles can be scaled as $|\hat T_s|^2 E_p$ or the particles can be randomly annihilated for $z > |\hat T_s|^2$.

\begin{figure}[h]
	\centering
	\includegraphics[width=0.6\linewidth]{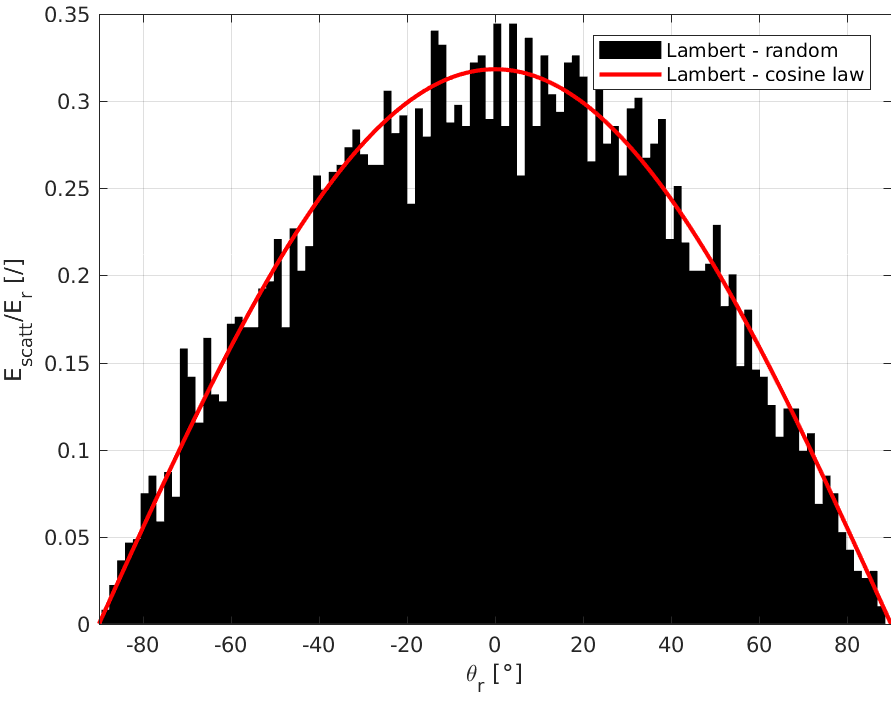}
	\caption{Angular distribution of the normalized scattered energy (for $z_1 < s_s$) according to Lambert's cosine law approximated with random outcomes of the upper part of eq.~(\ref{eq:Lambert2}). For easier comparison with the curve $s_s(\theta_i) = 1$ in Fig.~\ref{fig:Lambert_cosine_law}, the distribution is plotted as a histogram of $\theta_r = 2\arccos(\sqrt{z_2})-\pi/2$ in polar coordinates ($\phi_r = 0$) normalized with $N_p \pi^2/(2 N_{bins})$, where $N_p = 10000$ is the number of scattered particles and $N_{bins} = 100$ is the number of bins in the histogram.}
	\label{fig:Lambert_ray_tracing}
\end{figure}

\textbf{Scattering} is very commonly modelled using Lambert's cosine law from eq.~(\ref{eq:Lambert}) with the scattering coefficient $s_s$ independent from the angle of incidence. However, creating new scattered particles would dramatically increase the computational load for any non-zero value of $s_s$, especially if their number should be high enough to emulate accurately the angular distribution from Fig.~\ref{fig:Lambert_cosine_law}. As a reasonable alternative, no new particles are introduced. Instead, polar and azimuthal angles of each reflected particle's path (ray) to the scattering surface are not necessarily specular, but calculated as
\begin{flalign}\label{eq:Lambert2}
\begin{aligned}
\theta_r = |2\arccos(\sqrt{z_2})-\pi/2|, &\quad \phi_r = 2\pi z_3,  &\quad\text{ for } z_1 < s_s. &\\
\theta_r = \theta_i, &\quad \phi_r = \phi_i+\pi, &\quad\text{ for } z_1 \ge s_s. &
\end{aligned}
\end{flalign}
On a large sample of scattered particles, the upper part of the equation gives approximately the same angular distribution of scattered energy as Lambert's cosine law (with assumed uniform distribution of $\phi_r$). This is demonstrated in Fig.~\ref{fig:Lambert_ray_tracing} on the sample of 10000 incident (and scattered, for $z_1 < s_s$) particles. The lower part of eq.~(\ref{eq:Lambert2}) represents specularly reflected particles.

Local polar and azimuthal angles, $\theta_i$ and $\phi_i$, are calculated at the point of reflection from the known vectors $\boldsymbol e_r(\theta,\phi)$ of the incident particle and surface normal $\boldsymbol n$. Similarly, new values of global coordinates $\theta$ and $\phi$  and $\boldsymbol e_r$ of the reflected particle are calculated from the obtained $\theta_r$, $\phi_r$, and $\boldsymbol n$. For a transmitted particle, $\theta$ and $\phi$ are simply taken over from the incident particle. Values of the time, $t$, and particle energy (if energy scaling is applied), $E_p$, are updated after each reflection or transmission and the conditions $E_p < E_{p,min}$ and $t > t_{end}$ are checked. If any of the two conditions is met, tracing of the particle is stopped and the next particle can be observed. If not, tracing of the reflected or transmitted particle with the calculated direction $\boldsymbol e_r(\theta, \phi)$ can be continued with $\boldsymbol x_s$ becoming new $\boldsymbol x_0$ and $t$ becoming $t_0$.

Parallel to the inspection of possible surfaces crossing a particle's trajectory, a check should be made whether the particle hits one or more \textbf{receivers} on its path. Volumes of virtual receiver regions must be finite. For example, an omnidirectional receiver is represented by a spherical part of the room volume with radius $a_{rec}$, which is centred at the receiver location. The radius should not be too small, in order to ensure that a statistically valid number of particles hit the receiver\footnote{\label{ft:number_of_particles}Since finite number of randomly generated particles represent a continuous sound field in the room, too few received particles result in poor accuracy of the calculations (the statistical sample of the field is too small). For properly set $E_{p,min}$ and $t_{end}$, the number of particles which hit a receiving sphere depends primarily on its volume, $V_{rec}=4 a_{rec}^3 \pi/3$, the total number of generated particles, $N_p$, as well as the size of the room and absorption in it (when particles get annihilated or fully absorbed or dissipated). Smaller $V_{rec}$ should thus imply larger $N_p$ to maintain statistical validity. Larger values of $N_p$ are preferred especially for large or damped rooms. For example, reasonable values in large concert halls are $2a_{rec} \approx 1\text{\,m}$ and $N_p \sim \mathcal{O}(10^5)$ and in much smaller lecture halls $N_p \sim \mathcal{O}(10^3)$. The number of particles can be larger (for higher accuracy) as more powerful computational hardware and algorithms become available.}. For a receiver located at $\boldsymbol y_{rec}$, the check can be performed\footnote{Certain complications and inaccuracies can occur if a reflecting surface in the room partly covers or intersects the receiving volume. We neglect such cases.} by measuring the distance $d$ between $\boldsymbol y_{rec}$ and the closest point on the particle's path $\boldsymbol x_0 + R_{rec} \boldsymbol e_r$:
\begin{equation}
d = |\boldsymbol x_0 + R_{rec} \boldsymbol e_r - \boldsymbol y_{rec}|,
\end{equation}
where
\begin{equation}
R_{rec} = (\boldsymbol y_{rec} - \boldsymbol x_0) \cdot \boldsymbol e_r.
\end{equation}
If $d < a_{rec}$, $R_{rec}>0$, and $R_{rec}$ is smaller than $R$ of the next surface on the particle's trajectory (which has the smallest positive $R$ among all the surfaces), then the particle crosses the receiving volume. Its energy when it enters the volume (scaled according to eq.~(\ref{eq:ray_tracing_air_dissipation}) with $R_{rec}$ replacing $r$, if dissipation in air should be included) and time of arrival ($t_0 + R_{rec}/c_0$) are stored. Angles of arrival can also be stored (both for receivers and surfaces), since they provide valuable information on spaciousness and diffuseness of the field.

Energies of the received particles are summed for each receiver within certain non-overlapping time frames\footnote{While long time frames lower the temporal resolution of the results, short frames may increase the inaccuracy, due to fewer particles hitting the receivers within the short frames. An interval of 5-10\,ms can be chosen as a compromise, which also matches well the inertia of human hearing, as discussed in section~\ref{ch:sources_dynamics}.}, based on the assumption of broadband and mutually incoherent superposed waves. If energies of all particles are fixed and equal (for example, $E_p = 1$\,J/m$^3$) and energy decay is achieved only by means of annihilation, the summation comes down to counting the received particles in each time frame and multiplying the result with the initial energy per particle from eq.~(\ref{eq:ray_energy_omni_source2}). Storage of the energies of traced particles is not required, which is one of the advantages of the annihilation approach. The obtained distribution of the received sound energy over time, $E(t_i)$, where $i = 1,2,...$ is the number of each successive time frame, is an estimation of \textbf{echogram} (reflectogram) or energy impulse response averaged over the time frames.

Echograms are the crucial result of ray tracing simulations. They can be readily used for estimation of many energy-based descriptors of room acoustics from section~\ref{ch:descriptors_of_room_acoustics}. In addition to this, they can be used for \textbf{auralization} -- rendering sound fields at the receiver locations in simulated rooms. For this purpose, however, an additional information on phase has to be imposed. A set of obtained echograms in different frequency bands (typically octaves) is observed as a set of energy spectral densities in different time frames, with the frequency resolution dictated by the size of the frequency bands. Usually low frequency resolution can be increased by interpolating and extrapolating the values in narrower frequency bands with centre frequencies $f_j$ ($j = 1, 2,...$) from the values obtained with lower resolution, in the considered frequency range.

From an estimated energy spectral density $E(f_j,t_i)$ and eq.~(\ref{eq:intensity_energy_complex_time_average_plane_wave}), discrete frequency response in each time frame $i$ can be estimated as $\hat p_i(f_j) = \sqrt{2 E(f_j,t_i) \rho_0 c_0^2} e^{j\Phi(f_j,t_i)}$. The exact values of the phase $\Phi(f_j,t_i)$ cannot be extracted from the energy impulse responses. They are often modelled as stochastic and generated randomly\footnote{Still, phase must be an odd function of real-valued (both positive and negative) frequency for real LTI systems: $\Phi(f)=-\Phi(-f)$. Minimum phase function is often used, which also has the properties of causality and stability. Amplitude is in contrast an even function, $|\hat p (f)| = |\hat p (-f)|$.}. Discrete version of the inverse (fast) Fourier transform from eq.~(\ref{eq:inv_Fourier_p}) and merging the time signals obtained from different time frames provide an estimated impulse response of the room. The response can be used for auralization, for example, by convolving it with some anechoic (``dry'') input signal (discrete form of eq.~(\ref{eq:convolution})). The results is an estimated received signal for the given source and receiver locations in the analysed room, which can be subjectively assessed. Moreover, impulse responses obtained at two or more specific locations can be used for rendering binaural or ambisonic recordings with additional spatial information, which is not contained in a single impulse response.

It should be emphasized again that the algorithm considered above is by no means computationally optimized. Many adjustments can be made to make the computations faster with lower processing or memory requirements. For example, the normal vector $\boldsymbol n$ in eq.~(\ref{eq:supporting_plane_normal}) does not have to be normalized to a unit vector by the denominators for the rest of the algorithm to produce the same results. The conditions~(\ref{eq:intersection_point_inside_triangle}) can be formulated and checked more efficiently after expressing the vectors with two-dimensional coordinates, since all points belong to the same supporting plane. Many other optimization techniques are used in the development of software for ray tracing simulations, which are not in the focus here.

\subsubsection{Workflow}

Ray tracing algorithms, such as the one described above, are usually inaccessible to the users of software tools for room acoustic simulations. Therefore, it makes sense to describe a typical procedure for performing numerical room acoustics based on ray tracing. In practice, the workflow consists of the following several steps.

The first step is creating the three-dimensional \textbf{geometry} of the room with all relevant surfaces and objects in it. As a rule, this is the most time consuming step. The boundary surface of the room must be closed, so possible openings can be treated as surfaces with high absorption coefficient values. The geometrical model of the room should never be too detailed (which is often in contrast to architectural drawings and models and, unfortunately, hinders their direct use for acoustic simulations; some adaptation of the available three-dimensional CAD models is usually necessary). The reason is not only that numerous surfaces increase the computational load (for example, for identifying the surface of subsequent reflection), but accuracy of the results can also decrease. Since sound waves are treated as rays which do not diffract, unrealistic reflections and other acoustic effects can take place even at acoustically compact surfaces. Even if a surface scatters the incident sound due to its geometric irregularities, these irregularities should be represented by higher values of the scattering coefficient of one or several relatively large surfaces, rather than by many small surfaces. For example, curved surfaces should be only crudely divided into plane surfaces and additional diffusion due to curvature is modelled by means of the scattering coefficient.

The second step is specifying acoustically relevant properties of all \textbf{surfaces}. These are primarily absorption and scattering coefficients. It is advisable that the latter one is always at least 0.05, even for acoustically large flat surfaces, because finite surfaces inevitably scatter the sound close to their edges. Transparency coefficient can also be defined for inner surfaces, such as reflectors, to take into account the transmitted energy, especially at lower frequencies (recall eq.~(\ref{eq:absorption_coeff_plane_surface_Z_Zconst_free_membrane})). 

In the final step of setting up the calculations, the \textbf{sources and receivers} have to be defined, in particular their locations and directivity. The latter is more common for sources (receivers are rarely directional) and defined in local coordinates. Manufacturers of loudspeakers often offer files with three-dimensional radiation patterns which can be readily imported in the most frequently used software packages for acoustic ray tracing. In addition to this, sound power of the sources can be specified in each frequency band.

After the geometry, surfaces, sources, and receivers have been defined, the calculations can be executed. This is followed by the \textbf{post-processing} and interpretation of the results, including the extraction of relevant quantities. One of the major advantages of numerical simulations over measurements in real rooms is that different acoustic measures can be tested quickly and at smaller costs, for example, changes of the geometry (size and orientation of reflectors or boundary surfaces) or acoustic properties of surfaces (absorption and scattering). If measurement results are also available, the simulated model can be adjusted (especially regarding the less known parameters, such as scattering coefficient), such that the numerically obtained values match reasonably well the measured values. In this way the simulation procedure can be calibrated and validated before further acoustic measures are tested. Another advantage is that many more relevant quantities can be covered than with measurements or analytical formulas, such as angles of propagation of distinct reflections, free path length distribution (recall section~\ref{ch:ray-tracing_rectangular_room} and eq.~(\ref{eq:ray_path_length_average})), coverage with the direct sound or low-order reflections from surfaces of special interest, spatial distribution of acoustic quantities and descriptors, etc. Information on these can significantly help optimization of the acoustic design.

\subsection{Image sources}\label{ch:image_sources}

As discussed in section~\ref{ch:early_reflections_strategy}, strong low-order reflections are often critical for room acoustics but difficult for accurate predictions. Unfortunately, this applies for ray tracing calculations, as well. In extreme cases, important reflections may be completely left out from the calculation due to its stochastic nature and the limited number of traced particles. The traced particles may miss some of the surfaces or receivers on a reflection path and, as a result, the reflection is not present in the obtained echogram. Even if the reflection is captured, its energy can be estimated inaccurately because of too few particles or some of many approximations of the ray tracing method -- treating sound waves as rays (and summing time-averaged energy of broadband, mutually incoherent plane waves instead of more general superposition), neglecting diffraction, possibly inaccurate absorption and scattering coefficients as well as scattering models. For improved accuracy, the earliest (mostly 1$^\text{st}$- and 2$^\text{nd}$-order) reflections can be estimated with the aid of image sources introduced in section~\ref{ch:infinite_plane_surface}. In fact, image source technique is often combined with ray tracing in \textbf{hybrid} room acoustic calculations. The important low-order reflections are estimated more accurately with image sources, while stochastic ray tracing is used for less critical late reflections.

As pointed out in section~\ref{ch:infinite_plane_surface}, acoustically large flat and uniform surfaces reflect specularly and can be replaced with image sources, for calculations of sound fields in front of them. Each $n^\text{th}$-order ($n = 1,2,...$) image source is due to one $(n-1)^\text{th}$-order image source, with actual sources in the room having formally zero-order. Since the model is not based on sound rays, amplitude decay with the distance from the source (for a point source $|p| \sim 1/r$) has to be taken into account for all sources, including the image sources, and complex sound pressure (not energy) at a receiver point (not a receiving sphere) $\boldsymbol x$ can be estimated from eq.~(\ref{eq:solution_tailored_Green_wave_eq_free_space_compact_source_emission_time_amplitude}) for compact omnidirectional sources, starting with the lowest-order sources. (Amplitude of the source function, $|\hat Q|$, is related to the sound power of the source via eq.~(\ref{eq:acoustic_power_complex_time_average_spherical_wave}).) Contributions of different sources are summed according to wave superposition. If eq.~(\ref{eq:solution_tailored_Green_wave_eq_free_space_compact_directed_source_emission_time_freq}) is used for directional sources, an image source has to be oriented symmetrically to the corresponding lower-order source with respect to the supporting plane of the reflecting surface (as a mirror image). Notice that a mirroring surface can lie outside the physical boundaries of actual reflecting surface with finite dimensions, as long as sound waves from the image source reach a receiver point (or another surface) through the part of the supporting plane where the actual surface is located.

If the reflecting surface is uniform with angularly independent absorption coefficient, absorption of the incident sound can be included by multiplying the image source function with $\sqrt{1-\alpha_s}$. This is equivalent to replacing $\hat{R}_s(\theta)$ in eq.~(\ref{eq:p_plane_surface}) with $\sqrt{1-\alpha_s}$,  following from eq.~(\ref{eq:absorption_coeff_plane_surface}) and assuming no phase changes at the surface\footnote{Alternatively, minimum phase function can again be used for modelling phase of the reflection coefficient.}. It is also in agreement with eq.~(\ref{eq:SPL_dir-refl}). Similarly as in this equation, scattered energy can be excluded by additional multiplication with $\sqrt{1-s_s}$. In any case, image sources capture only specular reflections.

Compared to the ray tracing method, calculation of sound pressure rather than energy is more appropriate for estimation of superposition of direct and reflected waves. There is no need for simplified models of sound propagation, energy summation of incoherent waves, or plane waves assumption in the acoustic far field. In principle, the reflected sound amplitude can be calculated more accurately (if all relevant image sources are correctly defined), since it does not depend on the stochasticity of sound particles or size of the receiving sphere. Moreover, the sound is not required to be broadband. Still, validity of the image source technique is strongly limited to acoustically large flat surfaces which reflect essentially \textbf{specularly}. As demonstrated in section~\ref{ch:rectangular_surface}, even hard finite surfaces reflect non-specularly at the edges and act as diffusers in certain frequency ranges. More realistic irregular surfaces can reflect even more non-specularly. Finally, the accuracy depends also on the available absorption and scattering coefficients of the surfaces.

In addition to this, implementation of the image source technique introduces two further issues -- inaudible image sources and increase of the number of image sources (and therefore computational load) with their order. As already mentioned, a mirror plane behind which an image source is located can lie in the extension of actual surface of the room. Importantly, the sound waves from such a source reach some receiver points or other surfaces after propagating ``through'' the actual surface and thus represent the reflections from it. For all other receivers and surfaces, the same image source is \textbf{inaudible} and sound which it radiates unphysical. Therefore, whether an image source should be included or not depends not only on the lower-order sources and reflecting surface, but also on the associated higher-order image sources and ultimately receiver points in the end of the chain. These dependencies are not easy to determine in general, especially for higher-order image sources. Audibility of each image source has to be checked for each receiver independently, for example, by backtracing along the path from a receiver to the last image source in the chain. If the path does not cross an actual reflecting surface but its extension in the supporting plane, the image source is not audible and should be excluded for the particular receiver. If it does, the check is repeated for the path between the intersection point at the mirror surface of the image source and the next lower-order image source associated with the previously considered one (whether this path crosses another actual surface). This procedure is continued until a zero-order (actual) source in the room is reached. The initial highest-order image source is audible for the selected receiver only if all observed paths cross actual surfaces.

Checking audibility of each image source for each receiver point makes significant part of the computational load required by the image source technique. As we will see shortly, the number of image sources quickly becomes large for higher orders. Not only contributions of many sources have to be calculated and summed, but many checks of audibility have to be performed, as well. When image sources are used in combination with ray tracing (in hybrid schemes), results of the latter technique, namely the calculated particle trajectories, can be utilized to determine audibility of the image sources. If a particle hits a receiver after $n$ reflections, the $n^\text{th}$-order image source behind the last reflecting surface on the particle's trajectory is audible for the receiver. As before, the number of traced particles must be large enough to identify all important reflections. However, estimation of their strengths is now improved with the image source technique.

Another limitation of the method is that the \textbf{number} of image sources increases rapidly with their order, which makes the image source technique impractical for calculation of late reflections. For a single zero-order source and $N$ surfaces of the room, maximum $N$ first-order image sources can exist. If each image source is mirrored with respect to the other $N-1$ surfaces, the maximum of $N(N-1)$ 2$^{\text{nd}}$-order image sources is obtained. Continuing this process up to the order $n$, each higher orders introduces multiplication with $N-1$, which gives the total of
\begin{equation}
N_q = \sum_{i=1}^{n} N(N-1)^{i-1} = N \sum_{i=1}^{n} (N-1)^{i-1} = N \frac{(N-1)^n-1}{N-2}
\end{equation}
possible image sources. For the last equality, the sum was expressed from the equality
\begin{equation}
\begin{aligned}
\Sigma &= \sum_{i=1}^{n} (N-1)^{i-1} = 1 + (N-1)^1 + ... + (N-1)^{n-1} \\
&= \frac{1}{N-1} [ (N-1)^1 + (N-1)^2 + ... + (N-1)^n ] = \frac{\Sigma - 1 + (N-1)^n}{N-1}.
\end{aligned}
\end{equation}
Such a large growth with $n$ limits the application of image sources in practice to the first few orders, unless the room geometry is very simple with a small number of surfaces. Moreover, as the order of image sources and their distances from the actual room interior increase (which correspond to longer path lengths of later reflections), the fraction of inaudible image sources increases. For these reasons, diffuse late energy, starting with the reflections of, say, the 4$^\text{th}$- or 5$^\text{th}$-order, is estimated more efficiently and accurately with ray tracing.

\subsection{Scale models}

Ray tracing calculations of early reflections can be improved with the aid of image sources. However, this holds only for specular reflections from acoustically large and flat surfaces. Many room surfaces which provide important early reflections are not acoustically large (at least not in the entire frequency range) and scatter the sound. On the other hand, accurate calculation of scattering in rooms is computationally very expensive and practically unfeasible. If \textbf{early scattered sound} is critical (usually in rooms for music production, in which acoustics has a high priority, such as concert halls and opera houses), it can be estimated and optimized in scale models.

Scale models are built physical models of rooms with scaled geometry. Typical scaling ratio for relatively large halls is 1:10 (\textbf{scale factor} 10) or somewhat higher. Consequently, all distances and sound propagation intervals measured in a scale model should be multiplied with the scale factor before comparison with the actual room\footnote{We assume that the scaled room is filled with air with the same speed of sound as the original room. Other media (both gases and liquids) were used occasionally in the past, but nowadays very rarely.}, while the analysed frequencies should be divided by the same factor. In other words, the relevant Helmholtz number values are equal in the real room and its scale model (shorter sound wavelengths are adapted to the smaller geometry). While larger scale factors result in less cumbersome and cheaper models, restrictions are imposed by the acoustic phenomena at the scaled frequencies, which can significantly deviate from those in the actual room and thus decrease accuracy of the predictions.

The considered frequencies are higher than actual frequencies by the scale factor and this is associated with several complications in practice:
\begin{itemize}
	\item absorption and other acoustic properties of the surfaces should match those of the actual surfaces but shifted to higher frequencies, which can be achieved only approximately,
	\item if the medium (air) is the same as in the original room, attenuation constant ($m_{air}$) at scaled frequencies should be higher than at actual frequencies by the scale factor (since the propagation paths are shorter by the same factor), which is generally not the case,
	\item sources and receivers should usually be omnidirectional, even at very high scaled frequencies (often in ultrasound range), which implies their very small sizes,
	\item sources should generate accordingly very short impulses for direct impulse response measurements.
\end{itemize}
These issues limit the validity of measurements in scale models and their applicability mainly to low-order reflections from \textbf{highly reflecting} surfaces. With regard to this and considerably lower costs of production, partial scale models (rather than complete scaled rooms) can be used for inspecting particularly critical early reflections in certain parts of the room, typically close to the stage. For basic analyses, massive, fully reflecting (at scaled frequencies) materials can be used to represent all acoustically hard surfaces in the room and highly absorbing porous materials for all surfaces which essentially absorb, as well as for avoiding additional reflections from the surrounding. Dissipation in air is often neglected, especially for the earliest reflections with short propagation paths.

Omnidirectional radiation of a source in a scale model should be verified, as well as its dynamics over the entire (scaled) frequency range of interest (in order to ensure a sufficient signal-to-noise ratio during measurements). Spark generators are frequently used, since they provide very short impulses and fairly omnidirectional radiation patterns, due to the very small size of their tips. This is more difficult to achieve with small electro-acoustic transducers, for example, in the form of a dodecahedron. However, their advantage is that deterministic sequences from section~\ref{ch:acquisition_room_impulse_response} can be used, which allow more robust acquisition of impulse responses. Small (usually 1/4-inch or 1/8-inch) omnidirectional microphones are commonly used as receivers.

Due to their price and demanding process of production, scale models are used only in large projects, in which room acoustics is in focus. Even then, they are used in combination with more flexible and faster numerical simulations. Complementing the simulations, scale models account for \textbf{diffraction} and \textbf{scattering} at surfaces more accurately.

\section*{Literature}

\addcontentsline{toc}{section}{Literature}

L. Beranek: “Concert Halls and Opera Houses: Music, Acoustics, and Architecture” (2$^{\text{nd}}$ edition), Springer, 2004

T. J. Cox, P. D'Antonio: “Acoustic Absorbers and Diffusers” (2$^{\text{nd}}$ edition), Taylor \& Francis, 2009

W. Fasold, E. Veres: “Schallschutz und Raumakustik in der Praxis”, Verlag für Bauwesen Berlin, 1998

H. Kuttruff: “Room Acoustics” (6$^{\text{th}}$ edition), CRC Press, 2017

J. Meyer: “Acoustics and the Performance of Music” (5$^{\text{th}}$ edition), Springer, 2009

S.W. Rienstra, A. Hirschberg: “An Introduction to Acoustics”, Eindhoven University of Technology, 2017

M. Vorländer: “Auralization -- Fundamentals of Acoustics, Modelling, Simulation, Algorithms and Acoustic Virtual Reality”, Springer, 2008



\end{document}